\documentclass[11pt,letterpaper]{article}

\pdfoutput=1

\PassOptionsToPackage{nosort}{cite}

\usepackage{textcomp}
\usepackage[british]{chet}
\usepackage{mathtools}
\usepackage{amssymb,amsmath,amsfonts,mathrsfs,braket}
\usepackage[T1]{fontenc}
\usepackage{tikz}
\usepackage{diagbox}
\usepackage{tcolorbox}
\usepackage{bm}
\usepackage[colorinlistoftodos,textsize=tiny]{todonotes}
\usepackage{hyperref}
\usepackage{enumitem}
\usepackage{float}

\usetikzlibrary{external,calc}
\tikzset{external/figure name={ancb1-figure}}

\makeatletter
\renewcommand{\todo}[2][]{\tikzexternaldisable\@todo[#1]{#2}\tikzexternalenable}
\makeatother

\definecolor{tabblue}{RGB}{31,119,180}
\definecolor{tabred}{RGB}{214,39,40}

\makeatletter
\@ifundefined{zoomcircleradius}{\newlength{\zoomcircleradius}}{}
\makeatother
\tcbset{shield externalize}
\tcbuselibrary{theorems,skins,breakable}

\usepackage{subcaption}
\usepackage[font=small,format=hang,labelfont={sf,bf}]{caption}
\usepackage{pgfplots}
\usepackage{helvet}
\usepackage[eulergreek]{sansmath}
\pgfplotsset{compat=1.18,
    xlabel shift=-2pt,
    ylabel shift=-4pt,
    every picture/.append style={trim axis left, trim axis right},
    every axis label/.append style={font=\sansmath\sffamily\footnotesize},
    every axis title/.append style={font=\sansmath\sffamily\footnotesize, yshift=-0.75ex},
    every tick label/.append style={font=\sansmath\sffamily\footnotesize},
    every axis legend/.append style={font=\sansmath\sffamily\footnotesize}
}
\usepgfplotslibrary{fillbetween, statistics}

\definecolor{forestgreen}{rgb}{0.0, 0.27, 0.13}
\definecolor{darkblue}{rgb}{0.0, 0.0, 0.55}
\definecolor{darkred}{rgb}{0.55, 0.0, 0.0}
\hypersetup{colorlinks,
           linkcolor={forestgreen},
           citecolor={darkblue},
           urlcolor={darkred},
           pdftitle={Neural Spectral Bias and Conformal Correlators I: Introduction and Applications},
           pdfdisplaydoctitle
}

\newtcolorbox{cluster}[1][]{
  colback=blue!5!white,
  colframe=blue!75!black,
  title={$\triangleright$ Cluster runs},
  #1
}

\makeatletter
\g@addto@macro\bfseries{\boldmath}
\makeatother

\allowdisplaybreaks

\newcommand{\lsp}{\hspace{0.5pt}}

\renewcommand{\ge}{\geqslant}

\renewcommand{\geq}{\geqslant}
\renewcommand{\leq}{\leqslant}

\def\DD{{\cal D}}

\def\GG{{\cal G}}
\def\HH{{\cal H}}

\def\LL{{\cal L}}
\def\MM{{\cal M}}
\def\NN{{\cal N}}
\def\OO{{\cal O}}

\def\WW{{\cal W}}

\def\IR{{\mathbb R}}

\def\IC{{\mathbb C}}

\def\N{{\mathbb N}}

\newcommand{\Lp}{L^p}
\newcommand{\Wkp}{W^{k,p}}
\newcommand{\Hk}{H^k}
\newcommand{\Df}{{\Delta_\phi}}

\preprint{ITCP-2026-5\\CCTP-2-26-5}

\title{Neural Spectral Bias and Conformal Correlators I\\[5pt] 
Introduction and Applications}

\author{Kausik Ghosh,$^{\gamma,}$\email{kau.rock91@gmail.com}
Sidhaarth Kumar,$^{\gamma,}$\email{sidhaarth.kumar@kcl.ac.uk}
Vasilis Niarchos,$^{\delta,}$\email{niarchos@physics.uoc.gr} 
Andreas Stergiou$^{\gamma,}$\email{andreas.stergiou@kcl.ac.uk}}

\affiliation{$^{\gamma}$Department of Mathematics, King's College London, Strand, London WC2R 2LS, United Kingdom\\
$^\delta$Department of Physics, ITCP \& CCTP, University of Crete, 71003 Heraklion, Greece}

\abstract{We demonstrate that simple feed-forward neural networks (NNs) can accurately compute correlation functions of conformal field theories (CFTs) on a line. Strikingly, by optimising a NN solely on crossing symmetry and providing only the scaling dimension of the leading non-trivial operator and the correlator's value at a single ``anchor point'', we can reconstruct target physical correlators to within a few percent. We establish the robustness of this minimal-data approach across a broad class of theories and dimensions, including generalised free fields, contact and one-loop Witten diagrams in AdS$_2$, unitary and non-unitary 2d minimal models, the 3d Ising model, and half-BPS correlators in 4d $\mathcal{N}=4$ super-Yang--Mills theory, together with several thermal two-point functions, notably including those of the 3d Ising model. We argue that this remarkable alignment between NNs and CFTs stems from the spectral bias of gradient-based training, which heavily favours smooth functions. To ground this connection, we analyse the smoothness of conformal correlators using fractional Sobolev semi-norms, Chebyshev spectral decompositions, and a measure based on curvature. Finally, we establish the broader reconstructive power of this technique by extending it beyond the diagonal kinematics of the line.}

\date{April 2026}

\begin{document}

\maketitle

\hypersetup{pageanchor=true}

\setcounter{tocdepth}{2}

\toc

\section{Introduction}
\label{intro}

Conformal field theories (CFTs) encode universal information about critical phenomena, quantum gravity through holography, and the structure of quantum field theory (QFT) itself. A complete solution of the local properties of a given CFT amounts to a full determination of the spectrum of primary operators and their correlation functions at arbitrary points. Even for the best-studied models, such as the 3d Ising CFT, this programme remains far from complete, and the determination of the full functional forms of higher-point correlators at generic kinematic configurations is a largely open problem.

The conformal bootstrap programme \cite{Ferrara:1971zy, Polyakov:1974gs, Rattazzi:2008pe, Poland:2018epd} attacks this problem by exploiting the associativity of the operator product expansion (OPE) together with unitarity (positivity). For example, conformal invariance constrains the four-point function of identical scalar primaries $\phi$ (with scaling dimension $\Df$) to the form
\begin{equation}
    \label{introaa}
    \langle \phi(x_1) \phi(x_2) \phi(x_3) \phi(x_4) \rangle =
    \frac{1}{(x_{12}^2 x_{34}^2)^{\Df}}
    \,\GG(z,\bar z)\,,
\end{equation}
where $x_{ij}^2 = (x_i-x_j)^2$ and the cross-ratios are defined through 
\begin{equation}
    z\bar z=\frac{x_{12}^2 x_{34}^2}{x_{13}^2 x_{24}^2}\,,\qquad
    (1-z)(1-\bar z)=\frac{x_{14}^2 x_{23}^2}{x_{13}^2 x_{24}^2}\,.
\end{equation}
The reduced correlator $\GG(z,\bar z)$ is theory-dependent and has a decomposition in terms of conformal blocks \cite{Dolan:2000ut, Dolan:2003hv}. Symmetry under the exchange $x_2\leftrightarrow x_3$ in the correlation function \eqref{introaa} imposes the crossing equation
\begin{equation}
    \label{introab}
    \GG(z,\bar z)=\left(\frac{z\bar z}{(1-z)(1-\bar z)}\right)^{\Df}\GG(1-z,1-\bar z)\,.
\end{equation}
The standard bootstrap methodology leverages \eqref{introab}, along with unitarity, to derive rigorous bounds on isolated CFT data (scaling dimensions and OPE coefficients), which enter through the conformal-block decomposition of $\mathcal{G}$. However, the complete functional dependence of $\GG(z,\bar z)$ over its full kinematic domain, which encodes information about a countably infinite set of scaling dimensions and OPE coefficients, is typically not the direct target of these analyses and is rarely directly accessible, although in cases with enough data it can be reconstructed; see e.g.~\cite{Rychkov:2016mrc}.

On the real line, $z=\bar z$, the reduced correlator collapses to a single-variable function $\GG(z)=\GG(z,z)$ satisfying
\begin{equation}
    \label{introac}
    \GG(z)=\left(\frac{z}{1-z}\right)^{2\Df}\GG(1-z)\,.
\end{equation}
In these kinematics, the $s$-channel $(12)$-$(34)$ OPE converges in the interval $z\in [0,1)$. Setting $\bar z=z$ eliminates explicit spin dependence in the conformal block decomposition, yet the resulting single-variable function retains a significant part of the dynamical content of the theory. Line-restricted correlators have attracted independent attention within the bootstrap programme \cite{Hogervorst:2013sma, Mazac:2018mdx, Mazac:2019shk, Lanzetta:2025xfw} and arise naturally in 1d conformal systems such as line defects \cite{Ghosh:2021ruh, Cuomo:2021kfm, Cuomo:2021rkm, Pannell:2023pwz} and long range models \cite{Benedetti:2025nzp, Ghosh:2026gku, Carmi:2018qzm}.

The present work introduces a new computational strategy that recovers the function $\GG(z)$ on the interval $(0,1)$ from minuscule input: the scaling dimension of the external operator, $\Df$, the leading power-law behaviour of $\GG(z)$ near $z=0$ (equivalently, a gap in the OPE spectrum), and a single numerical value $\GG(z_0)$ at one reference point $z_0$ (an anchor). $\GG(z)$ is represented as a (typically small) multi-layer perceptron (MLP) and the crossing equation~\eqref{introac}, together with the anchor constraint, defines the loss functional for gradient-based (Adam \cite{Kingma:2014}) optimisation. The precise formulation is presented in Section~\ref{setup}. The proposed approach has been inspired heavily by the deep finite-temperature bootstrap in Ref.~\cite{Niarchos:2025cdg}, which combined NN-parametrisations of OPE tails with dispersion relations in thermal two-point functions.

Crossing symmetry together with the gap and anchor conditions constitute a severely under-determined problem that admits an infinite-dimensional family of solutions, the vast majority of which are not fully-fledged physical CFT correlators. A physical CFT correlator is a special solution of this problem, and it would be impossible to recover it uniquely from such minimal requirements without some additional guide. In fact, a significant part of the effort in the modern conformal bootstrap programme has been devoted to the discovery of efficient ways to overcome such inherent difficulties and to develop more constraining bootstrap strategies. So, how could a simple neural network (NN) representation bypass this obvious obstruction? 

It is well known that a NN does not explore arbitrary functions during its
optimisation. There is a well-documented phenomenon in machine learning \cite{rahaman2019spectral, Xu:2019frequency, Luo:2019theory}, called \emph{spectral bias} (or \emph{frequency principle}), whereby low-frequency, smooth target components are learned preferentially by NNs during gradient-based training. In our context, we observe that this subtle indirect bias singles out an almost unique smooth function. In all the examples we have explored (across different theories and spacetime dimensions), this output has been found to provide an unexpected, surprisingly good approximation of the target physical CFT correlator! This can only be true as a general fact if CFT correlators possess a hitherto unknown smoothness property that the anchored NNs pick up rather naturally.  

In Section~\ref{NNbias}, we review the relevant properties of NN optimisation drawing on the neural tangent kernel (NTK) framework \cite{Jacot:2018ntk} and the associated reproducing kernel Hilbert space (RKHS) implicit regularisation. This allows us to understand better what kind of functions we recover through optimisation with NNs in our context and under what conditions.

Section~\ref{smooth} develops this theme further through an independent analysis of correlator smoothness. The main goal here is to explore, within a number of concrete examples, whether 1d CFT correlators possess some special feature of smoothness that distinguishes them from other crossing-symmetric functions with the same gap and anchor-point properties, and whether they minimise an appropriate quantity. We introduce three complementary diagnostic tools---the fractional Sobolev (Gagliardo) semi-norm, Chebyshev spectral coefficients, and a measure based on curvature---and apply them to exact CFT correlators and one-parameter families of crossing-symmetric deformations. With a few minor differences, all three measures suggest that the physical correlators are among the smoothest functions, which corroborates the picture that the NN optimiser selects them precisely because they are within some smoothness attractor. This motivates the existence of a sharp measure in CFT, whose exact minima are physical 1d correlators. It would be very interesting to establish such a measure rigorously in CFT, if it indeed exists.

Another way to probe the role of smoothness in CFT correlators is by fitting them with Chebyshev polynomials in a Tikhonov scheme where higher Chebyshev coefficients are exponentially suppressed. In Section~\ref{chebyshev}, we demonstrate in a few examples that this alternative non-NN representation of CFT correlators in a Chebyshev basis can also work well, as long as one fits with a suitable suppression of the higher Chebyshev coefficients. The results in this section reinforce the importance of smoothness in CFT correlators, but also highlight what makes the NN representation so remarkable. In a Chebyshev--Tikhonov scheme the decay coefficients enforcing smoothness are a priori unknown, their optimal values can vary from case to case, and it is unclear how to determine them without knowing the target correlator. The NN takes care of this aspect automatically because of an implicit bias towards smoothness that aligns very well with the physics of CFT correlators.

The bulk of the paper is devoted to explicit applications that demonstrate how anchored NNs recover 1d correlation functions. In this work, we analyse the examples that follow.
\begin{enumerate}[label=(\arabic*)]
    \item \textbf{Generalised free fields} (Section~\ref{GFF}): We study bosonic and fermionic GFFs at various values of $\Df$. This is a simple, standard benchmark with a universal form across spacetime dimensions.

    \item \textbf{Witten diagrams for scalar fields in} $\mathbf{AdS}_{\boldsymbol{2}}$ (Section~\ref{ads}): This is an example of a 1d CFT within the AdS/CFT correspondence. We study the $\phi^4$ contact Witten diagram and the one-loop (bubble) diagram, both of which are known analytically and involve (poly-)logarithmic dependence. 
    
    \item \textbf{2d minimal models} (Section~\ref{minimal}): As warm-up to a set of non-trivial correlation functions on the line in $d>1$ dimensions, we study the $\phi_{1,2}$-field four-point functions in the family of unitary series $\MM(m,m+1)$. The corresponding functions involve sums of products of hypergeometric conformal blocks. We also study the non-unitary Lee--Yang model $\MM(2,5)$, which exhibits the potential applicability of the approach beyond the realm of unitarity.
    
    \item \textbf{The Ising model} (Section~\ref{ising}): In this case, we extend the analysis of the 2d Ising CFT (the $\MM(3,4)$ minimal model) beyond the four-point function of the $\phi_{1,2}=\sigma$ operator. We consider both the identical ($\sigma\sigma\sigma\sigma$, $\epsilon\epsilon\epsilon\epsilon$) and mixed ($\sigma\sigma\epsilon\epsilon$, $\sigma\epsilon\epsilon\sigma$) correlator systems and compare with the analytic results. This allows us to exhibit how the approach works in the context of a mixed-correlator bootstrap. Even less trivially, we also consider the case of the same system of correlators in the 3d Ising CFT, where exact answers for the corresponding correlation functions are unavailable with current techniques. We compare our predictions with independent conformal bootstrap and fuzzy-sphere regularisation data. In this context, our approach becomes genuinely predictive.

    \item \textbf{Wilson-Fisher fixed points in $4-\varepsilon$ dimensions} (Section~\ref{wf}): In the $\text{AdS}_2$ example, we demonstrated how anchored NNs can recover separately different-order contributions in the perturbative expansion of a correlator. Here we exhibit the same effect in the $\varepsilon$-expansion of Wilson-Fisher fixed points in non-integer $4-\varepsilon$ dimensions.

    \item \textbf{Half-BPS operators in 4d $\NN=4$ SYM} (Section~\ref{sym}): As an example of non-trivial four-point functions in a 4d CFT, we study correlators of half-BPS operators in 4d $\NN=4$ SYM theory in the large-$c$ limit.
    
    \item \textbf{Thermal two-point functions} (Section~\ref{thermal}): In this section, we change gears and move from four-point functions on $\IR^d$ to thermal two-point correlators on $\IR^{d-1}\times S^1_\beta$. The Kubo--Martin--Schwinger (KMS) condition at zero spatial separation reduces to the crossing equation~\eqref{introac}, enabling the application of the same methodology to finite-temperature two-point correlators. We study GFFs, generic 2d CFTs, and the $\langle\sigma\sigma\rangle_\beta$, $\langle\epsilon\epsilon\rangle_\beta$ thermal correlators of the 3d Ising model. The latter constitutes an interesting example where the exact correlators are unknown. To proceed, we deduce an approximate anchor point using as external input a single thermal OPE datum. The comparison with the recent numerical and analytic studies in \cite{Barrat:2025wbi, Barrat:2025nvu} is very encouraging.
    
    \item \textbf{From the line to the plane} (Section~\ref{plane}): The main focus of the discussion in this paper are CFT correlation functions on a line. It is natural to ask if we can use anchored NNs to go beyond the restricted kinematics on a line. For four-point functions on $\IR^d$, that entails the computation of the correlator for generic cross-ratio parameters $(z,\bar z)$ on the plane. For thermal two-point functions, it entails the computation of the correlator at non-zero spatial separation. We present a specific extension of the method in this direction that enforces crossing on concentric circles around the crossing-symmetric point $z=\bar z=\frac{1}{2}$, and preliminary successful tests in a few characteristic examples.
\end{enumerate}

Across all examples, we find that the anchored NN approximation reproduces exact or independently known numerical results with relative errors at the level of a few percent, using lightweight two- or three-layer MLPs trained in minutes on a modern laptop. 

In two companion papers~\cite{GKNS:2, GKNS:3}, we present further successful applications of the method to the 2d modular bootstrap on the torus and the annulus, as well as to superconformal correlators on the half-BPS Wilson line defect CFT in 4d $\NN=4$ SYM theory, and the holomorphic bootstrap in 6d $(2,0)$ SCFTs. An accompanying short paper~\cite{GKNS:0} provides a concise summary of the main results.

We conclude in Section~\ref{outlook} with a discussion of open problems and future directions. Useful facts and supplementary material are relegated to the appendices at the end of the paper.

\paragraph{Relation to previous work.} Previous applications of deep learning methods in conformal field theory include \cite{Kantor:2021kbx,Kantor:2021jpz,Kantor:2022epi,Niarchos:2023lot,Halverson:2024axc,Niarchos:2025cdg} (see also Refs.~\cite{Laio:2022ayq,Huang:2025qkk,Benjamin:2026lbj}, which involve similar methodologies). The present work is more closely related to \cite{Niarchos:2025cdg}, which expressed, for the first time, thermal two-point correlation functions in CFT in terms of NNs. Ref.~\cite{Niarchos:2025cdg} combined crossing symmetry with input from thermal dispersion relations.

\section{Setup}
\label{setup}

In this section, we describe the precise formulation of the computational problem we would like to solve. The objects we want to reconstruct are one-variable functions $\GG(z)$ on the interval $z\in(0,1)$ that satisfy the crossing symmetry constraint~\eqref{introac}. The external parameter $\Df$ takes positive real values in unitary theories but may be negative in non-unitary ones. Our framework does not assume unitarity.

\subsection{Statement of the problem}

We decompose the target correlator in two parts as
\begin{equation}
    \label{setupaa}
    \GG(z) = L(z) + H(z)\,.
\end{equation}
$L(z)$ (standing for `low') is a prescribed function that accounts for the dominant contributions to $\GG(z)$ in the limit $z\to 0$. In the simplest case (and most common in the ensuing applications), we set $L(z)=1$, which corresponds to the universal identity-operator contribution in the operator product expansion. More generally, we can use $L(z)$ to input further low-scaling dimension contributions to the OPE, or in cases like the $\text{AdS}_2$ examples in Section \ref{ads}, a non-trivial small-$z$ behaviour. In Section \ref{ads} this involves a function with logarithms. The remainder component, $H(z)$ (which stands for `high'), encodes all the higher-energy OPE data of the correlator. It is the unknown quantity to be determined.

On $H(z)$ we impose the following conditions:
\begin{enumerate}[label=(\alph*)]
    \item \textbf{Gap condition.} The leading small-$z$ asymptotics of $H(z)$ is a known power: $H(z)\sim z^\delta$ as $z\to 0$, with $\delta>0$ specified, but the overall coefficient left free. In the language of the OPE, $\delta$ corresponds to the scaling dimension of the lowest-lying operator (above those already included in $L(z)$) that appears in the $\phi\times\phi$ channel. We implement this by setting $H(z) = z^\delta\,h(z)$. 

    \item \textbf{Anchor condition.} The value of $H(z)$ at a single reference point $z_0\in(0,1)$ is prescribed: $H(z_0)=H_0$. From a CFT standpoint, this relation constitutes a sum-rule that packages the contributions of infinitely many operators into one number. We refer to the pair $(z_0,H_0)$ as the \textit{anchor point} of the problem. In practice, we typically choose $z_0$ in the range $0.3$--$0.4$; the precise value has a relatively mild effect on the results as long as $z_0$ is not chosen very close to 0 or 1.
\end{enumerate}

These data---$\Df$, $\delta$, and $(z_0,H_0)$---are the \emph{only} input to the computation. Remarkably, they suffice for the NN optimiser to reconstruct the full function $\GG(z)$ to percent-level accuracy.

\subsection{The degeneracy of crossing-symmetric solutions}

The conditions above are far from sufficient, in principle, to select a unique function. To see this explicitly, suppose $H_1(z)$ is some solution. Then,
\begin{equation}
    \label{setupba}
    H_2(z) = H_1(z) + (1-z)^{-2\Df}\,b(z)
\end{equation}
is also a solution whenever $b(z)$ is a symmetric function, $b(z)=b(1-z)$, that vanishes at the anchor point, $b(z_0)=0$, and is sub-leading relative to $z^\delta$ near $z=0$, i.e.\ $\lim_{z\to 0}z^{-\delta}b(z)=0$. Since the space of such functions $b$ is infinite-dimensional, the problem possesses a vast family of exact solutions.

Yet, physical CFT correlators are not expected to be generic members of this family. For example, a four-point function of local operators possesses an analytic continuation to the double-slit domain $\IC\setminus\{(-\infty,0]\cup[1,\infty)\}$ and obeys polynomial boundedness at infinity (Regge-type bounds). Restricted to the real segment $[0,1)$, these constraints endow the physical correlator with a high degree of smoothness and spectral regularity that is absent in a generic crossing-symmetric function. The question we are posing is whether a suitable optimisation algorithm, combined with an appropriate function-space parametrisation, can single out the physical correlator from the infinite-dimensional solution set of the above problem.

\subsection{Neural network implementation}
\label{setup:NN}

Our proposal represents the unknown function $H(z)$ through a fully connected feed-forward MLP. The detailed architecture and training protocol are as follows.

\paragraph{Architecture.} We consider a light network with $2$ or $3$ hidden layers and width $64$ or $128$ neurons per layer. The input is a single real number $z\in(0,1)$ and the output is a scalar. We compose the network output with the gap prefactor and a prefactor that captures the $z\to 1$ asymptotics to form
\begin{equation}
    \label{setupca}
    H(z) = z^\delta (1-z)^{-\delta'} \,\text{NN}_{\boldsymbol\theta}(z)\,,
\end{equation}
where $\boldsymbol\theta$ collectively denotes the trainable weights and biases of the MLP. The prefactor $z^\delta$, which hard-codes the correct leading small-$z$ behaviour, biases the network towards functions $\text{NN}_{\boldsymbol\theta}(z)$ that are regular at $z=0$ (when $\delta>0$). On the other hand, the factor $(1-z)^{-\delta'}$, which follows from crossing symmetry without further input, allows the NN to implement the crossing symmetry relation more efficiently near $z=1$. Typically, $\delta'=2\Delta_\phi$.

\paragraph{Activation functions.} We exclusively employ smooth activation functions---either $\tanh$ or GELU (Gaussian error linear unit),
\begin{equation}
    \label{gelu}
    {\rm GELU}(x) = \frac{x}{2}\bigg[ 1 + {\rm erf}\left( \frac{x}{\sqrt 2} \right) \bigg]
    \simeq \frac{x}{2} \bigg[ 1 + \tanh\left(\sqrt{\frac{2}{\pi}}(x+0.044715\, x^3)\right)\bigg]
    \,.
\end{equation}
Both produce smooth outputs. In contrast, piecewise-linear activations, such as ${\rm ReLU}(x)=\max(0,x)$ (rectified linear unit), were found to produce less smooth functions, degrading the quality of the results when we compare with analytic CFT correlators. The effects on the output of the tanh and GELU vs ReLU activation functions will be explained using standard arguments in the context of the neural tangent kernel framework in Section~\ref{NNbias}. The choice between tanh and GELU does not yield significant differences in general, but we found that it can lead to small problem-dependent gains.

\paragraph{Loss function.} The total loss function $\LL$ involves the sum of two contributions evaluated on a discrete training grid $\{z_i\}_{i=1}^N$ with $z_i\in (0,1)$,
\begin{equation}
    \label{setupcb}
    \LL(\boldsymbol\theta) = \LL_{\rm cross}(\boldsymbol\theta) + \lambda_{\rm anc}\,\LL_{\rm anc}(\boldsymbol\theta)\,,
\end{equation}
where $\lambda_{\rm anc}$ is a weighting coefficient (typically of order unity). The anchor loss $\LL_{\rm anc}$ is simply
\begin{equation}
    \label{setupcd}
    \LL_{\rm anc} = (H(z_0)-H_0)^2\,.
\end{equation}
The crossing loss $\LL_{\rm cross}$ is designed to enforce the crossing symmetry condition \eqref{introac}. 
There is a variety of implementations of this loss. In this paper, we use the \textit{mean relative square loss}
\begin{equation}
    \label{setupcc}
    \LL_{\rm cross} = \frac{1}{N}\sum_{i=1}^N \Big\{ \DD(z_i)^{-1} \Big[\GG(z_i) - \left(\tfrac{z_i}{1-z_i}\right)^{2\Df}\GG(1-z_i)\Big]\Big\}^2\,,
\end{equation}
with $\GG(z)=L(z)+H(z)$ as in \eqref{setupaa} and denominator $\DD(z_i)=1+|\GG(z_i)|+|\left(\tfrac{z_i}{1-z_i}\right)^{2\Df}\GG(1-z_i)|$. This particular choice of the loss function is motivated by its simplicity and the \emph{a posteriori} observation that it works sensibly in all the examples we analysed in this paper.

\paragraph{Optimiser and learning rate schedule.} We use the \texttt{Adam} optimiser \cite{Kingma:2014} with an initial learning rate typically in the range $5\times 10^{-4}$--$10^{-3}$, that decays by a {\tt StepLR} schedule with step size $500$ and multiplicative factor $\gamma=0.98$, down to a minimum learning rate of $10^{-5}$--$10^{-6}$. A small weight decay of $10^{-6}$ was added to improve stability, without introducing a significant impact on the results.

For each problem we ran 100 independent instances of the code on the cluster to collect statistics. Each run was programmed to terminate after 5K epochs of no improvement of the training loss. Typical runs of this type achieved their optimal loss within at most a couple of minutes.

\paragraph{Training grid.} The training was performed on a uniform grid of $N=200$--$400$ points in a subinterval of $(0,1)$. The endpoints were chosen slightly away from $0$ and $1$ to avoid endpoint singularities; typical choices were $[10^{-4},\,0.6]$ or $[10^{-4},\,0.9]$.

\paragraph{Reporting scheme.} In Sections~\ref{GFF}-\ref{plane} we analyse a variety of different theories and correlators. Unless stated otherwise, results are based on an ensemble of 100 independent runs with anchor point $z_0=0.3$. The plots are arranged as follows: $(a)$ top-left, $(b)$ top-right and $(c)$ bottom:
\begin{itemize}
\item[$(a)$] The mean predicted correlator on the line over the 100 cluster runs is shown as a solid blue curve, together with a shaded blue $\pm 1$ standard-deviation band.

\item[$(b)$] A plot of the relative difference between the prediction and the analytic result at each point of the domain of the functions. The mean relative error is shown as a solid blue curve with a shaded blue $\pm 1$ standard-deviation band.
\noindent We define the relative prediction error as
\begin{equation}
    \label{gff1o4a}
    \textrm{Prediction relative error} = \frac{({\rm Predicted~value}) - ({\rm Exact~value})}{1 + |({\rm Exact~value})|}\,.
\end{equation}
1 has been added to the denominator to regularise the divergence when the exact value becomes very small. In this plot, the prediction relative error vanishes by construction at the anchor point $z_0$, and almost vanishes at its crossed value at $1-z_0$, as dictated by the approximate crossing symmetry imposed by the training.

\item[$(c)$] A histogram of the distribution of the predictions at the crossing-symmetric point $z=0.5$ is shown, together with the histogram mean shown as a solid blue line; when available, the corresponding reference value is shown as a black dashed line. This plot gives a specific sense of the spread of the predictions and how they compare with the analytic correlators.

\end{itemize}

Additional details about the runs in each problem reported in this paper, as well as the explicit \texttt{Python} code used, can be found in a dedicated \href{https://github.com}{\tt GitHub} repository, \href{https://github.com/andstergiou/nn-cft}{\tt andstergiou/nn-cft}, which is archived on \href{https://zenodo.org}{\texttt{Zenodo}}~\cite{nn_cft_code}.

\section{Spectral bias in neural network optimisation}
\label{NNbias}

The setup of the previous section converts the search for crossing-symmetric correlators into a physics-informed neural network (PINN) optimisation. The novelty lies in the observation, borne out by every example we have examined, that the optimiser consistently converges to a \emph{physical} CFT correlator, even though infinitely many crossing-symmetric alternatives exist.

Two aspects of the setup conspire to produce this outcome. First, the spectral bias of gradient-based optimisation of NNs with smooth activations selects smooth, low-complexity functions to make an interpolation through the anchor point. Second, the anchor point itself has a strong stabilising effect: it constrains the overall normalisation of $H(z)$ and suppresses the large statistical fluctuations that could otherwise appear in an unanchored optimisation.

In this section we review the empirical and theoretical basis for the spectral bias phenomenon in NNs, and explain how it corroborates our observations and how it motivates a mechanism for the selection of physical CFT correlators.

\subsection{Spectral bias/frequency principle}

The spectral bias, or frequency principle, in gradient descent-based training of NNs has been documented in a series of studies by Xu et al.\ \cite{Xu:2019frequency} and Rahaman et al.\ \cite{rahaman2019spectral} (see also the references therein for related work). The setup is the following: a fully connected NN $f_{\boldsymbol\theta}:\IR\to\IR$ with trainable parameters $\boldsymbol\theta$ is trained via gradient descent on a mean-square loss $\LL=\frac{1}{N}\sum_{i=1}^N|f_{\boldsymbol\theta}(x_i)-y_i|^2$ over a data set $\{(x_i,y_i)\}$. Monitoring the Fourier spectrum of the residual $r(x)=f_{\boldsymbol\theta}(x)-y(x)$ at successive training epochs reveals a clear pattern: low-frequency Fourier components of $r$ decay to zero much faster than high-frequency ones. In many situations the high-frequency residual remains non-negligible even after the loss has effectively plateaued; Luo et al.\ \cite{Luo:2019theory}. The frequency principle has been observed across diverse architectures, activation functions, and datasets \cite{Xu:2019frequency}. It provides a partial explanation for why deep networks generalise well on natural data (which typically has most of its energy in low frequencies) but struggle with high-frequency patterns.

The strength of this implicit bias depends crucially on the activation function. Smooth activations ($\tanh$, GELU, sigmoid, etc.) produce network outputs that are themselves smooth. Piecewise-linear activations (ReLU, Leaky-ReLU) permit kinks at each neuron and can therefore represent non-smooth functions, weakening the spectral bias. This is consistent with our observation that by using tanh or GELU activation functions we achieve smoother predictions that approximate better 1d CFT correlators.

\subsection{The neural tangent kernel perspective}

A rigorous and quantitative understanding of spectral bias is afforded by the neural tangent kernel (NTK) framework of Jacot, Gabriel and Hongler \cite{Jacot:2018ntk}. Given a network $f_{\boldsymbol\theta}(x)$, the NTK is the positive-definite kernel
\begin{equation}
    \label{NTKaa}
    \Theta(x,x') = \sum_{p} \frac{\partial f_{\boldsymbol\theta}(x)}{\partial \theta_p}\,\frac{\partial f_{\boldsymbol\theta}(x')}{\partial \theta_p}\,.
\end{equation}
In the infinite-width limit, i.e.\ the limit where each hidden layer width goes to infinity, with random initialisation, it is argued that $\Theta$ converges to a deterministic kernel $\Theta^*$ that remains frozen during training. Continuous-time gradient descent on the loss then becomes
\begin{equation}
    \label{NTKaa2}
    \frac{\partial f_{\boldsymbol\theta}(x)}{\partial t} = -\int \Theta^*(x,x')\,\frac{\delta\LL}{\delta f(x')}\,dx'\,,
\end{equation}
which is a kernel regression problem with kernel $\Theta^*$. The training dynamics becomes linear for a mean square loss function $\LL$.

Denoting by $\{\phi_k, \lambda_k\}$ the eigenvectors and eigenvalues of $\Theta^*$, respectively, one finds that the training converges along the direction $\phi_k$ at an exponential rate with exponent proportional to $\lambda_k$, and so modes with large eigenvalues are learned quickly, while modes with small eigenvalues converge slowly. Bietti and Mairal \cite{bietti2019inductive} and subsequent works \cite{Basri:2019,Cao:2019,Geifman:2020eigendecay,murray:2023,Li:2024eigenvalue} demonstrated that for NTKs built from smooth activations (possibly with a finite number of non-smooth points like ReLU), the eigenvalues obey a polynomial decay law $\lambda_k\sim k^{-\alpha}$ with $\alpha$ increasing with the smoothness of the activation. For $C^\infty$ activations, like tanh or GELU, the decay is exponential. This directly implies a hierarchy of learnability: target function components aligned with the leading eigenvectors, which correspond to low-frequency, spatially smooth modes, are learned in $\OO(1/\lambda_k)$ gradient steps, while high-frequency components, associated with eigenvalues that are polynomially or exponentially smaller, require correspondingly more training time. For activations like ReLU the gap is polynomial in the frequency, whereas for $C^\infty$ activations the gap is exponential, making high-frequency learning prohibitively slow in the kernel regime.

Although the NTK framework applies to a specific regime of infinite width, it provides some general lessons, which are expected to carry over in more general practical situations. It suggests, for example, that the spectral bias will be more pronounced for smooth activation functions, small learning rates and optimisers based on standard stochastic gradient descent (SGD). Other features, like the form of the loss function, can also affect the outcome. We have confirmed these predictions in our runs and chose hyperparameters accordingly. 

A noteworthy aspect is the choice of the optimiser. Although vanilla SGD is preferable based on the above discussion, in practical applications it turns out to be inefficient in solving the crossing symmetry constraint. Adaptive methods with momentum, like \texttt{Adam} or \texttt{AdamW}, were very efficient in solving the crossing symmetry constraint. In principle, that comes with reduced suppression of the spectral bias, which can lead to less smooth functions. We did not observe this. In all our runs, which were almost exclusively based on the use of \texttt{Adam} on light-weight NNs, we found that the combination of \texttt{Adam} with small learning rates and smooth activation functions was naturally biased towards smooth outputs in accordance with the qualitative predictions of the NTK framework.

\subsection{Implicit regularisation: RKHS norm minimisation}

Beyond governing convergence rates, the NTK framework reveals an implicit regularisation mechanism. Among all functions that are consistent with the imposed constraints (in our case, the crossing equation evaluated on the training grid, together with the anchor), gradient descent converges (in the lazy-training regime, where the weights and biases evolve slowly) to the function of minimal RKHS norm $\|f\|_{\HH_{\Theta^*}}$ associated with the frozen kernel $\Theta^*$ \cite{bietti2019inductive} (see also \cite{gunasekar2018characterizing,arora2019fine} for the general minimum-norm implicit bias of gradient descent in over-parametrised models). Explicitly, the RKHS norm reads
\begin{equation}
    \label{NTKab}
    \|f\|_{\HH_{\Theta^*}}^2 = \sum_k \frac{|\hat f_k|^2}{\lambda_k}\,,
\end{equation}
where $\hat f_k=\langle f,\phi_k\rangle$ are the projections onto the eigenbasis. Because $\lambda_k$ decays for large $k$, high-frequency components are penalised with increasing severity. The minimum-norm interpolant is therefore the smoothest function (in the RKHS sense) that satisfies the constraints.

Translating to our context, the crossing equation~\eqref{introac} and the anchor constraint~\eqref{setupcd} define a convex feasible set in function space, and the NN optimiser selects from that set the element with the smallest RKHS norm. The key empirical finding of our paper is that, across all examples investigated, the resulting function coincides (to within a few percent) with a physical CFT correlator. In this manner, the anchored neural network reformulation of the conformal bootstrap reveals that physical CFT correlators are the smoothest crossing-symmetric functions sharing the same gap and anchor data. This implies a deep, powerful variational principle for 1d CFT correlators, which we try to test independently (with non-NN tools) in the next section.

\section{On the smoothness of correlation functions}
\label{smooth}

In this section, we explore the hypothesis that physical 1d CFT correlators are distinguishable from other crossing-symmetric functions with the same gap and anchor data by an independent measure of smoothness (formulated without the use of neural networks). We test this prediction using three complementary diagnostic tools: the fractional Sobolev semi-norm, Chebyshev spectral coefficients, and a curvature-based measure constructed directly out of first and second derivatives. A summary of useful background on Sobolev spaces is collected, for the convenience of the reader, in Appendix~\ref{sobolev}.

This section is independent of the evidence reported in the rest of the paper and can be safely skipped during a first reading.

\subsection{Diagnostic tools}
\label{smooth:measures}

\paragraph{Fractional Sobolev semi-norm.}
For $s\in(0,1)$, $p\in [1,\infty)$ and a function $f:\Omega \to\IR$, defined on an open domain $\Omega\subseteq \IR^n$, the fractional Sobolev semi-norm (also known as the Gagliardo semi-norm)
\begin{equation}
    \label{smoothaaa}
    [f]_{W^{s,p}}^p(\Omega) = \int_\Omega \int_\Omega \frac{|f(x)-f(y)|^p}{|x-y|^{n+ps}}\,dx\,dy
\end{equation}
quantifies global regularity at fractional order $s$: it penalises point-to-point differences weighted by an inverse-distance kernel that becomes more singular as $s\to 1$.

In this work we are interested in 1d functions. For that reason, we set $n=1$ in \eqref{smoothaaa}. Moreover, motivated by the fact that four-point CFT correlators on a line have an analytic continuation on the complex plane (away from two branch cuts on the real line) and a standard fact from harmonic analysis that relates the Dirichlet energy of a harmonic extension on the upper half plane to the fractional Sobolev semi-norm on the real axis with $s=\frac{1}{2}$, $p=2$, we consider the more specific case of semi-norms of the form 
\begin{equation}
    \label{smoothaa}
    [f]_{W^{\frac{1}{2},2}}^2(\Omega) = \int_\Omega \int_\Omega \frac{|f(x)-f(y)|^2}{|x-y|^{2}}\,dx\,dy
    \,.
\end{equation}
In harmonic analysis this particular integrand appears in the context of the harmonic extension from the whole real line $\Omega=\IR$ to the upper half-plane. The analogous extension from the finite interval $(0,1)$ to the double-slit domain $\IC/\{(-\infty,0]\cup [1,\infty)$ can be obtained by applying the conformal map $\Phi(z)=i\sqrt{z/(z-1)}$, which leads to a fractional semi-norm of the form
\begin{equation}
    \label{smoothaaK}
    \int_0^1 \int_0^1 K(x,y) |f(x)-f(y)|^2\, dx\,dy\,,
\end{equation}
with a modified kernel
\begin{equation}
    \label{smoothaaKa}
    K(x,y) = \frac{1}{2\pi}\,\frac{(x y)^{-1/4}\,(1-x)^{-3/4}\,(1-y)^{-3/4}}{\bigl(\sqrt{x/(1-x)}-\sqrt{y/(1-y)}\,\bigr)^2}
    \,.
\end{equation}
A review of relevant mathematical details can be found in Appendix \ref{sobolev}.

In practice, in our context, we compute the fractional Sobolev semi-norm of the functions $H(z)=\GG(z)-L(z)$, which are typically regular at $z=0$, but diverge at $z=1$. Since our functions are crossing symmetric, we chose to analyse their smoothness in the half-interval $[0,1/2]$ (where the functions are regular). For that purpose, we rescaled $z$ by setting $z=\frac{x}{2}$ and considered $H(x/2)$ in the interval $x\in [0,1]$. Accordingly, we define the following two measures of smoothness, which are inspired by the $s=1/2$, $p=2$ fractional Sobolev semi-norm
\begin{equation}
    \label{smoothaa2}
    [H]_1 = \int_{(0,1)^2}\frac{|H\left( \frac{x}{2}\right)-H\left( \frac{y}{2}\right)|^2}{|x-y|^2}\,dx\,dy\,,
\end{equation}
and
\begin{equation}
    \label{smoothaa3}
    [H]_2 = \int_{(0,1)^2}K(x,y)\left| H\left( \frac{x}{2}\right)-H\left( \frac{y}{2}\right) \right|^2 \,dx\,dy\,,
\end{equation}
with the kernel $K$ defined in Eq.\ \eqref{smoothaaKa}.
In numerical computations, both integrals were evaluated on a uniform grid of $5000$ points with analytic regularisation near the diagonal $x=y$ point.

In what follows, $[H]_1$ and $[H]_2$ will provide a preliminary basis of discussion of the smoothness properties of 1d CFT correlators and their crossing-symmetric deformations, but one should keep in mind that, eventually, other measures of smoothness may prove to be more appropriate in CFT.

\paragraph{Chebyshev spectral decomposition.} Another diagnostic tool of smoothness that we will examine here are Chebyshev spectral decompositions. We consider again $H(x/2)$ for $x\in [0,1]$ and study its expansion in Chebyshev polynomials, 
\begin{equation}
    \label{chebaa}
    H\left(\frac{x}{2}\right)=\sum_{n=0}^\infty c_n\,T_n(2x-1)
    \,.
\end{equation}
For concreteness, we focus on the first $n_{\rm max}=1000$ coefficients. The decay rate of $|c_n|$ at large $n$ encodes the smoothness class of $H$: analytic functions exhibit super-algebraic (typically geometric) decay, whereas finitely differentiable functions decay only polynomially. A related scalar measure of smoothness can be obtained from the logarithmic growth rate defined as
\begin{equation}
    \label{smoothab2}
    \gamma(s) = \frac{\displaystyle{\sum_{n=0}^\infty} \log(1+n^2)\,(1+n^2)^s\,|c_n|^2}{\displaystyle{\sum_{n=0}^{\infty}} (1+n^2)^s\,|c_n|^2}
    \,.
\end{equation}
In numerical computations, the infinite $n$-sums were estimated by truncating at $n_{\rm max}$. In that case, the plot of $\gamma(s)$ exhibits a sharp jump above some intermediate value of $s$, which is an artefact of the truncation. Below that value, smoother functions are characterised by lower values of $\gamma(s)$.

\paragraph{A differential diagnostic.}
A simpler measure of smoothness can be defined as the squared curvature functional,
\begin{equation}
    \label{curvature}
    K[H]=\int_{0}^{\frac{1}{2}} dz\, \frac{\bigl( H''(z)\bigr)^2}{\bigl(1+(H'(z))^2\bigr)^{5/2}}
\end{equation}
for the 1d curve $(z, H(z))$. Since the integrand depends on the first and second derivatives of \(H(z)\), this functional provides an integrated measure of the local variation of the crossing solution over the interval \(z\in[0,0.5]\). Once again, we restrict the integral to the interval \(z\in[0,0.5]\) because crossing symmetry fixes the behaviour on the complementary part of the Euclidean region. A word of caution is that, depending on \(\Delta_{\phi}\), taking two derivatives may introduce a singularity in the \(z\to 0\) limit. In such cases, we introduce a lower cutoff $0.05$ to regulate the integral.

\subsection{Crossing-symmetric deformations}
\label{smooth:examples}

To probe whether the physical correlator is the smoothest crossing-symmetric function (under a certain suitable measure of smoothness), we need a controlled family of crossing-symmetric functions that include a physical correlator and deformations thereof. We consider two types of crossing-symmetric functions.

\paragraph{I: Polynomial deformations.} As we noted previously in Section~\ref{setup}, given any solution to the crossing equation, $H_1(z)$, the function $H_2(z)=H_1(z)+(1-z)^{-2\Df}\,b(z)$ is also a solution that preserves the gap and the anchor point at $z=z_0$, whenever $b(z)=b(1-z)$, $b(z_0)=0$, and $\lim_{z\to 0}z^{-\delta} b(z)=0$. We will examine a concrete one-parameter family of such deformations
\begin{equation}
    \label{smoothba2}
    b(z;a) = a\Big[-z_0(1-z_0)\big(z(1-z)\big)^2 + \big(z(1-z)\big)^3\Big]\,,
\end{equation}
parametrised by the parameter $a$. The choice $a=0$ recovers the undeformed correlator.

\paragraph{II: Linear combinations of correlators.} A more interesting deformation arises when we consider linear combinations of physical CFT correlators. The only restrictions on the linear combination are the common external dimension, gap and anchor point. In what follows, we consider a linear combination of three physical CFT four-point correlation functions on a line: a bosonic GFF deformed by a linear combination with a fermionic GFF and a 2d minimal model. All the correlators have the same $\Df$, weighted to preserve the anchor point and the leading small-$z$ asymptotics. Such deformations respect the full analytic structure (branch points, boundedness) of CFT, yet generically differ from each of the constituent correlators at generic $z$. 

More concretely, we consider a linear combination of correlators with external scaling dimension
\begin{equation}
    \label{smoothbadelta}
    \Delta_\phi = \Delta_m = \frac{1}{2} - \frac{3}{2(1+m)}
    \,,
\end{equation}
for $m=3,4,\ldots.$ This  coincides with the scaling dimension of the lowest-dimension scalar primary operator $\phi_{1,2}$ in the 2d minimal model $\mathcal{M}(m,m+1)$. The one-parameter family of functions of interest is
\begin{equation}  \label{smoothbb}
   \GG(z;a, z_0, m) = \GG_{B}(z)+a\bigg(\GG_{F}(z)-\GG_V(z)\bigg)+\bigg(\frac{z}{z_0}\bigg)^{2\Delta_m}a\bigg(-\GG_{F}(z_0)+\GG_V(z_0)\bigg) ,
\end{equation}
where 
\begin{equation}
    \GG_{B/F}(z)=1+\eta^{B/F} z^{2\Delta_m}+\biggl(\frac{z}{1-z}\biggr)^{2\Delta_m},
\end{equation}
$\eta^{B}=1, \eta^F=-1$ and $\GG_V(z)$ represents the four-point correlator of the spin operator of 2d minimal model restricted on the line \eqref{minaa}. When the deformation parameter $a$ is equal to zero, this combination reduces to the bosonic GFF with $\Delta_\phi=\Delta_m$; for $a\neq 0$ it combines the 2d minimal model $\MM(m,m+1)$ correlator and the fermionic GFF correlator, while keeping the gap and the anchor point fixed.

We note that the leading order behaviour of $H(z)=\GG(z)-1$ near $z=0$ is $z^{2\Delta_\phi}$ for all the functions except for a special one with
\begin{align}
    \label{smoothcomboad}
    a = a^*(m) =& -2 \Biggl[ 1 + \frac{(1-z_0)^{\frac{3}{1+m}}}{z_0-1} - z_0^{-1+\frac{3}{1+m}} \notag\\
    &\quad\quad+ \bigl(z_0(1-z_0)\bigr)^{-1+\frac{3}{1+m}} \,{}_2F_1\!\left(\frac{1}{1+m}, -1+\frac{3}{1+m}; \frac{2}{1+m}; z_0\right)^{\!2} \notag\\
    &\quad\quad+ \frac{2^{-3+\frac{4}{1+m}}}{\pi\,\Gamma\!\left(\frac{3}{2}-\frac{1}{1+m}\right)^{\!2}} \bigl(z_0(1-z_0)\bigr)^{\frac{m}{1+m}} \left(\cos\frac{(m-3)\pi}{1+m} - \cos\frac{(1-m)\pi}{1+m}\right) \notag\\
    &\quad\quad\quad\times \Gamma\!\left(\frac{2}{1+m}\right)^{\!2} \Gamma\!\left(2-\frac{3}{1+m}\right)^{\!2} \,{}_2F_1\!\left(\frac{m}{1+m}, 2-\frac{3}{1+m}; \frac{2m}{1+m}; z_0\right)^{\!2} \Biggr]^{-1}\,,
\end{align}
where the coefficient of the leading $z^{2\Delta_\phi}$ power vanishes.

For quick reference, for the arbitrarily chosen value $z_0=0.4$ (that will be used momentarily) we get 
\begin{equation}
    \label{smoothcomboae}
    a^*(3) = -57.033\,, \qquad
    a^*(4) = -55.0473\,, \qquad
    a^*(5) = -62.279
    \,.
\end{equation}

\subsection{Examples}

\subsubsection{GFF with polynomial deformations}
\label{gffdefs}

To be concrete, we choose a specific random dimension, $\Df=0.8$. This number was selected to be less than 1 on purpose, to guarantee that the polynomial deformation \eqref{smoothba2} (which is quadratic at leading order at small $z$) does not alter the gap condition. For the corresponding bosonic GFF with $H(z)=\GG(z)-L(z)$, and $L(z)=1$ selected simply as the identity contribution, we compute the quantities $[H]_1$, $[H]_2$ and the Chebyshev spectra for the five values of the deformation parameter $a=-500,-100,\,0,\,100,\,500$. The results are collected in Table~\ref{tab:GFF_100_H} and the Chebyshev spectra are displayed in Fig.~\ref{fig:GFF_cheb}.
\begin{table}[H]
    \centering
    \begin{tabular}{c|c|c|c|c|c}
      semi-norm & $a=-500$& $a=-100$ & $a=0$ (GFF) & $a=100$ & $a=500$ \\ \hline\hline
      $[H]_1$   & 7.522 &  2.044 & 2.273 & 3.143 & 13.010 \\
      $[H]_2$   & 0.7867 & 0.217 & 0.180 & 0.187 & 0.636
    \end{tabular}
    \caption{Semi-norms $[H]_1$, $[H]_2$ for the bosonic GFF ($\Df=0.8$) and four polynomial deformations of the type \eqref{smoothba2} at $a=-500,-100,100,500$.}
    \label{tab:GFF_100_H}
\end{table}

\begin{figure*}[ht!]
\centering

\pgfplotsset{
  myaxis/.style={
    width=6.5cm,
    height=3cm,
    scale only axis,
    title style={font=\sffamily\sansmath\scriptsize},
    label style={font=\sffamily\sansmath\scriptsize},
    tick label style={font=\sffamily\sansmath\scriptsize},
    legend style={
      font=\sffamily\sansmath\scriptsize,
      cells={anchor=west},
      draw=black,
      fill=white,
      inner xsep=1pt,
      inner ysep=1pt
    },
    minor tick num=2,
    grid=both,
    grid style={line width=.1pt, draw=gray!25},
    major grid style={line width=.2pt, draw=gray!45}
  }
}

% Two stacked columns: left = a=-100, right = a=100
\begin{minipage}[t]{0.49\textwidth}
\centering

% --- Functions (a=-100) ---
\begin{tikzpicture}
\begin{axis}[myaxis,
  xlabel={$z$}, ylabel={$H(z)$},
  title={Functions on $[0,0.5]$},
  legend pos=north west]
\addplot[blue] table[x index=0,y index=1,col sep=space]{plot_data/GFF_pn100_functions.dat};
\addlegendentry{GFF}
\addplot[red]  table[x index=0,y index=2,col sep=space]{plot_data/GFF_pn100_functions.dat};
\addlegendentry{GFF($a=-100$)}
\end{axis}
\end{tikzpicture}

% --- Spectrum (a=-100) ---
{\centering\hspace*{-2.8mm}\begin{tikzpicture}
\begin{semilogyaxis}[myaxis,
  xlabel={Chebyshev mode $n$},
  ylabel={$|c_n|$},
  title={Chebyshev Spectrum Comparison},
  legend pos=north east,
  restrict x to domain=0:200]
\addplot[blue] table[x index=0,y index=1,col sep=space]{plot_data/GFF_pn100_spectrum.dat};
\addlegendentry{GFF}
\addplot[red]  table[x index=0,y index=2,col sep=space]{plot_data/GFF_pn100_spectrum.dat};
\addlegendentry{GFF($a=-100$)}
\end{semilogyaxis}
\end{tikzpicture}\par}

% --- Gamma (a=-100) ---
\begin{tikzpicture}
\begin{axis}[myaxis,
  xlabel={$s$}, ylabel={$\gamma(s)$},
  title={Logarithmic Growth Rate $\gamma(s)$},
  legend pos=north west]
\addplot[blue] table[x index=0,y index=1,col sep=space]{plot_data/GFF_pn100_gamma.dat};
\addlegendentry{GFF}
\addplot[red]  table[x index=0,y index=2,col sep=space]{plot_data/GFF_pn100_gamma.dat};
\addlegendentry{GFF($a=-100$)}
\end{axis}
\end{tikzpicture}
\end{minipage}
\hfill
\begin{minipage}[t]{0.49\textwidth}
\centering

% --- Functions (a=100) ---
\begin{tikzpicture}
\begin{axis}[myaxis,
  xlabel={$z$}, ylabel={$H(z)$},
  title={Functions on $[0,0.5]$},
  legend pos=north west]
\addplot[blue] table[x index=0,y index=1,col sep=space]{plot_data/GFF_p100_functions.dat};
\addlegendentry{GFF}
\addplot[red]  table[x index=0,y index=2,col sep=space]{plot_data/GFF_p100_functions.dat};
\addlegendentry{GFF($a=100$)}
\end{axis}
\end{tikzpicture}

% --- Spectrum (a=100) ---
{\centering\hspace*{-2.7mm}\begin{tikzpicture}
\begin{semilogyaxis}[myaxis,
  xlabel={Chebyshev mode $n$},
  ylabel={$|c_n|$},
  title={Chebyshev Spectrum Comparison},
  legend pos=north east,
  restrict x to domain=0:200]
\addplot[blue] table[x index=0,y index=1,col sep=space]{plot_data/GFF_p100_spectrum.dat};
\addlegendentry{GFF}
\addplot[red]  table[x index=0,y index=2,col sep=space]{plot_data/GFF_p100_spectrum.dat};
\addlegendentry{GFF($a=100$)}
\end{semilogyaxis}
\end{tikzpicture}\par}

% --- Gamma (a=100) ---
\begin{tikzpicture}
\begin{axis}[myaxis,
  xlabel={$s$}, ylabel={$\gamma(s)$},
  title={Logarithmic Growth Rate $\gamma(s)$},
  legend pos=north west]
\addplot[blue] table[x index=0,y index=1,col sep=space]{plot_data/GFF_p100_gamma.dat};
\addlegendentry{GFF}
\addplot[red]  table[x index=0,y index=2,col sep=space]{plot_data/GFF_p100_gamma.dat};
\addlegendentry{GFF($a=100$)}
\end{axis}
\end{tikzpicture}

\end{minipage}
\caption{\texorpdfstring{%
The first column presents (top to bottom) $H(z)$ (for $z\in[0,0.5]$), the absolute value of the first 200 Chebyshev coefficients, and the logarithmic growth rate $\gamma(s)$ for the bosonic GFF correlator ($\Delta_\phi=0.8$) comparing $a=-100$ (red) to $a=0$ (blue). The second column presents the same diagnostics comparing $a=100$ (red) to $a=0$ (blue).%
}{Left column: diagnostics for a=-100 vs a=0. Right column: diagnostics for a=100 vs a=0.}}
\label{fig:GFF_cheb}
\end{figure*}

The plots of the function $H(z)$ for $z\in [0,0.5]$ and the deformation values $a=-100$ (on the left) and $a=100$ (on the right) appear on the first row in Fig.\ \ref{fig:GFF_cheb}. Visually, it is evident that the undeformed function (blue curve) is more straight and smooth compared to the two deformations (red curves). The Chebyshev spectrum decay depicted on the second row does not exhibit significant differences between the deformed and undeformed functions in these two cases. However, a more distinctive difference appears in the plots of $\gamma(s)$ on the third row. There, we can see that the red curves of the deformed functions exhibit higher values than their undeformed versions for values of $s$ between the vicinity of 1 and below the jump around $s\simeq 2$.\footnote{Plotting $\gamma(s)$ for higher values of $s$ reveals a sharp jump that saturates around $s=4$ to a plateau around 14. This jump is due to the truncation of the sum over $n$ in the numerical estimation of $\gamma(s)$. Anything after the jump can be ignored in our analysis. Moreover, since we care about the large-$n$ decay of the spectrum we can also ignore the very small values of $s$.} The lower values of $\gamma(s)$ for the blue curve support the picture that the undeformed GFF correlator is a smoother function. 

When the size of the deformation parameter $|a|$ grows further, the picture arising from the Chebyshev spectra becomes cleaner. For both positive and negative $a$, the undeformed GFF correlator is clearly the smoother function according to the Chebyshev criteria of smoothness.

The semi-norms $[H]_1$, $[H]_2$ tell a similar story in Table \ref{tab:GFF_100_H}. $[H]_2$ assigns its smallest value closer to the undeformed correlator compared to $[H]_1$. For large $|a|$ both semi-norms unambiguously assign larger values to the deformed function. The values of $[H]_1$ and $[H]_2$ at $a=\pm 500$ exhibit this clearly.

\subsubsection{2d minimal models on a line with polynomial deformations}

As another example, we consider the polynomial deformations \eqref{smoothba2} of the four-point function $\langle \phi_{1,2}\phi_{1,2}\phi_{1,2}\phi_{1,2}\rangle$ on a line in the 2d minimal models $\mathcal{M}(m, m+1)$. For concreteness, we set $m=5$. Qualitatively similar results were observed also for several other values of $m$. Fig.\ \ref{fig:min_cheb} presents plots of the deformations and their corresponding Chebyshev spectra for deformation parameters $a=- 10,0,10$. In Table \ref{tab:min_100_H} we list the values of $[H]_1$, $[H]_2$ for deformation parameters $a=-100,-10,0,10,100$. 

\begin{table}[H]
    \centering
    \begin{tabular}{c|c|c|c|c|c}
      semi-norm & $a=-100$ & $a=-10$ & $a=0$ (GFF) & $a=10$ & $a=100$ \\ \hline\hline
      $[H]_1$   & 0.1921 & 0.09178 &  0.1072 & 0.128 & 0.5538\\ 
      $[H]_2$   & 0.01887 & 0.005487 &  0.005712 & 0.006279 & 0.02679 
    \end{tabular}
    \caption{Semi-norms $[H]_1$, $[H]_2$ for the minimal model $\MM(5,6)$ and four polynomial deformations at $a=-100,-10,10,100$.}
    \label{tab:min_100_H}
\end{table}

\begin{figure}[ht!]
\centering

\pgfplotsset{
  myaxis/.style={
    width=6.5cm,
    height=3cm,
    scale only axis,
    title style={font=\sffamily\sansmath\scriptsize},
    label style={font=\sffamily\sansmath\scriptsize},
    tick label style={font=\sffamily\sansmath\scriptsize},
    legend style={
      font=\sffamily\sansmath\scriptsize,
      cells={anchor=west},
      draw=black,
      fill=white,
      inner xsep=1pt,
      inner ysep=1pt
    },
    minor tick num=2,
    grid=both,
    grid style={line width=.1pt, draw=gray!25},
    major grid style={line width=.2pt, draw=gray!45}
  }
}

% Two stacked columns: left = a=-10, right = a=10
\begin{minipage}[t]{0.49\textwidth}
\centering

% --- Functions (a=-10) ---
\begin{tikzpicture}
\begin{axis}[myaxis,
  xlabel={$z$}, ylabel={$H(z)$},
  title={Functions on $[0,0.5]$},
  legend pos=north west]
\addplot[blue] table[x index=0,y index=1,col sep=space]{plot_data/minimal_m10_functions.dat};
\addlegendentry{$\mathcal{M}(5,6)$}
\addplot[red]  table[x index=0,y index=2,col sep=space]{plot_data/minimal_m10_functions.dat};
\addlegendentry{$\mathcal{M}(5,6)$($a=-10$)}
\end{axis}
\end{tikzpicture}

% --- Spectrum (a=-10) ---
{\centering\hspace*{-4mm}\begin{tikzpicture}
\begin{semilogyaxis}[myaxis,
  xlabel={Chebyshev mode $n$},
  ylabel={$|c_n|$},
  title={Chebyshev Spectrum Comparison},
  legend pos=north east,
  restrict x to domain=0:200]
\addplot[blue] table[x index=0,y index=1,col sep=space]{plot_data/minimal_m10_spectrum.dat};
\addlegendentry{$\mathcal{M}(5,6)$}
\addplot[red]  table[x index=0,y index=2,col sep=space]{plot_data/minimal_m10_spectrum.dat};
\addlegendentry{$\mathcal{M}(5,6)$($a=-10$)}
\end{semilogyaxis}
\end{tikzpicture}\par}

% --- Gamma (a=-10) ---
{\centering\hspace*{2.4mm}\begin{tikzpicture}
\begin{axis}[myaxis,
  xlabel={$s$}, ylabel={$\gamma(s)$},
  title={Logarithmic Growth Rate $\gamma(s)$},
  legend pos=north west,
  restrict x to domain=0:3.35]
\addplot[blue] table[x index=0,y index=1,col sep=space]{plot_data/minimal_m10_gamma.dat};
\addlegendentry{$\mathcal{M}(5,6)$}
\addplot[red]  table[x index=0,y index=2,col sep=space]{plot_data/minimal_m10_gamma.dat};
\addlegendentry{$\mathcal{M}(5,6)$($a=-10$)}
\end{axis}
\end{tikzpicture}\par}

\end{minipage}
\hfill
\begin{minipage}[t]{0.49\textwidth}
\centering

% --- Functions (a=10) ---
\begin{tikzpicture}
\begin{axis}[myaxis,
  xlabel={$z$}, ylabel={$H(z)$},
  title={Functions on $[0,0.5]$},
  legend pos=north west]
\addplot[blue] table[x index=0,y index=1,col sep=space]{plot_data/minimal_p10_functions.dat};
\addlegendentry{$\mathcal{M}(5,6)$}
\addplot[red]  table[x index=0,y index=2,col sep=space]{plot_data/minimal_p10_functions.dat};
\addlegendentry{$\mathcal{M}(5,6)$($a=10$)}
\end{axis}
\end{tikzpicture}

% --- Spectrum (a=10) ---
{\centering\hspace*{-4mm}\begin{tikzpicture}
\begin{semilogyaxis}[myaxis,
  xlabel={Chebyshev mode $n$},
  ylabel={$|c_n|$},
  title={Chebyshev Spectrum Comparison},
  legend pos=north east,
  restrict x to domain=0:200]
\addplot[blue] table[x index=0,y index=1,col sep=space]{plot_data/minimal_p10_spectrum.dat};
\addlegendentry{$\mathcal{M}(5,6)$}
\addplot[red]  table[x index=0,y index=2,col sep=space]{plot_data/minimal_p10_spectrum.dat};
\addlegendentry{$\mathcal{M}(5,6)$($a=10$)}
\end{semilogyaxis}
\end{tikzpicture}\par}

% --- Gamma (a=10) ---
\begin{tikzpicture}
\begin{axis}[myaxis,
  xlabel={$s$}, ylabel={$\gamma(s)$},
  title={Logarithmic Growth Rate $\gamma(s)$},
  legend pos=north west,
  restrict x to domain=0:3.35]
\addplot[blue] table[x index=0,y index=1,col sep=space]{plot_data/minimal_p10_gamma.dat};
\addlegendentry{$\mathcal{M}(5,6)$}
\addplot[red]  table[x index=0,y index=2,col sep=space]{plot_data/minimal_p10_gamma.dat};
\addlegendentry{$\mathcal{M}(5,6)$($a=10$)}
\end{axis}
\end{tikzpicture}

\end{minipage}
\caption{\texorpdfstring{%
The first column presents (top to bottom) $H(z)$ (for $z\in[0,0.5]$), the absolute value of the first 200 Chebyshev coefficients, and the logarithmic growth rate $\gamma(s)$ for the correlator of the 2d minimal model $\MM(5,6)$ with polynomial deformation parameters $a=-10$ (red curves) and $a=0$ (blue curves). The second column presents the same diagnostics comparing $a=10$ (red) to $a=0$ (blue).%
}{Left column: diagnostics for a=-10 vs a=0. Right column: diagnostics for a=10 vs a=0.}}
\label{fig:min_cheb}
\end{figure}

The qualitative features of these results are the same as those in the previous case of GFF deformations in Subsection~\ref{gffdefs}. When the deformation parameter is sufficiently large, all measures clearly show that the undeformed minimal model correlator is the smoothest function. When the deformation parameter is small (e.g.\ $a=\pm 10$ in Fig.\ \ref{fig:min_cheb} and Table \ref{tab:min_100_H}) the deformed functions have comparable measures of smoothness with the undeformed correlators and some measures may not assign strictly the largest degree of smoothness to the undeformed correlator. For example, we see in Table \ref{tab:min_100_H} that both $[H]_1$ and $[H]_2$ are slightly smaller for the deformed function at $a=-10$.

\subsubsection{Linear combinations of GFF and minimal models}
\label{linearcombo}

Next, we consider a second family of deformations: the linear combinations of bosonic GFF, fermionic GFF and 2d minimal model correlators in Eq.\ \eqref{smoothbb}. For concreteness, we discuss two representative cases, the $m=3$ case with $\Df=\frac{1}{8}$ and the $m=5$ case with $\Df=\frac{1}{4}$. For vanishing deformation parameter the function is a bosonic GFF correlator. When the deformation parameter is large the deformation is considerable and all the diagnostics of smoothness agree that the undeformed GFF correlator is comparatively smoother. For that reason, we focus the presentation at or around the critical values $a^*(m)$ (see Eq.\ \eqref{smoothcomboae}) where the leading small-$z$ behaviour of the deformation is affected most prominently by the deformation.

\begin{figure}[H]
\centering

\pgfplotsset{
  myaxis/.style={
    width=6.5cm,
    height=3cm,
    scale only axis,
    title style={font=\sffamily\sansmath\scriptsize},
    label style={font=\sffamily\sansmath\scriptsize},
    tick label style={font=\sffamily\sansmath\scriptsize},
    legend style={
      font=\sffamily\sansmath\scriptsize,
      draw=black,
      fill=white,
      inner xsep=1pt,
      inner ysep=1pt
    },
    minor tick num=2,
    grid=both,
    grid style={line width=.1pt, draw=gray!25},
    major grid style={line width=.2pt, draw=gray!45}
  }
}

% ===================== TOP ROW (2 PANELS) =====================

\begin{minipage}[t]{0.48\textwidth}
\centering
\begin{tikzpicture}
\begin{axis}[myaxis,
  xlabel={$z$},
  ylabel={$H(z)$},
  title={Functions on $[0,0.5]$},
  legend pos=south east]
\addplot[blue] table[x index=0,y index=1,col sep=space]{plot_data/linear_combo_m3_extra1_functions.dat};
\addlegendentry{GFF}
\addplot[red] table[x index=0,y index=2,col sep=space]{plot_data/linear_combo_m3_extra1_functions.dat};
\addlegendentry{Linear combo}
\end{axis}
\end{tikzpicture}
\end{minipage}
\hfill
\begin{minipage}[t]{0.48\textwidth}
\centering
\begin{tikzpicture}
\begin{semilogyaxis}[myaxis,
  xlabel={Mode $n$},
  ylabel={$|c_n|$},
  title={Chebyshev Spectrum Comparison},
  legend pos=north east,
  restrict x to domain=0:200]
\addplot[blue] table[x index=0,y index=1,col sep=space]{plot_data/linear_combo_m3_extra1_spectrum.dat};
\addlegendentry{GFF}
\addplot[red] table[x index=0,y index=2,col sep=space]{plot_data/linear_combo_m3_extra1_spectrum.dat};
\addlegendentry{Linear combo}
\end{semilogyaxis}
\end{tikzpicture}
\end{minipage}

% ===================== BOTTOM ROW (CENTERED) =====================

\begin{minipage}[t]{0.60\textwidth}
\centering
\begin{tikzpicture}
\begin{axis}[myaxis,
  xlabel={$s$},
  ylabel={$\gamma(s)$},
  title={Logarithmic Growth Rate $\gamma(s)$},
  legend pos=north west]
\addplot[blue] table[x index=0,y index=1,col sep=space]{plot_data/linear_combo_m3_extra1_gamma.dat};
\addlegendentry{GFF}
\addplot[red] table[x index=0,y index=2,col sep=space]{plot_data/linear_combo_m3_extra1_gamma.dat};
\addlegendentry{Linear combo}
\end{axis}
\end{tikzpicture}
\end{minipage}

\caption{Chebyshev spectra for the linear combinations of correlators \eqref{smoothbb} with parameters $(m,a)= (3,0),(3,a^*(3))$. The blue curves correspond to the undeformed GFF correlator and the red curves to the linear combination deformation. At $a^*$ the leading small-$z$ asymptotics of the deformation is visibly different in the top left plot.}
\label{fig:linear_combo_m3}
\end{figure}
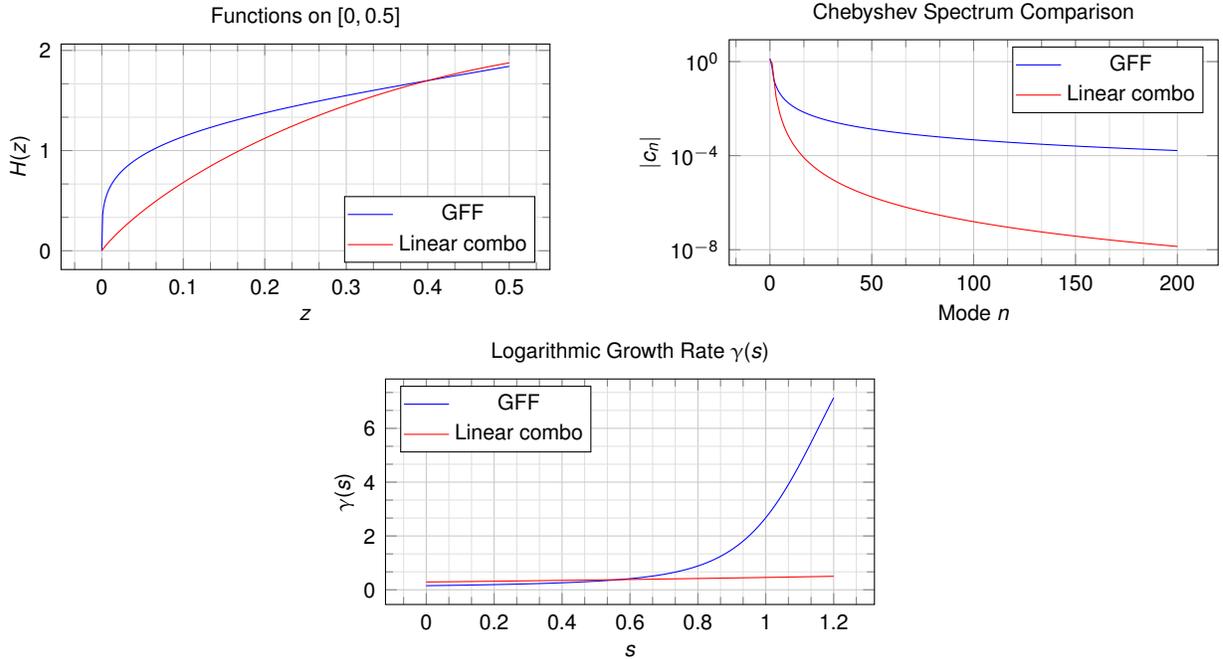

\paragraph{$\boldsymbol{m=3}$ deformations.}
The plot of the $a^*(3)$ deformation and the corresponding Chebyshev spectra appear in Fig.\ \ref{fig:linear_combo_m3}. The corresponding $[H]_1$, $[H]_2$ semi-norm values are summarised in Table \ref{tab:linear_combo_m3}.

\begin{table}[H]
    \centering
    \begin{tabular}{c|c|c}
      \textrm{semi-norm type} & $a=0$ & $a=a^*(3)=-57.033$ \\ \hline\hline
      $[H]_1$   &  1.864 & 3.473 \\ 
      $[H]_2$   &  0.3667 & 0.3761  
    \end{tabular}
    \caption{$[H]_1$, $[H]_2$ semi-norm values for the linear combinations $(m,a)= (3,0),(3,a^*(3))$.}
    \label{tab:linear_combo_m3}
\end{table}

\begin{figure}[H]
\centering

\pgfplotsset{
  myaxis/.style={
    width=6.5cm,
    height=3cm,
    scale only axis,
    title style={font=\sffamily\sansmath\scriptsize},
    label style={font=\sffamily\sansmath\scriptsize},
    tick label style={font=\sffamily\sansmath\scriptsize},
    legend style={
      font=\sffamily\sansmath\scriptsize,
      draw=black,
      fill=white,
      inner xsep=1pt,
      inner ysep=1pt
    },
    minor tick num=2,
    grid=both,
    grid style={line width=.1pt, draw=gray!25},
    major grid style={line width=.2pt, draw=gray!45}
  }
}

% ===================== TOP ROW (2 PANELS) =====================

\begin{minipage}[t]{0.48\textwidth}
\centering
\begin{tikzpicture}
\begin{axis}[myaxis,
  xlabel={$a$},
  ylabel={Value},
  title={$[H]_{1}$}]
\addplot[blue] table[x index=0,y index=1,col sep=space]{plot_data/seminorm_vs_aF_m_3.dat};
\end{axis}
\end{tikzpicture}
\end{minipage}
\hfill
\begin{minipage}[t]{0.48\textwidth}
\centering
\begin{tikzpicture}
\begin{axis}[myaxis,
  xlabel={a},
  ylabel={Value},
  title={$[H]_{2}$}]
\addplot[blue] table[x index=0,y index=2,col sep=space]{plot_data/seminorm_vs_aF_m_3.dat};
\end{axis}
\end{tikzpicture}
\end{minipage}
\caption{$[H]_1$, $[H]_2$ semi-norms as a function of the deformation parameter $a$ for fixed $m=3$.}
\label{fig:semi-norm_vs_a_m_3}
\end{figure}
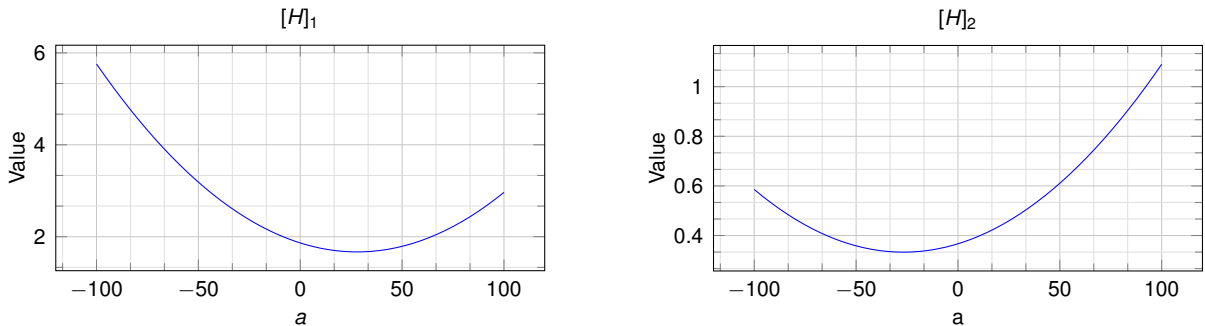

In this case, the plot of the functions creates the impression that the deformed correlator (red curve in the top left plot of Fig.\ \ref{fig:linear_combo_m3}) is smoother and the Chebyshev spectra confirm that. However, both semi-norm measures assign higher smoothness to the undeformed GFF correlator. To investigate this further we computed the $[H]_1$, $[H]_2$ semi-norms for a range of $a$ values that are summarised in Table \ref{tab:linear_combo_m3_several_a}. We observe (see also Fig.\ \ref{fig:semi-norm_vs_a_m_3}) that $[H]_1$ assigns a minimum around $a\simeq 25$ and $[H]_2$ a minimum around $a\simeq -25$. Both of these configurations are located in the vicinity of the undeformed GFF correlator and share with it similar Chebyshev spectra. Along the same lines, in Section~\ref{GFF} we observe that the NN optimisation favours functions in the vicinity of GFF, instead of functions in the vicinity of $a^*(3)$.
\begin{table}[H]
    \centering
    \scriptsize
\setlength{\tabcolsep}{4pt}
\begin{tabular}{c|c|c|c|c|c|c|c|c|c|c}
\diagbox{$[H]$}{$a$} 
 & $-75$ & $-57.033$ & $-45$ & $-25$ & $-10$ & $0$ & $5$ & $10$ & $50$ & $75$ \\ \hline\hline
$[H]_1$ 
 & 4.316 & 3.473 & 2.998 & 2.369 & 2.029 & 1.864 & 1.801 & 1.750 & 1.791 & 2.222 \\
$[H]_2$ 
 & 0.4425 & 0.3761 & 0.3486 & 0.3331 & 0.3462 & 0.3667 & 0.3805 & 0.3967 & 0.6107 & 0.8211 \\
\end{tabular}
    \caption{$[H]_1$, $[H]_2$ semi-norm values for linear combinations with varying deformation parameter $a$ at fixed $m=3$.}
    \label{tab:linear_combo_m3_several_a}
\end{table} 

\begin{figure}[H]
\centering

\pgfplotsset{
  myaxis/.style={
    width=6.5cm,
    height=3cm,
    scale only axis,
    title style={font=\sffamily\sansmath\scriptsize},
    label style={font=\sffamily\sansmath\scriptsize},
    tick label style={font=\sffamily\sansmath\scriptsize},
    legend style={
      font=\sffamily\sansmath\scriptsize,
      draw=black,
      fill=white,
      inner xsep=1pt,
      inner ysep=1pt
    },
    minor tick num=2,
    grid=both,
    grid style={line width=.1pt, draw=gray!25},
    major grid style={line width=.2pt, draw=gray!45}
  }
}

% ===================== TOP ROW (2 PANELS) =====================

\begin{minipage}[t]{0.48\textwidth}
\centering
\begin{tikzpicture}
\begin{axis}[myaxis,
  xlabel={$z$},
  ylabel={$H(z)$},
  title={Functions on $[0,0.5]$},
  legend pos=south east]
\addplot[blue] table[x index=0,y index=1,col sep=space]{plot_data/linear_combo_m5_extra1_functions.dat};
\addlegendentry{GFF}
\addplot[red] table[x index=0,y index=2,col sep=space]{plot_data/linear_combo_m5_extra1_functions.dat};
\addlegendentry{Linear combo}
\end{axis}
\end{tikzpicture}
\end{minipage}
\hfill
\begin{minipage}[t]{0.48\textwidth}
\centering
\begin{tikzpicture}
\begin{semilogyaxis}[myaxis,
  xlabel={Mode $n$},
  ylabel={$|c_n|$},
  title={Chebyshev Spectrum Comparison},
  legend pos=north east,
  restrict x to domain=0:200]
\addplot[blue] table[x index=0,y index=1,col sep=space]{plot_data/linear_combo_m5_extra1_spectrum.dat};
\addlegendentry{GFF}
\addplot[red] table[x index=0,y index=2,col sep=space]{plot_data/linear_combo_m5_extra1_spectrum.dat};
\addlegendentry{Linear combo}
\end{semilogyaxis}
\end{tikzpicture}
\end{minipage}

% ===================== BOTTOM ROW (CENTERED) =====================

\begin{minipage}[t]{0.60\textwidth}
\centering
\begin{tikzpicture}
\begin{axis}[myaxis,
  xlabel={$s$},
  ylabel={$\gamma(s)$},
  title={Logarithmic Growth Rate $\gamma(s)$},
  legend pos=north west]
\addplot[blue] table[x index=0,y index=1,col sep=space]{plot_data/linear_combo_m5_extra1_gamma.dat};
\addlegendentry{GFF}
\addplot[red] table[x index=0,y index=2,col sep=space]{plot_data/linear_combo_m5_extra1_gamma.dat};
\addlegendentry{Linear combo}
\end{axis}
\end{tikzpicture}
\end{minipage}

\caption{Chebyshev spectra for the linear combinations $(m,a)= (5,0),(5,a^*(5))$. The blue curves correspond to the undeformed GFF correlator and the red curves to the linear combination deformation.}
\label{fig:linear_combo_m5}
\end{figure}
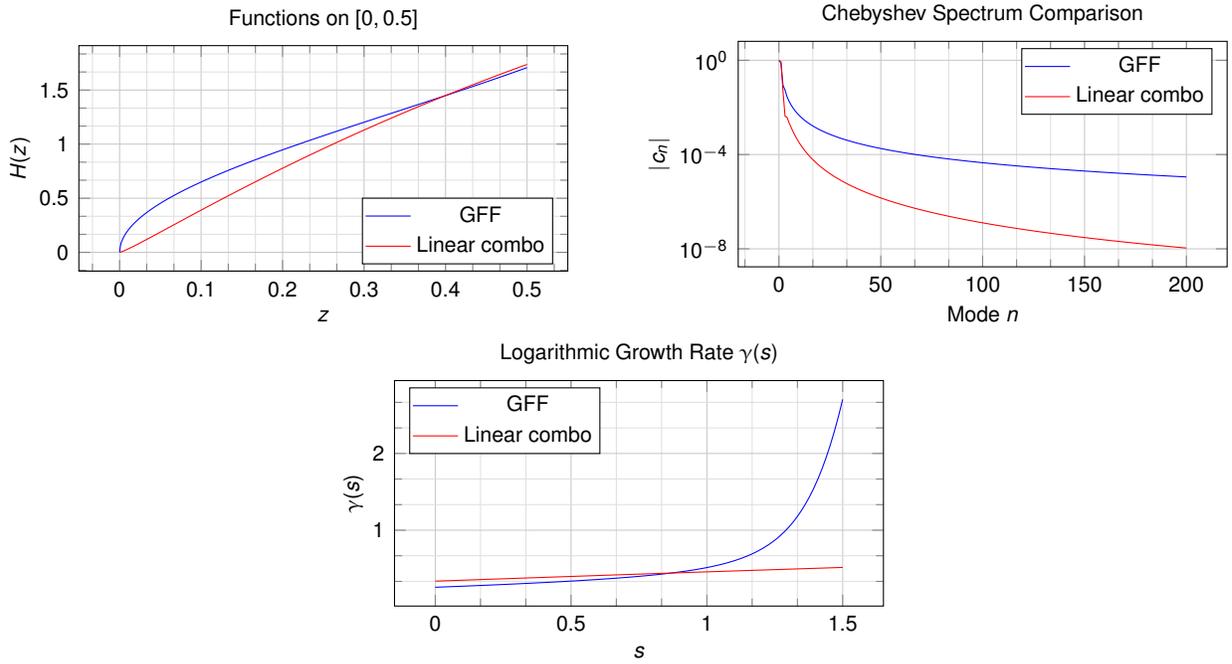

\paragraph{$\boldsymbol{m=5}$ deformations.}
An analogous analysis for $m=5$ yields a similar pattern. Both semi-norms are minimised in a neighbourhood of $a=0$ (see Table~\ref{tab:linear_combo_m5_several_aF} and Fig.\ \ref{fig:semi-norm_vs_a_m_5}), but the Chebyshev spectrum analysis favours configurations near $a^*(5)$ as the smoothest configurations (see Fig.\ \ref{fig:linear_combo_m5}). For the Sobolev semi-norms, it is apparent in Fig.\ \ref{fig:semi-norm_vs_a_m_5} that $[H]_2$ has a minimum closer to the undeformed correlator, around $a\simeq 25$, whereas $[H]_1$ has a minimum around $a\simeq 100$. We note in passing that the configuration favoured by the NN optimisation (see Fig.\ \ref{fig:gfb_1o4_summary} in Section~\ref{GFF}) lies also in the vicinity of the undeformed correlator.

\begin{table}[H]
    \centering
    \scriptsize
\setlength{\tabcolsep}{4pt}
\begin{tabular}{c|c|c|c|c|c|c|c|c|c|c}
\diagbox{$[H]$}{$a$} 
 & $-75$ & $-62.279$ & $-45$ & $-25$ & $-10$ & $0$ & $5$ & $10$ & $50$ & $75$ \\ \hline\hline
$[H]_1$ 
 & 3.345 & 3.113 & 2.825 & 2.532 & 2.340 & 2.225 & 2.172 & 2.121 & 1.811 & 1.704 \\
$[H]_2$ 
 & 0.2955 & 0.2806 & 0.2637 & 0.2487 & 0.2406 & 0.2368 & 0.2354 & 0.2342 & 0.2362 & 0.2475 \\
\end{tabular}
    \caption{$[H]_1$, $[H]_2$ semi-norm values for linear combinations with varying deformation parameter $a$ at fixed $m=5$.}
    \label{tab:linear_combo_m5_several_aF}
\end{table}

\begin{figure}[H]
\centering

\pgfplotsset{
  myaxis/.style={
    width=6.5cm,
    height=3cm,
    scale only axis,
    title style={font=\sffamily\sansmath\scriptsize},
    label style={font=\sffamily\sansmath\scriptsize},
    tick label style={font=\sffamily\sansmath\scriptsize},
    legend style={
      font=\sffamily\sansmath\scriptsize,
      draw=black,
      fill=white,
      inner xsep=1pt,
      inner ysep=1pt
    },
    minor tick num=2,
    grid=both,
    grid style={line width=.1pt, draw=gray!25},
    major grid style={line width=.2pt, draw=gray!45}
  }
}

% ===================== TOP ROW (2 PANELS) =====================

\begin{minipage}[t]{0.48\textwidth}
\centering
\begin{tikzpicture}
\begin{axis}[myaxis,
  xlabel={$a$},
  ylabel={Value},
  title={$[H]_{1}$}]
\addplot[blue] table[x index=0,y index=1,col sep=space]{plot_data/seminorm_vs_aF_m_5.dat};
\end{axis}
\end{tikzpicture}
\end{minipage}
\hfill
\begin{minipage}[t]{0.48\textwidth}
\centering
\begin{tikzpicture}
\begin{axis}[myaxis,
  xlabel={a},
  ylabel={Value},
  title={$[H]_{2}$}]
\addplot[blue] table[x index=0,y index=2,col sep=space]{plot_data/seminorm_vs_aF_m_5.dat};
\end{axis}
\end{tikzpicture}
\end{minipage}
\caption{$[H]_1$, $[H]_2$ semi-norms as a function of the deformation parameter $a$ for fixed $m=5$.}
\label{fig:semi-norm_vs_a_m_5}
\end{figure}
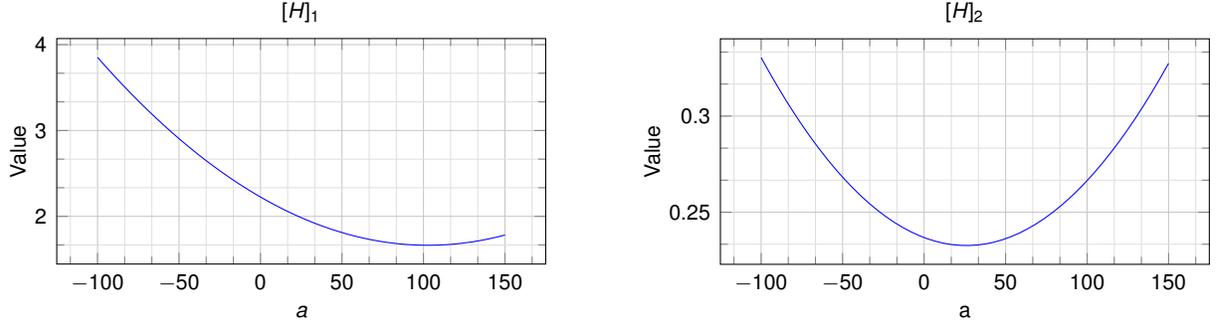

\subsection{Differential diagnostic of correlator smoothness}
In Eq.\ \eqref{curvature}, we defined a simple diagnostic of smoothness based on the first and second derivatives of $H$. We find that this diagnostic leads to conclusions consistent with those obtained from the other probes discussed above. Table~\ref{tab:kmin} lists the values \(a^{(K)}_{\min}\) at which the functional \(K[H]\) is minimised within the deformations \eqref{smoothbb} labelled by \(m\), together with the corresponding minimum value of \(K[H]\). The results exhibit substantial model dependence in the location of the minimum as a function of \(a\), with \(a^{(K)}_{\min}\) taking both negative and positive values across the examples considered. At the same time, the minimum values of \(K[H]\) remain within a relatively narrow range.

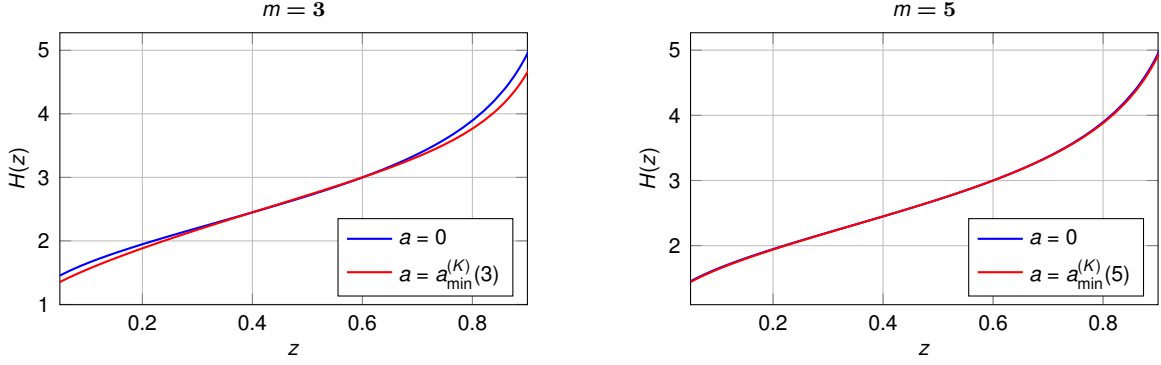
\begin{figure}[t!]
    \centering
    \begin{minipage}[t]{0.47\textwidth}
        \centering
        \begin{tikzpicture}
            \begin{axis}[
                width=\linewidth,
                height=5.2cm,
                xlabel={$z$},
                ylabel={$H(z)$},
                xmin=0.05, xmax=0.9,
                grid=major,
                legend pos=south east,
                legend cell align=left,
                title={$\boldsymbol{m=3}$},
                title style={font=\sffamily\sansmath\scriptsize},
                label style={font=\sffamily\sansmath\scriptsize},
                tick label style={font=\sffamily\sansmath\scriptsize},
                legend style={font=\sffamily\sansmath\scriptsize, draw=black, fill=white},
            ]
                \addplot[blue, thick] table[x index=0, y index=1, col sep=space]{plot_data/curvaturefunctional_Fdeform_a0_blue_from_a-23.5_show.dat};
                \addlegendentry{$a=0$}
                \addplot[red, thick] table[x index=0, y index=1, col sep=space]{plot_data/curvaturefunctional_Fdeform_a-23.5_red.dat};
                \addlegendentry{$a=a^{(K)}_{\min}(3)$}
            \end{axis}
        \end{tikzpicture}
    \end{minipage}
    \hspace{0.02\textwidth}
    \begin{minipage}[t]{0.47\textwidth}
        \centering
        \begin{tikzpicture}
            \begin{axis}[
                width=\linewidth,
                height=5.2cm,
                xlabel={$z$},
                ylabel={$H(z)$},
                xmin=0.05, xmax=0.9,
                grid=major,
                legend pos=south east,
                legend cell align=left,
                title={$\boldsymbol{m=5}$},
                title style={font=\sffamily\sansmath\scriptsize},
                label style={font=\sffamily\sansmath\scriptsize},
                tick label style={font=\sffamily\sansmath\scriptsize},
                legend style={font=\sffamily\sansmath\scriptsize, draw=black, fill=white},
            ]
                \addplot[blue, thick] table[x index=0, y index=1, col sep=space]{plot_data/curvaturefunctional_Fdeform_a0_blue_from_a-2.6_show.dat};
                \addlegendentry{$a=0$}
                \addplot[red, thick] table[x index=0, y index=1, col sep=space]{plot_data/curvaturefunctional_Fdeform_a-2.6_red.dat};
                \addlegendentry{$a=a^{(K)}_{\min}(5)$}
            \end{axis}
        \end{tikzpicture}
    \end{minipage}
    \caption{Functional comparison between the undeformed solution ($a=0$, solid blue) and the deformation at the curvature minimum ($a=a^{(K)}_{\min}(m)$, solid red) for the linear-combination family \eqref{smoothbb}, shown for $m=3$ (left) and $m=5$ (right).}
    \label{fig:curvaturefunctional_m3_m5}
\end{figure}

\begin{table}[ht]
\centering
\begin{tabular}{c c c}
\hline
$m$ & $a^{(K)}_{\min}$ & $K[H]\big|_{a=a^{(K)}_{\min}}$ \\
\hline
$3$ & $-23.5$ & $0.209053$ \\
$5$ & $-2.6$ & $0.147764$ \\
$-\frac{32}{10}$ & $-0.3$ & $3.47993$ \\
$23$ & $-81.5$ & $ 0.0174311$ \\
$-23$ & $247.7$ & $0.0637875$ \\
\hline
\end{tabular}
\caption{For each linear combination \eqref{smoothbb} labelled by $m$, we present the value of $a^{(K)}_{\min}$ at which the smoothness functional $K[H]$ is minimised, together with the corresponding minimum value of $K[H]$.}
\label{tab:kmin}
\end{table}

At first sight, this may appear surprising. For the functions under consideration, the undeformed generalised free boson correlator corresponds to \(a=0\), so one might have expected the smoothness diagnostic to select this value. Instead, the minimum of \(K[H]\) occurs at a non-zero value \(a^{(K)}_{\min}\). Interestingly, however, for the values of \(a^{(K)}_{\min}\) shown above, the corresponding correlators differ from the \(a=0\) solution only at the few-percent level (see Fig.\ \ref{fig:curvaturefunctional_m3_m5}). Thus, within the framework of our diagnostic, these correlators are not sharply distinguishable, and the smoothness functional is not sufficiently sensitive to uniquely isolate the undeformed generalised free boson solution when the differences are this small. This observation is consistent with the results shown in table \ref{tab:linear_combo_m3_several_a} and table \ref{tab:linear_combo_m5_several_aF} using other diagnostics of smoothness.

Nevertheless, this is already sufficient for our purposes. Even if the functional does not uniquely identify the exact physical correlator within such a narrow neighbourhood, it does succeed in selecting comparatively smoother solutions within the much larger space of crossing-symmetric candidates. In this way, the combination of crossing symmetry, anchoring conditions, and a preference for smoothness appears to guide the solution toward the physically relevant function space.

\subsection{A summary of emerging lessons}
\label{smooth:lessons}
What lessons can we distil from the above analysis? The main observations in the above examples are the following:
\begin{itemize}
    \item[(1)] In cases of `extreme' deformation, the Chebyshev spectra, the two fractional Sobolev semi-norm-inspired quantities $[H]_1$, $[H]_2$ and the curvature-based measure $K[H]$ provide the same ranking of smoothness and single out the undeformed physical CFT correlator as the smoothest function. In that sense, the above analysis supports the colloquial intuition that \textit{``CFT correlators on a line are smooth functions''}. 
    
    \item[(2)] When an undeformed CFT correlator is compared to a nearby deformation, the ranking of smoothness based on the diagnostics that we examined can disagree. The example of the linear combinations \eqref{smoothbb} exhibits this disagreement clearly. There is a range of the deformation parameters where the Chebyshev spectra and the two semi-norms point in opposite directions.
    
    \item[(3)] When we compare the quantities $[H]_1$ and $[H]_2$ (whose definition was based on the structure of the functions in the half interval $[0,0.5]$), we also detect subtle differences. An example was observed in Tables \ref{tab:GFF_100_H}, \ref{tab:min_100_H}, in the case of the polynomial GFF and minimal model deformations, respectively. We also observed subtle differences in the case of the $m=3$ and $m=5$ linear-combination deformations. It is unclear if any of these two measures is an appropriate measure of correlator smoothness in CFT. However, we note that in the $m=5$ example in Fig.\ \ref{fig:semi-norm_vs_a_m_5} the minimum of $[H]_2$ is closer to the configuration favoured by the NN optimisation. Recall that $[H]_2$ was motivated by the two-branch cut analytic structure of physical four-point correlation functions in unitary CFT. The picture emerging from the curvature-based measure $K[H]$ is comparable. Whether there is a variant of these quantities, whose exact minima are always physical CFT correlation functions on a line, remains an open question.
\end{itemize}

In what follows, we examine more closely what types of functions are favoured by the optimisation of anchored neural networks in a range of diverse examples across different CFTs in different spacetime dimensions and different types of observables. 

\begin{figure}[H]
    \centering
    \begin{tikzpicture}
        \begin{axis}[
            width=0.72\textwidth,
            height=5.5cm,
            xlabel={$z$},
            ylabel={Relative error},
            xmin=0.05, xmax=0.9,
            scaled y ticks=false,
            yticklabel style={/pgf/number format/fixed, /pgf/number format/precision=2},
            grid=major,
            legend pos=north west,
            legend cell align=left,
            title style={font=\sffamily\sansmath\scriptsize},
            label style={font=\sffamily\sansmath\scriptsize},
            tick label style={font=\sffamily\sansmath\scriptsize},
            legend style={font=\sffamily\sansmath\scriptsize, draw=black, fill=white},
        ]
            \addplot[red, thick] table[x=z, y=err_anchor_0p4, col sep=space]{plot_data/figure7_relative_error.dat};
            \addlegendentry{$z_0=0.4$}
            \addplot[blue, thick] table[x=z, y=err_anchor_0p3, col sep=space]{plot_data/figure7_relative_error.dat};
            \addlegendentry{$z_0=0.3$}
            \addplot[green!60!black, thick] table[x=z, y=err_anchor_0p2, col sep=space]{plot_data/figure7_relative_error.dat};
            \addlegendentry{$z_0=0.2$}
            \addplot[black, thick] table[x=z, y=err_anchor_0p1, col sep=space]{plot_data/figure7_relative_error.dat};
            \addlegendentry{$z_0=0.1$}
        \end{axis}
    \end{tikzpicture}
    \caption{Relative error between the correlator selected by minimizing \(K[H]\) and the undeformed (\(a=0\)) solution, shown for several choices of anchor point \(z_0\) in Eq.~\eqref{smoothbb} (here for the case \(m=3\)).}
    \label{fig:anchor_error}
\end{figure}
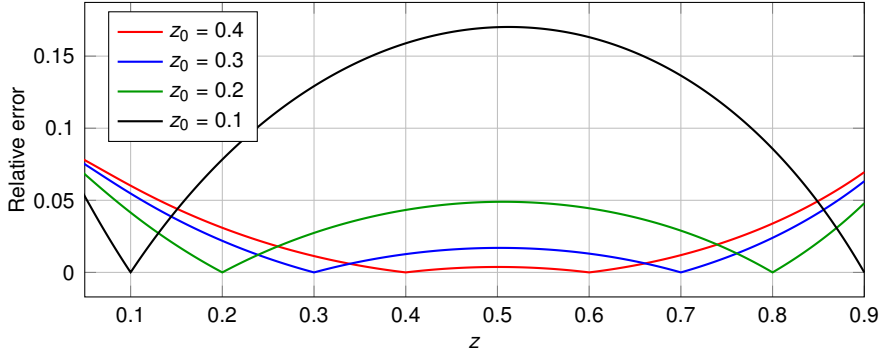

We note in passing the observation that in most cases the neural network prediction approximates the target physical correlator better when the anchor point is chosen in the intermediate Euclidean region, whereas the accuracy deteriorates as the anchor is moved closer to the \(z\to 0\) limit.\footnote{There are exceptions to this rule. In section~\ref{3dthermal} placing the anchor point as low as $z_0=0.05$ yields sensible, stable results.} In practice, the range \(z\sim 0.25\text{--}0.35\) appears to be the sweet spot. To illustrate this, in Fig.~\ref{fig:anchor_error} we plot the relative error between the correlator selected by minimizing \(K[H]\) and the \(a=0\) solution for several choices of anchor point, varying \(z\) from \(0.4\) to \(0.1\) for the case \(m=3\). 

\section{Generalised free fields}
\label{GFF}

Generalised free fields (GFFs) are some of the simplest conformal systems. For a scalar GFF $\phi$ with scaling dimension $\Delta_\phi$ the four-point function on a line reads (in any spacetime dimension)
\begin{equation}
    \label{1gffaa}
    \langle{\phi}(x_1) \phi(x_2) \phi(x_3) {\phi}(x_4) \rangle=\frac{1}{x_{12}^{2{\Delta}_{\phi}}x_{34}^{2{\Delta}_{\phi}}} \GG_{B/F}(z)\,,
\end{equation}
with
\begin{equation}
    \label{1gffab}
    \GG_{B/F}(z)=1+\eta^{B/F} z^{2\Delta_{\phi}}+\frac{z^{2\Delta_{\phi}}}{(1-z)^{2\Delta_{\phi}}}
    \,.
\end{equation}

For illustration, we will consider three explicit cases: 
\begin{itemize}
\item Bosonic GFFs with $\Delta_\phi=\frac{1}{4}$. This example, which is related to the $m=5$ family of linear combinations in \ref{linearcombo}, will also allow us to examine whether anchored neural networks can distinguish physical CFT correlators against linear combinations of CFT correlators.  
\item Bosonic and fermionic GFFs with the randomly chosen scaling dimension $\Delta_\phi = 1.618$.
\end{itemize}

In all cases, we set $L(z)=1$ (as very minimal input that acknowledges the contribution of the identity operator in the OPE) and write
\begin{equation}
    \label{1gffac}
    \GG_{B/F} (z) = L(z) + H(z)
\end{equation}
with 
\begin{equation}
    \label{1gffad}
    H(z) = z^{2\Delta_\phi} (1-z)^{-2\Delta_\phi} \, {\rm NN}_{\boldsymbol{\theta}}(z)
    \,.
\end{equation}
The first factor $z^{2\Delta_\phi}$ sets the gap parameter, $\delta=2\Delta_\phi$. The second factor $(1-z)^{-2\Delta_\phi}$ follows directly from the crossing symmetry constraint \eqref{introac} and does not entail any new information.
An anchor point value at $z_0=0.3$, $H(z_0)$, is selected using the exact equation \eqref{1gffab}.

With this minimal input, we ask whether optimisation of ${\rm NN}_{\boldsymbol{\theta}}(z)$ on crossing-symmetry can reconstruct numerically the complete exact correlator \eqref{1gffab}.

\subsection{Bosonic GFF at \texorpdfstring{$\Delta_\phi=\frac{1}{4}$}{Delta\_phi=1/4}}
\label{GFFb1_4}
First we consider the case of a bosonic GFF at $\Delta_\phi=\frac{1}{4}$. Deformations of this correlator were also analysed in Section \ref{linearcombo} for the $m=5$ case.

\begin{figure}[H]
    \centering
    % Subfigure 1: Comparison
    \begin{subfigure}[b]{0.49\textwidth}
        \centering
        \begin{tikzpicture}
            \begin{axis}[
                width=\linewidth, height=6cm,
                xlabel={$z$},
                ylabel={$\GG_{B}(z)$},
                title={Ensemble vs Exact},
                grid=major,
                ytick distance=0.5,
                legend pos=north west,
            ]
	                % Exact Solution
	                \addplot [black, dashed, thick] table [x=z, y=Exact, col sep=space] {plot_data/gfb_1o4_ensemble_comparison.dat};
	                \addlegendentry{Exact}

                % Ensemble Mean
                \addplot [blue, thick] table [x=z, y=Mean, col sep=space] {plot_data/gfb_1o4_ensemble_comparison.dat};
                \addlegendentry{Mean}

                % Linear Combination m=5
                \addplot [red, thick, densely dotted] table [x=z, y=LinCombo, col sep=space] {plot_data/gfb_1o4_ensemble_comparison.dat};
                \addlegendentry{Lin.\ Combo $m=5$}

                % Uncertainty Band
                \addplot [forget plot, name path=upper, draw=none] table [x=z, y=Mean_plus_Std, col sep=space] {plot_data/gfb_1o4_ensemble_comparison.dat};
                \addplot [forget plot, name path=lower, draw=none] table [x=z, y=Mean_minus_Std, col sep=space] {plot_data/gfb_1o4_ensemble_comparison.dat};
                \addplot [forget plot, fill=blue!30, fill opacity=0.5, draw=none] fill between [of=upper and lower];
                \addlegendimage{legend image code/.code={\fill[blue!30, draw=blue!50] (0cm,-0.1cm) rectangle (0.6cm,0.1cm);}}
                \addlegendentry{Mean $\pm$ 1 Std}
            \end{axis}
        \end{tikzpicture}
        %\caption{}
        %\label{fig:gfb_1o4_comparison}
    \end{subfigure}
    \hfill
    %
    % Subfigure 2: Percentage Error
    \begin{subfigure}[b]{0.49\textwidth}
        \centering
        \begin{tikzpicture}
            \begin{axis}[
                width=\linewidth, height=6cm,
                xlabel={$z$},
                ylabel={Error (\%)},
                ylabel style={at={(axis description cs:1.10,0.5)}, anchor=south},
                title={Prediction Error},
                grid=major,
                ytick distance=0.5,
                legend style={
                    at={(0.5,0.02)},
                    anchor=south,
                    }
            ]
                % Percentage Error
                \addplot [blue, thick] table [x=z, y=PctError, col sep=space] {plot_data/gfb_1o4_percentage_error.dat};
                \addlegendentry{Mean Error}

                % Uncertainty Band
                \addplot [forget plot, name path=upper, draw=none] table [x=z, y=PctError_plus_PctStd, col sep=space] {plot_data/gfb_1o4_percentage_error.dat};
                \addplot [forget plot, name path=lower, draw=none] table [x=z, y=PctError_minus_PctStd, col sep=space] {plot_data/gfb_1o4_percentage_error.dat};
                \addplot [forget plot, fill=blue!30, fill opacity=0.5, draw=none] fill between [of=upper and lower];
                \addlegendimage{legend image code/.code={\fill[blue!30, draw=blue!50] (0cm,-0.1cm) rectangle (0.6cm,0.1cm);}}
                \addlegendentry{Error $\pm$ 1 Std}
            \end{axis}
        \end{tikzpicture}
        %\caption{}
        %\label{fig:gfb_1o4_error}
    \end{subfigure}
    %
    %\vspace{1em}
    %
    % Subfigure 3: Histogram
    \begin{subfigure}[b]{0.65\textwidth}
        \centering
        \begin{tikzpicture}
            \begin{axis}[
                width=\linewidth, height=6.5cm,
                xlabel={$\mathcal{G}(z=0.5)$},
                ylabel={Count},
                title={Distribution at $z=0.5$},
                ybar,
                bar width=0.001369,
                xmin=2.704797, xmax=2.758273,
                ymin=0,
                ymajorgrids=true,
                xmajorgrids=false,
                legend pos=north west,
                legend style={xshift=16pt},
                scaled ticks=false,
                ytick distance=2,
                xticklabel style={
                    /pgf/number format/fixed,
                    /pgf/number format/precision=3,
                    font=\sansmath\sffamily\footnotesize,
                }
            ]
                \addplot [fill=blue!50, draw=black, opacity=0.7] table [x=BinCenter, y=Count, col sep=space] {plot_data/gfb_1o4_histogram_z0.500.dat};
                \addlegendentry{Model Predictions}
                \draw [black, dashed, thick] (axis cs:2.707227,\pgfkeysvalueof{/pgfplots/ymin}) -- (axis cs:2.707227,\pgfkeysvalueof{/pgfplots/ymax});
                \addlegendimage{legend image code/.code={\draw[black, dashed, thick] (0cm,0cm) -- (0.6cm,0cm);}}
                \addlegendentry{Exact ($2.707$)}
                \draw [blue, thick] (axis cs:2.738405,\pgfkeysvalueof{/pgfplots/ymin}) -- (axis cs:2.738405,\pgfkeysvalueof{/pgfplots/ymax});
                \addlegendimage{legend image code/.code={\draw[blue, thick] (0cm,0cm) -- (0.6cm,0cm);}}
                \addlegendentry{Mean ($2.738$)}
            \end{axis}
        \end{tikzpicture}
        %\caption{}
        %\label{fig:gfb_1o4_hist}
    \end{subfigure}

    \caption{NN-predicted $\GG_{B}(z)$ for a bosonic GFF at $\Delta_\phi=\frac{1}{4}$. In the top left panel the dashed red line corresponds to $(m,a)=(5,a^{*}(5))$ in the family of linear combinations discussed in Section~\ref{linearcombo}. These results are compared to the exact correlator \eqref{1gffab}. The NN prediction at $z=0.5$ is $2.738\pm 0.005$.}
    \label{fig:gfb_1o4_summary}
\end{figure}
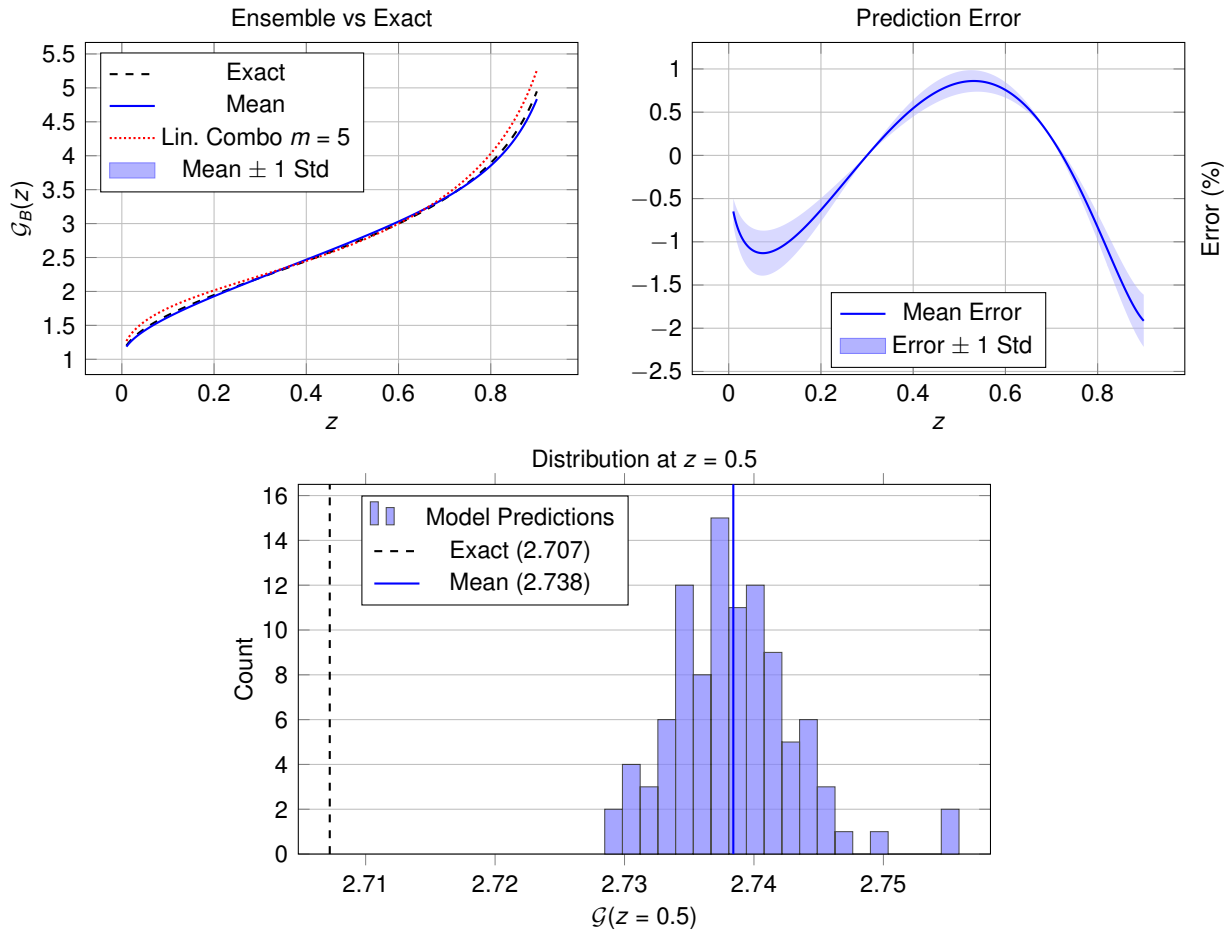

Our results for this theory are presented in Fig.\ \ref{fig:gfb_1o4_summary}; refer to the reporting scheme in Section~\ref{setup:NN}. On 100 independent runs, the crossing equation was satisfied with a mean square (MS) training loss of $(2.56\pm 0.371)\times 10^{-6}$. We observe that the NN prediction approximates the bosonic GFF correlator within a relative error that is mostly within 1\%. 

Returning to the discussion in Section~\ref{linearcombo}, we notice that the $[H]_2$ minimum at deformation parameter $a=25$ in Eq.\ \eqref{smoothbb} (which is represented by the dotted red curve on the top left plot in Fig.\ \ref{fig:gfb_1o4_summary}) lies in the vicinity of the undeformed bosonic GFF correlator, but not as close as the NN prediction. The more drastic deformation at $a^*(5)=-62.279$ (that appeared to be a smoother function according to the Chebyshev spectrum analysis), is further away. 

In summary, we observe that the anchored neural network optimisation favours a configuration in the close vicinity of the pure bosonic GFF correlator over the many possible deformations around it.

\subsection{Bosonic GFF at \texorpdfstring{$\Delta_\phi=1.618$}{Delta\_phi=1.618}}
\label{gffbosonic1618}

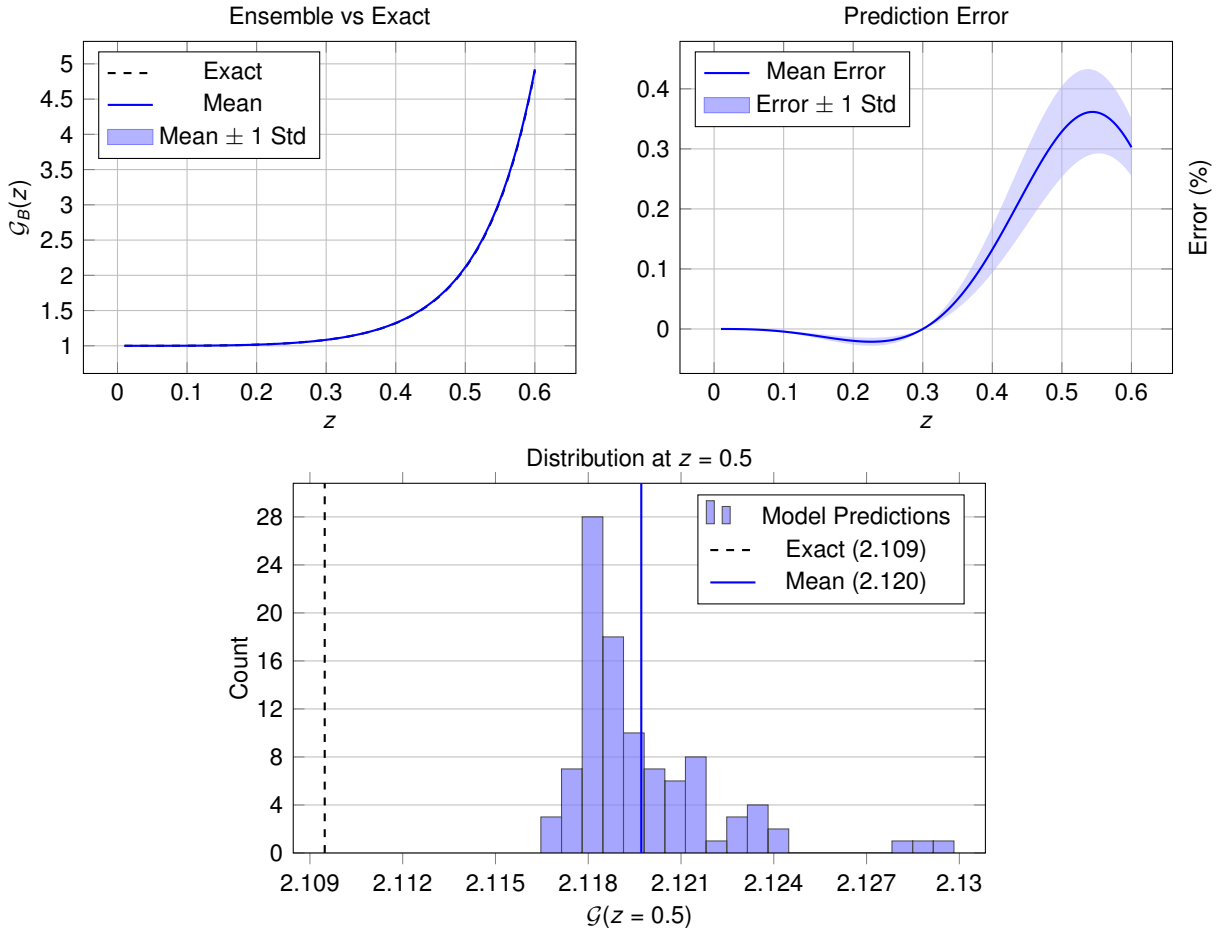
\begin{figure}[ht!]
    \centering
    % Subfigure 1: Comparison
    \begin{subfigure}[b]{0.49\textwidth}
        \centering
        \begin{tikzpicture}
            \begin{axis}[
                width=\linewidth, height=6cm,
                xlabel={$z$},
                ylabel={$\GG_{B}(z)$},
                title={Ensemble vs Exact},
                grid=major,
                ytick distance=0.5,
                legend pos=north west,
            ]
	                % Exact Solution
	                \addplot [black, dashed, thick] table [x=z, y=Exact, col sep=space] {plot_data/gfb_1.618_ensemble_comparison.dat};
	                \addlegendentry{Exact}

                % Ensemble Mean
                \addplot [blue, thick] table [x=z, y=Mean, col sep=space] {plot_data/gfb_1.618_ensemble_comparison.dat};
                \addlegendentry{Mean}

                % Uncertainty Band
                \addplot [forget plot, name path=upper, draw=none] table [x=z, y=Mean_plus_Std, col sep=space] {plot_data/gfb_1.618_ensemble_comparison.dat};
                \addplot [forget plot, name path=lower, draw=none] table [x=z, y=Mean_minus_Std, col sep=space] {plot_data/gfb_1.618_ensemble_comparison.dat};
                \addplot [forget plot, fill=blue!30, fill opacity=0.5, draw=none] fill between [of=upper and lower];
                \addlegendimage{legend image code/.code={\fill[blue!30, draw=blue!50] (0cm,-0.1cm) rectangle (0.6cm,0.1cm);}}
                \addlegendentry{Mean $\pm$ 1 Std}
            \end{axis}
        \end{tikzpicture}
        %\caption{}
        %\label{fig:gfb_1.618_comparison}
    \end{subfigure}
    \hfill
    %
    % Subfigure 2: Percentage Error
    \begin{subfigure}[b]{0.49\textwidth}
        \centering
        \begin{tikzpicture}
            \begin{axis}[
                width=\linewidth, height=6cm,
                xlabel={$z$},
                ylabel={Error (\%)},
                ylabel style={at={(axis description cs:1.10,0.5)}, anchor=south},
                title={Prediction Error},
                grid=major,
                ytick distance=0.1,
                legend pos=north west,
            ]
                % Percentage Error
                \addplot [blue, thick] table [x=z, y=PctError, col sep=space] {plot_data/gfb_1.618_percentage_error.dat};
                \addlegendentry{Mean Error}

                % Uncertainty Band
                \addplot [forget plot, name path=upper, draw=none] table [x=z, y=PctError_plus_PctStd, col sep=space] {plot_data/gfb_1.618_percentage_error.dat};
                \addplot [forget plot, name path=lower, draw=none] table [x=z, y=PctError_minus_PctStd, col sep=space] {plot_data/gfb_1.618_percentage_error.dat};
                \addplot [forget plot, fill=blue!30, fill opacity=0.5, draw=none] fill between [of=upper and lower];
                \addlegendimage{legend image code/.code={\fill[blue!30, draw=blue!50] (0cm,-0.1cm) rectangle (0.6cm,0.1cm);}}
                \addlegendentry{Error $\pm$ 1 Std}
            \end{axis}
        \end{tikzpicture}
        %\caption{}
        %\label{fig:gfb_1.618_error}
    \end{subfigure}
    %
    %\vspace{1em}
    %
    % Subfigure 3: Histogram
    \begin{subfigure}[b]{0.65\textwidth}
        \centering
        \begin{tikzpicture}
            \begin{axis}[
                width=\linewidth, height=6.5cm,
                xlabel={$\mathcal{G}(z=0.5)$},
                ylabel={Count},
                title={Distribution at $z=0.5$},
                ybar,
                bar width=0.000668,
                xmin=2.108460, xmax=2.130843,
                ymin=0,
                ymajorgrids=true,
                xmajorgrids=false,
                legend pos=north east,
                scaled ticks=false,
                ytick distance=4,
                xtick distance=0.003,
                xticklabel style={
                    /pgf/number format/fixed,
                    /pgf/number format/precision=3,
                    font=\sansmath\sffamily\footnotesize,
                }
            ]
                \addplot [fill=blue!50, draw=black, opacity=0.7] table [x=BinCenter, y=Count, col sep=space] {plot_data/gfb_1.618_histogram_z0.500.dat};
                \addlegendentry{Model Predictions}
                \draw [black, dashed, thick] (axis cs:2.109478,\pgfkeysvalueof{/pgfplots/ymin}) -- (axis cs:2.109478,\pgfkeysvalueof{/pgfplots/ymax});
                \addlegendimage{legend image code/.code={\draw[black, dashed, thick] (0cm,0cm) -- (0.6cm,0cm);}}
                \addlegendentry{Exact ($2.109$)}
                \draw [blue, thick] (axis cs:2.119715,\pgfkeysvalueof{/pgfplots/ymin}) -- (axis cs:2.119715,\pgfkeysvalueof{/pgfplots/ymax});
                \addlegendimage{legend image code/.code={\draw[blue, thick] (0cm,0cm) -- (0.6cm,0cm);}}
                \addlegendentry{Mean ($2.120$)}
            \end{axis}
        \end{tikzpicture}
        %\caption{}
        %\label{fig:gfb_1.618_hist}
    \end{subfigure}

    \caption{NN-predicted $\GG_{B}(z)$ for a bosonic GFF at $\Delta_\phi=1.618$. These results are compared to the exact correlator \eqref{1gffab}. The NN prediction at $z=0.5$ is $2.120\pm 0.002$.}
    \label{fig:gfb_1.618_summary}
\end{figure}

The next example concerns the four-point function on a line of a bosonic GFF at $\Delta_\phi=1.618$. The corresponding results are reported in Fig.\ \ref{fig:gfb_1.618_summary}. Since the correlator attains quickly relatively large values in the vicinity of $z=1$ we present plots in the range $[0,0.6]$. On 100 independent runs, the crossing equation was satisfied with MS training loss $(2.93 \pm 1.80)\times 10^{-7}$. We observe that the relative error between the NN prediction and the corresponding analytic correlator is less than 1\% in this case.

\subsection{Fermionic GFF at \texorpdfstring{$\Delta_\phi=1.618$}{Delta\_phi=1.618}}

Analogous results for the fermionic GFF at $\Delta_\phi=1.618$ are reported in Fig.\ \ref{fig:gff_1.618_summary}. The MS training loss in this case was $(6.58\pm 4.59)\times 10^{-8}$ and the comparison with the analytic correlator exhibits, again, a relative error below 1\%.

\begin{figure}[ht!]
    \centering
    % Subfigure 1: Comparison
    \begin{subfigure}[b]{0.49\textwidth}
        \centering
        \begin{tikzpicture}
            \begin{axis}[
                width=\linewidth, height=6cm,
                xlabel={$z$},
                ylabel={$\GG_{F}(z)$},
                title={Ensemble vs Exact},
                grid=major,
                ytick distance=0.5,
                legend pos=north west,
            ]
                % Exact Solution
                \addplot [black, dashed, thick] table [x=z, y=Exact, col sep=space] {plot_data/gff_1.618_ensemble_comparison.dat};
                \addlegendentry{Exact}

                % Ensemble Mean
                \addplot [blue, thick] table [x=z, y=Mean, col sep=space] {plot_data/gff_1.618_ensemble_comparison.dat};
                \addlegendentry{Mean}

                % Uncertainty Band
                \addplot [forget plot, name path=upper, draw=none] table [x=z, y=Mean_plus_Std, col sep=space] {plot_data/gff_1.618_ensemble_comparison.dat};
                \addplot [forget plot, name path=lower, draw=none] table [x=z, y=Mean_minus_Std, col sep=space] {plot_data/gff_1.618_ensemble_comparison.dat};
                \addplot [forget plot, fill=blue!30, fill opacity=0.5, draw=none] fill between [of=upper and lower];
                \addlegendimage{legend image code/.code={\fill[blue!30, draw=blue!50] (0cm,-0.1cm) rectangle (0.6cm,0.1cm);}}
                \addlegendentry{Mean $\pm$ 1 Std}
            \end{axis}
        \end{tikzpicture}
        %\caption{}
        %\label{fig:gff_1.618_comparison}
    \end{subfigure}
    \hfill
    %
    % Subfigure 2: Percentage Error
    \begin{subfigure}[b]{0.49\textwidth}
        \centering
        \begin{tikzpicture}
            \begin{axis}[
                width=\linewidth, height=6cm,
                xlabel={$z$},
                ylabel={Error (\%)},
                ylabel style={at={(axis description cs:1.10,0.5)}, anchor=south},
                title={Prediction Error},
                grid=major,
                ytick distance=0.1,
                legend pos=south west,
            ]
                % Percentage Error
                \addplot [blue, thick] table [x=z, y=PctError, col sep=space] {plot_data/gff_1.618_percentage_error.dat};
                \addlegendentry{Mean Error}

                % Uncertainty Band
                \addplot [forget plot, name path=upper, draw=none] table [x=z, y=PctError_plus_PctStd, col sep=space] {plot_data/gff_1.618_percentage_error.dat};
                \addplot [forget plot, name path=lower, draw=none] table [x=z, y=PctError_minus_PctStd, col sep=space] {plot_data/gff_1.618_percentage_error.dat};
                \addplot [forget plot, fill=blue!30, fill opacity=0.5, draw=none] fill between [of=upper and lower];
                \addlegendimage{legend image code/.code={\fill[blue!30, draw=blue!50] (0cm,-0.1cm) rectangle (0.6cm,0.1cm);}}
                \addlegendentry{Error $\pm$ 1 Std}
            \end{axis}
        \end{tikzpicture}
        %\caption{}
        %\label{fig:gff_1.618_error}
    \end{subfigure}
    %
    %\vspace{1em}
    %
    % Subfigure 3: Histogram
    \begin{subfigure}[b]{0.65\textwidth}
        \centering
        \begin{tikzpicture}
            \begin{axis}[
                width=\linewidth, height=6.5cm,
                xlabel={$\mathcal{G}(z=0.5)$},
                ylabel={Count},
                title={Distribution at $z=0.5$},
                ybar,
                bar width=0.000733,
	                xmin=1.870665, xmax=1.888984,
	                ymin=0,
	                ymajorgrids=true,
	                xmajorgrids=false,
	                ytick distance=2,
	                legend pos=north east,
	                scaled ticks=false,
	                xticklabel style={
	                    /pgf/number format/fixed,
	                    /pgf/number format/precision=3,
	                    font=\sansmath\sffamily\footnotesize,
	                }
	            ]
	                \addplot [fill=blue!50, draw=black, opacity=0.7] table [x=BinCenter, y=Count, col sep=space] {plot_data/gff_1.618_histogram_z0.500.dat};
	                \addlegendentry{Model Predictions}
                \draw [black, dashed, thick] (axis cs:1.888151,\pgfkeysvalueof{/pgfplots/ymin}) -- (axis cs:1.888151,\pgfkeysvalueof{/pgfplots/ymax});
                \addlegendimage{legend image code/.code={\draw[black, dashed, thick] (0cm,0cm) -- (0.6cm,0cm);}}
                \addlegendentry{Exact ($1.888$)}
                \draw [blue, thick] (axis cs:1.878741,\pgfkeysvalueof{/pgfplots/ymin}) -- (axis cs:1.878741,\pgfkeysvalueof{/pgfplots/ymax});
                \addlegendimage{legend image code/.code={\draw[blue, thick] (0cm,0cm) -- (0.6cm,0cm);}}
                \addlegendentry{Mean ($1.879$)}
            \end{axis}
        \end{tikzpicture}
        %\caption{}
        %\label{fig:gff_1.618_hist}
    \end{subfigure}

    \caption{NN-predicted $\GG_{F}(z)$ for a fermionic GFF at $\Delta_\phi=1.618$. These results are compared to the exact correlator \eqref{1gffab}. The NN prediction at $z=0.5$ is $1.879\pm 0.003$.}
    \label{fig:gff_1.618_summary}
\end{figure}

\subsection{Fermionic GFF at \texorpdfstring{$\Delta_\phi=1.618$}{Delta\_phi=1.618}}

Analogous results for the fermionic GFF at $\Delta_\phi=1.618$ are reported in Fig.\ \ref{fig:gff_1.618_summary}. The MS training loss in this case was $(6.58\pm 4.59)\times 10^{-8}$ and the comparison with the analytic correlator exhibits, again, a relative error below 1\%.

\section{\texorpdfstring{CFT$_1$}{CFT1}: Witten diagrams in \texorpdfstring{$\text{AdS}_2$}{AdS2}}
\label{ads}

The next class of applications involves Witten-diagram correlators of a $\phi^4$ scalar field theory in $\text{AdS}_2$. These correlators capture holographically the correlators of a dual 1d CFT. Exact expressions are available in closed form, providing another controlled benchmark for our method. These examples are distinctive, because the leading behaviour near $z=0$ involves logarithmic (rather than purely power-law) behaviour, testing the flexibility of the NN parametrisation in a situation that differs qualitatively from the generic OPE setting.

\subsection{Contact diagram}
\label{ads:contact}

\begin{figure}[ht]
    \centering
    % Subfigure 1: Comparison
    \begin{subfigure}[b]{0.49\textwidth}
        \centering
        \begin{tikzpicture}
            \begin{axis}[
                width=\linewidth, height=6cm,
                xlabel={$z$},
                ylabel={$\mathcal{G}_{\text{contact}}(z)$},
                title={Ensemble vs Exact},
                grid=major,
                ytick distance=1,
                legend pos=south west,
            ]
                % Exact Solution
                \addplot [black, dashed, thick] table [x=z, y=Exact, col sep=space] {plot_data/AdS2_contact_ensemble_comparison.dat};
                \addlegendentry{Exact}

                % Ensemble Mean
                \addplot [blue, thick] table [x=z, y=Mean, col sep=space] {plot_data/AdS2_contact_ensemble_comparison.dat};
                \addlegendentry{Mean}

                % Uncertainty Band
                \addplot [forget plot, name path=upper, draw=none] table [x=z, y=Mean_plus_Std, col sep=space] {plot_data/AdS2_contact_ensemble_comparison.dat};
                \addplot [forget plot, name path=lower, draw=none] table [x=z, y=Mean_minus_Std, col sep=space] {plot_data/AdS2_contact_ensemble_comparison.dat};
                \addplot [forget plot, fill=blue!30, fill opacity=0.5, draw=none] fill between [of=upper and lower];
                \addlegendimage{legend image code/.code={\fill[blue!30, draw=blue!50] (0cm,-0.1cm) rectangle (0.6cm,0.1cm);}}
                \addlegendentry{Mean $\pm$ 1 Std}
            \end{axis}
        \end{tikzpicture}
        %\caption{}
        %\label{fig:AdS2_contact_comparison}
    \end{subfigure}
    \hfill
    %
    % Subfigure 2: Percentage Error
    \begin{subfigure}[b]{0.49\textwidth}
        \centering
        \begin{tikzpicture}
	            \begin{axis}[
	                width=\linewidth, height=6cm,
	                xlabel={$z$},
	                ylabel={Error (\%)},
	                ylabel style={at={(axis description cs:1.10,0.5)}, anchor=south},
	                title={Prediction Error},
	                grid=major,
		                ymin=-1.6, ymax=0.4,
		                ytick={-1.6,-1.2,-0.8,-0.4,0,0.4},
			                legend pos=south west,
			            ]
	                % Percentage Error
	                \addplot [blue, thick] table [x=z, y=PctError, col sep=space] {plot_data/AdS2_contact_percentage_error.dat};
	                \addlegendentry{Mean Error}

                % Uncertainty Band
                \addplot [forget plot, name path=upper, draw=none] table [x=z, y=PctError_plus_PctStd, col sep=space] {plot_data/AdS2_contact_percentage_error.dat};
                \addplot [forget plot, name path=lower, draw=none] table [x=z, y=PctError_minus_PctStd, col sep=space] {plot_data/AdS2_contact_percentage_error.dat};
                \addplot [forget plot, fill=blue!30, fill opacity=0.5, draw=none] fill between [of=upper and lower];
                \addlegendimage{legend image code/.code={\fill[blue!30, draw=blue!50] (0cm,-0.1cm) rectangle (0.6cm,0.1cm);}}
                \addlegendentry{Error $\pm$ 1 Std}
            \end{axis}
        \end{tikzpicture}
        %\caption{}
        %\label{fig:AdS2_contact_error}
    \end{subfigure}
    %
    %\vspace{1em}
    %
    % Subfigure 3: Histogram
    \begin{subfigure}[b]{0.65\textwidth}
        \centering
        \begin{tikzpicture}
	            \begin{axis}[
	                width=\linewidth, height=6.5cm,
	                xlabel={$\mathcal{G}(z=0.5)$},
	                ylabel={Count},
	                title={Distribution at $z=0.5$},
	                ybar,
	                bar width=0.000163,
	                xmin=-1.389000, xmax=-1.382303,
	                ymin=0,
	                ymajorgrids=true,
		                xmajorgrids=false,
		                ytick distance=2,
		                legend pos=north west,
		                scaled ticks=false,
		                xticklabel style={
		                    /pgf/number format/fixed,
		                    /pgf/number format/precision=3,
                    font=\sansmath\sffamily\footnotesize,
                }
            ]
                \addplot [fill=blue!50, draw=black, opacity=0.7] table [x=BinCenter, y=Count, col sep=space] {plot_data/AdS2_contact_histogram_z0.500.dat};
                \addlegendentry{Model Predictions}
	                \draw [black, dashed, thick] (axis cs:-1.388054,\pgfkeysvalueof{/pgfplots/ymin}) -- (axis cs:-1.388054,\pgfkeysvalueof{/pgfplots/ymax});
	                \addlegendimage{legend image code/.code={\draw[black, dashed, thick] (0cm,0cm) -- (0.6cm,0cm);}}
	                \addlegendentry{Exact ($-1.388$)}
	                \draw [blue, thick] (axis cs:-1.384226,\pgfkeysvalueof{/pgfplots/ymin}) -- (axis cs:-1.384226,\pgfkeysvalueof{/pgfplots/ymax});
	                \addlegendimage{legend image code/.code={\draw[blue, thick] (0cm,0cm) -- (0.6cm,0cm);}}
	                \addlegendentry{Mean ($-1.384$)}
	            \end{axis}
	        \end{tikzpicture}
	        %\caption{}
	        %\label{fig:AdS2_contact_hist}
	    \end{subfigure}

	    \caption{NN-predicted $\GG_{\rm contact}(z)$ for the $\phi^4$ tree-level contact diagram in $\text{AdS}_2$ with $\Df=1$. These results are compared to the exact correlator \eqref{adsaa}. The NN prediction at $z=0.5$ is $-1.3842\pm 0.0006$.}
	    \label{fig:AdS2_contact_summary}
\end{figure}
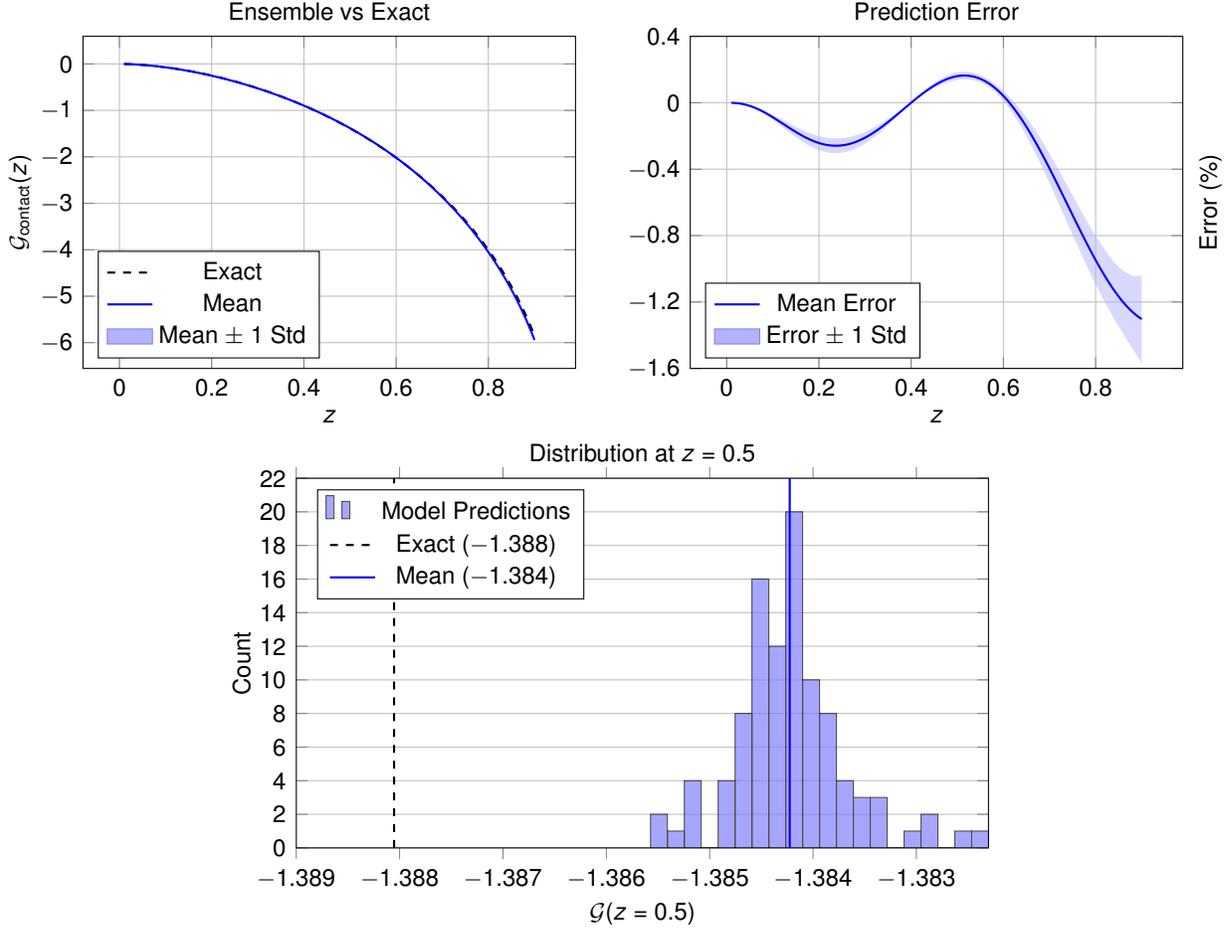

First, we consider the tree-level $\phi^4$ contact Witten diagram in $\text{AdS}_2$ at external dimension $\Df=1$, whose exact expression is
\begin{equation}
    \label{adsaa}
    \GG_{\rm contact}(z) = 2z^2\left(\frac{\log(1-z)}{z}+\frac{\log z}{1-z}\right).
\end{equation}
It is straightforward to check that this function satisfies the crossing symmetry equation \eqref{introac}. The dominant contribution in the expansion around $z=0$ is
\begin{equation}
    \label{adsac}
    \GG_{\rm contact}(z) = 2(-1+\log z)\,z^2 + \OO(z^3)\,,
\end{equation}
which involves a $z^2\log z$ term. We absorb this term into the ``known'' piece,
\begin{equation}
    \label{adsad}
    L(z) = 2(-1+\log z)\,z^2\,,
\end{equation}
and define the remainder $H(z)=\GG_{\rm contact}(z)-L(z)$. The behaviour of $H$ near $z=0$ starts at $\OO(z^3\log z)$, while crossing implies $\GG_{\rm contact}(z)\sim 2\log(1-z)$ near $z=1$. It is useful to bake this asymptotics into the ansatz for the NN parametrisation of $H$ by setting
\begin{equation}
    \label{adsae}
    H(z) = \big( z^3 \log z + z^3\log(1-z)\big)\,\text{NN}_{\boldsymbol\theta}(z)\,.
\end{equation}
With this ansatz we trained an anchored NN on the crossing equation and an anchor point at $z_0=0.4$ derived from the exact expression \eqref{adsaa}. The results are presented in Fig.\ \eqref{fig:AdS2_contact_summary}. The MS of the training loss on 100 runs was $(7.19\pm 2.82)\times 10^{-8}$. The observed deviation of the NN predictions from the exact result is below 1\% for most of the range of $z$, reaching about 1.3\% at $z=0.9$, which is related to the degree at which the NN managed to satisfy the crossing equation.

\subsection{One-loop (bubble) diagram}
\label{ads:oneloop}

\begin{figure}[H]
    \centering
    % Subfigure 1: Comparison
    \begin{subfigure}[b]{0.49\textwidth}
        \centering
        \begin{tikzpicture}
            \begin{axis}[
                width=\linewidth, height=6cm,
                xlabel={$z$},
                ylabel={$\mathcal{G}_{\text{bubble}}(z)$},
                ytick distance=20,
                title={Ensemble vs Exact},
                grid=major,
                legend pos=north west,
            ]
                % Exact Solution
                \addplot [black, dashed, thick] table [x=z, y=Exact, col sep=space] {plot_data/AdS2_one_loop_ensemble_comparison.dat};
                \addlegendentry{Exact}

                % Ensemble Mean
                \addplot [blue, thick] table [x=z, y=Mean, col sep=space] {plot_data/AdS2_one_loop_ensemble_comparison.dat};
                \addlegendentry{Mean}

                % Uncertainty Band
                \addplot [forget plot, name path=upper, draw=none] table [x=z, y=Mean_plus_Std, col sep=space] {plot_data/AdS2_one_loop_ensemble_comparison.dat};
                \addplot [forget plot, name path=lower, draw=none] table [x=z, y=Mean_minus_Std, col sep=space] {plot_data/AdS2_one_loop_ensemble_comparison.dat};
                \addplot [forget plot, fill=blue!30, fill opacity=0.5, draw=none] fill between [of=upper and lower];
                \addlegendimage{legend image code/.code={\fill[blue!30, draw=blue!50] (0cm,-0.1cm) rectangle (0.6cm,0.1cm);}}
                \addlegendentry{Mean $\pm$ 1 Std}
            \end{axis}
        \end{tikzpicture}
        %\caption{}
        %\label{fig:AdS2_one_loop_comparison}
    \end{subfigure}
    \hfill
    %
    % Subfigure 2: Percentage Error
    \begin{subfigure}[b]{0.49\textwidth}
        \centering
        \begin{tikzpicture}
		            \begin{axis}[
		                width=\linewidth, height=6cm,
		                xlabel={$z$},
		                ylabel={Error (\%)},
		                ylabel style={at={(axis description cs:1.10,0.5)}, anchor=south},
		                ymin=-1.2, ymax=1.2,
		                ytick={-1.2,-0.8,-0.4,0,0.4,0.8,1.2},
		                title={Prediction Error},
		                grid=major,
		                legend pos=north west,
		            ]
                % Percentage Error
                \addplot [blue, thick] table [x=z, y=PctError, col sep=space] {plot_data/AdS2_one_loop_percentage_error.dat};
                \addlegendentry{Mean Error}

                % Uncertainty Band
                \addplot [forget plot, name path=upper, draw=none] table [x=z, y=PctError_plus_PctStd, col sep=space] {plot_data/AdS2_one_loop_percentage_error.dat};
                \addplot [forget plot, name path=lower, draw=none] table [x=z, y=PctError_minus_PctStd, col sep=space] {plot_data/AdS2_one_loop_percentage_error.dat};
                \addplot [forget plot, fill=blue!30, fill opacity=0.5, draw=none] fill between [of=upper and lower];
                \addlegendimage{legend image code/.code={\fill[blue!30, draw=blue!50] (0cm,-0.1cm) rectangle (0.6cm,0.1cm);}}
                \addlegendentry{Error $\pm$ 1 Std}
            \end{axis}
        \end{tikzpicture}
        %\caption{}
        %\label{fig:AdS2_one_loop_error}
    \end{subfigure}
    %
    %\vspace{1em}
    %
    % Subfigure 3: Histogram
    \begin{subfigure}[b]{0.65\textwidth}
        \centering
        \begin{tikzpicture}
	            \begin{axis}[
	                width=\linewidth, height=6.5cm,
	                xlabel={$\mathcal{G}(z=0.5)$},
	                ylabel={Count},
	                title={Distribution at $z=0.5$},
	                ybar,
	                bar width=0.024507,
	                xmin=27.002750, xmax=27.492891,
	                ymin=0,
	                ytick distance=3,
	                ymajorgrids=true,
	                xmajorgrids=false,
	                legend pos=north west,
                scaled ticks=false,
                xticklabel style={
                    /pgf/number format/fixed,
                    /pgf/number format/precision=3,
                    font=\sansmath\sffamily\footnotesize,
                }
            ]
                \addplot [fill=blue!50, draw=black, opacity=0.7] table [x=BinCenter, y=Count, col sep=space] {plot_data/AdS2_one_loop_histogram_z0.500.dat};
                \addlegendentry{Model Predictions}
	                \draw [black, dashed, thick] (axis cs:27.123687,\pgfkeysvalueof{/pgfplots/ymin}) -- (axis cs:27.123687,\pgfkeysvalueof{/pgfplots/ymax});
	                \addlegendimage{legend image code/.code={\draw[black, dashed, thick] (0cm,0cm) -- (0.6cm,0cm);}}
	                \addlegendentry{Exact ($27.124$)}
	                \draw [blue, thick] (axis cs:27.299574,\pgfkeysvalueof{/pgfplots/ymin}) -- (axis cs:27.299574,\pgfkeysvalueof{/pgfplots/ymax});
	                \addlegendimage{legend image code/.code={\draw[blue, thick] (0cm,0cm) -- (0.6cm,0cm);}}
	                \addlegendentry{Mean ($27.300$)}
	            \end{axis}
	        \end{tikzpicture}
	        %\caption{}
	        %\label{fig:AdS2_one_loop_hist}
	    \end{subfigure}

	    \caption{NN-predicted $\GG_{\rm bubble}(z)$ for the $\phi^4$ one-loop (bubble) diagram in $\text{AdS}_2$ with $\Df=1$. These results are compared to the exact correlator \eqref{adsba}-\eqref{adsbc}. The NN prediction at $z=0.5$ is $27.300\pm 0.107$.}
	    \label{fig:AdS2_one_loop_summary}
	\end{figure}
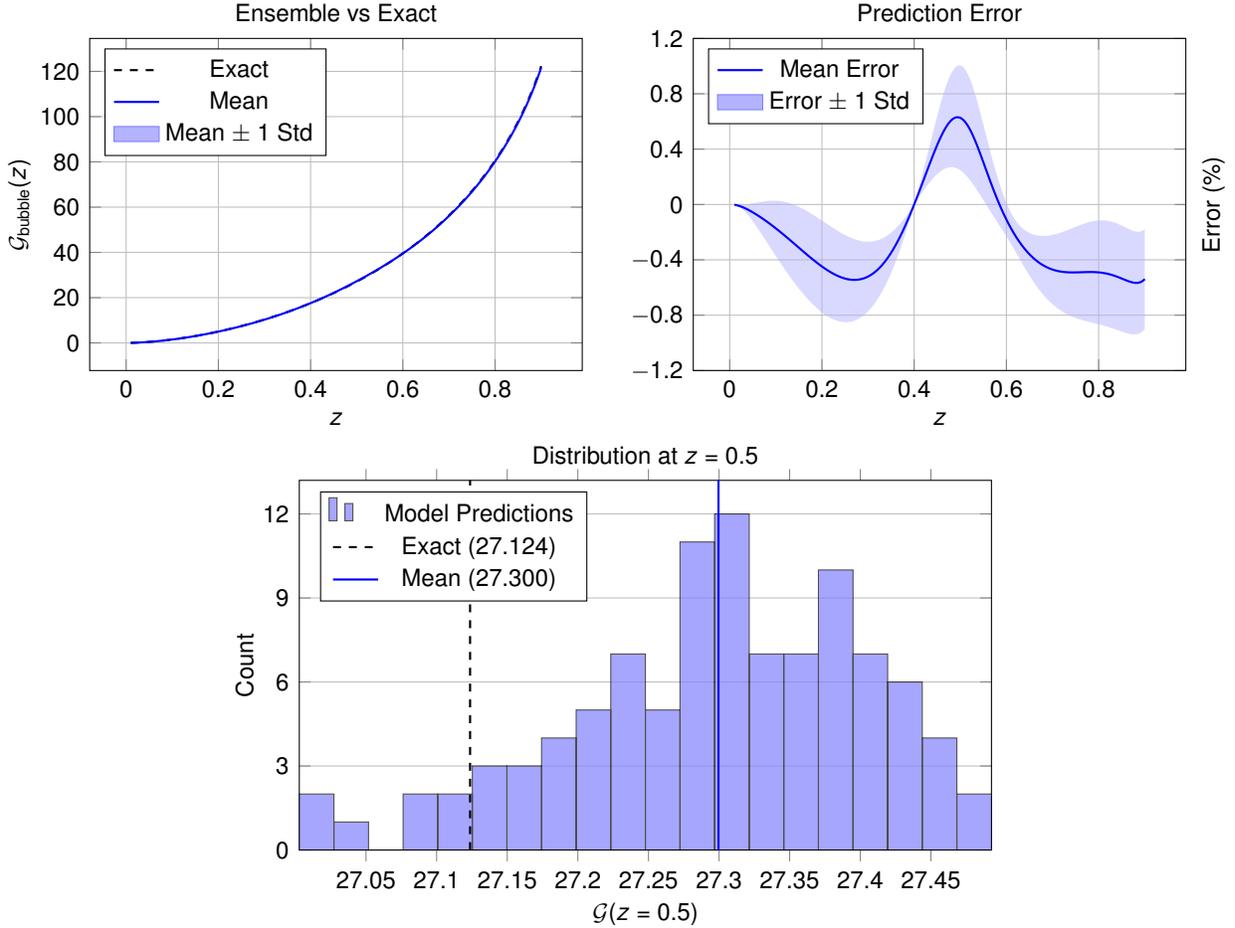

At one loop, the scalar bubble diagram in $\text{AdS}_2$ yields a considerably more complex correlator involving polylogarithms up to $\text{Li}_4$ \cite{Mazac:2018ycv,Ferrero:2019luz}. The exact function is
\begin{equation}
    \label{adsba}
    \GG_{\rm bubble}(z) = L(z) + H_{\rm bubble}(z)
\end{equation}
with
\begin{equation}
    \label{adsbb}
    L(z) = \frac{27}{250} \bigg( 95 + 2\pi^4 - 80 \log z + 30 (\log z)^2 + 80 \zeta(3) - 200\zeta(3) \log z  \bigg) z^2
    \,,
\end{equation}

\begin{align}
\label{adsbc}
& H_{\rm bubble}(z)=-\frac{27}{250}\, z^{2}
\Bigl(95 + 2\pi^{4} - 80\log z + 30\log^{2} z + 80\zeta(3) - 200\log z\,\zeta(3)\Bigr)
\nonumber\\
&+\frac{9}{250\,(z-1)^{2}}\Bigg[
\pi^{4} z^{2}\bigl(6-(z-2)z\bigr)
+15\Bigl(
4 z^{3}\operatorname{arctanh}(1-2z)
+(2z-1)\log^{4}(1-z)
+2 z^{2}\log z
\nonumber\\
& +4(1-2z)\log^{3}(1-z)\log z
+2\log(1-z)\Bigl(z-2z^{2}+\pi^{2}(z-1)^{2}(2+z^{2})\log z\Bigr)+2\log^{2}(1-z)
\nonumber\\
&\Bigl(\pi^{2}(2z-1)-3(z-1)^{2}(1+z^{2})\log^{2} z\Bigr)-180 (z-1)^{2}\Bigl((z^{2}-1)\log(1-z)+(2+z^{2})\log z\Bigr)\operatorname{Li}_{3}(1-z)
\Bigr)
\nonumber\\
&-180 z^{2}\Bigl((3+(z-2)z)\log(1-z)+(z-2)z\log z\Bigr)\operatorname{Li}_{3}(z)
\nonumber\\
&+360 (z-2) z^{3}\operatorname{Li}_{4}(1-z)
+360 (z-1)^{3}(1+z)\operatorname{Li}_{4}(z)
+360(2z-1)\operatorname{Li}_{4}\!\left(\frac{z}{z-1}\right)
\nonumber\\
&+60\Bigl(
-8 z^{3}\operatorname{arctanh}(1-2z)
+(-3+z(2+11z))\log(1-z)
-(-6+z(12+z))\log z
\Bigr)\zeta(3)
\Bigg].
\end{align}
Up to overall numerical factors, near $z=0$
\begin{equation}
    \label{adsbd}
    H_{\rm bubble}(z) \sim z^3 (\log z)^2
    ~.
\end{equation}
In the vicinity of $z=1$, we can deduce, using the crossing equation \eqref{introac}, that the correlator behaves as $(\log(1-z))^2$, which implies that
\begin{equation}
    \label{adsbe}
    H_{\rm bubble}(z) \sim (\log(1-z))^2
    \,.
\end{equation}
	Accordingly, we set
	\begin{equation}
	    \label{adsbe2}
	    H(z) = \big(z^3 \log z - z^2(\log(1-z))^2\big)\,\text{NN}_{\boldsymbol\theta}(z)
	    \,.
	\end{equation}
The results of the corresponding anchored neural network optimisation appear in Fig.\ \ref{fig:AdS2_one_loop_summary}. 

The corresponding MS training loss on 100 runs was higher in this case compared to the contact case of the previous subsection, namely $(1.19\pm0.64)\times 10^{-6}$. The deviation of the NN predictions from the exact result is below 0.7\% across the range of $z$ shown.

\paragraph*{Note on asymptotics.} To construct the NN prefactor, we cannot simply add the $z\to 0$ and $z\to 1$ asymptotics (as $\log(1-z)$ dominates both limits) nor multiply them (which over-suppresses the leading behaviour). Instead, we sum them but dress the $z\to 1$ term with a factor $z^n$ to suppress it near $z=0$. For the bubble diagram, however, a relative plus sign ($z^3\log z+z^n(\log(1-z))^2$) introduces an artificial root on $(0,1)$ for any $n\ge 1$, forcing the NN to vanish there. To avoid this (an issue absent in the contact diagram prefactor) we use a relative minus sign in \eqref{adsbe2}, preserving the correct endpoint behaviour up to overall normalisation without introducing spurious zeros.

\section{\texorpdfstring{CFT$_2$}{CFT2}: unitary and non-unitary minimal models}
\label{minimal}

The 2d minimal models $\MM(p,q)$ ($p,q\geq 2$) furnish an infinite family of exactly solvable CFTs. Their four-point functions are expressible in terms of hypergeometric functions, making them non-trivial benchmarks for our method's performance as the scaling dimension $\Df$ varies.

In this section, we consider the four-point function of the conformal primary field $\phi_{1,2}$ in the unitary minimal models $\MM(m,m+1)$ $(m=3,4,5,\ldots)$ and the non-unitary Lee--Yang model $\MM(2,5)$. The latter model will allow us to probe the applicability of the method in an explicit non-unitary setting. 

Let us quickly review some of the relevant exact results for these models.

\subsection{Useful exact results}

\subsubsection{Unitary models}
\label{minimalunitary}

In the unitary minimal models $\MM(m,m+1)$ the conformal primary $\phi_{1,2}$ has scaling dimension\footnote{By scaling dimension we mean here the full scaling dimension, $\Delta=h+\bar h$, which is a sum of the left- and right-moving conformal weights.}
\begin{equation}
    \label{minaa0}
    \Df = \frac{1}{2} - \frac{3}{2(m+1)}\,.
\end{equation}
For example, the case of the 2d Ising model corresponds to $m=3$ and $\phi_{1,2}$ is in that context the spin conformal primary $\sigma$, whose scaling dimension is $\frac{1}{8}$.

On the full $(z,\bar z)$ plane, the reduced four-point function of $\phi_{1,2}$ factorises into holomorphic and anti-holomorphic conformal blocks
\begin{equation}
    \label{minaa}
    \GG_V(z,\bar z) = \GG_1(z)\,\GG_1(\bar z) + N(\Df)\, \GG_2(z)\, \GG_2(\bar z)\,,
\end{equation}
with
\begin{align}
    \label{minab}
    \GG_1(z) &= (1-z)^{-\Df}\,{}_2F_1\big(\tfrac13(1-2\Df),\,-2\Df;\,\tfrac23(1-2\Df);\,z\big)\,,\nonumber\\
    \GG_2(z) &= (1-z)^{\frac{1+\Df}{3}}\,z^{\frac{1+4\Df}{3}}\,{}_2F_1\big(\tfrac23(1+\Df),\,1+2\Df;\,\tfrac43(1+\Df);\,z\big)\,.
\end{align}
The relative normalisation reads
\begin{equation}
    \label{minac}
    N(\Df) = \frac{2^{1-\frac{8(\Df+1)}{3}}\,\Gamma\bigl(\frac{2}{3}-\frac{4\Df}{3}\bigr)^2\,\Gamma(2\Df+1)^2\,\bigl(\sin\frac{\pi(16\Df+1)}{6}-\cos\frac{\pi(4\Df+1)}{3}\bigr)}{\pi\,\Gamma\bigl(\frac{2\Df}{3}+\frac{7}{6}\bigr)^2}\,.
\end{equation}
Restricting on the diagonal $z=\bar z$ gives a one-variable function $\GG_V(z)=\GG(z,z)$ that satisfies \eqref{introac}.

\subsubsection{A non-unitary model}

The Lee--Yang model, $\MM(2,5)$, has two conformal primaries---the identity and $\phi_{1,2}$. The latter has scaling dimension $\Delta_\phi=-\frac{2}{5}$.

On the full $(z,\bar z)$ plane, the reduced four-point function of $\phi_{1,2}$ takes the form \eqref{minaa} with 
\begin{equation}
    \label{minad}
    \GG_1(z) = (1-z)^{2/5} \, _2F_1\left(\frac{3}{5},\frac{4}{5};\frac{6}{5};z\right),
\end{equation}
\begin{equation}
    \label{minae}
    \GG_2(z) =\frac{\sqrt[5]{1-z} \, _2F_1\left(\frac{1}{5},\frac{2}{5};\frac{4}{5};z\right)}{\sqrt[5]{z}}\,,
\end{equation}
and the relative factor
\begin{equation}
    \label{minaf}
    N(\Delta_\phi) = -\frac{25\ 2^{2/5} \sqrt{5} \Gamma \left(\frac{1}{5}\right)^2 \Gamma \left(\frac{6}{5}\right)^2}{\pi  \Gamma \left(-\frac{1}{10}\right)^2}
    \,.
\end{equation}

\subsection{NN reconstruction}

In standard fashion, we set $\GG(z)=L(z)+H(z)$. For the unitary models we also set $L(z)=1$ to separate the trivial contribution of the identity operator and represent $H(z)$ in neuromorphic form by setting
\begin{equation}
    \label{minba}
    H(z) = z^\delta (1-z)^{-2\Delta_\phi} \, {\rm NN}_{\boldsymbol{\theta}}(z)
    \,.
\end{equation}
For the unitary minimal models $\MM(m,m+1)$ the exponent $\delta$ is related to the contribution of the $\phi_{1,3}$ primary, which implies $\delta=\frac{2(m-1)}{m+1}$. 

For the Lee--Yang model, $\delta=-\frac{2}{5}$. In this case, we set $L(z)=(1-z)^{\frac{4}{5}}$, corresponding to the identity-block contribution in the crossed ($t$-)channel, and parametrise the remainder $H(z)=\GG(z)-L(z)$ as
\begin{equation}
    \label{minbaextra}
    H(z) = z^{-\frac{2}{5}} (1-z)^{\frac{2}{5}} \, {\rm NN}_{\boldsymbol{\theta}}(z)
    \,.
\end{equation}

\subsubsection{\texorpdfstring{$\MM(3,4)$}{M(3,4)}: Ising model}
\label{2dising1}
\begin{figure}[H]
    \centering
    % Subfigure 1: Comparison
    \begin{subfigure}[b]{0.49\textwidth}
        \centering
        \begin{tikzpicture}
            \begin{axis}[
                width=\linewidth, height=6cm,
                xlabel={$z$},
                ylabel={$\mathcal{G}(z)$},
                title={Ensemble vs Exact},
                grid=major,
                ytick distance=0.1,
                legend pos=north west,
            ]
                % Exact Solution
                \addplot [black, dashed, thick] table [x=z, y=Exact, col sep=space] {plot_data/ising_ensemble_comparison.dat};
                \addlegendentry{Exact}

                % Ensemble Mean
                \addplot [blue, thick] table [x=z, y=Mean, col sep=space] {plot_data/ising_ensemble_comparison.dat};
                \addlegendentry{Mean}

                % Uncertainty Band
                \addplot [forget plot, name path=upper, draw=none] table [x=z, y=Mean_plus_Std, col sep=space] {plot_data/ising_ensemble_comparison.dat};
                \addplot [forget plot, name path=lower, draw=none] table [x=z, y=Mean_minus_Std, col sep=space] {plot_data/ising_ensemble_comparison.dat};
                \addplot [forget plot, fill=blue!30, fill opacity=0.5, draw=none] fill between [of=upper and lower];
                \addlegendimage{legend image code/.code={\fill[blue!30, draw=blue!50] (0cm,-0.1cm) rectangle (0.6cm,0.1cm);}}
                \addlegendentry{Mean $\pm$ 1 Std}
            \end{axis}
        \end{tikzpicture}
        %\caption{}
        %\label{fig:ising_comparison}
    \end{subfigure}
    \hfill
    %
    % Subfigure 2: Percentage Error
    \begin{subfigure}[b]{0.49\textwidth}
        \centering
        \begin{tikzpicture}
            \begin{axis}[
                width=\linewidth, height=6cm,
                xlabel={$z$},
                ylabel={Error (\%)},
                ylabel style={at={(axis description cs:1.10,0.5)}, anchor=south},
                title={Prediction Error},
                grid=major,
                ytick distance=0.2,
                legend pos=north west,
            ]
                % Percentage Error
                \addplot [blue, thick] table [x=z, y=PctError, col sep=space] {plot_data/ising_percentage_error.dat};
                \addlegendentry{Mean Error}

                % Uncertainty Band
                \addplot [forget plot, name path=upper, draw=none] table [x=z, y=PctError_plus_PctStd, col sep=space] {plot_data/ising_percentage_error.dat};
                \addplot [forget plot, name path=lower, draw=none] table [x=z, y=PctError_minus_PctStd, col sep=space] {plot_data/ising_percentage_error.dat};
                \addplot [forget plot, fill=blue!30, fill opacity=0.5, draw=none] fill between [of=upper and lower];
                \addlegendimage{legend image code/.code={\fill[blue!30, draw=blue!50] (0cm,-0.1cm) rectangle (0.6cm,0.1cm);}}
                \addlegendentry{Error $\pm$ 1 Std}
            \end{axis}
        \end{tikzpicture}
        %\caption{}
        %\label{fig:ising_error}
    \end{subfigure}
    %
    %\vspace{1em}
    %
    % Subfigure 3: Histogram
    \begin{subfigure}[b]{0.65\textwidth}
        \centering
        \begin{tikzpicture}
            \begin{axis}[
                width=\linewidth, height=6.5cm,
                xlabel={$\mathcal{G}(z=0.5)$},
                ylabel={Count},
                title={Distribution at $z=0.5$},
                ybar,
                bar width=0.003821,
                xmin=1.167719, xmax=1.201340,
                ymin=0,
                ymajorgrids=true,
                xmajorgrids=false,
                legend pos=north west,
                scaled ticks=false,
                ytick distance=3, 
                xticklabel style={
                    /pgf/number format/fixed,
                    /pgf/number format/precision=3,
                    font=\sansmath\sffamily\footnotesize,
                }
            ]
                \addplot [fill=blue!50, draw=black, opacity=0.7] table [x=BinCenter, y=Count, col sep=space] {plot_data/ising_histogram_z0.500.dat};
                \addlegendentry{Model Predictions}
                \draw [black, dashed, thick] (axis cs:1.189201,\pgfkeysvalueof{/pgfplots/ymin}) -- (axis cs:1.189201,\pgfkeysvalueof{/pgfplots/ymax});
                \addlegendimage{legend image code/.code={\draw[black, dashed, thick] (0cm,0cm) -- (0.6cm,0cm);}}
                \addlegendentry{Exact ($1.189$)}
                \draw [blue, thick] (axis cs:1.193946,\pgfkeysvalueof{/pgfplots/ymin}) -- (axis cs:1.193946,\pgfkeysvalueof{/pgfplots/ymax});
                \addlegendimage{legend image code/.code={\draw[blue, thick] (0cm,0cm) -- (0.6cm,0cm);}}
                \addlegendentry{Mean ($1.194$)}
            \end{axis}
        \end{tikzpicture}
        %\caption{}
        %\label{fig:ising_hist}
    \end{subfigure}

    \caption{NN-predicted reduced four-point function $\GG(z)$ of the correlator $\langle \phi_{1,2}\phi_{1,2}\phi_{1,2}\phi_{1,2}\rangle$ for the 2d Ising model, $\MM(3,4)$. These results were based on 43 independent runs filtered at training loss below $10^{-5}$ and are compared to the exact correlator \eqref{minaa0}-\eqref{minac} with $m = 3$. The NN prediction at $z=0.5$ is $1.194\pm 0.006$.}
    \label{fig:ising_summary}
\end{figure}
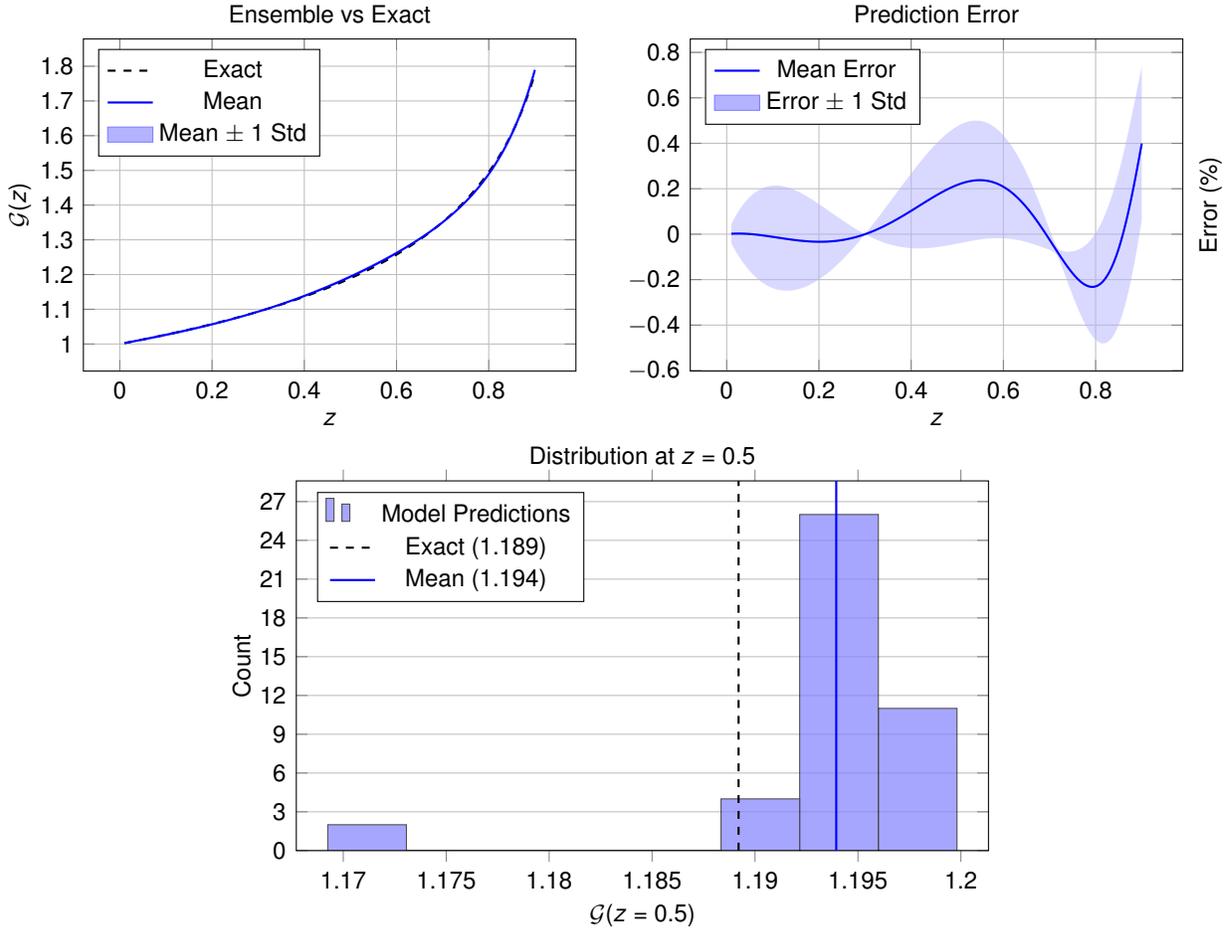

The results for the 2d Ising model are summarised in Fig.~\ref{fig:ising_summary}. They are based on 100 independent runs, retaining only those whose training loss falls below a chosen cutoff. In this case, we impose a threshold of $10^{-5}$ in order to discard runs that have not learned the crossing equation \eqref{introac} to an adequate degree. The point is not that $10^{-5}$ has any special significance in itself, but rather that, for this ensemble, the training outcomes are clearly split between well-converged runs and poorly converged ones, so the cutoff serves to isolate the former. With this criterion, 57 runs are rejected, leaving the set displayed in Fig.~\ref{fig:ising_summary}. The retained runs have an MS training loss of $(2.94\pm0.691)\times10^{-6}$.

We observe that the predicted correlator is within less than 1\% percent difference from the exact analytic result throughout the $z$-interval.

\subsubsection{\texorpdfstring{$\MM(4,5)$}{M(4,5)}: Tricritical Ising model}

\begin{figure}[ht!]
    \centering
    % Subfigure 1: Comparison
    \begin{subfigure}[b]{0.49\textwidth}
        \centering
        \begin{tikzpicture}
            \begin{axis}[
                width=\linewidth, height=6cm,
                xlabel={$z$},
                ylabel={$\mathcal{G}(z)$},
                title={Ensemble vs Exact},
                grid=major,
                ytick distance=0.2,
                legend pos=north west,
            ]
                % Exact Solution
                \addplot [black, dashed, thick] table [x=z, y=Exact, col sep=space] {plot_data/mm_4_ensemble_comparison.dat};
                \addlegendentry{Exact}

                % Ensemble Mean
                \addplot [blue, thick] table [x=z, y=Mean, col sep=space] {plot_data/mm_4_ensemble_comparison.dat};
                \addlegendentry{Mean}

                % Uncertainty Band
                \addplot [forget plot, name path=upper, draw=none] table [x=z, y=Mean_plus_Std, col sep=space] {plot_data/mm_4_ensemble_comparison.dat};
                \addplot [forget plot, name path=lower, draw=none] table [x=z, y=Mean_minus_Std, col sep=space] {plot_data/mm_4_ensemble_comparison.dat};
                \addplot [forget plot, fill=blue!30, fill opacity=0.5, draw=none] fill between [of=upper and lower];
                \addlegendimage{legend image code/.code={\fill[blue!30, draw=blue!50] (0cm,-0.1cm) rectangle (0.6cm,0.1cm);}}
                \addlegendentry{Mean $\pm$ 1 Std}
            \end{axis}
        \end{tikzpicture}
        %\caption{}
        %\label{fig:mm_4_comparison}
    \end{subfigure}
    \hfill
    %
    % Subfigure 2: Percentage Error
    \begin{subfigure}[b]{0.49\textwidth}
        \centering
        \begin{tikzpicture}
            \begin{axis}[
                width=\linewidth, height=6cm,
                xlabel={$z$},
                ylabel={Error (\%)},
                ylabel style={at={(axis description cs:1.10,0.5)}, anchor=south},
                title={Prediction Error},
                grid=major,
                ytick distance=0.2,
                legend pos=north west,
            ]
                % Percentage Error
                \addplot [blue, thick] table [x=z, y=PctError, col sep=space] {plot_data/mm_4_percentage_error.dat};
                \addlegendentry{Mean Error}

                % Uncertainty Band
                \addplot [forget plot, name path=upper, draw=none] table [x=z, y=PctError_plus_PctStd, col sep=space] {plot_data/mm_4_percentage_error.dat};
                \addplot [forget plot, name path=lower, draw=none] table [x=z, y=PctError_minus_PctStd, col sep=space] {plot_data/mm_4_percentage_error.dat};
                \addplot [forget plot, fill=blue!30, fill opacity=0.5, draw=none] fill between [of=upper and lower];
                \addlegendimage{legend image code/.code={\fill[blue!30, draw=blue!50] (0cm,-0.1cm) rectangle (0.6cm,0.1cm);}}
                \addlegendentry{Error $\pm$ 1 Std}
            \end{axis}
        \end{tikzpicture}
        %\caption{}
        %\label{fig:mm_4_error}
    \end{subfigure}
    %
    %\vspace{1em}
    %
    % Subfigure 3: Histogram
    \begin{subfigure}[b]{0.65\textwidth}
        \centering
        \begin{tikzpicture}
            \begin{axis}[
                width=\linewidth, height=6.5cm,
                xlabel={$\mathcal{G}(z=0.5)$},
                ylabel={Count},
                title={Distribution at $z=0.5$},
                ybar,
                bar width=0.002223,
                xmin=1.255859, xmax=1.280315,
                ymin=0,
                ymajorgrids=true,
                xmajorgrids=false,
                legend pos=north west,
                scaled ticks=false,
                xtick distance=0.003,
                ytick distance=3,
                xticklabel style={
                    /pgf/number format/fixed,
                    /pgf/number format/precision=3,
                    font=\sansmath\sffamily\footnotesize,
                }
            ]
                \addplot [fill=blue!50, draw=black, opacity=0.7] table [x=BinCenter, y=Count, col sep=space] {plot_data/mm_4_histogram_z0.500.dat};
                \addlegendentry{Model Predictions}
                \draw [black, dashed, thick] (axis cs:1.270243,\pgfkeysvalueof{/pgfplots/ymin}) -- (axis cs:1.270243,\pgfkeysvalueof{/pgfplots/ymax});
                \addlegendimage{legend image code/.code={\draw[black, dashed, thick] (0cm,0cm) -- (0.6cm,0cm);}}
                \addlegendentry{Exact ($1.270$)}
                \draw [blue, thick] (axis cs:1.272023,\pgfkeysvalueof{/pgfplots/ymin}) -- (axis cs:1.272023,\pgfkeysvalueof{/pgfplots/ymax});
                \addlegendimage{legend image code/.code={\draw[blue, thick] (0cm,0cm) -- (0.6cm,0cm);}}
                \addlegendentry{Mean ($1.272$)}
            \end{axis}
        \end{tikzpicture}
        %\caption{}
        %\label{fig:mm_4_hist}
    \end{subfigure}

    \caption{NN-predicted reduced four-point function $\GG(z)$ of the correlator $\langle \phi_{1,2}\phi_{1,2}\phi_{1,2}\phi_{1,2}\rangle$ for the tricritical Ising model, $\MM(4,5)$. These results were based on 51 independent runs filtered at training loss below $10^{-5}$ and are compared to the exact correlator \eqref{minaa0}-\eqref{minac} with $m = 4$. The NN prediction at $z=0.5$ is $1.272 \pm 0.007$.}
    \label{fig:mm_4_summary}
\end{figure}
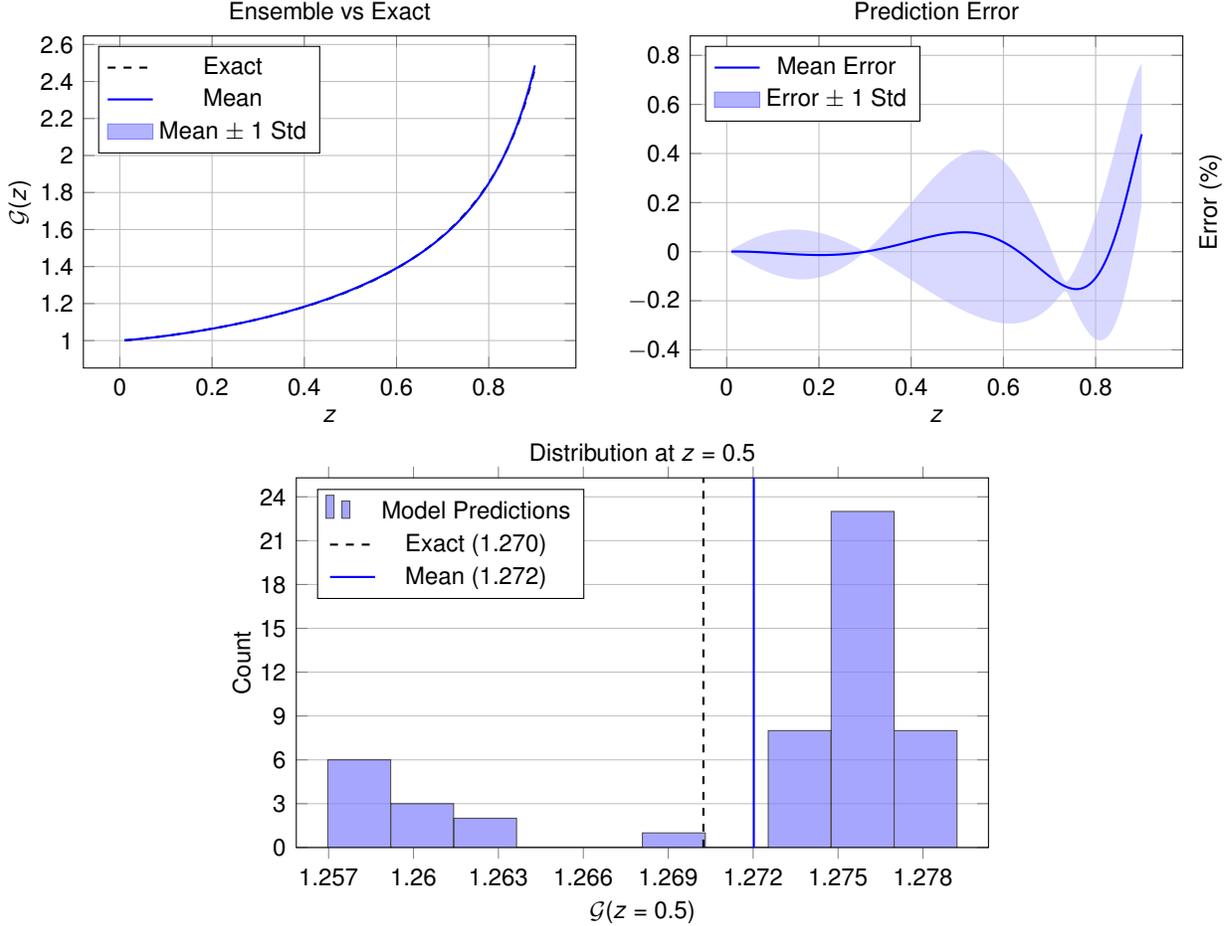

The results of the NN predictions for the tricritical Ising model are collected in the plots of Fig.\ \ref{fig:mm_4_summary}. After filtering 100 independent runs to a training loss below $10^{-5}$ we obtained 51 runs with MS training loss $(4.72\pm 3.17) \times 10^{-6}$. The mean prediction agrees with the expected analytical result within a relative error that is again below 1\%.

\subsubsection{\texorpdfstring{$\MM(7,8)$}{M(7,8)} and \texorpdfstring{$\MM(13,14)$}{M(13,14)}}

Analogous results have been collected for the minimal models $\MM(7,8)$ and $\MM(13,14)$. The relevant plots can be found in Appendix \ref{app:moreminimal}. We have also checked correlators for which more than two Virasoro primaries contribute, such as the spin-field correlator in the tricritical Ising model, and found that our method reconstructs the correlator with excellent accuracy.

\subsubsection{\texorpdfstring{$\MM(2,5)$}{M(2,5)}: Lee--Yang model}

\begin{figure}[H]
    \centering
    % Subfigure 1: Comparison
    \begin{subfigure}[b]{0.49\textwidth}
        \centering
        \begin{tikzpicture}
            \begin{axis}[
                width=\linewidth, height=6cm,
                xlabel={$z$},
                ylabel={$\mathcal{G}(z)$},
                title={Ensemble vs Exact},
                grid=major,
                ytick distance=2,
                legend pos=south east,
            ]
                % Exact Solution
                \addplot [black, dashed, thick] table [x=z, y=Exact, col sep=space] {plot_data/lee-yang_whole_correlator_ensemble_comparison.dat};
                \addlegendentry{Exact}

                % Ensemble Mean
                \addplot [blue, thick] table [x=z, y=Mean, col sep=space] {plot_data/lee-yang_whole_correlator_ensemble_comparison.dat};
                \addlegendentry{Mean}

                % Uncertainty Band
                \addplot [forget plot, name path=upper, draw=none] table [x=z, y=Mean_plus_Std, col sep=space] {plot_data/lee-yang_whole_correlator_ensemble_comparison.dat};
                \addplot [forget plot, name path=lower, draw=none] table [x=z, y=Mean_minus_Std, col sep=space] {plot_data/lee-yang_whole_correlator_ensemble_comparison.dat};
                \addplot [forget plot, fill=blue!30, fill opacity=0.5, draw=none] fill between [of=upper and lower];
                \addlegendimage{legend image code/.code={\fill[blue!30, draw=blue!50] (0cm,-0.1cm) rectangle (0.6cm,0.1cm);}}
                \addlegendentry{Mean $\pm$ 1 Std}
            \end{axis}
        \end{tikzpicture}
        %\caption{}
        %\label{fig:lee-yang_whole_correlator_comparison}
    \end{subfigure}
    \hfill
    %
    % Subfigure 2: Percentage Error
    \begin{subfigure}[b]{0.49\textwidth}
        \centering
        \begin{tikzpicture}
            \begin{axis}[
                width=\linewidth, height=6cm,
                xlabel={$z$},
                ylabel={Error (\%)},
                ylabel style={at={(axis description cs:1.10,0.5)}, anchor=south},
	                title={Prediction Error},
	                grid=major,
	                ytick distance=1,
	                legend style={xshift=45pt},
	                legend pos=south west,
	            ]
                % Percentage Error
                \addplot [blue, thick] table [x=z, y=PctError, col sep=space] {plot_data/lee-yang_whole_correlator_percentage_error.dat};
                \addlegendentry{Mean Error}

                % Uncertainty Band
                \addplot [forget plot, name path=upper, draw=none] table [x=z, y=PctError_plus_PctStd, col sep=space] {plot_data/lee-yang_whole_correlator_percentage_error.dat};
                \addplot [forget plot, name path=lower, draw=none] table [x=z, y=PctError_minus_PctStd, col sep=space] {plot_data/lee-yang_whole_correlator_percentage_error.dat};
                \addplot [forget plot, fill=blue!30, fill opacity=0.5, draw=none] fill between [of=upper and lower];
                \addlegendimage{legend image code/.code={\fill[blue!30, draw=blue!50] (0cm,-0.1cm) rectangle (0.6cm,0.1cm);}}
                \addlegendentry{Error $\pm$ 1 Std}
            \end{axis}
        \end{tikzpicture}
        %\caption{}
        %\label{fig:lee-yang_whole_correlator_error}
    \end{subfigure}
    %
    %\vspace{1em}
    %
    % Subfigure 3: Histogram
    \begin{subfigure}[b]{0.65\textwidth}
        \centering
        \begin{tikzpicture}
            \begin{axis}[
                width=\linewidth, height=6.5cm,
                xlabel={$\mathcal{G}(z=0.5)$},
                ylabel={Count},
                title={Distribution at $z=0.5$},
                ybar,
                bar width=0.004539,
                xmin=-3.228338, xmax=-3.130584,
                ymin=0,
                ymajorgrids=true,
                xmajorgrids=false,
                ytick distance=2,
                legend pos=north west,
                scaled ticks=false,
                xticklabel style={
                    /pgf/number format/fixed,
                    /pgf/number format/precision=3,
                    font=\sansmath\sffamily\footnotesize,
                }
            ]
                \addplot [fill=blue!50, draw=black, opacity=0.7] table [x=BinCenter, y=Count, col sep=space] {plot_data/lee-yang_whole_correlator_histogram_z0.500.dat};
                \addlegendentry{Model Predictions}
                \draw [black, dashed, thick] (axis cs:-3.180222,\pgfkeysvalueof{/pgfplots/ymin}) -- (axis cs:-3.180222,\pgfkeysvalueof{/pgfplots/ymax});
                \addlegendimage{legend image code/.code={\draw[black, dashed, thick] (0cm,0cm) -- (0.6cm,0cm);}}
                \addlegendentry{Exact ($-3.180$)}
                \draw [blue, thick] (axis cs:-3.171302,\pgfkeysvalueof{/pgfplots/ymin}) -- (axis cs:-3.171302,\pgfkeysvalueof{/pgfplots/ymax});
                \addlegendimage{legend image code/.code={\draw[blue, thick] (0cm,0cm) -- (0.6cm,0cm);}}
                \addlegendentry{Mean ($-3.171$)}
            \end{axis}
        \end{tikzpicture}
        %\caption{}
        %\label{fig:lee-yang_whole_correlator_hist}
    \end{subfigure}

    \caption{NN-predicted reduced four-point function $\GG(z)$ of the correlator $\langle \phi_{1,2}\phi_{1,2}\phi_{1,2}\phi_{1,2}\rangle$ for the Lee--Yang model, $\MM(2,5)$. The anchor point is taken to be $z_{0}=0.7$. These results are compared to the exact correlator \eqref{minad}-\eqref{minaf}. Our predictions at $z=0.5$ is $-3.171\pm 0.017$.}
    \label{fig:lee-yang_whole_correlator_summary}
\end{figure}
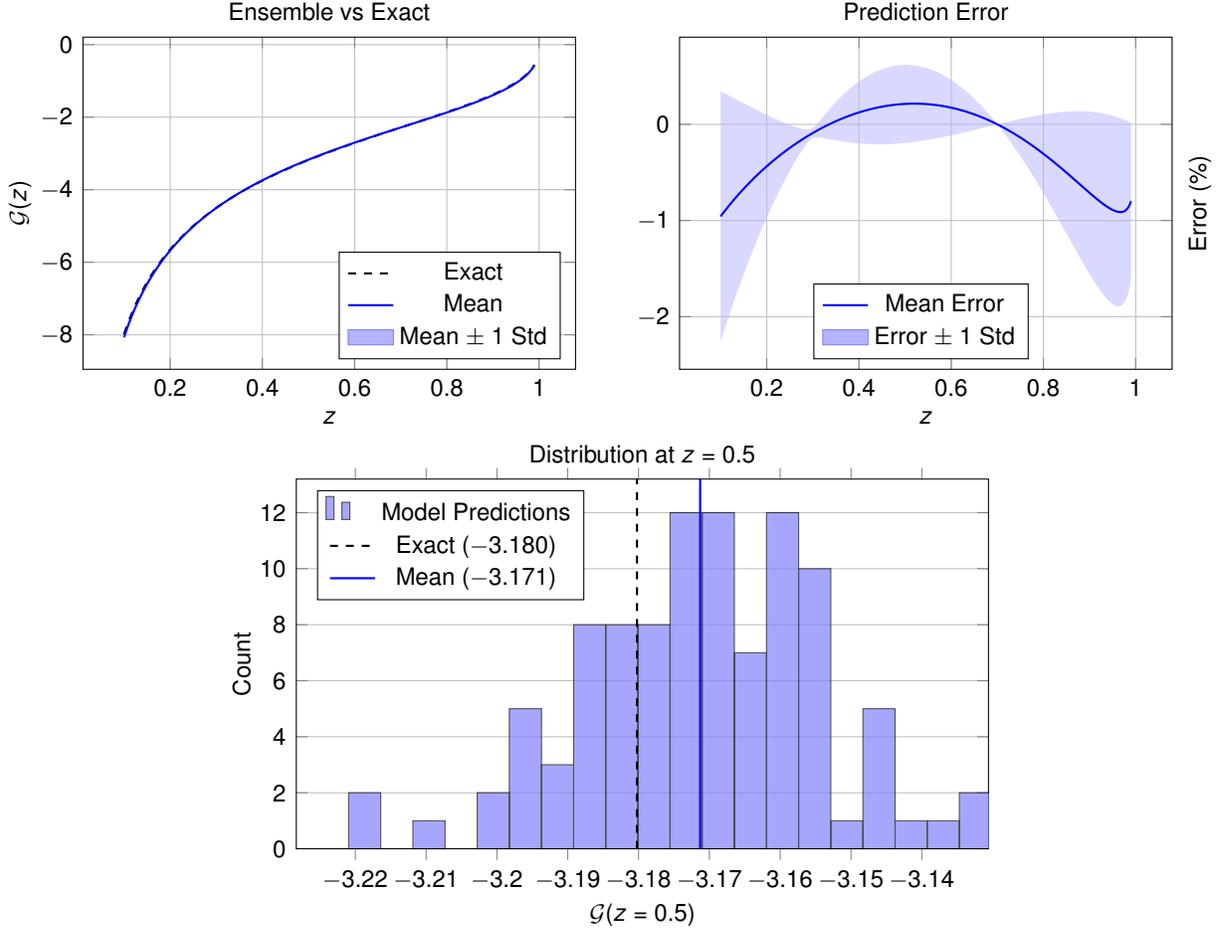

The results for the non-unitary Lee--Yang minimal model, $\MM(2,5)$, are summarised in Fig.\ \ref{fig:lee-yang_whole_correlator_summary}. Here we set $L(z)=(1-z)^{\frac{4}{5}}$ and parametrise the remainder $H(z)$ as in Eq.~\eqref{minbaextra}, with an anchor point at $z_0=0.7$. Since $\Delta_\phi<0$, the correlator diverges as $z\to 0$; our choice of ansatz and anchor is therefore designed to bias the network toward a representation controlled by the crossed ($t$-)channel expansion. Over 100 independent runs, the corresponding MS training loss was $(2.63\pm2.35)\times 10^{-7}$.

\section{2d and 3d Ising models: more correlators}
\label{ising}

The Ising model---in both 2d and 3d---represents the most physically significant application treated in this paper. In 2d the exact correlators are available and serve as precision benchmarks; in 3d no closed-form expressions exist, so the anchored neural network predictions become genuinely predictive. We have already discussed the $\langle \sigma \sigma \sigma \sigma \rangle$ correlator in the 2d Ising model in Section \ref{2dising1}. Besides the extension to the 3d Ising model, the novelty of this section is that it provides an instructive example of a mixed-correlator system. 

\subsection{Crossing equations for mixed systems}

Consider four scalar primary operators $\phi_i,\phi_j,\phi_k,\phi_l$ with corresponding scaling dimensions $\Delta_i,\Delta_j,\Delta_k$, $\Delta_l$. On the line, we adopt the normalisation convention
\begin{equation}
    \label{isingaa}
    \GG^{ijkl}(x_1,x_2,x_3,x_4) =
    \langle \phi_i(x_1)\phi_j(x_2)\phi_k(x_3)\phi_l(x_4)\rangle = \frac{1}{x_{12}^{\Delta_i+\Delta_j}\,x_{34}^{\Delta_k+\Delta_l}}\left(\frac{x_{24}}{x_{14}}\right)^{\Delta_{ij}}\left(\frac{x_{14}}{x_{13}}\right)^{\Delta_{kl}}\,\GG(z)
    ~,
\end{equation}
where $\Delta_{ji}=\Delta_j-\Delta_i$. When the operators are identical operators the crossing equation reduces to \eqref{introac}. For the mixed $\sigma$-$\epsilon$ system, the $s$-$t$ channel crossing relates two distinct reduced correlators
\begin{equation}
    \label{isingac}
    \GG^{\sigma\epsilon\epsilon\sigma}(z) = \frac{z^{\Delta_\sigma+\Delta_\epsilon}}{(1-z)^{2\Delta_\epsilon}}\,\GG^{\sigma\sigma\epsilon\epsilon}(1-z)\,.
\end{equation}
Since the left- and right-hand sides involve different functions, the NN implementation employs two independent networks (or one network with two scalar outputs), each with its own gap prefactor and anchor constraint. 

Alternatively, notice that by defining the linear combinations
\begin{equation}
    \label{isingadaa}
    \GG_\pm(z) = z^{-(\Delta_\sigma +\Delta_\epsilon)}\GG^{\sigma \epsilon\epsilon\sigma}(z) \pm (1-z)^{-2\Delta_\epsilon} \GG^{\sigma\sigma \epsilon \epsilon}(1-z)\,,
\end{equation}
Eq.\ \eqref{isingac} can be reformulated as two independent crossing equations,
\begin{equation}
    \label{isigngadab}
    \GG_\pm (z) = \pm \GG_\pm (1-z)\,,
\end{equation}
which one could attempt to solve with the independent optimisation of two anchored neural networks. In what follows, we will not adopt this approach. We will use, instead, NNs to directly parametrise the correlators $\GG^{\sigma \epsilon\epsilon\sigma}(z)$, $\GG^{\sigma\sigma \epsilon \epsilon}(z)$, and solve the original crossing equation \eqref{isingac}. We exhibit the performance of this approach in the 2d Ising model. Anchored neural networks optimised in this manner can produce in a single run an approximation of two mixed correlators at the same level of accuracy as the one obtained for correlators of identical operators.

\subsection{2d Ising model}
\label{ising:2d}

The 2d Ising CFT has $\Delta_\sigma=1/8$ and $\Delta_\epsilon=1$. The exact line-restricted correlators read
\begin{gather}
    \label{isingba}
    \GG^{\sigma\sigma\sigma\sigma}(z) = \frac{1}{(1-z)^{1/4}}\,,\qquad
    \GG^{\epsilon\epsilon\epsilon\epsilon}(z) = \Bigl(1-z+\frac{z}{1-z}\Bigr)^{\!2}\,,
    \\
    \label{isingbamix}
    \GG^{\sigma\sigma\epsilon\epsilon}(z) = \frac{(z-2)^2}{4(1-z)}\,,\qquad
    \GG^{\sigma\epsilon\epsilon\sigma}(z) = \frac{z^{1/8}(z+1)^2}{4(z-1)^2}\,.    
\end{gather}

\subsubsection{Identical correlators}

\begin{figure}[H]
    \centering
    % Subfigure 1: Comparison
    \begin{subfigure}[b]{0.49\textwidth}
        \centering
        \begin{tikzpicture}
            \begin{axis}[
                width=\linewidth, height=6cm,
                xlabel={$z$},
                ylabel={$\mathcal{G}(z)$},
                title={Ensemble vs Exact},
                grid=major,
                legend pos=north west,
            ]
                % Exact Solution
                \addplot [black, dashed, thick] table [x=z, y=Exact, col sep=space] {plot_data/ising_eps_ensemble_comparison.dat};
                \addlegendentry{Exact}

                % Ensemble Mean
                \addplot [blue, thick] table [x=z, y=Mean, col sep=space] {plot_data/ising_eps_ensemble_comparison.dat};
                \addlegendentry{Mean}

                % Uncertainty Band
                \addplot [forget plot, name path=upper, draw=none] table [x=z, y=Mean_plus_Std, col sep=space] {plot_data/ising_eps_ensemble_comparison.dat};
                \addplot [forget plot, name path=lower, draw=none] table [x=z, y=Mean_minus_Std, col sep=space] {plot_data/ising_eps_ensemble_comparison.dat};
                \addplot [forget plot, fill=blue!30, fill opacity=0.5, draw=none] fill between [of=upper and lower];
                \addlegendimage{legend image code/.code={\fill[blue!30, draw=blue!50] (0cm,-0.1cm) rectangle (0.6cm,0.1cm);}}
                \addlegendentry{Mean $\pm$ 1 Std}
            \end{axis}
        \end{tikzpicture}
        %\caption{}
        %\label{fig:ising_eps_comparison}
    \end{subfigure}
    \hfill
    %
    % Subfigure 2: Percentage Error
    \begin{subfigure}[b]{0.49\textwidth}
        \centering
        \begin{tikzpicture}
            \begin{axis}[
                width=\linewidth, height=6cm,
                xlabel={$z$},
                ylabel={Error (\%)},
                ylabel style={at={(axis description cs:1.10,0.5)}, anchor=south},
                title={Prediction Error},
                grid=major,
                ytick distance=0.1,
                legend pos=north west,
            ]
                % Percentage Error
                \addplot [blue, thick] table [x=z, y=PctError, col sep=space] {plot_data/ising_eps_percentage_error.dat};
                \addlegendentry{Mean Error}

                % Uncertainty Band
                \addplot [forget plot, name path=upper, draw=none] table [x=z, y=PctError_plus_PctStd, col sep=space] {plot_data/ising_eps_percentage_error.dat};
                \addplot [forget plot, name path=lower, draw=none] table [x=z, y=PctError_minus_PctStd, col sep=space] {plot_data/ising_eps_percentage_error.dat};
                \addplot [forget plot, fill=blue!30, fill opacity=0.5, draw=none] fill between [of=upper and lower];
                \addlegendimage{legend image code/.code={\fill[blue!30, draw=blue!50] (0cm,-0.1cm) rectangle (0.6cm,0.1cm);}}
                \addlegendentry{Error $\pm$ 1 Std}
            \end{axis}
        \end{tikzpicture}
        %\caption{}
        %\label{fig:ising_eps_error}
    \end{subfigure}
    %
    %\vspace{1em}
    %
    % Subfigure 3: Histogram
    \begin{subfigure}[b]{0.65\textwidth}
        \centering
        \begin{tikzpicture}
            \begin{axis}[
                width=\linewidth, height=6.5cm,
                xlabel={$\mathcal{G}(z=0.5)$},
                ylabel={Count},
                title={Distribution at $z=0.5$},
                ybar,
                bar width=0.001380,
                xmin=2.241421, xmax=2.271772,
                ymin=0,
                ymajorgrids=true,
                xmajorgrids=false,
                legend pos=north east,
                scaled ticks=false,
                ytick distance=2,
                xticklabel style={
                    /pgf/number format/fixed,
                    /pgf/number format/precision=3,
                    font=\sansmath\sffamily\footnotesize,
                }
            ]
                \addplot [fill=blue!50, draw=black, opacity=0.7] table [x=BinCenter, y=Count, col sep=space] {plot_data/ising_eps_histogram_z0.500.dat};
                \addlegendentry{Model Predictions}
                \draw [black, dashed, thick] (axis cs:2.250401,\pgfkeysvalueof{/pgfplots/ymin}) -- (axis cs:2.250401,\pgfkeysvalueof{/pgfplots/ymax});
                \addlegendimage{legend image code/.code={\draw[black, dashed, thick] (0cm,0cm) -- (0.6cm,0cm);}}
                \addlegendentry{Exact ($2.250$)}
                \draw [blue, thick] (axis cs:2.253236,\pgfkeysvalueof{/pgfplots/ymin}) -- (axis cs:2.253236,\pgfkeysvalueof{/pgfplots/ymax});
                \addlegendimage{legend image code/.code={\draw[blue, thick] (0cm,0cm) -- (0.6cm,0cm);}}
                \addlegendentry{Mean ($2.253$)}
            \end{axis}
        \end{tikzpicture}
        %\caption{}
        %\label{fig:ising_eps_hist}
    \end{subfigure}

    \caption{NN-predicted reduced four-point function $\GG(z)$ of the correlator $\langle \epsilon\epsilon\epsilon\epsilon\rangle$ for the 2d Ising model.  These results are compared to the exact correlator given in \eqref{isingba}. The NN prediction at $z=0.5$ is $2.253\pm 0.007$. }
    \label{fig:ising_eps_summary}
\end{figure}
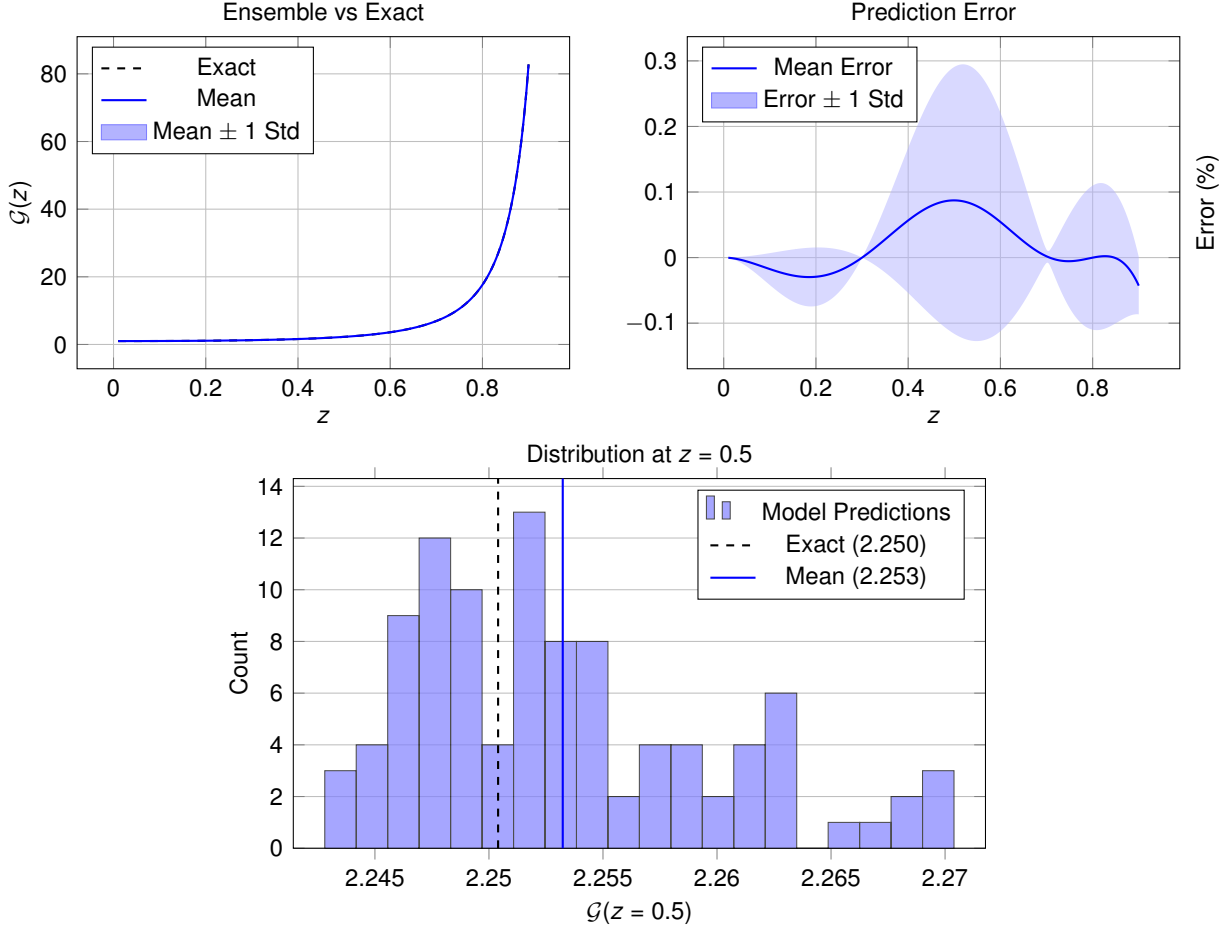

We have already discussed the case of the $\langle \sigma \sigma \sigma \sigma\rangle$ correlator in Section \ref{2dising1} with corresponding results appearing in Fig.\ \ref{fig:ising_summary}. For the $\langle \epsilon \epsilon\epsilon\epsilon\rangle$ correlator the anchored NN predictions are summarised in Fig.\ \ref{fig:ising_eps_summary}. Over 100 independent runs, the corresponding MS training loss was $(7.10\pm 5.80)\times10^{-8}$. The mean prediction agrees with the analytic result within a relative error well below 1\%.

\subsubsection{Mixed correlators}

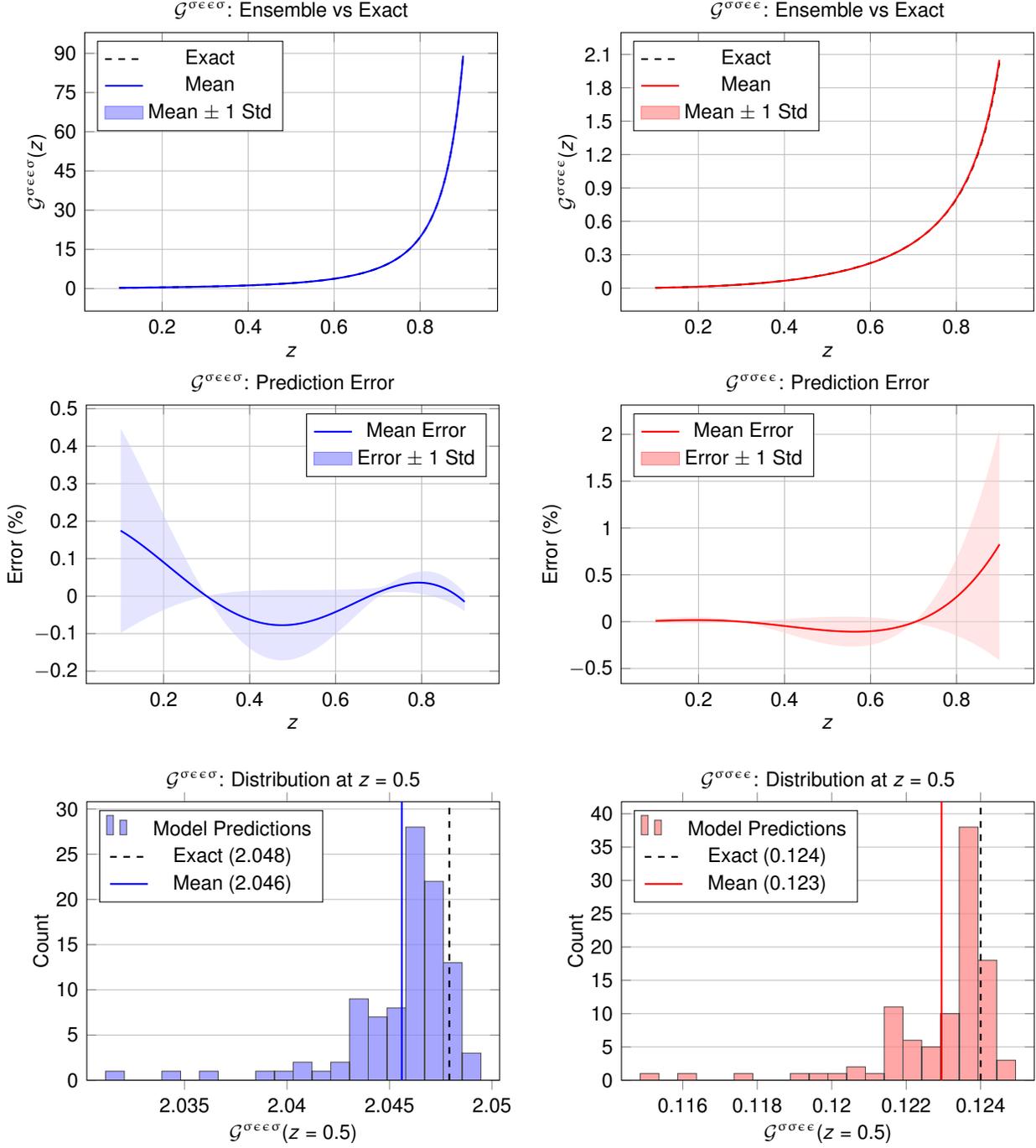
\begin{figure}[ht!]
    \centering

    %% Subfigure: Comparison for f
    \begin{subfigure}[b]{0.49\textwidth}
        \centering
        \begin{tikzpicture}
            \begin{axis}[
                width=\linewidth, height=6cm,
                xlabel={$z$},
                ylabel={$\mathcal{G}^{\sigma\epsilon\epsilon\sigma}(z)$},
                title={$\mathcal{G}^{\sigma\epsilon\epsilon\sigma}$: Ensemble vs Exact},
                grid=major,
                ytick distance=15,
                legend pos=north west,
            ]
                % Exact Solution
                \addplot [black, dashed, thick] table [x=z, y=Exact, col sep=space] {plot_data/ising_mixed_correlator_ensemble_comparison_f.dat};
                \addlegendentry{Exact}

                % Ensemble Mean
                \addplot [blue, thick] table [x=z, y=Mean, col sep=space] {plot_data/ising_mixed_correlator_ensemble_comparison_f.dat};
                \addlegendentry{Mean}

                % Uncertainty Band
                \addplot [forget plot, name path=upper, draw=none] table [x=z, y=Mean_plus_Std, col sep=space] {plot_data/ising_mixed_correlator_ensemble_comparison_f.dat};
                \addplot [forget plot, name path=lower, draw=none] table [x=z, y=Mean_minus_Std, col sep=space] {plot_data/ising_mixed_correlator_ensemble_comparison_f.dat};
                \addplot [forget plot, fill=blue!30, fill opacity=0.5, draw=none] fill between [of=upper and lower];
                \addlegendimage{legend image code/.code={\fill[blue!30, draw=blue!50] (0cm,-0.1cm) rectangle (0.6cm,0.1cm);}}
                \addlegendentry{Mean $\pm$ 1 Std}
            \end{axis}
        \end{tikzpicture}
    \end{subfigure}
    \hfill
    %% Subfigure: Comparison for g
    \begin{subfigure}[b]{0.49\textwidth}
        \centering
        \begin{tikzpicture}
            \begin{axis}[
                width=\linewidth, height=6cm,
                xlabel={$z$},
                ylabel={$\mathcal{G}^{\sigma\sigma\epsilon\epsilon}(z)$},
                title={$\mathcal{G}^{\sigma\sigma\epsilon\epsilon}$: Ensemble vs Exact},
                grid=major,
                ytick distance=0.3,
                legend pos=north west,
            ]
                % Exact Solution
                \addplot [black, dashed, thick] table [x=z, y=Exact, col sep=space] {plot_data/ising_mixed_correlator_ensemble_comparison_g.dat};
                \addlegendentry{Exact}

                % Ensemble Mean
                \addplot [red, thick] table [x=z, y=Mean, col sep=space] {plot_data/ising_mixed_correlator_ensemble_comparison_g.dat};
                \addlegendentry{Mean}

                % Uncertainty Band
                \addplot [forget plot, name path=upper, draw=none] table [x=z, y=Mean_plus_Std, col sep=space] {plot_data/ising_mixed_correlator_ensemble_comparison_g.dat};
                \addplot [forget plot, name path=lower, draw=none] table [x=z, y=Mean_minus_Std, col sep=space] {plot_data/ising_mixed_correlator_ensemble_comparison_g.dat};
                \addplot [forget plot, fill=red!30, fill opacity=0.5, draw=none] fill between [of=upper and lower];
                \addlegendimage{legend image code/.code={\fill[red!30, draw=red!50] (0cm,-0.1cm) rectangle (0.6cm,0.1cm);}}
                \addlegendentry{Mean $\pm$ 1 Std}
            \end{axis}
        \end{tikzpicture}
    \end{subfigure}
    \hspace{-0.1mm}
    \vspace{1em}
    %% Subfigure: Percentage Error for f
    \hspace*{-4mm}
    \begin{subfigure}[b]{0.49\textwidth}
        \centering
        \begin{tikzpicture}
            \begin{axis}[
                width=\linewidth, height=6cm,
                xlabel={$z$},
                ylabel={Error (\%)},
                title={$\mathcal{G}^{\sigma\epsilon\epsilon\sigma}$: Prediction Error},
                grid=major,
                ytick distance=0.1,
                legend pos=north east,
            ]
                % Percentage Error
                \addplot [blue, thick] table [x=z, y=PctError, col sep=space] {plot_data/ising_mixed_correlator_percentage_error_f.dat};
                \addlegendentry{Mean Error}

                % Uncertainty Band
                \addplot [forget plot, name path=upper, draw=none] table [x=z, y=PctError_plus_PctStd, col sep=space] {plot_data/ising_mixed_correlator_percentage_error_f.dat};
                \addplot [forget plot, name path=lower, draw=none] table [x=z, y=PctError_minus_PctStd, col sep=space] {plot_data/ising_mixed_correlator_percentage_error_f.dat};
                \addplot [forget plot, fill=blue!20, fill opacity=0.5, draw=none] fill between [of=upper and lower];
                \addlegendimage{legend image code/.code={\fill[blue!30, draw=blue!50] (0cm,-0.1cm) rectangle (0.6cm,0.1cm);}}
                \addlegendentry{Error $\pm$ 1 Std}
            \end{axis}
        \end{tikzpicture}
    \end{subfigure}
    %% Subfigure: Percentage Error for g
    \hspace*{1mm}
    \begin{subfigure}[b]{0.49\textwidth}
        \centering
        \begin{tikzpicture}
            \begin{axis}[
                width=\linewidth, height=6cm,
                xlabel={$z$},
                ylabel={Error (\%)},
                title={$\mathcal{G}^{\sigma\sigma\epsilon\epsilon}$: Prediction Error},
                grid=major,
                ytick distance=0.5,
                legend pos=north west,
            ]
                % Percentage Error
                \addplot [red, thick] table [x=z, y=PctError, col sep=space] {plot_data/ising_mixed_correlator_percentage_error_g.dat};
                \addlegendentry{Mean Error}

                % Uncertainty Band
                \addplot [forget plot, name path=upper, draw=none] table [x=z, y=PctError_plus_PctStd, col sep=space] {plot_data/ising_mixed_correlator_percentage_error_g.dat};
                \addplot [forget plot, name path=lower, draw=none] table [x=z, y=PctError_minus_PctStd, col sep=space] {plot_data/ising_mixed_correlator_percentage_error_g.dat};
                \addplot [forget plot, fill=red!20, fill opacity=0.5, draw=none] fill between [of=upper and lower];
                \addlegendimage{legend image code/.code={\fill[red!30, draw=red!50] (0cm,-0.1cm) rectangle (0.6cm,0.1cm);}}
                \addlegendentry{Error $\pm$ 1 Std}
            \end{axis}
        \end{tikzpicture}
    \end{subfigure}
    \vspace{1em}
    %% Subfigure: Histogram for f
    \hspace*{1mm}
    \begin{subfigure}[b]{0.49\textwidth}
        \centering
        \begin{tikzpicture}
            \begin{axis}[
                width=\linewidth, height=6cm,
                xlabel={$\mathcal{G}^{\sigma\epsilon\epsilon\sigma}(z=0.5)$},
                ylabel={Count},
                title={$\mathcal{G}^{\sigma\epsilon\epsilon\sigma}$: Distribution at $z=0.5$},
                ybar,
                bar width=0.000915,
                xmin=2.030236, xmax=2.050371,
                ymin=0,
                ymajorgrids=true,
                xmajorgrids=false,
                ytick distance=5,
                legend pos=north west,
                scaled ticks=false,
                xticklabel style={
                    /pgf/number format/fixed,
                    /pgf/number format/precision=3,
                    font=\sansmath\sffamily\footnotesize,
                }
            ]
                \addplot [fill=blue!50, draw=black, opacity=0.7] table [x=BinCenter, y=Count, col sep=space] {plot_data/ising_mixed_correlator_histogram_f_z0.499.dat};
                \addlegendentry{Model Predictions}
                \draw [black, dashed, thick] (axis cs:2.047917,\pgfkeysvalueof{/pgfplots/ymin}) -- (axis cs:2.047917,\pgfkeysvalueof{/pgfplots/ymax});
                \addlegendimage{legend image code/.code={\draw[black, dashed, thick] (0cm,0cm) -- (0.6cm,0cm);}}
                \addlegendentry{Exact ($2.048$)}
                \draw [blue, thick] (axis cs:2.045599,\pgfkeysvalueof{/pgfplots/ymin}) -- (axis cs:2.045599,\pgfkeysvalueof{/pgfplots/ymax});
                \addlegendimage{legend image code/.code={\draw[blue, thick] (0cm,0cm) -- (0.6cm,0cm);}}
                \addlegendentry{Mean ($2.046$)}
            \end{axis}
        \end{tikzpicture}
    \end{subfigure}
    %% Subfigure: Histogram for g
    \begin{subfigure}[b]{0.49\textwidth}
        \centering
        \begin{tikzpicture}
            \begin{axis}[
                width=\linewidth, height=6cm,
                xlabel={$\mathcal{G}^{\sigma\sigma\epsilon\epsilon}(z=0.5)$},
                ylabel={Count},
                title={$\mathcal{G}^{\sigma\sigma\epsilon\epsilon}$: Distribution at $z=0.5$},
                ybar,
                bar width=0.000504,
                xmin=0.114352, xmax=0.125441,
                ymin=0,
                ymajorgrids=true,
                xmajorgrids=false,
                ytick distance=5,
                legend pos=north west,
                scaled ticks=false,
                xticklabel style={
                    /pgf/number format/fixed,
                    /pgf/number format/precision=3,
                    font=\sansmath\sffamily\footnotesize,
                }
            ]
                \addplot [fill=red!50, draw=black, opacity=0.7] table [x=BinCenter, y=Count, col sep=space] {plot_data/ising_mixed_correlator_histogram_g_z0.499.dat};
                \addlegendentry{Model Predictions}
                \draw [black, dashed, thick] (axis cs:0.124000,\pgfkeysvalueof{/pgfplots/ymin}) -- (axis cs:0.124000,\pgfkeysvalueof{/pgfplots/ymax});
                \addlegendimage{legend image code/.code={\draw[black, dashed, thick] (0cm,0cm) -- (0.6cm,0cm);}}
                \addlegendentry{Exact ($0.124$)}
                \draw [red, thick] (axis cs:0.122949,\pgfkeysvalueof{/pgfplots/ymin}) -- (axis cs:0.122949,\pgfkeysvalueof{/pgfplots/ymax});
                \addlegendimage{legend image code/.code={\draw[red, thick] (0cm,0cm) -- (0.6cm,0cm);}}
                \addlegendentry{Mean ($0.123$)}
            \end{axis}
        \end{tikzpicture}
        \hspace*{-2.9mm}
    \end{subfigure}

    \caption{NN-predicted mixed correlators $\GG^{\sigma \epsilon\epsilon\sigma}(z)$, $\GG^{\sigma\sigma \epsilon \epsilon}(z)$ in the 2d Ising model that obey the crossing equation \eqref{isingac}. The left column (in blue) contains the data for $\GG^{\sigma \epsilon\epsilon\sigma}(z)$. The NN prediction at $z=0.5$ is $2.046\pm 0.003$ for $\GG^{\sigma \epsilon\epsilon\sigma}(z)$. The right column (in red) contains the data for $\GG^{\sigma\sigma \epsilon \epsilon}(z)$. The NN prediction at $z=0.5$ is $0.123\pm 0.002$ for $\GG^{\sigma\sigma \epsilon \epsilon}(z)$. Both results are compared to the exact correlators given in \eqref{isingbamix}.}
    \label{fig:ising_mixed_correlator_summary}
\end{figure}

The two mixed correlators $\GG^{\sigma \epsilon\epsilon\sigma}(z)$, $\GG^{\sigma\sigma \epsilon \epsilon}(z)$ are coupled by the crossing Eq.\ \eqref{isingac}. Each of these functions is parametrised by a separate network that receives its own anchor value and gap prefactor. The results of the corresponding optimisation are summarised in Fig.~\ref{fig:ising_mixed_correlator_summary}. Over 100 independent runs, the corresponding MS of the training loss was $(7.68\pm 5.22)\times 10^{-9}$. We observe a similar quality of matching to exact results as in single-field correlators with relative prediction errors below 1\%.

\subsection{3d Ising model}
\label{ising:3d}

The 3d Ising CFT is a more interesting case. High-precision scaling dimensions have been determined by the numerical conformal bootstrap \cite{Kos:2014bka,Kos:2016ysd}
\begin{equation}
    \label{isingca}
    \Delta_\sigma = 0.518148806(24)\,,\qquad \Delta_\epsilon = 1.41262528(29)\, ,
\end{equation}
but no exact analytic expressions for the four-point functions are known. As a result, this is a context where our approach can provide genuinely new numerical data. 

In what follows, we consider the $\langle \sigma \sigma \sigma \sigma \rangle$ and $\langle \epsilon \epsilon \epsilon \epsilon\rangle$ correlators. The crossing equations are set up precisely as in the 2d case (with the external scaling dimensions replaced by their 3d values), but the anchor values cannot be obtained now from an analytic solution. We approximate them using the fast convergence of the OPE expansion and the OPE data obtained by bootstrap methods in \cite{Simmons-Duffin:2016wlq}.

\subsubsection{\texorpdfstring{$\langle\sigma\sigma\sigma\sigma\rangle$}{<ssss>}}
\label{sec:ising3d_sigma}
\begin{figure}[ht]
    \centering
    % Subfigure 1: Comparison
    \begin{subfigure}[b]{0.49\textwidth}
        \centering
        \begin{tikzpicture}
            \begin{axis}[
                width=\linewidth, height=6cm,
                xlabel={$z$},
                ylabel={$\mathcal{G}(z)$},
                title={Ensemble vs Bootstrap},
                grid=major,
                ytick distance=0.3,
                legend pos=north west,
            ]
                % Exact Solution
                \addplot [black, dashed, thick] table [x=z, y=Bootstrap, col sep=space] {plot_data/sigma_four_pt_function_anchor_0.3_ensemble_comparison.dat};
                \addlegendentry{BT}

                % Ensemble Mean
                \addplot [blue, thick] table [x=z, y=Mean, col sep=space] {plot_data/sigma_four_pt_function_anchor_0.3_ensemble_comparison.dat};
                \addlegendentry{Mean}

                % Uncertainty Band
                \addplot [forget plot, name path=upper, draw=none] table [x=z, y=Mean_plus_Std, col sep=space] {plot_data/sigma_four_pt_function_anchor_0.3_ensemble_comparison.dat};
                \addplot [forget plot, name path=lower, draw=none] table [x=z, y=Mean_minus_Std, col sep=space] {plot_data/sigma_four_pt_function_anchor_0.3_ensemble_comparison.dat};
                \addplot [forget plot, fill=blue!30, fill opacity=0.5, draw=none] fill between [of=upper and lower];
                \addlegendimage{legend image code/.code={\fill[blue!30, draw=blue!50] (0cm,-0.1cm) rectangle (0.6cm,0.1cm);}}
                \addlegendentry{Mean $\pm$ 1 Std}
            \end{axis}
        \end{tikzpicture}
        %\caption{}
        %\label{fig:sigma_four_pt_function_anchor_0.3_comparison}
    \end{subfigure}
    \hfill
    %
    % Subfigure 2: Percentage Error
    \begin{subfigure}[b]{0.49\textwidth}
        \centering
        \begin{tikzpicture}
            \begin{axis}[
                width=\linewidth, height=6cm,
                xlabel={$z$},
                ylabel={Error (\%)},
                ylabel style={at={(axis description cs:1.10,0.5)}, anchor=south},
                title={Prediction Error},
                grid=major,
                ytick distance=0.02,
                legend pos=south west,
                yticklabel style={
                    /pgf/number format/fixed,
                    /pgf/number format/precision=2,
                    %/pgf/number format/fixed zerofill % To ensure it always shows .00
                },
            ]
                % Percentage Error
                \addplot [blue, thick] table [x=z, y=PctError, col sep=space] {plot_data/sigma_four_pt_function_anchor_0.3_percentage_error.dat};
                \addlegendentry{Mean Error}

                % Uncertainty Band
                \addplot [forget plot, name path=upper, draw=none] table [x=z, y=PctError_plus_PctStd, col sep=space] {plot_data/sigma_four_pt_function_anchor_0.3_percentage_error.dat};
                \addplot [forget plot, name path=lower, draw=none] table [x=z, y=PctError_minus_PctStd, col sep=space] {plot_data/sigma_four_pt_function_anchor_0.3_percentage_error.dat};
                \addplot [forget plot, fill=blue!30, fill opacity=0.5, draw=none] fill between [of=upper and lower];
                \addlegendimage{legend image code/.code={\fill[blue!30, draw=blue!50] (0cm,-0.1cm) rectangle (0.6cm,0.1cm);}}
                \addlegendentry{Error $\pm$ 1 Std}
            \end{axis}
        \end{tikzpicture}
        %\caption{}
        %\label{fig:sigma_four_pt_function_anchor_0.3_error}
    \end{subfigure}
    %
    %\vspace{1em}
    %
    % Subfigure 3: Histogram
    \begin{subfigure}[b]{0.65\textwidth}
        \centering
        \begin{tikzpicture}
            \begin{axis}[
                width=\linewidth, height=6.5cm,
                xlabel={$\mathcal{G}(z=0.5)$},
                ylabel={Count},
                title={Distribution at $z=0.5$},
                ybar,
                bar width=0.000197,
                xmin=1.763536, xmax=1.767875,
                ymin=0,
                ymajorgrids=true,
                xmajorgrids=false,
                ytick distance=5,
                legend pos=north east,
                scaled ticks=false,
                xticklabel style={
                    /pgf/number format/fixed,
                    /pgf/number format/precision=3,
                    font=\sansmath\sffamily\footnotesize,
                }
            ]
                \addplot [fill=blue!50, draw=black, opacity=0.7] table [x=BinCenter, y=Count, col sep=space] {plot_data/sigma_four_pt_function_anchor_0.3_histogram_z0.500.dat};
                \addlegendentry{Model Predictions}
                \draw [black, dashed, thick] (axis cs:1.766958,\pgfkeysvalueof{/pgfplots/ymin}) -- (axis cs:1.766958,\pgfkeysvalueof{/pgfplots/ymax});
                \addlegendimage{legend image code/.code={\draw[black, dashed, thick] (0cm,0cm) -- (0.6cm,0cm);}}
                \addlegendentry{BT ($1.76696$)}
                \draw [blue, thick] (axis cs:1.765066,\pgfkeysvalueof{/pgfplots/ymin}) -- (axis cs:1.765066,\pgfkeysvalueof{/pgfplots/ymax});
                \addlegendimage{legend image code/.code={\draw[blue, thick] (0cm,0cm) -- (0.6cm,0cm);}}
                \addlegendentry{Mean ($1.76507$)}
            \end{axis}
        \end{tikzpicture}
        %\caption{}
        %\label{fig:sigma_four_pt_function_anchor_0.3_hist}
    \end{subfigure}

	    \caption{NN-predicted correlator reduced four-point function $\GG(z)$ of the correlator $\langle \sigma \sigma \sigma \sigma\rangle$ in the 3d Ising model. The dashed black curve labelled BT (bootstrap truncation) is obtained by evaluating the $s$-channel OPE truncated to the operator content described in Section~\ref{sec:ising3d_sigma}, using bootstrap-extracted CFT data from \cite{Simmons-Duffin:2016wlq}. The anchor point at $z_0=0.3$ is fixed using this same BT approximation. The NN prediction at $z= 0.5$ is $1.76507\pm 0.00059$.}
    \label{fig:sigma_four_pt_function_anchor_0.3_summary}
\end{figure}
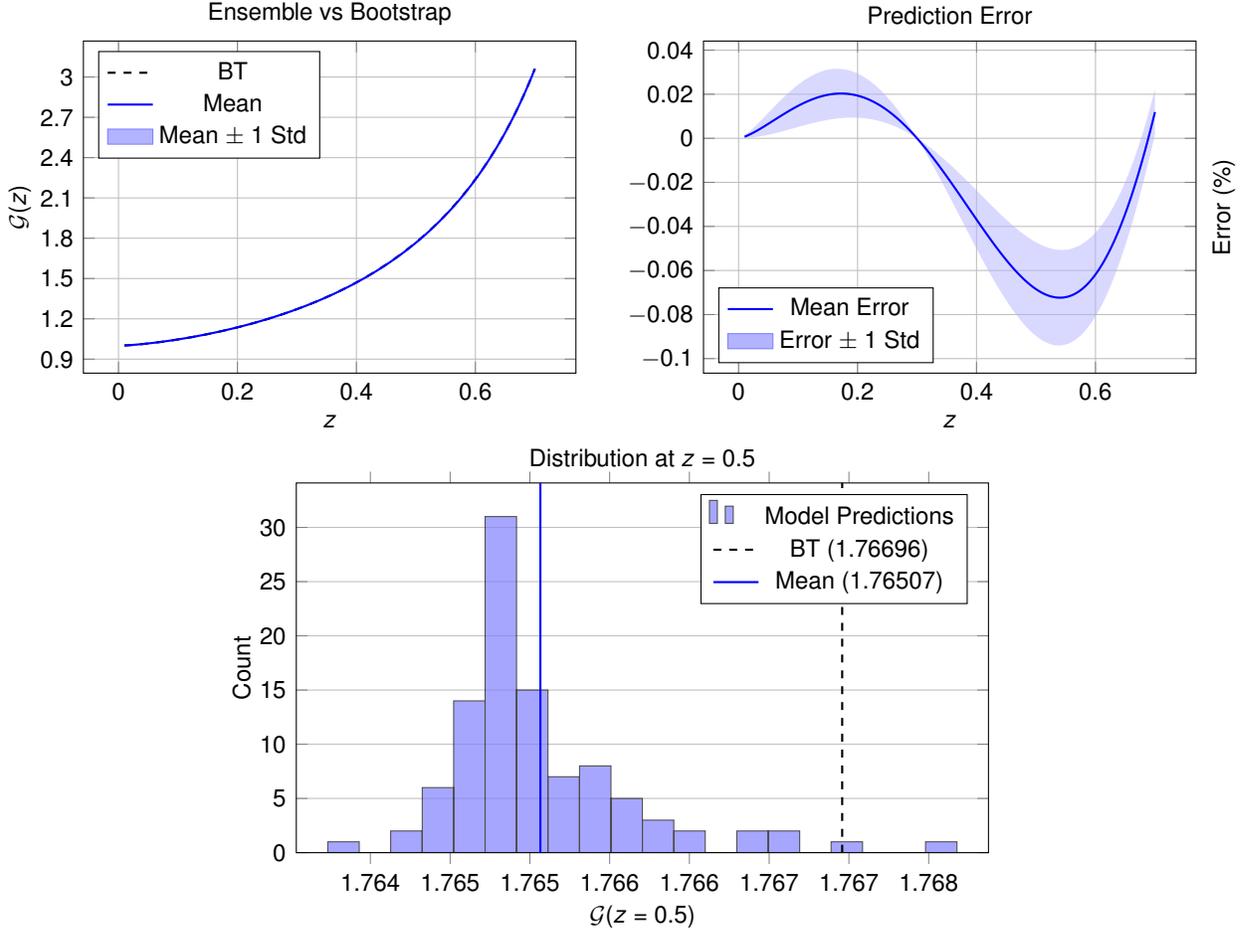

Results based on 100 independent runs for the $\langle\sigma\sigma\sigma\sigma\rangle$ correlator are summarised in Fig.\ \ref{fig:sigma_four_pt_function_anchor_0.3_summary}. The anchor point was set at $z_0=0.3$ and the corresponding anchor value was approximated with a truncation on the $s$-channel OPE importing from~\cite{Simmons-Duffin:2016wlq} the first four spin-$0$ operators, two spin-$2$ operators, two spin-$4$ operators, and one operator for each spin in the range $6$--$12$. By adding more operators we checked that the OPE convergence had already saturated with this ansatz. We set $L(z)=1$ and parametrise the remainder by
\begin{equation}
    H(z) = z^{\Delta_{\epsilon}}(1-z)^{-2\Delta_{\sigma}}\text{NN}_{\textbf{$\theta$}}(z).
\end{equation}
The predicted correlator is compared here with the truncated OPE expression, which is expected to fail sufficiently close to $z=1$. In Fig.\ \ref{fig:sigma_four_pt_function_anchor_0.3_summary} we compare up to $z=0.7$. The MS training loss for the results presented was $(1.25\pm 0.15)\times 10^{-7}$. 

\subsubsection{\texorpdfstring{$\langle\epsilon\epsilon\epsilon\epsilon\rangle$}{<eeee>}}
\label{sec:ising3d_epsilon}

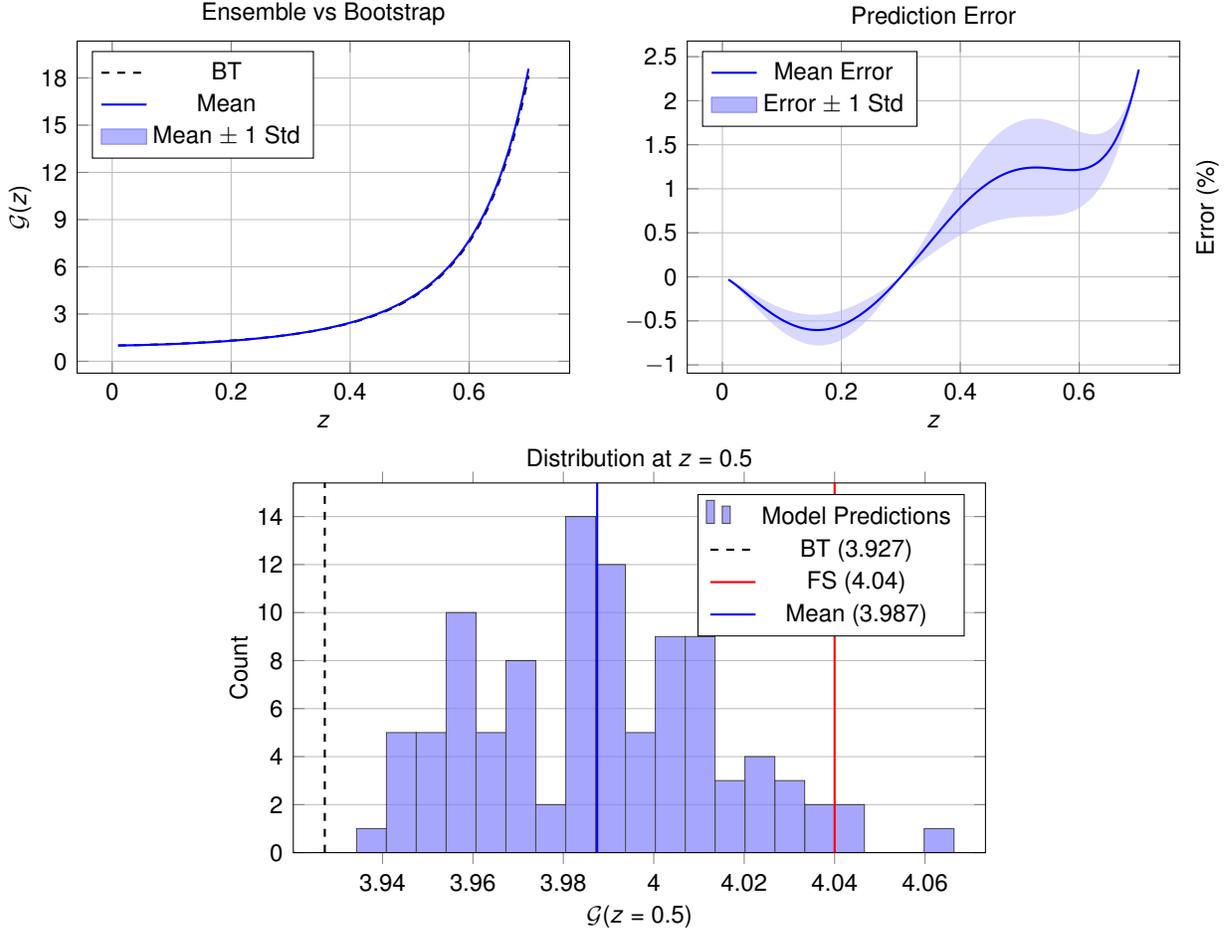
\begin{figure}[t!]
    \centering
    % Subfigure 1: Comparison
    \begin{subfigure}[b]{0.49\textwidth}
        \centering
        \begin{tikzpicture}
            \begin{axis}[
                width=\linewidth, height=6cm,
                xlabel={$z$},
                ylabel={$\mathcal{G}(z)$},
                title={Ensemble vs Bootstrap},
                grid=major,
                ytick distance=3,
                legend pos=north west,
            ]
                % Exact Solution
                \addplot [black, dashed, thick] table [x=z, y=Bootstrap, col sep=space] {plot_data/epsilon_four_pt_function_anchor_0.3_ensemble_comparison.dat};
                \addlegendentry{BT}

                % Ensemble Mean
                \addplot [blue, thick] table [x=z, y=Mean, col sep=space] {plot_data/epsilon_four_pt_function_anchor_0.3_ensemble_comparison.dat};
                \addlegendentry{Mean}

                % Uncertainty Band
                \addplot [forget plot, name path=upper, draw=none] table [x=z, y=Mean_plus_Std, col sep=space] {plot_data/epsilon_four_pt_function_anchor_0.3_ensemble_comparison.dat};
                \addplot [forget plot, name path=lower, draw=none] table [x=z, y=Mean_minus_Std, col sep=space] {plot_data/epsilon_four_pt_function_anchor_0.3_ensemble_comparison.dat};
                \addplot [forget plot, fill=blue!30, fill opacity=0.5, draw=none] fill between [of=upper and lower];
                \addlegendimage{legend image code/.code={\fill[blue!30, draw=blue!50] (0cm,-0.1cm) rectangle (0.6cm,0.1cm);}}
                \addlegendentry{Mean $\pm$ 1 Std}
            \end{axis}
        \end{tikzpicture}
        %\caption{}
        %\label{fig:epsilon_four_pt_function_anchor_0.3_comparison}
    \end{subfigure}
    \hfill
    %
    % Subfigure 2: Percentage Error
    \begin{subfigure}[b]{0.49\textwidth}
        \centering
        \begin{tikzpicture}
            \begin{axis}[
                width=\linewidth, height=6cm,
                xlabel={$z$},
                ylabel={Error (\%)},
                ylabel style={at={(axis description cs:1.10,0.5)}, anchor=south},
                title={Prediction Error},
                grid=major,
                ytick distance=0.5,
                legend pos=north west,
            ]
                % Percentage Error
                \addplot [blue, thick] table [x=z, y=PctError, col sep=space] {plot_data/epsilon_four_pt_function_anchor_0.3_percentage_error.dat};
                \addlegendentry{Mean Error}

                % Uncertainty Band
                \addplot [forget plot, name path=upper, draw=none] table [x=z, y=PctError_plus_PctStd, col sep=space] {plot_data/epsilon_four_pt_function_anchor_0.3_percentage_error.dat};
                \addplot [forget plot, name path=lower, draw=none] table [x=z, y=PctError_minus_PctStd, col sep=space] {plot_data/epsilon_four_pt_function_anchor_0.3_percentage_error.dat};
                \addplot [forget plot, fill=blue!30, fill opacity=0.5, draw=none] fill between [of=upper and lower];
                \addlegendimage{legend image code/.code={\fill[blue!30, draw=blue!50] (0cm,-0.1cm) rectangle (0.6cm,0.1cm);}}
                \addlegendentry{Error $\pm$ 1 Std}
            \end{axis}
        \end{tikzpicture}
        %\caption{}
        %\label{fig:epsilon_four_pt_function_anchor_0.3_error}
    \end{subfigure}
    %
    %\vspace{1em}
    %
    % Subfigure 3: Histogram
    \begin{subfigure}[b]{0.65\textwidth}
        \centering
        \begin{tikzpicture}
            \begin{axis}[
                width=\linewidth, height=6.5cm,
                xlabel={$\mathcal{G}(z=0.5)$},
                ylabel={Count},
                title={Distribution at $z=0.5$},
                ybar,
                bar width=0.006607,
                xmin=3.920273, xmax=4.073368,
                ymin=0,
                ymajorgrids=true,
                xmajorgrids=false,
                legend pos=north east,
                ytick distance=2,
                scaled ticks=false,
                xticklabel style={
                    /pgf/number format/fixed,
                    /pgf/number format/precision=3,
                    font=\sansmath\sffamily\footnotesize,
                }
            ]
                \addplot [fill=blue!50, draw=black, opacity=0.7] table [x=BinCenter, y=Count, col sep=space] {plot_data/epsilon_four_pt_function_anchor_0.3_histogram_z0.500.dat};
                \addlegendentry{Model Predictions}
                \draw [black, dashed, thick] (axis cs:3.927232,\pgfkeysvalueof{/pgfplots/ymin}) -- (axis cs:3.927232,\pgfkeysvalueof{/pgfplots/ymax});
                \addlegendimage{legend image code/.code={\draw[black, dashed, thick] (0cm,0cm) -- (0.6cm,0cm);}}
                \addlegendentry{BT ($3.927$)}
                \draw [red, thick] (axis cs:4.04,\pgfkeysvalueof{/pgfplots/ymin}) -- (axis cs:4.04,\pgfkeysvalueof{/pgfplots/ymax});
                \addlegendimage{legend image code/.code={\draw[red, thick] (0cm,0cm) -- (0.6cm,0cm);}}
	                \addlegendentry{FS ($4.04$)}
                \draw [blue, thick] (axis cs:3.987486,\pgfkeysvalueof{/pgfplots/ymin}) -- (axis cs:3.987486,\pgfkeysvalueof{/pgfplots/ymax});
                \addlegendimage{legend image code/.code={\draw[blue, thick] (0cm,0cm) -- (0.6cm,0cm);}}
                \addlegendentry{Mean ($3.987$)}
            \end{axis}
        \end{tikzpicture}
        %\caption{}
        %\label{fig:epsilon_four_pt_function_anchor_0.3_hist}
    \end{subfigure}

	    \caption{NN-predicted reduced four-point function $\GG(z)$ of the correlator $\langle \epsilon \epsilon \epsilon \epsilon \rangle$ in the 3d Ising model. The dashed black curve labelled BT (bootstrap truncation) is obtained by evaluating the $s$-channel OPE truncated to the operator content described in Section~\ref{sec:ising3d_epsilon}, using bootstrap-extracted CFT data from \cite{Simmons-Duffin:2016wlq}. The anchor point at $z_0=0.3$ is fixed using this same BT approximation. The NN prediction at $z=0.5$ is $3.987\pm 0.027$. The red line in the histogram, labelled FS,  denoted the fuzzy sphere bootstrap prediction at $z=0.5$ detailed in Appendix \ref{fuzzy}.}
    \label{fig:epsilon_four_pt_function_anchor_0.3_summary}
\end{figure}

Results based on 100 independent runs for the $\langle\epsilon\epsilon\epsilon\epsilon\rangle$ correlator are summarised in Fig.\ \ref{fig:epsilon_four_pt_function_anchor_0.3_summary}. The anchor value at $z_0=0.3$ was approximated using a truncation of the OPE based on the first four spin-$0$ operators, three spin-$2$ operators, three spin-$4$ operators, and two operators for each spin in the range $6$--$12$. These OPE data were borrowed from \cite{Simmons-Duffin:2016wlq}. We set $L(z)=1$ and parametrise the remainder by
\begin{equation}
    H(z) = z^{\Delta_{\epsilon}}(1-z)^{-2\Delta_{\epsilon}}\text{NN}_{\textbf{$\theta$}}(z).
\end{equation}
We observe that the predicted correlator is tracking the truncated OPE results within 2\% up to $z=0.7$. The truncated OPE is expected to fail in the vicinity of $z=1$. Accordingly, we see the disagreement growing in that region. The observed training loss MS was $(2.42 \pm 0.789) \times 10^{-6}$.

The higher scaling dimension of $\epsilon$ (compared to $\sigma$) makes the $\epsilon$-correlator more demanding. In order to examine more closely the validity of the NN prediction, we performed a further, non-trivial computation of the value of the correlator at the crossing symmetric point $z=0.5$ using fuzzy-sphere techniques \cite{Zhu:2022gjc,Han:2023yyb}. The pertinent details of that calculation, which involves Hamiltonian and DMRG \cite{White:1992zz} methods, are presented in Appendix \ref{fuzzy}. The fuzzy-sphere methods yield the value $4.04$. The NN prediction, $3.987 \pm 0.027$, is visibly closer to the fuzzy-sphere result.

\section{\texorpdfstring{CFT$_{4-\varepsilon}$}{CFT(4-epsilon)}: Wilson--Fisher fixed points}
\label{wf}

In Section \ref{ads}, we observed in the context of $\text{AdS}_2$ Witten diagrams that the anchored NNs can recover separately different terms in a parametric expansion of a correlator. Here, we explore this aspect further in a higher-dimensional example: Wilson--Fisher fixed points in $d=4-\varepsilon$ dimensions.

\subsection{Analytic results}

We consider the four-point correlator $\langle \phi\phi\phi\phi\rangle$ in the Wilson-Fisher fixed point of $\phi^4$ theory in $d=4-\varepsilon$ dimensions for $\varepsilon\ll 1$ (for related literature we refer the reader to \cite{Bissi:2019kkx,Bertucci:2022ptt}). Up to second order in $\varepsilon$ the external scaling dimension is
\begin{equation}
    \label{wfaa}
    \Delta_\phi = 1 - \tfrac{1}{2} \varepsilon + \tfrac{1}{108} \varepsilon^2 + \OO(\varepsilon^3)\,,
\end{equation}
and the reduced four-point function on the line is 
\begin{equation}
    \label{wfab}
    \GG(z)=\GG^{(0)}(z)+\varepsilon\lsp \GG^{(1)}(z)+\varepsilon^2 \GG^{(2)}(z)+\OO(\varepsilon^3)\,,
\end{equation}
with 
\begin{equation}
\begin{split}
    & \GG^{(1)}(z)=\frac{2}{3} z \left(\log (1-z)-\frac{z \log (z)}{z-1}\right),\\
    & \GG^{(2)}(z)=\frac{z \left(9 (z-1) \log ^2(1-z)+(34 (z-1)-36 \log (z)) \log (1-z)-z \log (z) (9 \log (z)+34)\right)}{81 (z-1)}\,.
    \end{split}
\end{equation}
We have defined the correlator with an extra power of $z^2$ so that all orders $\GG^{(i)}(z)$ of the $\varepsilon$ expansion of $\GG(z)$ obey the same crossing equation of the form
\begin{equation}
    \label{wfae}
    \GG^{(i)}(z) = \left(\frac{z}{1-z}\right)^2 \GG^{(i)}(1-z)\,.
\end{equation}

The zeroth order GFF contribution $\GG^{(0)}(z)$ was analysed with anchored NNs in a previous section. Here we focus on the reconstruction of $\GG^{(1)}(z)$ and $\GG^{(2)}(z)$. At leading order in the small $z$-expansion,
\begin{equation}
    \label{wfai}
    \GG^{(1)}(z) =\tfrac{2}{3} z^2 (-1+\log(z)) + \OO(z^3\log(z))\,,
\end{equation}
\begin{equation}
    \label{wfaj}
    \GG^{(2)}(z) = \tfrac{1}{81} z^2 \left(9 \log ^2(z)-2 \log (z)-34\right)+ \OO(z^3 \log(z)^2)
    \,.
\end{equation}
Accordingly, we assume knowledge of these leading order behaviours and set
\begin{equation}
    \label{wfak}
    L^{(1)}(z) =\tfrac{2}{3} z^2 (-1+\log(z))\,,\qquad
    L^{(2)}(z) =  \tfrac{1}{81} z^2 \left(9 \log ^2(z)-2 \log (z)-34\right).
\end{equation}
Our goal is to reconstruct the remaining part of the perturbative correlators $\GG^{(1)}, \GG^{(2)}$. 
\subsection{NN reconstruction}
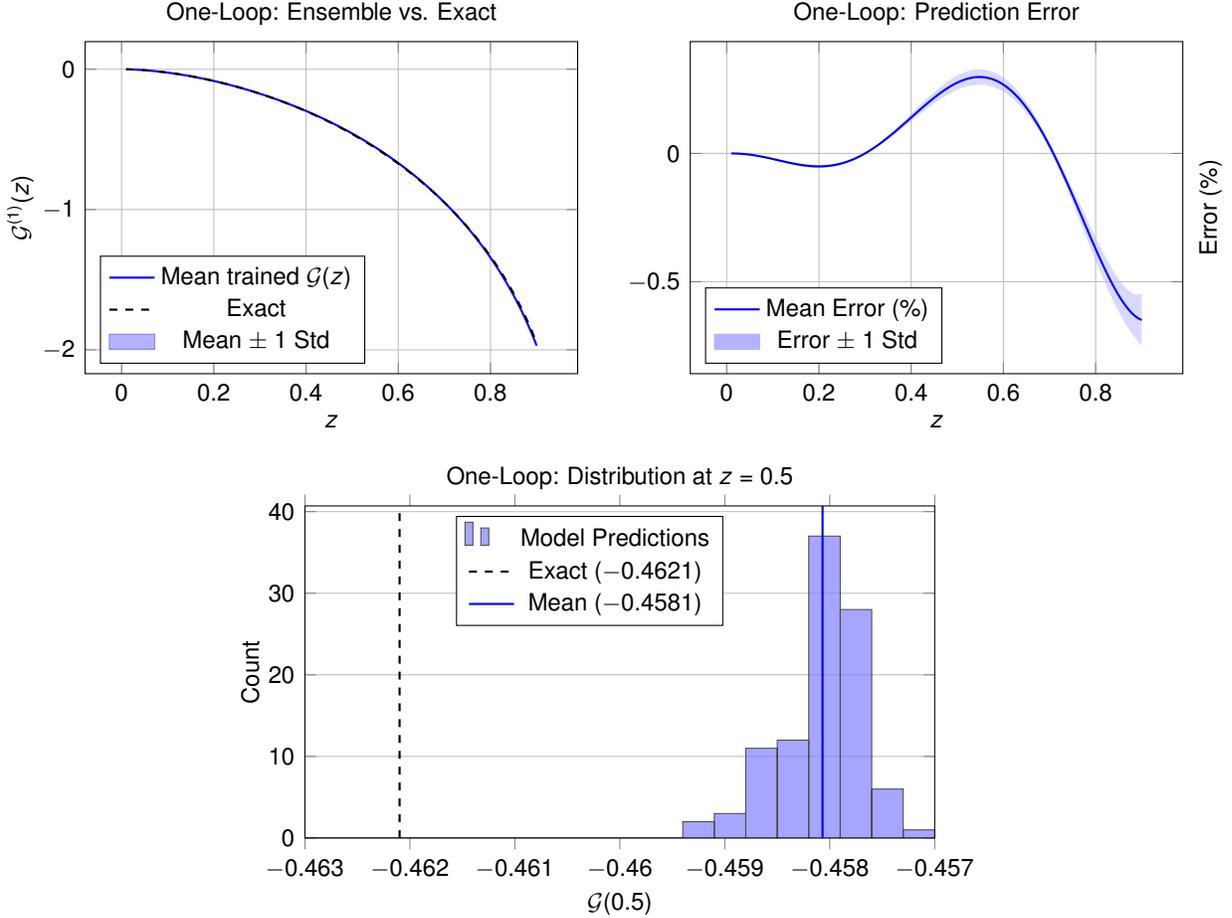
\begin{figure}[ht]
    \centering

    \begin{subfigure}[b]{0.49\textwidth}
        \centering
        \begin{tikzpicture}
	            \begin{axis}[
	                width=\linewidth, height=6cm,
	                xlabel={$z$},
	                ylabel={$\GG^{(1)}(z)$},
	                title={One-Loop: Ensemble vs. Exact},
	                grid=major,
	                legend pos=south west,
	            ]
                \addplot [blue, thick]
                table [x=z, y=mean, col sep=space]
                {plot_data/oneloop_mean_std.dat};
                \addlegendentry{Mean trained $\mathcal{G}(z)$}

                \addplot [black, dashed, thick]
                table [x=z, y=exact, col sep=space]
                {plot_data/oneloop_mean_std.dat};
                \addlegendentry{Exact}

                \addplot [forget plot, name path=upper, draw=none]
                table [x=z, y expr=\thisrow{mean}+\thisrow{std}, col sep=space]
                {plot_data/oneloop_mean_std.dat};

                \addplot [forget plot, name path=lower, draw=none]
                table [x=z, y expr=\thisrow{mean}-\thisrow{std}, col sep=space]
                {plot_data/oneloop_mean_std.dat};

                \addplot [forget plot, fill=blue!30, fill opacity=0.5, draw=none]
                fill between [of=upper and lower];
                \addlegendimage{legend image code/.code={\fill[blue!30, draw=blue!50] (0cm,-0.1cm) rectangle (0.6cm,0.1cm);}}
                \addlegendentry{Mean $\pm$ 1 Std}
            \end{axis}
        \end{tikzpicture}
    \end{subfigure}
    \hfill
    \begin{subfigure}[b]{0.49\textwidth}
        \centering
        \begin{tikzpicture}
	            \begin{axis}[
	                width=\linewidth, height=6cm,
	                xlabel={$z$},
	                ylabel={Error (\%)},
	                title={One-Loop: Prediction Error},
	                grid=major,
                    ylabel style={at={(axis description cs:1.10,0.5)}, anchor=south},
	                legend pos=south west,
	            ]
                \addplot [blue, thick]
                table [x=z, y=percent_error_mean, col sep=space]
                {plot_data/oneloop_mean_std.dat};
                \addlegendentry{Mean Error (\%)}

                \addplot [forget plot, name path=upper, draw=none]
                table [x=z, y expr=\thisrow{percent_error_mean}+\thisrow{percent_error_std}, col sep=space]
                {plot_data/oneloop_mean_std.dat};

                \addplot [forget plot, name path=lower, draw=none]
                table [x=z, y expr=\thisrow{percent_error_mean}-\thisrow{percent_error_std}, col sep=space]
                {plot_data/oneloop_mean_std.dat};

                \addplot [forget plot, fill=blue!30, fill opacity=0.5, draw=none]
                fill between [of=upper and lower];
                \addlegendimage{legend image code/.code={\fill[blue!30] (0cm,-0.1cm) rectangle (0.6cm,0.1cm);}}
                \addlegendentry{Error $\pm$ 1 Std}
            \end{axis}
        \end{tikzpicture}
    \end{subfigure}

    \vspace{0.6em}

    \begin{subfigure}[b]{0.60\textwidth}
        \centering
        \begin{tikzpicture}
            \begin{axis}[
	    width=\linewidth, height=6cm,
	    xlabel={$\mathcal{G}(0.5)$},
	    ylabel={Count},
	    title={One-Loop: Distribution at $z=0.5$},
	    xmin=-0.463, xmax=-0.457,
	    xtick={-0.463,-0.462,-0.461,-0.460,-0.459,-0.458,-0.457},
	    scaled x ticks=false,
    xticklabel style={
        /pgf/number format/fixed,
        /pgf/number format/precision=4
    },
    ymin=0,
    ybar,
    ymajorgrids=true,
    xmajorgrids=false,
    legend pos=north west,
    legend style={xshift=50pt},
]
                \addplot [forget plot, fill=blue!50, draw=black, opacity=0.7,
                    hist={bins=20}]
                table [y=prediction_at_z0.5, col sep=space]
                {plot_data/oneloop_raw_z0.5.dat};
                \addlegendimage{ybar,ybar legend,fill=blue!50, opacity=0.7, draw=black}
                \addlegendentry{Model Predictions}

	                \draw[black, dashed, thick]
	                (axis cs:-0.4620981412292,0) -- (axis cs:-0.4620981412292,\pgfkeysvalueof{/pgfplots/ymax});
	                \addlegendimage{legend image code/.code={\draw[black, dashed, thick] (0cm,0cm)--(0.6cm,0cm);}}
	                \addlegendentry{Exact ($-0.4621$)}

	                \draw[blue, thick]
	                (axis cs:-4.580683754224e-01,0) -- (axis cs:-4.580683754224e-01,\pgfkeysvalueof{/pgfplots/ymax});
	                \addlegendimage{legend image code/.code={\draw[blue, thick] (0cm,0cm)--(0.6cm,0cm);}}
	                \addlegendentry{Mean ($-0.4581$)}
            \end{axis}
        \end{tikzpicture}
	    \end{subfigure}

	    \caption{Plots depicting the NN-predicted results for the one-loop correlator, $\GG^{(1)}(z)$, in the Wilson--Fisher model in $4-\varepsilon$ using the crossing equation \eqref{wfae}. The NN prediction at z = 0.5 is $-0.4581 \pm 0.0004$.}
	    \label{fig:wf_oneloop_summary}
\end{figure}

We parametrise $H(z)$ as follows,
\begin{equation}
    H^{(1)}(z)=z^3\bigl(\log(z)+\log(1-z)\bigr)\mathrm{NN}_{\boldsymbol{\theta}}(z),\qquad H^{(2)}(z)=\bigl(z^3 \log(z)^2+\log(1-z)^2\bigr)\mathrm{NN}_{\boldsymbol{\theta}}(z).
\end{equation}
The anchor value is fixed at $z_0=0.3$ and the corresponding anchored NN predictions are summarised in Fig.\ \ref{fig:wf_oneloop_summary} for the one-loop correction $\GG^{(1)}$, and  Fig.~\ref{fig:wf_twoloop_summary} for the two-loop correction $\GG^{(2)}$. These results are based on 100 independent runs with MS training loss $(2.52\pm 1.31)\times 10^{-8}$ and $(6.87\pm 3.22)\times 10^{-9}$ for $\GG^{(1)}$ and $\GG^{(2)}$ respectively. 

% =====================
% Figure 2: Two-loop
% =====================
\begin{figure}[ht]
    \centering

    \begin{subfigure}[b]{0.49\textwidth}
        \centering
        \begin{tikzpicture}
	            \begin{axis}[
	                width=\linewidth, height=6cm,
	                xlabel={$z$},
	                ylabel={$\GG^{(2)}(z)$},
	                title={Two-Loop Case 1: Ensemble vs. Exact},
	                grid=major,
	                legend pos=north west,
	            ]
                \addplot [blue, thick]
                table [x=z, y=mean, col sep=space]
                {plot_data/twoloop_mean_std.dat};
                \addlegendentry{Mean trained $\mathcal{G}(z)$}

                \addplot [black, dashed, thick]
                table [x=z, y=exact, col sep=space]
                {plot_data/twoloop_mean_std.dat};
                \addlegendentry{Exact}

                \addplot [forget plot, name path=upper, draw=none]
                table [x=z, y expr=\thisrow{mean}+\thisrow{std}, col sep=space]
                {plot_data/twoloop_mean_std.dat};

                \addplot [forget plot, name path=lower, draw=none]
                table [x=z, y expr=\thisrow{mean}-\thisrow{std}, col sep=space]
                {plot_data/twoloop_mean_std.dat};

                \addplot [forget plot, fill=blue!30, fill opacity=0.5, draw=none]
                fill between [of=upper and lower];
                \addlegendimage{legend image code/.code={\fill[blue!30, draw=blue!50] (0cm,-0.1cm) rectangle (0.6cm,0.1cm);}}
                \addlegendentry{Mean $\pm$ 1 Std}
            \end{axis}
        \end{tikzpicture}
    \end{subfigure}
    \hfill
    \begin{subfigure}[b]{0.49\textwidth}
        \centering
        \begin{tikzpicture}
	            \begin{axis}[
	                width=\linewidth, height=6cm,
	                xlabel={$z$},
	                ylabel={Error (\%)},
	                title={Two-Loop Case 1: Prediction Error},
	                grid=major,
                    ylabel style={at={(axis description cs:1.10,0.5)}, anchor=south},
	                legend pos=north west,
	            ]
                \addplot [blue, thick]
                table [x=z, y=percent_error_mean, col sep=space]
                {plot_data/twoloop_mean_std.dat};
                \addlegendentry{Mean Error (\%)}

                \addplot [forget plot, name path=upper, draw=none]
                table [x=z, y expr=\thisrow{percent_error_mean}+\thisrow{percent_error_std}, col sep=space]
                {plot_data/twoloop_mean_std.dat};

                \addplot [forget plot, name path=lower, draw=none]
                table [x=z, y expr=\thisrow{percent_error_mean}-\thisrow{percent_error_std}, col sep=space]
                {plot_data/twoloop_mean_std.dat};

                \addplot [forget plot, fill=blue!30, fill opacity=0.5, draw=none]
                fill between [of=upper and lower];
                \addlegendimage{legend image code/.code={\fill[blue!30] (0cm,-0.1cm) rectangle (0.6cm,0.1cm);}}
                \addlegendentry{Error $\pm$ 1 Std}
            \end{axis}
        \end{tikzpicture}
    \end{subfigure}

    \vspace{0.6em}

    \begin{subfigure}[b]{0.60\textwidth}
        \centering
        \begin{tikzpicture}
            \begin{axis}[
	    width=\linewidth, height=6cm,
	    xlabel={$\mathcal{G}(0.5)$},
	    ylabel={Count},
	    title={Two-Loop Case 1: Distribution at $z=0.5$},
	    xmin=-0.027, xmax=-0.023,
	    xtick={-0.027,-0.026,-0.025,-0.024,-0.023},,
	    scaled x ticks=false,
    xticklabel style={
        /pgf/number format/fixed,
        /pgf/number format/precision=4
    },
    ymin=0,
    ybar,
    ymajorgrids=true,
    xmajorgrids=false,
    legend pos=north west,
    legend style={xshift=60pt},
]
                \addplot [forget plot, fill=blue!50, draw=black, opacity=0.7,
                    hist={bins=20}]
                table [y=prediction_at_z0.5, col sep=space]
                {plot_data/twoloop_raw_z0.5.dat};
                \addlegendimage{ybar,ybar legend,fill=blue!50, opacity=0.7, draw=black}
                \addlegendentry{Model Predictions}

	                \draw[black, dashed, thick]
	                (axis cs:-2.403232520767e-02,0) -- (axis cs:-2.403232520767e-02,\pgfkeysvalueof{/pgfplots/ymax});
	                \addlegendimage{legend image code/.code={\draw[black, dashed, thick] (0cm,0cm)--(0.6cm,0cm);}}
	                \addlegendentry{Exact ($-0.0240$)}

	                \draw[blue, thick]
	                (axis cs:-2.615662121129e-02,0) -- (axis cs:-2.615662121129e-02,\pgfkeysvalueof{/pgfplots/ymax});
	                \addlegendimage{legend image code/.code={\draw[blue, thick] (0cm,0cm)--(0.6cm,0cm);}}
	                \addlegendentry{Mean ($-0.0262$)}
            \end{axis}
        \end{tikzpicture}
	    \end{subfigure}

	    \caption{Plots depicting the NN-predicted results for the two-loop correlator, $\GG^{(2)}(z)$, in the Wilson-Fisher model in $4-\varepsilon$ dimensions using the crossing equation \eqref{wfae}. The NN prediction at $z=0.5$ is $-0.0262\pm0.0002$.}
	    \label{fig:wf_twoloop_summary}
\end{figure}

We note that the multiplication of the perturbative corrections $\GG^{(i)}$ by the overall factor $z^2$ is not strictly necessary, and we could instead have worked with functions obeying the crossing equation 
\begin{equation} \label{eq:crossing2}
    \GG^{(i)}(z)=\GG^{(i)}(1-z)\,.
\end{equation}
Since these functions are perturbative contributions to the full correlator, they do not need to be strictly monotonic in the interval $z\in (0,1)$. Moreover, the asymptotic behaviour supplied explicitly to the neural network near $z=0$ is rather intricate, involving both $\log z$ and $\log^2 z$ terms. This is qualitatively different from the nonperturbative correlators considered earlier, whose leading behaviour near $z=0$ is typically governed by a power law. In those cases, once the relevant power-law dependence is factored out, the neural network appears to capture the remaining structure rather efficiently. Here, by contrast, the additional logarithmic terms complicate the asymptotic structure and render the reconstruction problem significantly more subtle. More precisely, if we consider $\GG^{(2)}(z)$ without the overall $z^2$ factor and define the corresponding $L^{(2)}(z)$ simply by removing this factor from the previous definition, the neural network struggles to reproduce the correct asymptotic slope as $z\to 0$. This difficulty hints that, at the perturbative level, the smoothness property of the target correlator is significantly weakened. From the perspective of neural-network optimisation, this naturally leads to a broader and less selective landscape of admissible functions, making it harder for the training dynamics to isolate the physical solution.

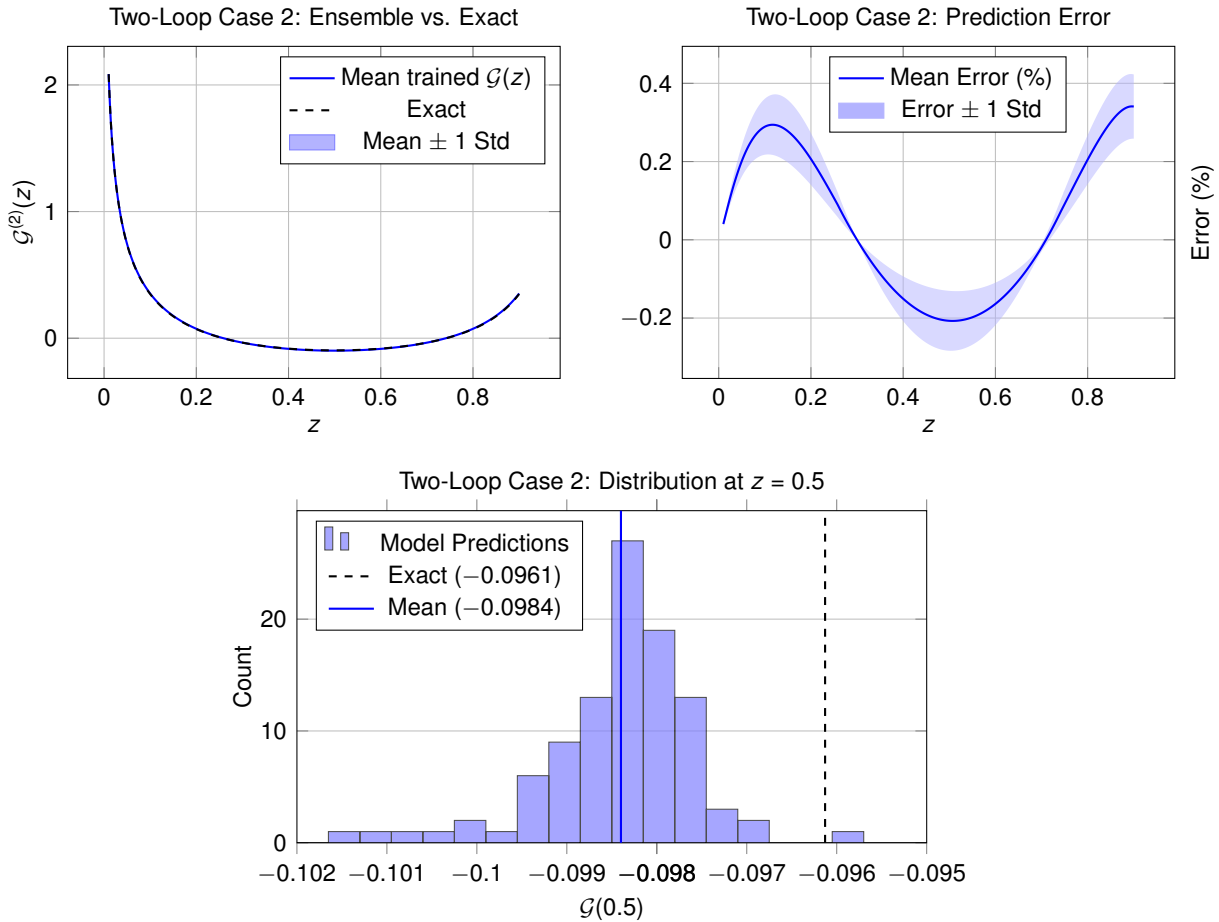
\begin{figure}[H]
    \centering

    \begin{subfigure}[b]{0.49\textwidth}
        \centering
        \begin{tikzpicture}
	            \begin{axis}[
	                width=\linewidth, height=6cm,
	                xlabel={$z$},
	                ylabel={$\GG^{(2)}(z)$},
	                title={Two-Loop Case 2: Ensemble vs. Exact},
	                grid=major,
	                legend pos=north east,
	            ]
                \addplot [blue, thick]
                table [x=z, y=mean, col sep=space]
                {plot_data/curvy2_mean_std.dat};
                \addlegendentry{Mean trained $\mathcal{G}(z)$}

                \addplot [black, dashed, thick]
                table [x=z, y=exact, col sep=space]
                {plot_data/curvy2_mean_std.dat};
                \addlegendentry{Exact}

                \addplot [forget plot, name path=upper, draw=none]
                table [x=z, y expr=\thisrow{mean}+\thisrow{std}, col sep=space]
                {plot_data/curvy2_mean_std.dat};

                \addplot [forget plot, name path=lower, draw=none]
                table [x=z, y expr=\thisrow{mean}-\thisrow{std}, col sep=space]
                {plot_data/curvy2_mean_std.dat};

                \addplot [forget plot, fill=blue!30, fill opacity=0.5, draw=none]
                fill between [of=upper and lower];
                \addlegendimage{legend image code/.code={\fill[blue!30, draw=blue!50] (0cm,-0.1cm) rectangle (0.6cm,0.1cm);}}
                \addlegendentry{Mean $\pm$ 1 Std}
            \end{axis}
        \end{tikzpicture}
    \end{subfigure}
    \hfill
    \begin{subfigure}[b]{0.49\textwidth}
        \centering
        \begin{tikzpicture}
	            \begin{axis}[
	                width=\linewidth, height=6cm,
	                xlabel={$z$},
	                ylabel={Error (\%)},
	                title={Two-Loop Case 2: Prediction Error},
	                grid=major,
	                legend pos=north west,
                    ylabel style={at={(axis description cs:1.10,0.5)}, anchor=south},
	                legend style={xshift=50pt}
	            ]
                \addplot [blue, thick]
                table [x=z, y=percent_error_mean, col sep=space]
                {plot_data/curvy2_mean_std.dat};
                \addlegendentry{Mean Error (\%)}

                \addplot [forget plot, name path=upper, draw=none]
                table [x=z, y expr=\thisrow{percent_error_mean}+\thisrow{percent_error_std}, col sep=space]
                {plot_data/curvy2_mean_std.dat};

                \addplot [forget plot, name path=lower, draw=none]
                table [x=z, y expr=\thisrow{percent_error_mean}-\thisrow{percent_error_std}, col sep=space]
                {plot_data/curvy2_mean_std.dat};

                \addplot [forget plot, fill=blue!30, fill opacity=0.5, draw=none]
                fill between [of=upper and lower];
                \addlegendimage{legend image code/.code={\fill[blue!30] (0cm,-0.1cm) rectangle (0.6cm,0.1cm);}}
                \addlegendentry{Error $\pm$ 1 Std}
            \end{axis}
        \end{tikzpicture}
    \end{subfigure}

    \vspace{0.6em}

    \begin{subfigure}[b]{0.60\textwidth}
        \centering
        \begin{tikzpicture}
            \begin{axis}[
	    width=\linewidth, height=6cm,
	    xlabel={$\mathcal{G}(0.5)$},
	    ylabel={Count},
	    title={Two-Loop Case 2: Distribution at $z=0.5$},
	    xmin=-0.102, xmax=-0.095,
	    xtick={-0.102,-0.101,-0.10,-0.099,-0.098,-0.098,-0.097,-0.096,-0.095},,
	    scaled x ticks=false,
    xticklabel style={
        /pgf/number format/fixed,
        /pgf/number format/precision=4
    },
    ymin=0,
    ybar,
    ymajorgrids=true,
    xmajorgrids=false,
    legend pos=north west,
]
                \addplot [forget plot, fill=blue!50, draw=black, opacity=0.7,
                    hist={bins=20}]
                table [y=prediction_at_z0.5, col sep=space]
                {plot_data/curvy2_raw_z0.5.dat};
                \addlegendimage{ybar,ybar legend,fill=blue!50, opacity=0.7, draw=black}
                \addlegendentry{Model Predictions}

	                \draw[black, dashed, thick]
	                (axis cs:-9.612929748124e-02,0) -- (axis cs:-9.612929748124e-02,\pgfkeysvalueof{/pgfplots/ymax});
	                \addlegendimage{legend image code/.code={\draw[black, dashed, thick] (0cm,0cm)--(0.6cm,0cm);}}
	                \addlegendentry{Exact ($-0.0961$)}

	                \draw[blue, thick]
	                (axis cs:-9.839833949510e-02,0) -- (axis cs:-9.839833949510e-02,\pgfkeysvalueof{/pgfplots/ymax});
	                \addlegendimage{legend image code/.code={\draw[blue, thick] (0cm,0cm)--(0.6cm,0cm);}}
	                \addlegendentry{Mean ($-0.0984$)}
            \end{axis}
        \end{tikzpicture}
	    \end{subfigure}

	    \caption{Plots depicting the NN-predicted results for the two-loop correlator, $\GG^{(2)}(z)$ (without the overall $z^2$), in the Wilson--Fisher model in $4-\varepsilon$ dimensions using the crossing equation \eqref{eq:crossing2}. The NN prediction at $z=0.5$ is $-0.0984\pm0.0008$.}
	    \label{fig:wf_curvy2_summary}
\end{figure}

To overcome this issue, a simple ansatz in which the unknown part is written as $z\,\log^2(z)\mathrm{NN}_{\boldsymbol{\theta}}(z)$ is not flexible enough to capture the correct asymptotic structure efficiently. We therefore absorb the leading logarithmic singular behaviour into the input asymptotic function,
\begin{equation}
L^{(2)}(z)=\frac{1}{81} z \left(9 \log ^2(z)-20 \log (z)\right)+\frac{1}{81} \left(9 \log ^2(z)-2 \log (z)-34\right)+\OO(z)\,,
\end{equation}
and parametrise the remaining part as
\begin{equation}
    H^{(2)}(z)=z\bigl(1+\log^2(1-z)\bigr)\,\mathrm{NN}_{\boldsymbol{\theta}}(z)\,.
\end{equation}
In this way, the network is no longer forced to reproduce the nontrivial logarithmic structure from scratch, and can instead focus on learning the remaining scaling behaviour as $z\to 0$. With this modification, we recover the expected results shown in Fig.~\ref{fig:wf_curvy2_summary}. The observed MS training loss was $(8.01\pm 2.55)\times 10^{-8}$. This example illustrates the important role played by the smoothness of the target function in the performance of our approach.

% =====================
% Figure 3: Curvy2
% =====================

\section{\texorpdfstring{CFT$_4$}{CFT4}: Half-BPS operators in 4d \texorpdfstring{$\NN=4$}{N=4} SYM theory}
\label{sym}

In this section, we consider a non-trivial example of a four-point correlation function in a four-dimensional gauge theory---the 4d $\NN=4$ SYM. We consider further examples of supersymmetric correlation functions in superconformal field theories in \cite{GKNS:3}.

\subsection{A short introduction to the correlators of interest}

The four-dimensional $\NN=4$ SYM theory is a superconformal field theory with $SO(6)_R$ R-symmetry. It contains scalar half-BPS superconformal primary operators $S_{IJ}$ in the $\boldsymbol{20}'$ traceless-symmetric representation of $SO(6)_R$. It is convenient to introduce null $SO(6)_R$ polarisation vectors $y^I$ $(I=1,2,\ldots,6)$ and denote
\begin{equation}
    \label{symaa}
    S_2(x,y) = S_{IJ}(x) y^I y^J
    ~.
\end{equation}
Here $x$ expresses the four-dimensional spacetime dependence and $y$ the R-symmetry polarisation. The scalars $S_2$ are in the same supermultiplet as the energy-momentum tensor.

We are interested in the four-point correlation functions
\begin{equation}
\label{symab}
    \langle S_2(x_1,y_1) S_2(x_2,y_2) S_2(x_3,y_3) S_2(x_4,y_4) \rangle=\bigg(\frac{(y_1\cdot y_2)  (y_3\cdot y_4)}{x_{12}^2 x_{34}^2}\bigg)^2 \GG(u,v,\sigma,\tau)\,.
\end{equation}
To proceed, it is useful to re-express the cross-ratios in the form
\begin{equation}
\label{symac}
\begin{split}
   & u=z\bar{z},\qquad v=(1-z)(1-\bar{z})\,,\\
    & \sigma=\alpha\bar{\alpha}, \qquad \tau=(1-\alpha)(1-\bar{\alpha})\,.
    \end{split}
\end{equation}
In this notation, one can show, \cite{Dolan:2004mu,Nirschl:2004pa}, that the superconformal Ward identities imply the decomposition
\begin{equation}
    \GG(u,v,\sigma,\tau)=
    \GG_{\rm free}(u,v,\sigma,\tau)
    +(\alpha z-1)(\alpha \bar{z}-1)(\bar{\alpha} z-1)(\bar{\alpha} \bar{z}-1) \HH(u,v)\,,
\end{equation}
where $\GG_{\rm free}$ is a function fully determined in free field theory and $\HH(u,v)$ a function that contains all the non-trivial dynamics of the interactions. In addition, crossing symmetry requires 
\begin{equation}
\label{symad}
    v^2 \HH(u,v)-u^2 \HH(v,u)+\frac{u-v}{c}+\left(u^2-v^2\right)=0\,.
\end{equation}
In the large central charge limit, $c\gg 1$, the function $\HH(u,v)$ admits a $1/c$ expansion of the form
\begin{equation}
\label{symae}
    \HH(u,v)=\HH^{(0)}(u,v)+\frac{1}{c} \HH^{(1)}(u,v)+\ldots
    ~,
\end{equation}
where
\begin{equation}
\label{symaf}
   \HH^{(0)}(u,v)= \frac{1}{v^2}+1\,,\qquad \HH^{(1)}(u,v)=\frac{1}{v}-u^2 \bar{D}_{2422}(u,v)\,.
\end{equation}
$\bar D_{2422}(u,v)$ is a suitable function, \cite{Dolan:2001tt}, that appears in Witten diagram computations in AdS/CFT. After setting $z=\bar{z}$, at $\mathcal{O}(1/c)$ the crossing equation implies
\begin{equation}
\label{symag}
    (1-z)^4 \HH^{(1)}(z)-z^4 \HH^{(1)}(1-z)-(1-2z)=0
\end{equation}
and the exact solution in $\NN=4$ SYM theory gives
\begin{equation}
\label{symai}
    \begin{split}
        \bar{D}_{2422}(z)= \frac{1}{35 (z-1)^5 z^5} \Bigl[ & 4 (z (2 z-7)+7) z^5 \tanh ^{-1}(1-2 z) \\
        & + 4 (z-1) ((z-1) z+1)^2 z - 2 (7 (z-1) z+2) \log (1-z) \Bigr]\,.
    \end{split}
\end{equation}
In what follows, we use this expression to set the anchor point and, with this input, attempt to re-derive numerically the full expression of the perturbative correction $\HH^{(1)}(z)$ in \eqref{symaf}, \eqref{symai} using the crossing equation \eqref{symag}. We set $L(z)=1$ and parametrise the remainder as $H(z) = z(1-z)^{-2}NN_{\textbf{$\theta$}}(z)$.

\subsection{NN reconstruction}

\begin{figure}[ht]
    \centering
    % Subfigure 1: Comparison
    \begin{subfigure}[b]{0.49\textwidth}
        \centering
        \begin{tikzpicture}
            \begin{axis}[
                width=\linewidth, height=6cm,
                xlabel={$z$},
                ylabel={$\mathcal{H}^{(1)}(z)$},
                title={Ensemble vs Exact},
                grid=major,
                ytick distance=10,
                legend pos=north west,
            ]
                % Exact Solution
                \addplot [black, dashed, thick] table [x=z, y=Exact, col sep=space] {plot_data/o20p_ensemble_comparison.dat};
                \addlegendentry{Exact}

                % Ensemble Mean
                \addplot [blue, thick] table [x=z, y=Mean, col sep=space] {plot_data/o20p_ensemble_comparison.dat};
                \addlegendentry{Mean}

                % Uncertainty Band
                \addplot [forget plot, name path=upper, draw=none] table [x=z, y=Mean_plus_Std, col sep=space] {plot_data/o20p_ensemble_comparison.dat};
                \addplot [forget plot, name path=lower, draw=none] table [x=z, y=Mean_minus_Std, col sep=space] {plot_data/o20p_ensemble_comparison.dat};
                \addplot [forget plot, fill=blue!30, fill opacity=0.5, draw=none] fill between [of=upper and lower];
                \addlegendimage{legend image code/.code={\fill[blue!30, draw=blue!50] (0cm,-0.1cm) rectangle (0.6cm,0.1cm);}}
                \addlegendentry{Mean $\pm$ 1 Std}
            \end{axis}
        \end{tikzpicture}
        %\caption{}
        %\label{fig:o20p_comparison}
    \end{subfigure}
    \hfill
    %
    % Subfigure 2: Percentage Error
    \begin{subfigure}[b]{0.49\textwidth}
        \centering
        \begin{tikzpicture}
            \begin{axis}[
                width=\linewidth, height=6cm,
                xlabel={$z$},
                ylabel={Error (\%)},
                ylabel style={at={(axis description cs:1.10,0.5)}, anchor=south},
                title={Prediction Error},
                grid=major,
                ytick distance=0.5,
                legend pos=north west,
            ]
                % Percentage Error
                \addplot [blue, thick] table [x=z, y=PctError, col sep=space] {plot_data/o20p_percentage_error.dat};
                \addlegendentry{Mean Error}

                % Uncertainty Band
                \addplot [forget plot, name path=upper, draw=none] table [x=z, y=PctError_plus_PctStd, col sep=space] {plot_data/o20p_percentage_error.dat};
                \addplot [forget plot, name path=lower, draw=none] table [x=z, y=PctError_minus_PctStd, col sep=space] {plot_data/o20p_percentage_error.dat};
                \addplot [forget plot, fill=blue!30, fill opacity=0.5, draw=none] fill between [of=upper and lower];
                \addlegendimage{legend image code/.code={\fill[blue!30, draw=blue!50] (0cm,-0.1cm) rectangle (0.6cm,0.1cm);}}
                \addlegendentry{Error $\pm$ 1 Std}
            \end{axis}
        \end{tikzpicture}
        %\caption{}
        %\label{fig:o20p_error}
    \end{subfigure}
    %
    %\vspace{1em}
    %
    % Subfigure 3: Histogram
    \begin{subfigure}[b]{0.65\textwidth}
        \centering
        \begin{tikzpicture}
            \begin{axis}[
                width=\linewidth, height=6.5cm,
                xlabel={$\mathcal{H}^{(1)}(z=0.5)$},
                ylabel={Count},
                title={Distribution at $z=0.5$},
                ybar,
                bar width=0.002144,
                xmin=3.566947, xmax=3.614116,
                ymin=0,
                ytick distance=3,
                ymajorgrids=true,
                xmajorgrids=false,
                legend pos=north west,
                scaled ticks=false,
                xticklabel style={
                    /pgf/number format/fixed,
                    /pgf/number format/precision=3,
                    font=\sansmath\sffamily\footnotesize,
                }
            ]
                \addplot [fill=blue!50, draw=black, opacity=0.7] table [x=BinCenter, y=Count, col sep=space] {plot_data/o20p_histogram_z0.500.dat};
                \addlegendentry{Model Predictions}
                \draw [black, dashed, thick] (axis cs:3.605735,\pgfkeysvalueof{/pgfplots/ymin}) -- (axis cs:3.605735,\pgfkeysvalueof{/pgfplots/ymax});
                \addlegendimage{legend image code/.code={\draw[black, dashed, thick] (0cm,0cm) -- (0.6cm,0cm);}}
                \addlegendentry{Exact ($3.606$)}
                \draw [blue, thick] (axis cs:3.601434,\pgfkeysvalueof{/pgfplots/ymin}) -- (axis cs:3.601434,\pgfkeysvalueof{/pgfplots/ymax});
                \addlegendimage{legend image code/.code={\draw[blue, thick] (0cm,0cm) -- (0.6cm,0cm);}}
                \addlegendentry{Mean ($3.601$)}
            \end{axis}
        \end{tikzpicture}
        %\caption{}
        %\label{fig:o20p_hist}
    \end{subfigure}

    \caption{NN-predicted results for the leading $1/c$ correction to the correlator $\langle S_2 S_2 S_2 S_2 \rangle$ in the large-$c$ limit of the 4d $\NN=4$ SYM theory, $\HH^{(1)}(z)$. in \eqref{symaf}, \eqref{symai}. The NN prediction at $z=0.5$ is $3.601\pm 0.007$.}
    \label{fig:o20p_summary}
\end{figure}
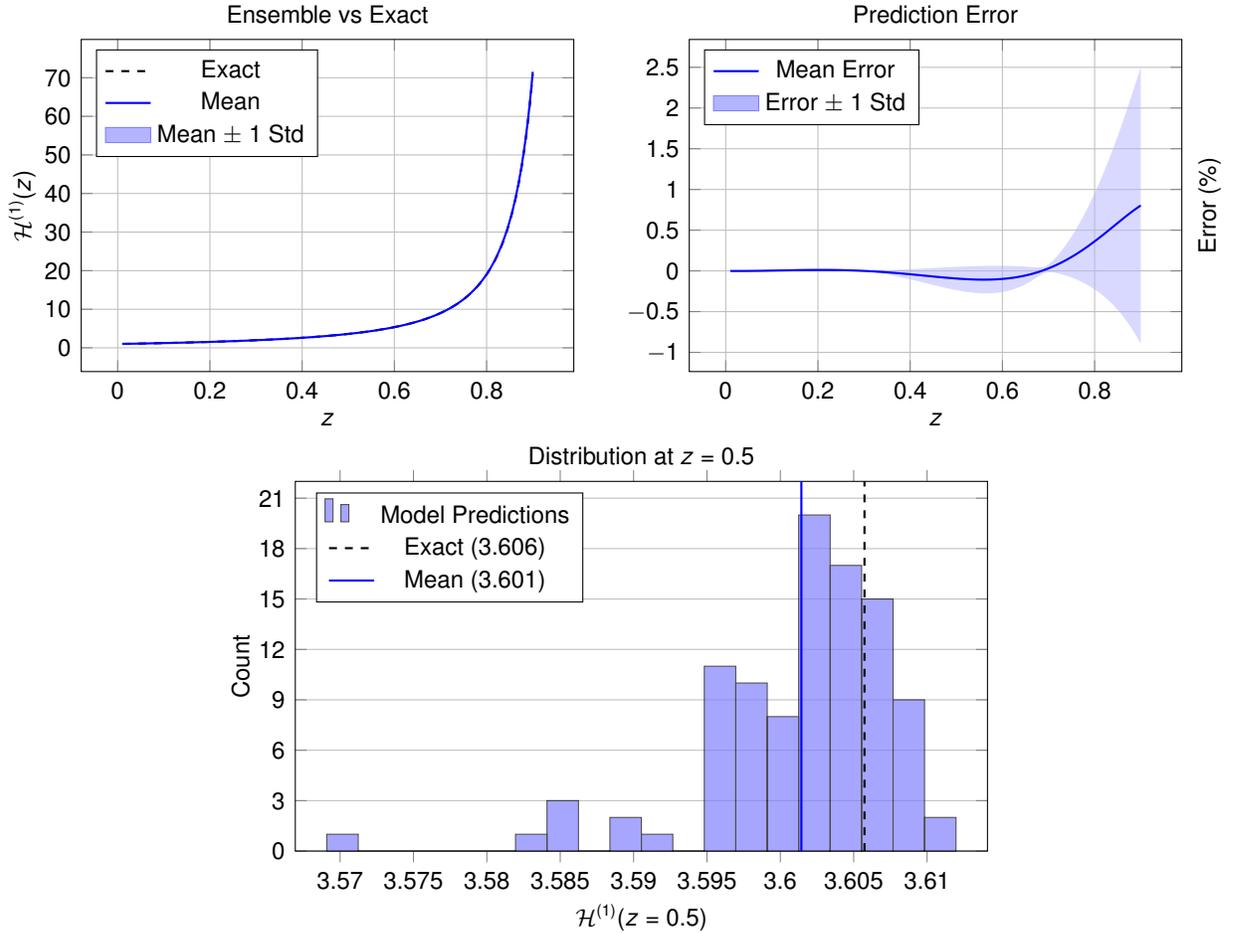

The results for $\HH^{(1)}$ of 100 independent runs are summarised in Fig.\ \ref{fig:o20p_summary}. The anchor point was set at $z_0=0.3$ and the observed training loss MS was $(8.06 \pm 6.99) \times  10^{-8}$. We observe agreement with the analytic result \eqref{symaf}, \eqref{symai}, which is significantly below 1\% for most of the interval of the $z$-variable. For example, at the crossing-symmetric point $z=0.5$ the prediction is $3.601 \pm 0.006$, which compares well with the analytic value $3.606$.

In this example, we are computing a very specific $1/c$ correction in a non-trivial part of the full correlator. The exact derivation of the result either employs Witten diagrams in AdS/CFT or (bootstrap) arguments purely in $\NN=4$ SYM theory, see e.g.\ \cite{Alday:2014tsa}. At the level of numerics, the anchored neural network optimisation effectively reduces many of the non-trivial ingredients in these computations to the input of a single number!

\section{Thermal two-point functions at zero spatial separation}
\label{thermal}

Finite-temperature two-point correlators at zero spatial separation in CFT obey the crossing symmetry equation~\eqref{introac} by virtue of the Kubo--Martin--Schwinger (KMS) condition. We provide evidence that the scope of the anchored NN approach extends to thermal two-point functions as well. Using an identical PINN architecture to the one employed for vacuum four-point correlators on a line, we show in a variety of different examples how NNs reconstruct efficiently thermal two-point functions at zero spatial separation. These observations suggest that an underlying principle of smoothness extends beyond the vacuum four-point correlators in CFT.

\subsection{Useful facts about the thermal bootstrap}
\label{introthermal}

Consider a $d$-dimensional CFT on $S^1_\beta\times\IR^{d-1}$, with Euclidean time $\tau\in[0,\beta)$ compactified on a circle of circumference $\beta$ and thermal boundary conditions. In this context, one can study thermal properties of the theory at infinite spatial volume, with the parameter $\beta$ expressing the inverse temperature. The spacetime coordinates are split naturally to the time and space parts, $x=(\tau, \vec x)$.

In what follows, we focus on the thermal two-point function of a generic scalar primary $\phi$ with scaling dimension $\Df$, which is denoted as
\begin{equation}
    \label{thermaa}
    g(\tau,|x|) = \langle\phi(x)\,\phi(0)\rangle_\beta
    \,,
\end{equation}
where
\begin{equation}
    \label{thermaaa}
    |x|=\sqrt{\tau^2 + \vec x^{\lsp 2}}
    \,.
\end{equation}
We can use the $SO(d-1)$ symmetry of the theory on this background to set $\vec x=(\rho, 0,\ldots,0)$ for the spatial separation. In this frame, we define the following set of variables that are used interchangeably in the main text:
\begin{equation}
    \label{thermaab}
    z=\tau + i \rho\,, ~~ \bar z=\tau -i \rho\,,
\end{equation}
and
\begin{equation}
    \label{thermaac}
    z=rw\,, ~~ \bar z=rw^{-1}\,,
\end{equation}
where $w$ is a phase. In this notation, $|x|=r$ and the two-point function depends on $(\tau, r)$, or equivalently $(z,\bar z)$.

The KMS periodicity condition demands $g(\tau,r)=g(\beta-\tau,r)$. Henceforth, we set $\beta=1$ (without loss of generality), and by using the invariance of the two-point function under the parity symmetry $\rho \to -\rho$, we can recast the KMS condition as a crossing equation in the standard form:
\begin{equation}
    \label{thermaad}
    g(z,\bar z) = g(1-z, 1-\bar z)
    \,.
\end{equation}

\subsection*{OPE expansions of finite-temperature scalar two-point functions}
\label{ope}

It will be useful to recall a few basic facts about the properties of the OPE of thermal two-point functions, although we will not use its detailed structure heavily in our computation.

In a two-point function $\langle \phi(x) \phi(0) \rangle_\beta$, we can use the OPE to express the correlation function as a series over thermal one-point functions, which are typically the main computational target in thermal bootstrap studies. For an operator with scaling dimension $\Delta$, the one-point function is proportional to the temperature raised to the power $\Delta$, namely $\beta^{-\Delta}$. Moreover, since conformal descendants have vanishing thermal one-point functions, only the leading terms in a conformal family within an OPE contribute. This fact simplifies the form of the thermal conformal blocks. Putting everything together, one ends up with the following conformal block expansion of thermal two-point functions of identical scalar operators\footnote{We are setting $\beta=1$ in this formula.} \cite{Iliesiu:2018fao, Petkou:2018ynm}
\begin{equation}
    \label{opeaa}
    g(rw, rw^{-1}) = \sum_{\OO \in \phi\times \phi} a_\OO \, C_{J}^{(\nu)}\left(\frac{1}{2}(w+w^{-1})\right) r^{\Delta-2\Delta_\phi}
    \,,
\end{equation}
where
\begin{equation}
    \label{opeab}
    a_\OO = \frac{f_{\phi\phi \OO}b_\OO}{c_\OO} \frac{J !}{2^{J} (\nu)_{J}}
    \,,\qquad
    \nu = \frac{d-2}{2}\,,
\end{equation}
and $C_J^{(\nu)}(\eta)$ are Gegenbauer polynomials. The coefficients $C_{\phi\phi\OO}$ are three-point function coefficients in the zero-temperature theory, and $b_\OO$ are the thermal one-point coefficients. $\Delta, J$ represent the scaling dimension and spin of each contributing operator $\OO$ in the OPE. This expansion is convergent for $r<1$.

\subsubsection*{Zero spatial separation}

At zero spatial separation, ($\rho =0$, or equivalently $z=\tau \in[0,1)$), we set
\begin{equation}
    \label{zerospaa}
    \GG(z) = z^{2\Delta_\phi} g(z,z)
\end{equation}
and the OPE \eqref{opeaa} becomes
\begin{equation}
    \label{zerospab}
    \GG(z) = \sum_{\Delta} a_\Delta z^{\Delta}
\end{equation}
with
\begin{equation}
    \label{zerospac}
    a_\Delta = \sum_{\OO \in \phi\times \phi}^{\Delta \, \text{fixed}} a_\OO C_J^{(\nu)}(1)
    \,.
\end{equation}

We will examine if we can reconstruct such one-dimensional functions in $(0,1)$ by using an anchor point and the KMS condition, which now takes the form 
\begin{equation}
    \label{zerospad}
    \GG(z) = \left( \frac{z}{1-z} \right)^{2\Delta_\phi} \GG(1-z)
    \,.
\end{equation}
We implement the same training setup as in previous sections. In all cases, we set $L(z)=1$ to capture the identity contribution to the thermal OPE.

\subsection{Thermal GFFs}
\label{gffthermal}

For a scalar GFF at inverse temperature $\beta=1$, the exact thermal two-point function at zero spatial separation is
\begin{equation}
    \label{Tgffaa}
    \GG(z) = z^{2\Df}\Big(\zeta_H(2\Df,z)+\zeta_H(2\Df,1-z)\Big)\,,
\end{equation}
where $\zeta_H(s,a)=\sum_{n=0}^\infty(n+a)^{-s}$ is the Hurwitz zeta function. 

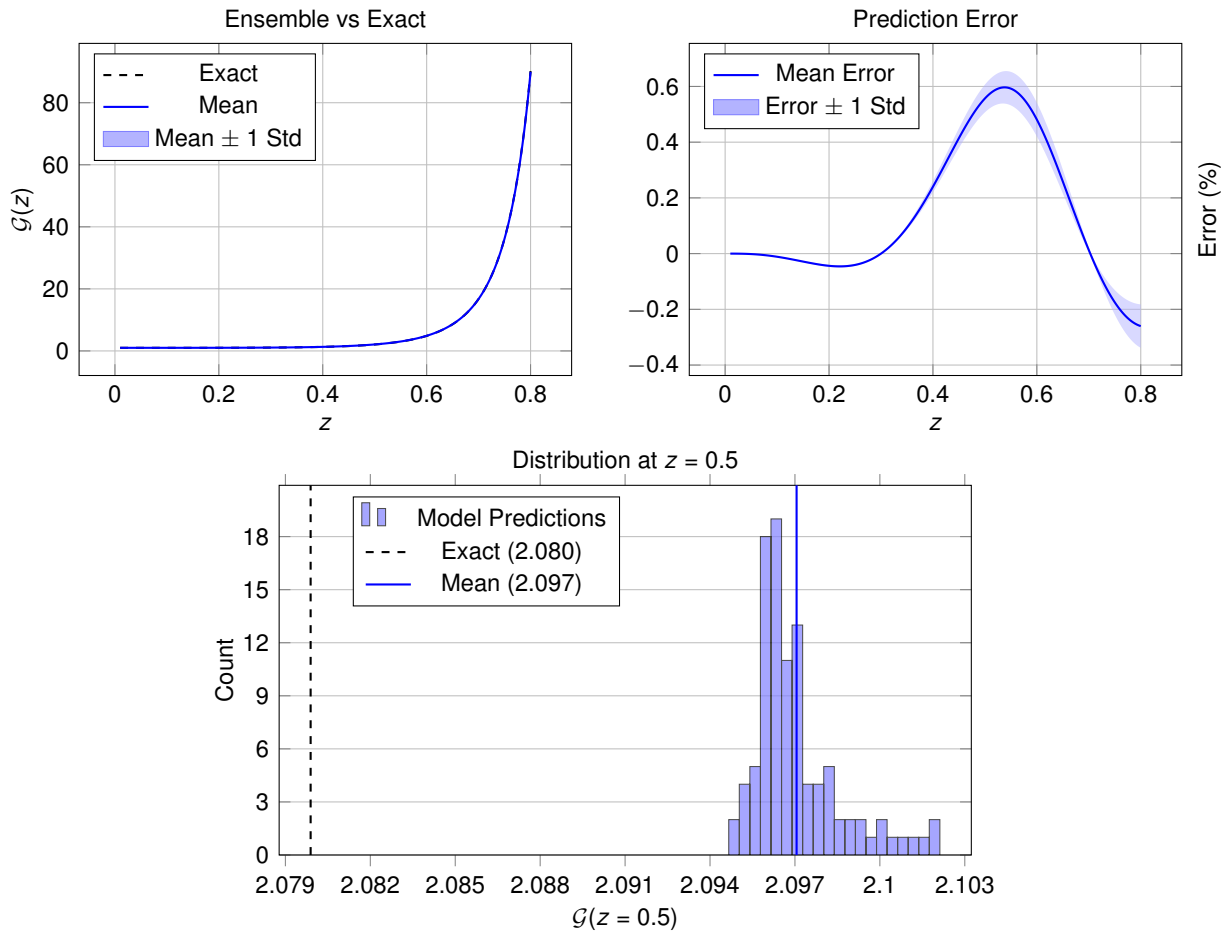
\begin{figure}[H]
    \centering
    % Subfigure 1: Comparison
    \begin{subfigure}[b]{0.49\textwidth}
        \centering
        \begin{tikzpicture}
            \begin{axis}[
                width=\linewidth, height=6cm,
                xlabel={$z$},
                ylabel={$\mathcal{G}(z)$},
                title={Ensemble vs Exact},
                grid=major,
                ytick distance=20,
                legend pos=north west,
            ]
                % Exact Solution
                \addplot [black, dashed, thick] table [x=z, y=Exact, col sep=space] {plot_data/gff_T_1.618_ensemble_comparison.dat};
                \addlegendentry{Exact}

                % Ensemble Mean
                \addplot [blue, thick] table [x=z, y=Mean, col sep=space] {plot_data/gff_T_1.618_ensemble_comparison.dat};
                \addlegendentry{Mean}

                % Uncertainty Band
                \addplot [forget plot, name path=upper, draw=none] table [x=z, y=Mean_plus_Std, col sep=space] {plot_data/gff_T_1.618_ensemble_comparison.dat};
                \addplot [forget plot, name path=lower, draw=none] table [x=z, y=Mean_minus_Std, col sep=space] {plot_data/gff_T_1.618_ensemble_comparison.dat};
                \addplot [forget plot, fill=blue!30, fill opacity=0.5, draw=none] fill between [of=upper and lower];
                \addlegendimage{legend image code/.code={\fill[blue!30, draw=blue!50] (0cm,-0.1cm) rectangle (0.6cm,0.1cm);}}
                \addlegendentry{Mean $\pm$ 1 Std}
            \end{axis}
        \end{tikzpicture}
        %\caption{}
        %\label{fig:gff_T_1.618_comparison}
    \end{subfigure}
    \hfill
    %
    % Subfigure 2: Percentage Error
    \begin{subfigure}[b]{0.49\textwidth}
        \centering
        \begin{tikzpicture}
            \begin{axis}[
                width=\linewidth, height=6cm,
                xlabel={$z$},
                ylabel={Error (\%)},
                ylabel style={at={(axis description cs:1.10,0.5)}, anchor=south},
                title={Prediction Error},
                grid=major,
                ytick distance=0.2,
                legend pos=north west,
            ]
                % Percentage Error
                \addplot [blue, thick] table [x=z, y=PctError, col sep=space] {plot_data/gff_T_1.618_percentage_error.dat};
                \addlegendentry{Mean Error}

                % Uncertainty Band
                \addplot [forget plot, name path=upper, draw=none] table [x=z, y=PctError_plus_PctStd, col sep=space] {plot_data/gff_T_1.618_percentage_error.dat};
                \addplot [forget plot, name path=lower, draw=none] table [x=z, y=PctError_minus_PctStd, col sep=space] {plot_data/gff_T_1.618_percentage_error.dat};
                \addplot [forget plot, fill=blue!30, fill opacity=0.5, draw=none] fill between [of=upper and lower];
                \addlegendimage{legend image code/.code={\fill[blue!30, draw=blue!50] (0cm,-0.1cm) rectangle (0.6cm,0.1cm);}}
                \addlegendentry{Error $\pm$ 1 Std}
            \end{axis}
        \end{tikzpicture}
        %\caption{}
        %\label{fig:gff_T_1.618_error}
    \end{subfigure}
    %
    %\vspace{1em}
    %
    % Subfigure 3: Histogram
    \begin{subfigure}[b]{0.65\textwidth}
        \centering
        \begin{tikzpicture}
            \begin{axis}[
                width=\linewidth, height=6.5cm,
                xlabel={$\mathcal{G}(z=0.5)$},
                ylabel={Count},
                title={Distribution at $z=0.5$},
                ybar,
                bar width=0.000373,
                xmin=2.078780, xmax=2.103230,
                ymin=0,
                ymajorgrids=true,
                xmajorgrids=false,
                xtick distance=0.003,
                ytick distance=3,
                legend pos=north west,
                legend style={xshift=20pt},
                scaled ticks=false,
                xticklabel style={
                    /pgf/number format/fixed,
                    /pgf/number format/precision=3,
                    font=\sansmath\sffamily\footnotesize,
                }
            ]
                \addplot [fill=blue!50, draw=black, opacity=0.7] table [x=BinCenter, y=Count, col sep=space] {plot_data/gff_T_1.618_histogram_z0.500.dat};
                \addlegendentry{Model Predictions}
                \draw [black, dashed, thick] (axis cs:2.079892,\pgfkeysvalueof{/pgfplots/ymin}) -- (axis cs:2.079892,\pgfkeysvalueof{/pgfplots/ymax});
                \addlegendimage{legend image code/.code={\draw[black, dashed, thick] (0cm,0cm) -- (0.6cm,0cm);}}
                \addlegendentry{Exact ($2.080$)}
                \draw [blue, thick] (axis cs:2.097058,\pgfkeysvalueof{/pgfplots/ymin}) -- (axis cs:2.097058,\pgfkeysvalueof{/pgfplots/ymax});
                \addlegendimage{legend image code/.code={\draw[blue, thick] (0cm,0cm) -- (0.6cm,0cm);}}
                \addlegendentry{Mean ($2.097$)}
            \end{axis}
        \end{tikzpicture}
        %\caption{}
        %\label{fig:gff_T_1.618_hist}
    \end{subfigure}

    \caption{NN-predicted results for the thermal two-point function of a GFF at $\Delta_\phi=1.618$. These results are compared to the analytic GFF correlator \eqref{Tgffaa}. The NN prediction at $z=0.5$ is $2.097\pm0.002$.}
    \label{fig:gff_T_1.618_summary}
\end{figure}

As an illustration, we consider the specific case of $\Delta_\phi=1.618$. Optimising a two-layer NN with an anchor point at $z_0=0.3$ deduced from the exact equation \eqref{Tgffaa}, and the ansatz 
\begin{equation}
    \label{Tgffab}
    H(z) = z^{2\Df}\,(1-z)^{-2\Df}\,\text{NN}_{\boldsymbol\theta}(z)
    \,,
\end{equation}
we obtain the results presented in Fig.\ \ref{fig:gff_T_1.618_summary}.

The quality of the prediction is comparable to that for zero-temperature four-point correlators ---compare, for example, with the corresponding GFF four-point correlator in Fig.\ \ref{fig:gfb_1.618_summary}. The MS training loss was $(3.20\pm2.37)\times10^{-7}$.

It is amusing to note that both here (for the case of the thermal two-point function $\langle \phi\phi\rangle_\beta$) and in section \ref{gffbosonic1618} (for the zero-temperature four-point function $\langle \phi\phi\phi\phi\rangle$ on a line) we are solving with NNs exactly the same problem. Only the anchor point is different and this is enough to differentiate between two functions whose analytic form is very different.

\subsection{Generic 2d Virasoro primaries}
\label{2dthermal}

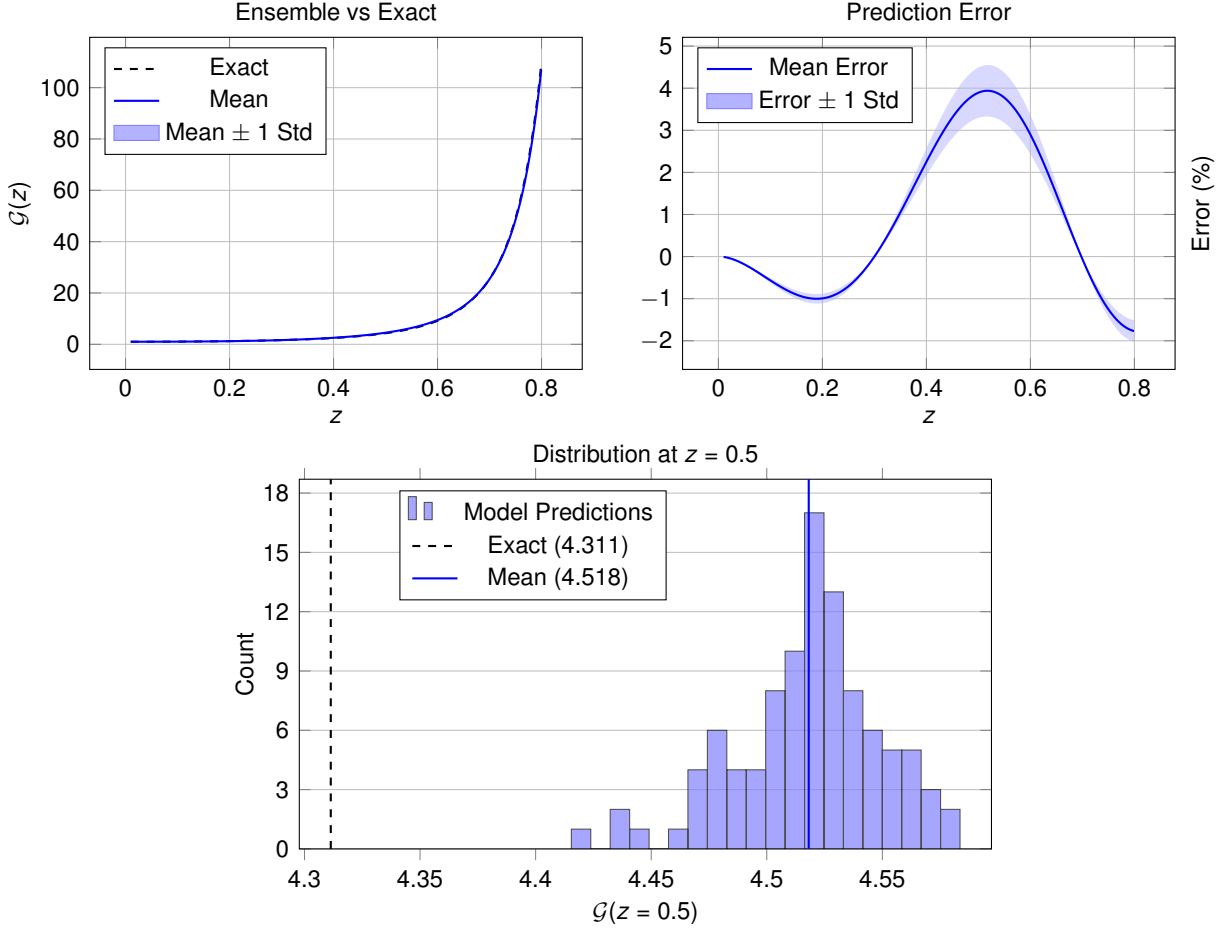
\begin{figure}[ht]
    \centering
    % Subfigure 1: Comparison
    \begin{subfigure}[b]{0.49\textwidth}
        \centering
        \begin{tikzpicture}
            \begin{axis}[
                width=\linewidth, height=6cm,
                xlabel={$z$},
                ylabel={$\mathcal{G}(z)$},
                title={Ensemble vs Exact},
                grid=major,
                ytick distance=20,
                legend pos=north west,
            ]
                % Exact Solution
                \addplot [black, dashed, thick] table [x=z, y=Exact, col sep=space] {plot_data/2d_virasoro_primaries_1.618_ensemble_comparison.dat};
                \addlegendentry{Exact}

                % Ensemble Mean
                \addplot [blue, thick] table [x=z, y=Mean, col sep=space] {plot_data/2d_virasoro_primaries_1.618_ensemble_comparison.dat};
                \addlegendentry{Mean}

                % Uncertainty Band
                \addplot [forget plot, name path=upper, draw=none] table [x=z, y=Mean_plus_Std, col sep=space] {plot_data/2d_virasoro_primaries_1.618_ensemble_comparison.dat};
                \addplot [forget plot, name path=lower, draw=none] table [x=z, y=Mean_minus_Std, col sep=space] {plot_data/2d_virasoro_primaries_1.618_ensemble_comparison.dat};
                \addplot [forget plot, fill=blue!30, fill opacity=0.5, draw=none] fill between [of=upper and lower];
                \addlegendimage{legend image code/.code={\fill[blue!30, draw=blue!50] (0cm,-0.1cm) rectangle (0.6cm,0.1cm);}}
                \addlegendentry{Mean $\pm$ 1 Std}
            \end{axis}
        \end{tikzpicture}
        %\caption{}
        %\label{fig:2d_virasoro_primaries_1.618_comparison}
    \end{subfigure}
    \hfill
    %
    % Subfigure 2: Percentage Error
    \begin{subfigure}[b]{0.49\textwidth}
        \centering
        \begin{tikzpicture}
            \begin{axis}[
                width=\linewidth, height=6cm,
                xlabel={$z$},
                ylabel={Error (\%)},
                ylabel style={at={(axis description cs:1.10,0.5)}, anchor=south},
                title={Prediction Error},
                grid=major,
                ytick distance=1,
                legend pos=north west,
            ]
                % Percentage Error
                \addplot [blue, thick] table [x=z, y=PctError, col sep=space] {plot_data/2d_virasoro_primaries_1.618_percentage_error.dat};
                \addlegendentry{Mean Error}

                % Uncertainty Band
                \addplot [forget plot, name path=upper, draw=none] table [x=z, y=PctError_plus_PctStd, col sep=space] {plot_data/2d_virasoro_primaries_1.618_percentage_error.dat};
                \addplot [forget plot, name path=lower, draw=none] table [x=z, y=PctError_minus_PctStd, col sep=space] {plot_data/2d_virasoro_primaries_1.618_percentage_error.dat};
                \addplot [forget plot, fill=blue!30, fill opacity=0.5, draw=none] fill between [of=upper and lower];
                \addlegendimage{legend image code/.code={\fill[blue!30, draw=blue!50] (0cm,-0.1cm) rectangle (0.6cm,0.1cm);}}
                \addlegendentry{Error $\pm$ 1 Std}
            \end{axis}
        \end{tikzpicture}
        %\caption{}
        %\label{fig:2d_virasoro_primaries_1.618_error}
    \end{subfigure}
    %
    %\vspace{1em}
    %
    % Subfigure 3: Histogram
    \begin{subfigure}[b]{0.65\textwidth}
        \centering
        \begin{tikzpicture}
            \begin{axis}[
                width=\linewidth, height=6.5cm,
                xlabel={$\mathcal{G}(z=0.5)$},
                ylabel={Count},
                title={Distribution at $z=0.5$},
                ybar,
                bar width=0.008407,
                xmin=4.297764, xmax=4.597202,
                ymin=0,
                ymajorgrids=true,
                xmajorgrids=false,
                legend pos=north west,
                ytick distance=3,
                legend style={xshift=30pt},
                scaled ticks=false,
                xticklabel style={
                    /pgf/number format/fixed,
                    /pgf/number format/precision=3,
                    font=\sansmath\sffamily\footnotesize,
                }
            ]
                \addplot [fill=blue!50, draw=black, opacity=0.7] table [x=BinCenter, y=Count, col sep=space] {plot_data/2d_virasoro_primaries_1.618_histogram_z0.500.dat};
                \addlegendentry{Model Predictions}
                \draw [black, dashed, thick] (axis cs:4.311375,\pgfkeysvalueof{/pgfplots/ymin}) -- (axis cs:4.311375,\pgfkeysvalueof{/pgfplots/ymax});
                \addlegendimage{legend image code/.code={\draw[black, dashed, thick] (0cm,0cm) -- (0.6cm,0cm);}}
                \addlegendentry{Exact ($4.311$)}
                \draw [blue, thick] (axis cs:4.518152,\pgfkeysvalueof{/pgfplots/ymin}) -- (axis cs:4.518152,\pgfkeysvalueof{/pgfplots/ymax});
                \addlegendimage{legend image code/.code={\draw[blue, thick] (0cm,0cm) -- (0.6cm,0cm);}}
                \addlegendentry{Mean ($4.518$)}
            \end{axis}
        \end{tikzpicture}
        %\caption{}
        %\label{fig:2d_virasoro_primaries_1.618_hist}
    \end{subfigure}

    \caption{NN-predicted results for a generic thermal two-point function of identical Virasoro primaries in 2d CFTs on the cylinder. For concreteness, here we set $\Delta_\phi=1.618$. These results are compared to the analytic GFF correlator \eqref{T2daa}. The NN prediction at $z=0.5$ is $4.518\pm 0.032$.}
    \label{fig:2d_virasoro_primaries_1.618_summary}
\end{figure}

Another standard example, with well-known analytic expressions, is provided by thermal two-point functions in 2d CFTs on the cylinder. At zero spatial separation, the two-point function of a generic Virasoro primary with total dimension $\Df=h_\phi+\bar h_\phi$ is fixed exactly by conformal invariance
\begin{equation}
    \label{T2daa}
    \GG(z) = \left(\frac{\pi z}{\sin(\pi z)}\right)^{2\Df}\,.
\end{equation}
As an illustration, we set again $\Delta_\phi=1.618$. In this case, we make the ansatz 
\begin{equation}
    \label{T2daaextra}
    H(z) = z^2\,(1-z)^{-2\Df}\,\text{NN}_{\boldsymbol\theta}(z)
    \,,
\end{equation}
to capture the appropriate gap above the identity. With an anchor point at $z_0=0.3$ the predictions of the anchored NN are as depicted in Fig.\ \ref{fig:2d_virasoro_primaries_1.618_summary}. The MS training loss was $(2.14\pm2.34)\times10^{-6}$.

\subsection{3d Ising thermal correlators}
\label{3dthermal}

Next we consider the 3d Ising CFT at finite temperature and infinite spatial volume. This is a more demanding theory, where exact thermal correlators are currently unavailable. Since we cannot infer the anchor point $H(z_0)$ from an analytic expression (as we did in other examples for benchmarking purposes), here we explore a different strategy. We use the leading non-trivial thermal OPE coefficient $a_{\Delta_\epsilon}$ of the energy operator as input from another method, e.g.\ Monte Carlo (MC), and evaluate at low $z=0.05$ so that OPE truncation is trustworthy. In this section, we explain how this is implemented and how the results compare with independent computation based on analytical approximations and numerical MC calculations.

As we mentioned previously in Section~\ref{smooth:lessons}, setting the anchor point close to $z=0$ can yield unstable results because a low anchor value may be unable to provide a numerically significant guide towards the target function. In the present example this problem does not seem to arise. Alternative strategies to determine the anchor point from low-lying CFT data will be outlined briefly in the outlook Section \ref{outlook}.

\subsubsection{Useful background}

We briefly remind the reader that the low-lying spectrum of the 3d Ising model contains the operators
\begin{alignat}{2}
    \label{Tisingaa}
    \sigma\lsp:~ \Delta_\sigma &= 0.518148806(24) \,, \qquad 
    &\epsilon\lsp:&~ \Delta_\epsilon = 1.41262528(29)\,,\nonumber\\
    T_{\mu\nu}\lsp:~ \Delta_T &= 3\,, \qquad 
    &\epsilon'\lsp:&~ \Delta_{\epsilon'} = 3.82951(61) 
    \,.
\end{alignat}
We will consider two thermal two-point functions at zero spatial separation
\begin{equation}
    \label{Tisingab}
    \langle \sigma(\tau)\sigma(0) \rangle_\beta \,, ~~
    \langle \epsilon(\tau) \epsilon(0) \rangle_\beta
    \,.
\end{equation}
The corresponding OPEs are
\begin{equation}
    \label{Tisingac}
    \sigma \times \sigma = {\bf 1} + \epsilon + T_{\mu\nu} + \epsilon' + \ldots
\end{equation}
and a similar expression for $\epsilon\times \epsilon$.

Ref.\ \cite{Barrat:2025wbi} studied the $\langle \sigma \sigma\rangle_\beta$ correlator using the combination of truncation and Tauberian theorem methods. Ref.\ \cite{Barrat:2025nvu} proposed an analytic approximation of both $\langle \sigma \sigma\rangle_\beta$, $\langle \epsilon \epsilon\rangle_\beta$ correlators using thermal dispersion arguments. The analytic expressions used in \cite{Barrat:2025nvu} are
\begin{equation}
    \label{Tisingad}
    \langle \sigma(\tau)\sigma(0) \rangle_{\beta=1} \simeq \sum_{\OO \in \{{\bf 1}, \epsilon, T_{\mu\nu}, \epsilon' \}} a^{(\sigma)}_{\Delta_\OO} \bigg( 
    \zeta_H(2\Delta_\sigma - \Delta_\OO, \tau) + \zeta_H(2\Delta_\sigma -\Delta_\OO, 1-\tau) 
    \bigg) + \kappa_\sigma
    \,,
\end{equation}
\begin{equation}
    \label{Tisingae}
    \langle \epsilon(\tau)\epsilon(0) \rangle_{\beta=1} \simeq \sum_{\OO \in \{{\bf 1}, \epsilon, T_{\mu\nu}\}} a^{(\epsilon)}_{\Delta_\OO} \bigg( 
    \zeta_H(2\Delta_\sigma - \Delta_\OO, \tau) + \zeta_H(2\Delta_\sigma -\Delta_\OO, 1-\tau) 
    \bigg) + \kappa_\epsilon
    \,.
\end{equation}
Using
\begin{equation}
    \label{Tisingaf}
    a^{(\sigma)}_{\Delta_\epsilon} = 0.75\,, ~~
    a^{(\sigma)}_{\Delta_T} = 1.97\,, ~~
    a^{(\sigma)}_{\Delta_{\epsilon'}} = 0.19\,, ~~
    \kappa_\sigma = -55.872165883552114
    \,,
\end{equation}
and
\begin{equation}
    \label{Tisingag}
    a^{(\epsilon)}_{\Delta_\epsilon} = 1.09\,, ~~
    a^{(\epsilon)}_{\Delta_T} = 5.37\,, ~~
    \kappa_\sigma = 1.81
\end{equation}
Ref.~\cite{Barrat:2025nvu} found experimentally that the resulting expressions are numerically close to MC results and the results obtained in \cite{Barrat:2025wbi}.

Notice that the correlator $\langle \epsilon\epsilon\rangle$ is more demanding in these approaches. It requires more than the leading correction to the Tauberian approximation and was, therefore, not analysed in \cite{Barrat:2025wbi}. Because of related difficulties, the sum in \eqref{Tisingae} was truncated very drastically at the $T_{\mu\nu}$ contribution and is expected to be a worse approximation to the actual thermal correlator compared to $\langle \sigma \sigma\rangle$.

In what follows, we analyse separately the correlators
\begin{equation}
    \label{Tisingai}
    \GG_\sigma(z) = z^{2\Delta_\sigma} \langle \sigma (z) \sigma(0) \rangle_{\beta = 1}
    \,,\qquad
    \GG_\epsilon(z) = z^{2\Delta_\epsilon} \langle \epsilon (z) \epsilon(0) \rangle_{\beta = 1}
\end{equation}
for $z=\tau \in [0,1)$ and compare them with the approximate predictions in \eqref{Tisingad}, \eqref{Tisingae}.

\subsubsection{\texorpdfstring{$\langle\sigma\sigma\rangle_\beta$}{<ss>\_beta}}
\label{sigsig}

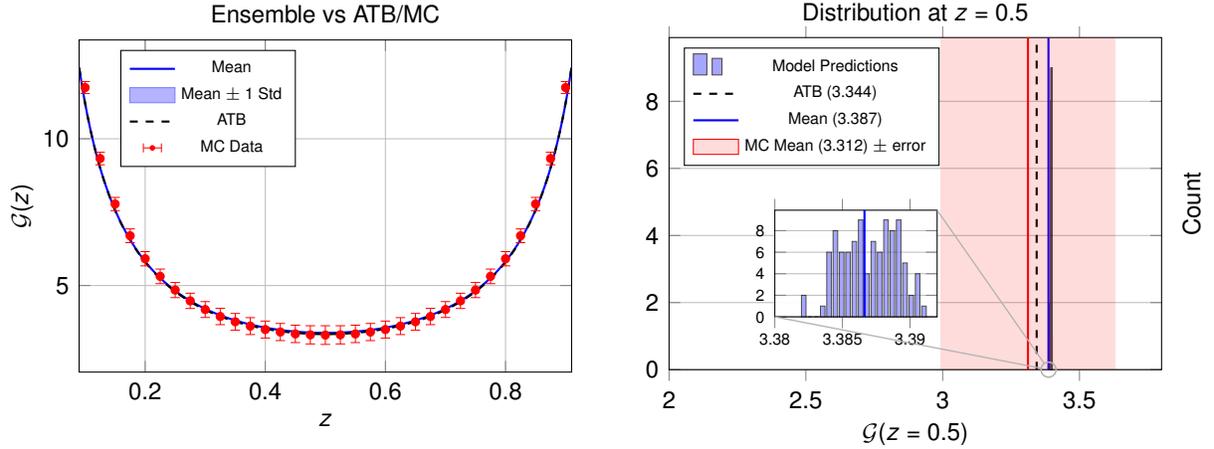
\begin{figure}[ht]
    \centering
    \begin{subfigure}[b]{0.49\textwidth}
        \centering
        \begin{tikzpicture}[remember picture]
	            \begin{axis}[
	                width=\linewidth, height=6cm,
	                xlabel={$z$},
	                ylabel={$\mathcal{G}(z)$},
		                title={Ensemble vs ATB/MC},
	                grid=major,
	                axis on top=false,
	                xmin=0.09, xmax=0.91,
                legend pos=north west,
                legend style={xshift=10pt, font=\sansmath\sffamily\tiny},
            ]
                \addplot [forget plot, name path=upper, draw=none] table [x=z, y=Mean_plus_Std, col sep=space] {plot_data/thermal_sigma_ensemble_comparison.dat};
                \addplot [forget plot, name path=lower, draw=none] table [x=z, y=Mean_minus_Std, col sep=space] {plot_data/thermal_sigma_ensemble_comparison.dat};
                \addplot [forget plot, fill=blue!30, fill opacity=0.5, draw=none] fill between [of=upper and lower];
                \addplot [forget plot, blue, thick] table [x=z, y=Mean, col sep=space] {plot_data/thermal_sigma_ensemble_comparison.dat};
                \addplot [forget plot, black, dashed, thick] table [x=z, y=Exact, col sep=space] {plot_data/thermal_sigma_ensemble_comparison.dat};
                \begin{pgfonlayer}{axis foreground}
                \addplot+[
                    forget plot,
                    only marks,
                    mark=*,
                        mark size=1.55pt,
                    mark options={fill=red, draw=red, line width=0.15pt},
                        fill opacity=1,
                        draw opacity=1,
                        opacity=1,
                    ] table [x=z, y=MC_Mean, col sep=space] {plot_data/thermal_sigma_mc_comparison.dat};
                \addplot+[
                    forget plot,
                    red,
                    draw=none,
                    no markers,
                    error bars/.cd,
                    y dir=both,
                    y explicit,
                        error mark=-,
                        error mark options={red, rotate=90, line width=0.35pt, mark size=1.8pt},
                        error bar style={line width=0.35pt, draw=red, draw opacity=1},
                    ] table [x=z, y=MC_Mean, y error=MC_Std, col sep=space] {plot_data/thermal_sigma_mc_comparison.dat};
                \end{pgfonlayer}
                \addlegendimage{legend image code/.code={\draw[blue, thick] (0cm,0cm) -- (0.6cm,0cm);}}
                \addlegendentry{Mean}
                \addlegendimage{legend image code/.code={\fill[blue!30, draw=blue!50] (0cm,-0.1cm) rectangle (0.6cm,0.1cm);}}
	                \addlegendentry{Mean $\pm$ 1 Std}
	                \addlegendimage{legend image code/.code={\draw[black, dashed, thick] (0cm,0cm) -- (0.6cm,0cm);}}
	                \addlegendentry{ATB}
	                \addlegendimage{legend image code/.code={%
	                    \draw[red, line width=0.35pt] (0.18cm,0cm) -- (0.42cm,0cm);
	                    \draw[red, line width=0.35pt] (0.18cm,-0.06cm) -- (0.18cm,0.06cm);
	                    \draw[red, line width=0.35pt] (0.42cm,-0.06cm) -- (0.42cm,0.06cm);
	                    \path[draw=red, fill=red] (0.3cm,0cm) circle[radius=0.03cm];
	                }}
		                \addlegendentry{MC Data}
            \end{axis}
        \end{tikzpicture}
    \end{subfigure}
    \hfill
    \begin{subfigure}[b]{0.49\textwidth}
        \centering
        \vspace{0.8cm}
        \raisebox{-0.25cm}{%
        \begin{tikzpicture}
            \begin{axis}[
                clip=false,
                name=mainhist,
                width=\linewidth, height=6cm,
                xlabel={$\mathcal{G}(z=0.5)$},
                ylabel={Count},
                title={Distribution at $z=0.5$},
                ybar,
                bar width=0.0004679270015035542,
                xmin=2, xmax=3.8,
                ymin=0,
                ymajorgrids=true,
                xmajorgrids=false,
                legend pos=north west,
                axis on top=false,
                legend style={font=\sansmath\sffamily\tiny},
                scaled ticks=false,
                ylabel style={at={(axis description cs:1.10,0.5)}, anchor=south},
                xticklabel style={
                    /pgf/number format/fixed,
                    /pgf/number format/precision=2,
                    font=\sansmath\sffamily\footnotesize,
                }
            ]
                \addplot [fill=blue!50, opacity=0.7, draw=black, forget plot] table [x=BinCenter, y=Count, col sep=space] {plot_data/thermal_sigma_histogram_z0.500.dat};
                \addlegendimage{legend image code/.code={\fill[blue!50, opacity=0.7, draw=black] (0cm,-0.1cm) rectangle (0.18cm,0.18cm); \fill[blue!50, opacity=0.7, draw=black] (0.25cm,-0.1cm) rectangle (0.38cm,0.12cm);}}
                \addlegendentry{Model Predictions}
	                \draw [black, dashed, thick] (axis cs:3.34352831402591,\pgfkeysvalueof{/pgfplots/ymin}) -- (axis cs:3.34352831402591,\pgfkeysvalueof{/pgfplots/ymax});
	                \addlegendimage{legend image code/.code={\draw[black, dashed, thick] (0cm,0cm) -- (0.6cm,0cm);}}
	                \addlegendentry{ATB (3.344)}
                \draw [blue, thick] (axis cs:3.386629959572441,\pgfkeysvalueof{/pgfplots/ymin}) -- (axis cs:3.386629959572441,\pgfkeysvalueof{/pgfplots/ymax});
                \addlegendimage{legend image code/.code={\draw[blue, thick] (0cm,0cm) -- (0.6cm,0cm);}}
                \addlegendentry{Mean (3.387)}
                \draw [fill=red, fill opacity=0.14, draw=none] (axis cs:2.993153710794585,\pgfkeysvalueof{/pgfplots/ymin}) rectangle (axis cs:3.630055130379896,\pgfkeysvalueof{/pgfplots/ymax});
                \addlegendimage{legend image code/.code={\fill[red, fill opacity=0.14, draw=red] (0cm,-0.1cm) rectangle (0.6cm,0.1cm);}}
                \addlegendentry{MC Mean (3.312) $\pm$ error}
                \draw [red, thick] (axis cs:3.311604420587241,\pgfkeysvalueof{/pgfplots/ymin}) -- (axis cs:3.311604420587241,\pgfkeysvalueof{/pgfplots/ymax});
                \coordinate (zoomcircle) at (axis cs:3.387,0);
                \draw[gray!60, line width=0.6pt] (zoomcircle) circle (\zoomcircleradius);
            \end{axis}
            \begin{axis}[
                at={(mainhist.south west)},
                anchor=south west,
                xshift=40pt,
                yshift=20pt,
                width=0.46\linewidth,
                height=3cm,
                ybar,
                bar width=0.000336907441082559,
                xmin=3.38, xmax=3.392,
                ymin=0,
                ymajorgrids=true,
                xmajorgrids=false,
                scaled ticks=false,
                axis background/.style={fill=white, fill opacity=0.95},
                legend style={
                    at={(0.90,0.98)},
                    anchor=north east,
                    draw=none,
                    fill=none,
                    fill opacity=0,
                    text opacity=1,
                    font=\sansmath\sffamily\tiny,
                    inner sep=0.8pt,
                },
                tick label style={font=\sansmath\sffamily\tiny},
                xticklabel style={
                    /pgf/number format/fixed,
                    /pgf/number format/precision=4,
                    font=\sansmath\sffamily\tiny,
                    yshift=2pt,
                },
                yticklabel style={font=\sansmath\sffamily\tiny},
	                name=zoomaxis,
	            ]
	                \addplot [fill=blue!50, opacity=0.7, draw=black, line width=0.35pt, forget plot] table [x=BinCenter, y=Count, col sep=space] {plot_data/thermal_sigma_histogram_z0.500.dat};
	                \draw [black, dashed, thick] (axis cs:3.34352831402591,\pgfkeysvalueof{/pgfplots/ymin}) -- (axis cs:3.34352831402591,\pgfkeysvalueof{/pgfplots/ymax});
	                \draw [blue, thick] (axis cs:3.386629959572441,\pgfkeysvalueof{/pgfplots/ymin}) -- (axis cs:3.386629959572441,\pgfkeysvalueof{/pgfplots/ymax});
	                \coordinate (targetNE) at (axis cs:3.392,\pgfkeysvalueof{/pgfplots/ymax});
	                \coordinate (targetSW) at (axis cs:3.38,\pgfkeysvalueof{/pgfplots/ymin});
	            \end{axis}
            \path let \p1=(zoomcircle), \p2=(targetNE) in coordinate (connNE) at ($(\p1)!\zoomcircleradius!(\p2)$);
            \path let \p1=(zoomcircle), \p2=(targetSW) in coordinate (connSW) at ($(\p1)!\zoomcircleradius!(\p2)$);
            \draw[gray!60, line width=0.5pt] (connNE) -- (targetNE);
            \draw[gray!60, line width=0.5pt] (connSW) -- (targetSW);
            \end{tikzpicture}%
        }
    \end{subfigure}
	    \caption{NN-predicted results for thermal two-point function $\langle\sigma\sigma\rangle_\beta$ in the 3d Ising model with anchor point fixed at $z_0=0.05$. The left panel compares the ensemble mean against the analytic thermal bootstrap (ATB) approximation of Barrat et al.\ \cite{Barrat:2025nvu} and Monte Carlo (MC) data points with vertical error bars. The right panel shows the histogram of NN outputs at $z = 0.5$, together with the ATB prediction and MC mean and error band. A zoomed inset highlights the prediction cluster. The NN prediction at $z = 0.5$ is $3.387\pm 0.002$.}
	    \label{fig:3d_Ising_T_sigma}
	\end{figure}

To study the thermal two-point correlator $ \langle \sigma(0) \sigma(z) \rangle_\beta$ we set $\GG_\sigma(z) = 1 + H_\sigma(z)$ and make the ansatz $H_\sigma(z) = z^{\Delta_\epsilon} (1-z)^{-2\Delta_\sigma} \rm{NN}_{{\boldsymbol \theta},\sigma}(z)$. The results, based on 100 independent runs with an MS training loss of $(2.25\pm 0.32) \times 10^{-7}$, are summarised in Fig.~\ref{fig:3d_Ising_T_sigma}. The prediction is exceptionally stable (despite the low anchor point set at $z_0=0.05$) and compares very well with the analytic approximation and MC mean values of Ref.~\cite{Barrat:2025nvu}. For example, at the crossing symmetric point $z=0.5$, the anchored NN prediction is $3.387\pm 0.002$, when the analytic approximation of \cite{Barrat:2025nvu} is 3.344 and the MC mean 3.312.

\subsubsection{\texorpdfstring{$\langle\epsilon\epsilon\rangle_\beta$}{<ee>\_beta}}
\label{epep}

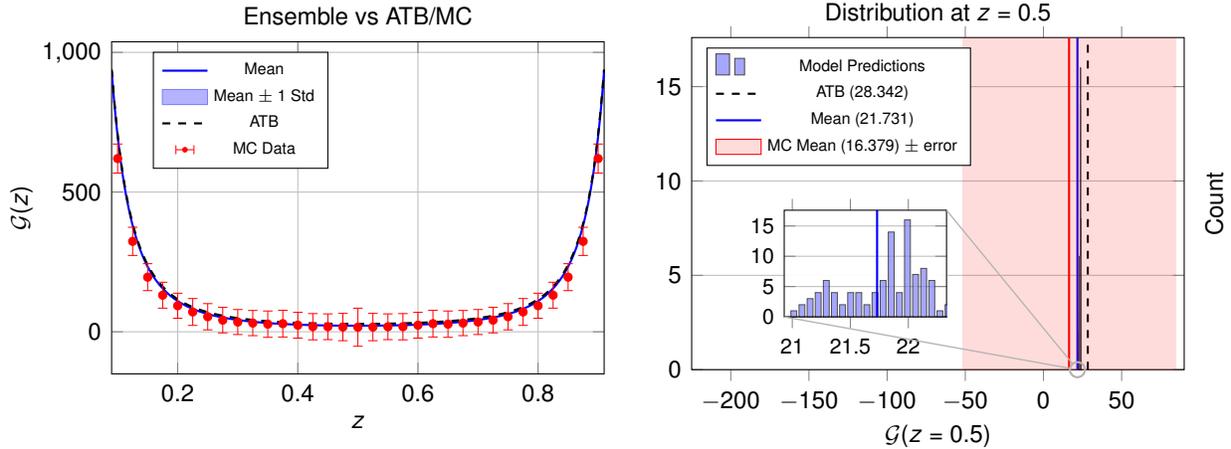
\begin{figure}[ht]
    \centering
    \begin{subfigure}[b]{0.49\textwidth}
        \centering
	        \begin{tikzpicture}[remember picture]
	            \begin{axis}[
	                width=\linewidth, height=6cm,
	                xlabel={$z$},
	                ylabel={$\mathcal{G}(z)$},
		                title={Ensemble vs ATB/MC},
	                grid=major,
	                xmin=0.09, xmax=0.91,
	                legend pos=north west,
                legend style={xshift=10pt, font=\sansmath\sffamily\tiny},
            ]
                \addplot [forget plot, name path=upper, draw=none] table [x=z, y=Mean_plus_Std, col sep=space] {plot_data/thermal_epsilon_ensemble_comparison.dat};
                \addplot [forget plot, name path=lower, draw=none] table [x=z, y=Mean_minus_Std, col sep=space] {plot_data/thermal_epsilon_ensemble_comparison.dat};
                \addplot [forget plot, fill=blue!30, fill opacity=0.5, draw=none] fill between [of=upper and lower];
                \addplot [forget plot, blue, thick] table [x=z, y=Mean, col sep=space] {plot_data/thermal_epsilon_ensemble_comparison.dat};
                \addplot [forget plot, black, dashed, thick] table [x=z, y=Exact, col sep=space] {plot_data/thermal_epsilon_ensemble_comparison.dat};
                \begin{pgfonlayer}{axis foreground}
                \addplot+[
                    forget plot,
                    only marks,
                    mark=*,
                        mark size=1.55pt,
                    mark options={fill=red, draw=red, line width=0.15pt},
                        fill opacity=1,
                        draw opacity=1,
                        opacity=1,
                    ] table [x=z, y=MC_Mean, col sep=space] {plot_data/thermal_epsilon_mc_comparison.dat};
                \addplot+[
                    forget plot,
                    red,
                    draw=none,
                    no markers,
                    error bars/.cd,
                    y dir=both,
                    y explicit,
                        error mark=-,
                        error mark options={red, rotate=90, line width=0.35pt, mark size=1.8pt},
                        error bar style={line width=0.35pt, draw=red, draw opacity=1},
                    ] table [x=z, y=MC_Mean, y error=MC_Std, col sep=space] {plot_data/thermal_epsilon_mc_comparison.dat};
                \end{pgfonlayer}
                \addlegendimage{legend image code/.code={\draw[blue, thick] (0cm,0cm) -- (0.6cm,0cm);}}
                \addlegendentry{Mean}
                \addlegendimage{legend image code/.code={\fill[blue!30, draw=blue!50] (0cm,-0.1cm) rectangle (0.6cm,0.1cm);}}
	                \addlegendentry{Mean $\pm$ 1 Std}
	                \addlegendimage{legend image code/.code={\draw[black, dashed, thick] (0cm,0cm) -- (0.6cm,0cm);}}
	                \addlegendentry{ATB}
		                \addlegendimage{legend image code/.code={%
		                    \draw[red, line width=0.35pt] (0.18cm,0cm) -- (0.42cm,0cm);
		                    \draw[red, line width=0.35pt] (0.18cm,-0.06cm) -- (0.18cm,0.06cm);
		                    \draw[red, line width=0.35pt] (0.42cm,-0.06cm) -- (0.42cm,0.06cm);
		                    \path[draw=red, fill=red] (0.3cm,0cm) circle[radius=0.03cm];
		                }}
		                \addlegendentry{MC Data}
            \end{axis}
        \end{tikzpicture}
    \end{subfigure}
    \hfill
    \begin{subfigure}[b]{0.49\textwidth}
        \centering
        \vspace{0.8cm}
        \raisebox{-0.25cm}{%
        \begin{tikzpicture}
            \begin{axis}[
                name=mainhist,
                width=\linewidth, height=6cm,
                xlabel={$\mathcal{G}(z=0.5)$},
                ylabel={Count},
                title={Distribution at $z=0.5$},
                ybar,
                bar width=0.06992879882973568,
                xmin=-225, xmax=90,
                ymin=0,
                ymajorgrids=true,
                xmajorgrids=false,
                legend pos=north west,
                clip=false,
                ylabel style={at={(axis description cs:1.10,0.5)}, anchor=south},
                axis on top=false,
                legend style={font=\sansmath\sffamily\tiny},
                scaled ticks=false,
                xticklabel style={
                    /pgf/number format/fixed,
                    /pgf/number format/precision=2,
                    font=\sansmath\sffamily\footnotesize,
                }
            ]
                \addplot [fill=blue!50, opacity=0.7, draw=black, forget plot] table [x=BinCenter, y=Count, col sep=space] {plot_data/thermal_epsilon_histogram_z0.500.dat};
                \addlegendimage{legend image code/.code={\fill[blue!50, opacity=0.7, draw=black] (0cm,-0.1cm) rectangle (0.18cm,0.18cm); \fill[blue!50, opacity=0.7, draw=black] (0.25cm,-0.1cm) rectangle (0.38cm,0.12cm);}}
                \addlegendentry{Model Predictions}
	                \draw [black, dashed, thick] (axis cs:28.34236328117764,\pgfkeysvalueof{/pgfplots/ymin}) -- (axis cs:28.34236328117764,\pgfkeysvalueof{/pgfplots/ymax});
	                \addlegendimage{legend image code/.code={\draw[black, dashed, thick] (0cm,0cm) -- (0.6cm,0cm);}}
	                \addlegendentry{ATB (28.342)}
                \draw [blue, thick] (axis cs:21.73089833365271,\pgfkeysvalueof{/pgfplots/ymin}) -- (axis cs:21.73089833365271,\pgfkeysvalueof{/pgfplots/ymax});
                \addlegendimage{legend image code/.code={\draw[blue, thick] (0cm,0cm) -- (0.6cm,0cm);}}
                \addlegendentry{Mean (21.731)}
                \draw [fill=red, fill opacity=0.14, draw=none] (axis cs:-51.73947663909554,\pgfkeysvalueof{/pgfplots/ymin}) rectangle (axis cs:84.4967449161641,\pgfkeysvalueof{/pgfplots/ymax});
                \addlegendimage{legend image code/.code={\fill[red, fill opacity=0.14, draw=red] (0cm,-0.1cm) rectangle (0.6cm,0.1cm);}}
                \addlegendentry{MC Mean (16.379) $\pm$ error}
                \draw [red, thick] (axis cs:16.37863413853428,\pgfkeysvalueof{/pgfplots/ymin}) -- (axis cs:16.37863413853428,\pgfkeysvalueof{/pgfplots/ymax});
                \coordinate (zoomcircle) at (axis cs:21.731,0);
            \end{axis}
            \begin{axis}[
                at={(mainhist.south west)},
                anchor=south west,
                xshift=35pt,
                yshift=20pt,
                width=0.46\linewidth,
                height=3cm,
                ybar,
                bar width=0.05034873515740969,
                xmin=20.93047331043197, xmax=22.32904928702671,
                ymin=0,
                ymajorgrids=true,
                xmajorgrids=false,
                scaled ticks=false,
                axis background/.style={fill=white, fill opacity=0.95},
                tick label style={font=\sansmath\sffamily\scriptsize},
                xticklabel style={
                    /pgf/number format/fixed,
                    /pgf/number format/precision=4,
                    font=\sansmath\sffamily\scriptsize,
                    yshift=0pt,
                },
	                yticklabel style={font=\sansmath\sffamily\scriptsize},
	                name=zoomaxis,
	            ]
	                \addplot [fill=blue!50, opacity=0.7, draw=black, line width=0.35pt, forget plot] table [x=BinCenter, y=Count, col sep=space] {plot_data/thermal_epsilon_histogram_z0.500.dat};
	                \draw [black, dashed, thick] (axis cs:28.34236328117764,\pgfkeysvalueof{/pgfplots/ymin}) -- (axis cs:28.34236328117764,\pgfkeysvalueof{/pgfplots/ymax});
	                \draw [blue, thick] (axis cs:21.73089833365271,\pgfkeysvalueof{/pgfplots/ymin}) -- (axis cs:21.73089833365271,\pgfkeysvalueof{/pgfplots/ymax});
	                \coordinate (targetNE) at (axis cs:22.32904928702671,\pgfkeysvalueof{/pgfplots/ymax});
	                \coordinate (targetSW) at (axis cs:20.93047331043197,\pgfkeysvalueof{/pgfplots/ymin});
	            \end{axis}
            \draw[gray!60, thick] (zoomcircle) circle (\zoomcircleradius);
            \path let \p1=(zoomcircle), \p2=(targetNE) in coordinate (connNE) at ($(\p1)!\zoomcircleradius!(\p2)$);
            \path let \p1=(zoomcircle), \p2=(targetSW) in coordinate (connSW) at ($(\p1)!\zoomcircleradius!(\p2)$);
            \draw[gray!60, line width=0.5pt] (connNE) -- (targetNE);
            \draw[gray!60, line width=0.5pt] (connSW) -- (targetSW);
        \end{tikzpicture}%
        }
    \end{subfigure}
	    \caption{NN-predicted results for thermal two-point function $\langle\epsilon\epsilon\rangle_\beta$ in the 3d Ising model with anchor point fixed at $z_0=0.05$. The left panel compares the ensemble mean against the analytic thermal bootstrap (ATB) approximation of Barrat et al.\ \cite{Barrat:2025nvu} and Monte Carlo (MC) data points with vertical error bars. The right panel shows the histogram of NN outputs at $z = 0.5$, together with the ATB prediction and MC mean and error band. A zoomed inset highlights the prediction cluster.  The NN prediction at $z = 0.5$ is $21.731\pm 0.337$.}
	    \label{fig:thermal_epsilon_summary}
	\end{figure}

The case of the thermal correlator $\langle \epsilon(0) \epsilon(z) \rangle_\beta$ can be studied with similar methods by setting $\mathcal{G}_\epsilon(z) = 1 + H_\epsilon(z)$ with $H_\epsilon(z) = z^{\Delta_\epsilon} (1-z)^{-2\Delta_\epsilon} \text{NN}_{\boldsymbol{\theta},\epsilon}(z)$. The results of the Anchored NN optimisation are summarised in Fig.~\ref{fig:thermal_epsilon_summary}. They are based on the mean and standard deviation of 100 runs with MS training loss of $(1.89\pm2.23)\times 10^{-4}$. Again, we observe that the NN optimisation yields sensible stable results that compare well with existing results in the literature. As an illustration, in this case, the analytic approximation and mean MC prediction, \cite{Barrat:2025nvu}, of the correlator at $z=0.5$ is 28.342 and 16.379 respectively. The anchored NN prediction is $21.731\pm 0.337$.  

\section{From the line to the plane}
\label{plane}

All examples discussed so far have been restricted to the diagonal kinematics $z=\bar z$, where the four-point function reduces to a single-variable problem. A natural and important question is whether the application of anchored NNs can be extended to the full kinematic domain $(z,\bar z)$, thereby recovering the complete four-point function on the plane. Here we propose, and test, such an extension, which does not require any further input beyond what is already needed for the anchored NN on the real line.

\subsection{Crossing on concentric circles}

On the full complex plane, the correlator $\GG(z,\bar z)$ depends on two real variables (or one complex variable and its conjugate) and the crossing equation~\eqref{introab} is a two-dimensional functional constraint. The most naive extension---training a two-input NN on a two-dimensional grid on the plane---is a more complex optimisation process where it is not immediately obvious if and how certain key aspects of the previous 1d methodology extend. For example, it is unclear how to choose anchor points and whether there is a prescription that yields stable predictions. 

A more structured approach, that stays close to the spirit of the previously employed 1d functional optimisation, exploits the fact that circles centred at the crossing-symmetric point $z=\bar z=\frac{1}{2}$ are invariant under the transformation $z\to 1-z$. More specifically, for any family of concentric circles
\begin{equation}
    \label{planeaa}
    z = \tfrac{1}{2} + R\,e^{i\alpha}\,,\qquad \bar z = \tfrac{1}{2} + R\,e^{-i\alpha}\,,\qquad \alpha\in[0,2\pi)\,,
\end{equation}
parametrised by the radius $R>0$, the crossing transformation $z\to 1-z$ translates to an angular shift $\alpha \to \alpha + \pi$ leaving the circle invariant. Therefore, on each such circle, the crossing equation becomes a constraint on a 1d function depending only on the angular variable $\alpha$. We can therefore explore the possibility of training an anchored NN on the circle. A nice property of this approach is that it does not require any new independent input of anchor point values beyond the anchor point needed for the bootstrap on the real line. As long as $0<R<\frac{1}{2}$, the circles \eqref{planeaa} intersect the real interval $(0,1)$ at two antipodal points (for $\alpha=0,\pi$). Assuming we have already solved for the correlator on $(0,1)$, we can now anchor the bootstrap on the circle on any of these two intersection points on the real interval. In this manner, we have an approach that decomposes the 2d problem of the correlator on a disc of radius $\frac{1}{2}$ centred at the crossing-symmetric point $z=\frac{1}{2}$ into a continuous family of 1d problems at separate values of $R$. Whether the spectral bias continues to reproduce CFT correlation functions is a priori unclear. In what follows, we provide encouraging preliminary evidence that it does!

The ensuing examples are not exhaustive. At this stage they are only meant to convey the promise of the approach. We can use the proposed method to make new predictions in all the cases analysed in the paper and more. We plan to return to a systematic exploration of this approach in the future and to study its implications in systems of physical interest.

\subsection{A technical point on implementation}

Having fixed a specific arbitrary value of the radius $R$ in $(0,\frac{1}{2})$, we are looking to predict the reduced correlator $\GG(z,\bar z)$ on the circle \eqref{planeaa}. Since $R$ is fixed, $\GG$ is effectively a 1d function of the angle $\alpha$, and we set accordingly $\GG(z,\bar z)=\GG(\alpha)$ keeping the value of $R$ implicit. In analogy to previous implementations on the real line, we set
\begin{equation}
    \label{planeba}
    \GG(\alpha) = L(\alpha) + H(\alpha)
\end{equation}
and solve crossing equations of the form
\begin{equation}
    \label{planebb}
    \GG(\alpha) = \left| \frac{z(\alpha)}{1-z(\alpha)}\right|^{2\Delta_\phi} \GG(\alpha + \pi)
    ~,
\end{equation}
where $\Delta_\phi$ is, as before, the scaling dimension of the external operator. $H(\alpha)$ is parametrised by a NN.

In many cases the correlator exhibits values that range over a few orders of magnitude. This can affect the quality of training, but it can be treated efficiently by training on $\log H$ (if $H$ is a strictly positive function), or ${\rm arcsinh}(\frac{H}{2})$ more generally. 

Moreover, since we are dealing with functions on a circle, the standard fully-connected architecture of an MLP with a tanh or GELU activation function is not very efficient. A key modification, which was found to produce very good results, involves replacing the raw input of the angle $\alpha$ by an explicit Fourier embedding before the first linear layer. For $N$ harmonics the embedding reads
\begin{equation}
  \beta(\alpha) = \bigl(\sin\alpha,\,\cos\alpha,\,\sin 2\alpha,\,\cos 2\alpha,\,
    \ldots,\,\sin N\alpha,\,\cos N\alpha\bigr) \in \mathbb{R}^{2N}
    ~.
  \label{eq:fourier-features}
\end{equation}
In this manner, the first linear layer has input dimension $2N$ instead of $1$ and all other layers are identical to a standard MLP. This architecture encodes efficiently the periodic structure of the problem and leads to faster implementation of the crossing symmetry. Practically, we have observed that better, more smooth outputs are produced for a single harmonic, $N = 1$, with tanh activation function, and this is the choice we implement in the examples below.

\subsection{GFF four-point functions}

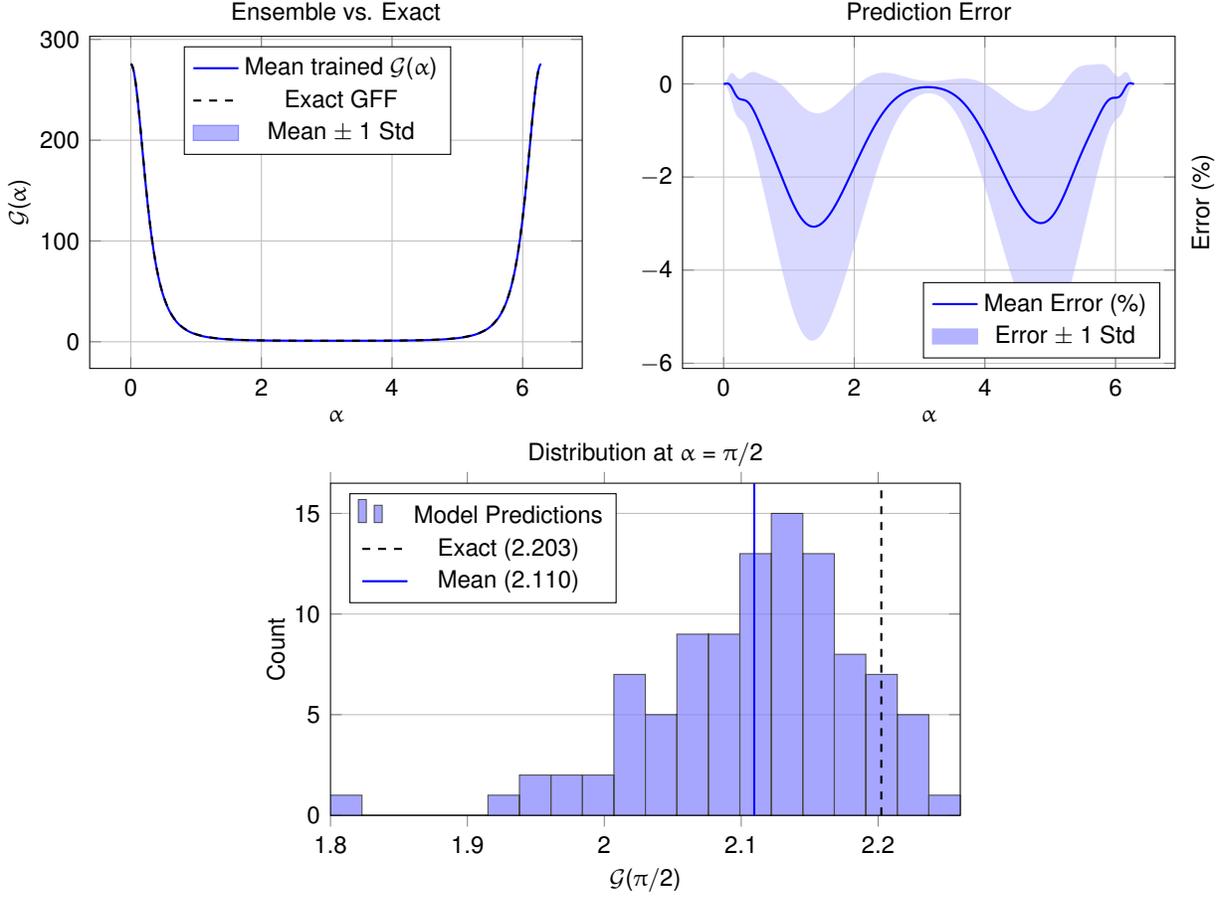
\begin{figure}[t!]
    \centering

    % Subfigure 1: Mean vs exact with std band
    \begin{subfigure}[b]{0.49\textwidth}
        \centering
        \begin{tikzpicture}
            \begin{axis}[
                width=\linewidth, height=6cm,
                xlabel={$\alpha$},
                ylabel={$\mathcal{G}(\alpha)$},
                title={Ensemble vs. Exact},
                grid=major,
                legend pos=north west,  legend style={xshift=30pt}
            ]

                % Mean
                \addplot [blue, thick]
                table [x=theta, y=mean_g, col sep=space]
                {plot_data/GFF_circle_means_1.618.dat};
                \addlegendentry{Mean trained $\mathcal{G}(\alpha)$}
                % Exact
                \addplot [black, dashed, thick]
                table [x=theta, y=exact_g, col sep=space]
                {plot_data/GFF_circle_means_1.618.dat};
                \addlegendentry{Exact GFF}
                % Uncertainty band
                \addplot [forget plot, name path=upper, draw=none]
                table [
                    x=theta,
                    y expr=\thisrow{mean_g}+\thisrow{std_g},
                    col sep=space
                ] {plot_data/GFF_circle_means_1.618.dat};

                \addplot [forget plot, name path=lower, draw=none]
                table [
                    x=theta,
                    y expr=\thisrow{mean_g}-\thisrow{std_g},
                    col sep=space
                ] {plot_data/GFF_circle_means_1.618.dat};

                \addplot [forget plot, fill=blue!30, fill opacity=0.5, draw=none]
                fill between [of=upper and lower];
                \addlegendimage{legend image code/.code={\fill[blue!30, draw=blue!50] (0cm,-0.1cm) rectangle (0.6cm,0.1cm);}}
                \addlegendentry{Mean $\pm$ 1 Std}

            \end{axis}
        \end{tikzpicture}
    \end{subfigure}
    \hfill
    % Subfigure 2: Percent error
    \begin{subfigure}[b]{0.49\textwidth}
        \centering
        \begin{tikzpicture}
            \begin{axis}[
                width=\linewidth, height=6cm,
                xlabel={$\alpha$},
                ylabel={Error (\%)},
                title={Prediction Error},
                grid=major,
                ylabel style={at={(axis description cs:1.10,0.5)}, anchor=south},
                legend pos=south east,
            ]
                % Mean percent error
                \addplot [blue, thick]
                table [x=theta, y expr =100*((\thisrow{mean_g}-\thisrow{exact_g})/(1+abs(\thisrow{exact_g})), col sep=space]
                {plot_data/GFF_circle_means_1.618.dat};
                \addlegendentry{Mean Error (\%)}

                % Error band
                \addplot [forget plot, name path=upper, draw=none]
                table [
                    x=theta,
                    y expr=100*((\thisrow{mean_g}+\thisrow{std_g})-\thisrow{exact_g})/(1+abs(\thisrow{exact_g})),
                    col sep=space
                ] {plot_data/GFF_circle_means_1.618.dat};

                \addplot [forget plot, name path=lower, draw=none]
                table [
                    x=theta,
                    y expr=100*((\thisrow{mean_g}-\thisrow{std_g})-\thisrow{exact_g})/(1+abs(\thisrow{exact_g})),
                    col sep=space
                ] {plot_data/GFF_circle_means_1.618.dat};

                \addplot [forget plot, fill=blue!30, fill opacity=0.5, draw=none]
                fill between [of=upper and lower];
                \addlegendimage{legend image code/.code={\fill[blue!30] (0cm,-0.1cm) rectangle (0.6cm,0.1cm);}}
                \addlegendentry{Error $\pm$ 1 Std}
            \end{axis}
        \end{tikzpicture}
    \end{subfigure}
    \hfill
    % Subfigure 3: Histogram at pi/2
\begin{subfigure}[b]{0.60\textwidth}
    \centering
    \begin{tikzpicture}
        \begin{axis}[
            width=\linewidth, height=6cm,
            xlabel={$\mathcal{G}(\pi/2)$},
            ylabel={Count},
            title={Distribution at $\alpha=\pi/2$},
            ymin=0,
            xmin=1.80, xmax=2.26,
            ybar,
            ymajorgrids=true,
            xmajorgrids=false,
            legend pos=north west,
            scaled ticks=false,
            xtick={1.8,1.9,2.0,2.1,2.2},
            xticklabel style={
                /pgf/number format/fixed,
                /pgf/number format/precision=1,
                font=\sansmath\sffamily\footnotesize,
            }
        ]

            % True histogram from raw samples
            \addplot [forget plot,
                fill=blue!50,
                draw=black,
                opacity=0.7,
                hist={
                    bins=20,
                    data min=1.80,
                    data max=2.26
                }
            ] table[y=pi2_prediction_g, col sep=space]
            {plot_data/GFF_circle_pi2_preds_1.618.dat};
            \addlegendimage{ybar,ybar legend,fill=blue!50, opacity  = 0.7, draw=black}
            \addlegendentry{Model Predictions};
            % Exact vertical line
            \draw[black, dashed, thick]
(axis cs:2.202340,0) -- (axis cs:2.202340,\pgfkeysvalueof{/pgfplots/ymax});
            \addlegendimage{legend image code/.code={\draw[black, dashed, thick] (0cm,0cm)--(0.6cm,0cm);}}
\addlegendentry{Exact ($2.203$)}
            \draw[blue, thick]
(axis cs:2.10951,0) -- (axis cs:2.10951,\pgfkeysvalueof{/pgfplots/ymax});

\addlegendimage{legend image code/.code={\draw[blue, thick] (0cm,0cm)--(0.6cm,0cm);}}
\addlegendentry{Mean ($2.110$)}
        \end{axis}
    \end{tikzpicture}
\end{subfigure}
    \caption{NN-predicted results for the GFF four-point function on the circle $z=0.5+0.35e^{i\alpha}$. The anchor point is fixed at $\alpha_{0} = \pi$. These results are compared to the exact correlator \eqref{planeca}. The bottom plot presents a histogram of the NN predictions at $\alpha=\pi/2$ with mean $2.110\pm 0.075$.}
    \label{fig:gff_circle_summary}
\end{figure}

As a first example, the concentric-circle approach was implemented and tested on the bosonic GFF, where the exact answer is 
\begin{equation}
    \label{planeca}
    \GG(z,\bar z)=1+(z\bar z)^{\Df}+\left(\frac{z\bar z}{(1-z)(1-\bar z)}\right)^{\Df}
    ~.
\end{equation}
A set of representative results are presented in Fig.\ \ref{fig:gff_circle_summary}. In this set of runs we set $R=0.35$. The highest deviation from the exact result occurs at $\alpha=\frac{\pi}{2}$, where the difference is still small (2.203 from the exact correlator and $2.110\pm 0.075$ from the anchored NN prediction). Similar results were obtained for other values of $R$ as well.

\subsection{2d minimal model four-point functions}

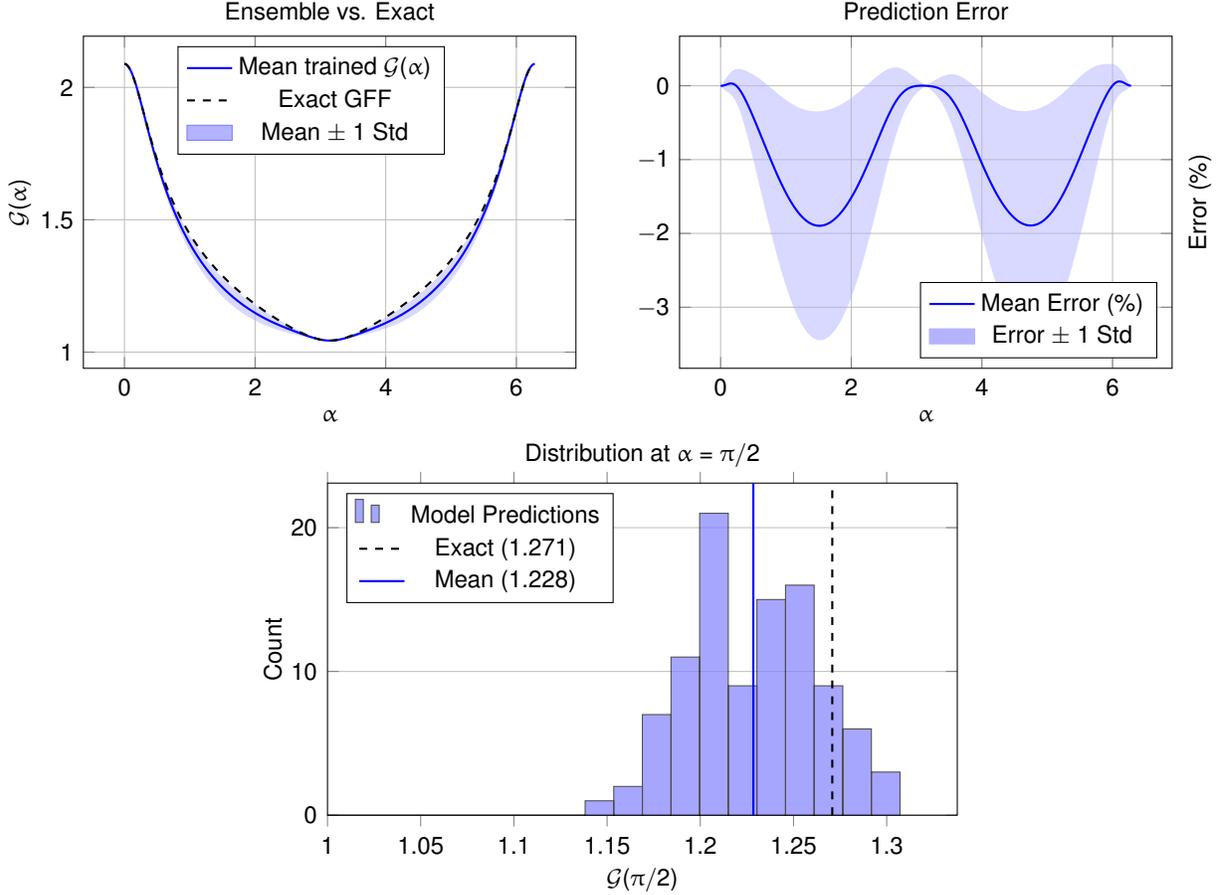
\begin{figure}[ht!]
    \centering

    % Subfigure 1: Mean vs exact with std band
    \begin{subfigure}[b]{0.49\textwidth}
        \centering
        \begin{tikzpicture}
            \begin{axis}[
                width=\linewidth, height=6cm,
                xlabel={$\alpha$},
                ylabel={$\mathcal{G}(\alpha)$},
                title={Ensemble vs. Exact},
                grid=major,
                legend pos=north west, legend style={xshift=30pt}
            ]

                % Mean
                \addplot [blue, thick]
                table [x=theta, y=mean_g, col sep=space]
                {plot_data/MinimalModel_M4_5_circle_means.dat};
                \addlegendentry{Mean trained $\mathcal{G}(\alpha)$}
                % Exact
                \addplot [black, dashed, thick]
                table [x=theta, y=exact_g, col sep=space]
                {plot_data/MinimalModel_M4_5_circle_means.dat};
                \addlegendentry{Exact GFF}
                % Uncertainty band
                \addplot [forget plot, name path=upper, draw=none]
                table [
                    x=theta,
                    y expr=\thisrow{mean_g}+\thisrow{std_g},
                    col sep=space
                ] {plot_data/MinimalModel_M4_5_circle_means.dat};

                \addplot [forget plot, name path=lower, draw=none]
                table [
                    x=theta,
                    y expr=\thisrow{mean_g}-\thisrow{std_g},
                    col sep=space
                ] {plot_data/MinimalModel_M4_5_circle_means.dat};

                \addplot [forget plot, fill=blue!30, fill opacity=0.5, draw=none]
                fill between [of=upper and lower];
                \addlegendimage{legend image code/.code={\fill[blue!30, draw=blue!50] (0cm,-0.1cm) rectangle (0.6cm,0.1cm);}}
                \addlegendentry{Mean $\pm$ 1 Std}
            \end{axis}
        \end{tikzpicture}
    \end{subfigure}
    \hfill
    % Subfigure 2: Percent error
    \begin{subfigure}[b]{0.49\textwidth}
        \centering
        \begin{tikzpicture}
            \begin{axis}[
                width=\linewidth, height=6cm,
                xlabel={$\alpha$},
                ylabel={Error (\%)},
                title={Prediction Error},
                grid=major,
                ylabel style={at={(axis description cs:1.10,0.5)}, anchor=south},
                legend pos=south east
            ]
                % Mean percent error
                \addplot [blue, thick]
                table [x=theta, y=prediction_error_pct, col sep=space]
                {plot_data/MinimalModel_M4_5_circle_means.dat};
                \addlegendentry{Mean Error (\%)}

                % Error band
                \addplot [forget plot, name path=upper, draw=none]
                table [
                    x=theta,
y expr={\thisrow{prediction_error_pct} + 100*\thisrow{std_g}/(1 + abs(\thisrow{mean_g}))},
                    col sep=space
                ] {plot_data/MinimalModel_M4_5_circle_means.dat};

                \addplot [forget plot, name path=lower, draw=none]
                table [
                    x=theta,
y expr={\thisrow{prediction_error_pct} - 100*\thisrow{std_g}/(1 + abs(\thisrow{mean_g}))},
                    col sep=space
                ] {plot_data/MinimalModel_M4_5_circle_means.dat};

                \addplot [forget plot, fill=blue!30, fill opacity=0.5, draw=none]
                fill between [of=upper and lower];
                \addlegendimage{legend image code/.code={\fill[blue!30] (0cm,-0.1cm) rectangle (0.6cm,0.1cm);}}
                \addlegendentry{Error $\pm$ 1 Std}

            \end{axis}
        \end{tikzpicture}
    \end{subfigure}
    \hfill
    % Subfigure 3: Histogram at pi/2
\begin{subfigure}[b]{0.60\textwidth}
    \centering
    \begin{tikzpicture}
        \begin{axis}[
            width=\linewidth, height=6cm,
            xlabel={$\mathcal{G}(\pi/2)$},
            ylabel={Count},
            title={Distribution at $\alpha=\pi/2$},
            ymin=0,
            ybar,
            xmin = 1,
            ymajorgrids=true,
            xmajorgrids=false,
            legend pos=north west,
            scaled ticks=false,
            xticklabel style={
                /pgf/number format/fixed,
                /pgf/number format/precision=2,
                font=\sansmath\sffamily\footnotesize,
            }
        ]

            % True histogram from raw samples
            \addplot [forget plot,
                fill=blue!50,
                draw=black,
                opacity=0.7,
                hist={
                    bins=20
                }
            ] table[y=pi2_prediction_g, col sep=space]
            {plot_data/MinimalModel_M4_5_circle_pi2_preds.dat};
            \addlegendimage{ybar,ybar legend,fill=blue!50, opacity  = 0.7, draw=black}
            \addlegendentry{Model Predictions};
            % Exact vertical line
            \draw[black, dashed, thick]
(axis cs:1.270811,0) -- (axis cs:1.270811,\pgfkeysvalueof{/pgfplots/ymax});
            \addlegendimage{legend image code/.code={\draw[black, dashed, thick] (0cm,0cm)--(0.6cm,0cm);}}
\addlegendentry{Exact ($1.271$)}
            \draw[blue, thick]
(axis cs:1.22846,0) -- (axis cs:1.22846,\pgfkeysvalueof{/pgfplots/ymax});

\addlegendimage{legend image code/.code={\draw[blue, thick] (0cm,0cm)--(0.6cm,0cm);}}
\addlegendentry{Mean ($1.228$)}
        \end{axis}
    \end{tikzpicture}
\end{subfigure}
    \caption{NN-predicted results for reduced four-point function $\langle \phi_{1,2}\phi_{1,2}\phi_{1,2}\phi_{1,2}\rangle$ in 2d minimal models on the circle $z=0.5+0.35e^{i\alpha}$. The anchor point is fixed at $\alpha_{0} = \pi$. These results are compared to the exact correlator \eqref{minaa}-\eqref{minac}. The bottom plot presents a histogram of the NN predictions at $\alpha=\pi/2$ with mean $1.228\pm 0.035$.}
    \label{fig:mmT_circle_summary}
\end{figure}

The exact reduced four-point function $\langle \phi_{1,2}\phi_{1,2}\phi_{1,2}\phi_{1,2}\rangle$ in 2d minimal models appears in Eqs.\ \eqref{minaa}-\eqref{minac} in Section \ref{minimalunitary}. As an example, here we apply the concentric-circle approach to the unitary minimal model $\MM(4,5)$ at $R=0.35$. The results of 100 independent runs are summarised in Fig.\ \ref{fig:mmT_circle_summary}.

\subsection{GFF thermal two-point functions}

\begin{figure}[ht]
    \centering

    % Subfigure 1: Mean vs exact with std band
    \begin{subfigure}[b]{0.49\textwidth}
        \centering
        \begin{tikzpicture}
            \begin{axis}[
                width=\linewidth, height=6cm,
                xlabel={$\alpha$},
                ylabel={$\mathcal{G}(\alpha)$},
                title={Ensemble vs. Exact},
                grid=major,
                legend pos=north west, legend style={xshift=30pt}
            ]

                % Mean
                \addplot [blue, thick]
                table [x=theta, y=mean_g, col sep=space]
                {plot_data/FiniteT_GFF_circle_means_deltaphi1.618_Ncut1000.dat};
                \addlegendentry{Mean trained $\mathcal{G}(\alpha)$}
                % Exact
                \addplot [black, dashed, thick]
                table [x=theta, y=exact_g, col sep=space]
                {plot_data/FiniteT_GFF_circle_means_deltaphi1.618_Ncut1000.dat};
                \addlegendentry{Exact GFF}
                % Uncertainty band
                \addplot [forget plot, name path=upper, draw=none]
                table [
                    x=theta,
                    y expr=\thisrow{mean_g}+\thisrow{std_g},
                    col sep=space
                ] {plot_data/FiniteT_GFF_circle_means_deltaphi1.618_Ncut1000.dat};

                \addplot [forget plot, name path=lower, draw=none]
                table [
                    x=theta,
                    y expr=\thisrow{mean_g}-\thisrow{std_g},
                    col sep=space
                ] {plot_data/FiniteT_GFF_circle_means_deltaphi1.618_Ncut1000.dat};

                \addplot [forget plot, fill=blue!30, fill opacity=0.5, draw=none]
                fill between [of=upper and lower];
                \addlegendimage{legend image code/.code={\fill[blue!30, draw=blue!50] (0cm,-0.1cm) rectangle (0.6cm,0.1cm);}}
                \addlegendentry{Mean $\pm$ 1 Std}

            \end{axis}
        \end{tikzpicture}
    \end{subfigure}
    \hfill
    % Subfigure 2: Percent error
    \begin{subfigure}[b]{0.49\textwidth}
        \centering
        \begin{tikzpicture}
            \begin{axis}[
                width=\linewidth, height=6cm,
                xlabel={$\alpha$},
                ylabel={Error (\%)},
                title={Prediction Error},
                grid=major,
                ylabel style={at={(axis description cs:1.10,0.5)}, anchor=south},
                legend pos=south east,
            ]
                % Mean percent error
                \addplot [blue, thick]
                table [x=theta, y expr =100*((\thisrow{mean_g}-\thisrow{exact_g})/(1+abs(\thisrow{exact_g})), col sep=space]
                {plot_data/FiniteT_GFF_circle_means_deltaphi1.618_Ncut1000.dat};
                \addlegendentry{Mean Error (\%)}

                % Error band
                \addplot [forget plot, name path=upper, draw=none]
                table [
                    x=theta,
                    y expr=100*((\thisrow{mean_g}+\thisrow{std_g})-\thisrow{exact_g})/(1+abs(\thisrow{exact_g})),
                    col sep=space
                ] {plot_data/FiniteT_GFF_circle_means_deltaphi1.618_Ncut1000.dat};

                \addplot [forget plot, name path=lower, draw=none]
                table [
                    x=theta,
                    y expr=100*((\thisrow{mean_g}-\thisrow{std_g})-\thisrow{exact_g})/(1+abs(\thisrow{exact_g})),
                    col sep=space
                ] {plot_data/FiniteT_GFF_circle_means_deltaphi1.618_Ncut1000.dat};

                \addplot [forget plot, fill=blue!30, fill opacity=0.5, draw=none]
                fill between [of=upper and lower];
                \addlegendimage{legend image code/.code={\fill[blue!30] (0cm,-0.1cm) rectangle (0.6cm,0.1cm);}}
                \addlegendentry{Error $\pm$ 1 Std}

            \end{axis}
        \end{tikzpicture}
    \end{subfigure}
    \hfill
    % Subfigure 3: Histogram at pi/2
\begin{subfigure}[b]{0.60\textwidth}
    \centering
    \begin{tikzpicture}
        \begin{axis}[
            width=\linewidth, height=6cm,
            xlabel={$\mathcal{G}(\pi/2)$},
            ylabel={Count},
            title={Distribution at $\alpha=\pi/2$},
            ymin=0,
            xmin=1.80, xmax=2.26,
            ybar,
            ymajorgrids=true,
            xmajorgrids=false,
            legend pos=north west,
            scaled ticks=false,
            xtick={1.8,1.9,2.0,2.1,2.2},
            xticklabel style={
                /pgf/number format/fixed,
                /pgf/number format/precision=1,
                font=\sansmath\sffamily\footnotesize,
            }
        ]

            % True histogram from raw samples
            \addplot [forget plot,
                fill=blue!50,
                draw=black,
                opacity=0.7,
                hist={
                    bins=20,
                    data min=1.80,
                    data max=2.26
                }
            ] table[y=pi2_prediction_g, col sep=space]
            {plot_data/FiniteT_GFF_circle_pi2_preds_deltaphi1.618_Ncut1000.dat};
            \addlegendimage{ybar,ybar legend,fill=blue!50, opacity  = 0.7, draw=black}
            \addlegendentry{Model Predictions};
            % Exact vertical line
            \draw[black, dashed, thick]
(axis cs:2.135092,0) -- (axis cs:2.135092,\pgfkeysvalueof{/pgfplots/ymax});
            \addlegendimage{legend image code/.code={\draw[black, dashed, thick] (0cm,0cm)--(0.6cm,0cm);}}
\addlegendentry{Exact ($2.135$)}
            \draw[blue, thick]
(axis cs:2.10896897,0) -- (axis cs:2.10896897,\pgfkeysvalueof{/pgfplots/ymax});

\addlegendimage{legend image code/.code={\draw[blue, thick] (0cm,0cm)--(0.6cm,0cm);}}
\addlegendentry{Mean ($2.109$)}
        \end{axis}
    \end{tikzpicture}
\end{subfigure}
    \caption{NN-predicted results for the GFF thermal two-point function on the circle $z=0.5+0.35e^{i\alpha}$ at $\Delta_\phi=1.618$. The anchor point is fixed at $\alpha_{0} = \pi$. These results are compared to the exact correlator \eqref{planeea}. The bottom plot presents a histogram of the NN predictions at $\alpha=\pi/2$ with mean $2.109\pm 0.075$.}
    \label{fig:gffT_circle_summary}
\end{figure}
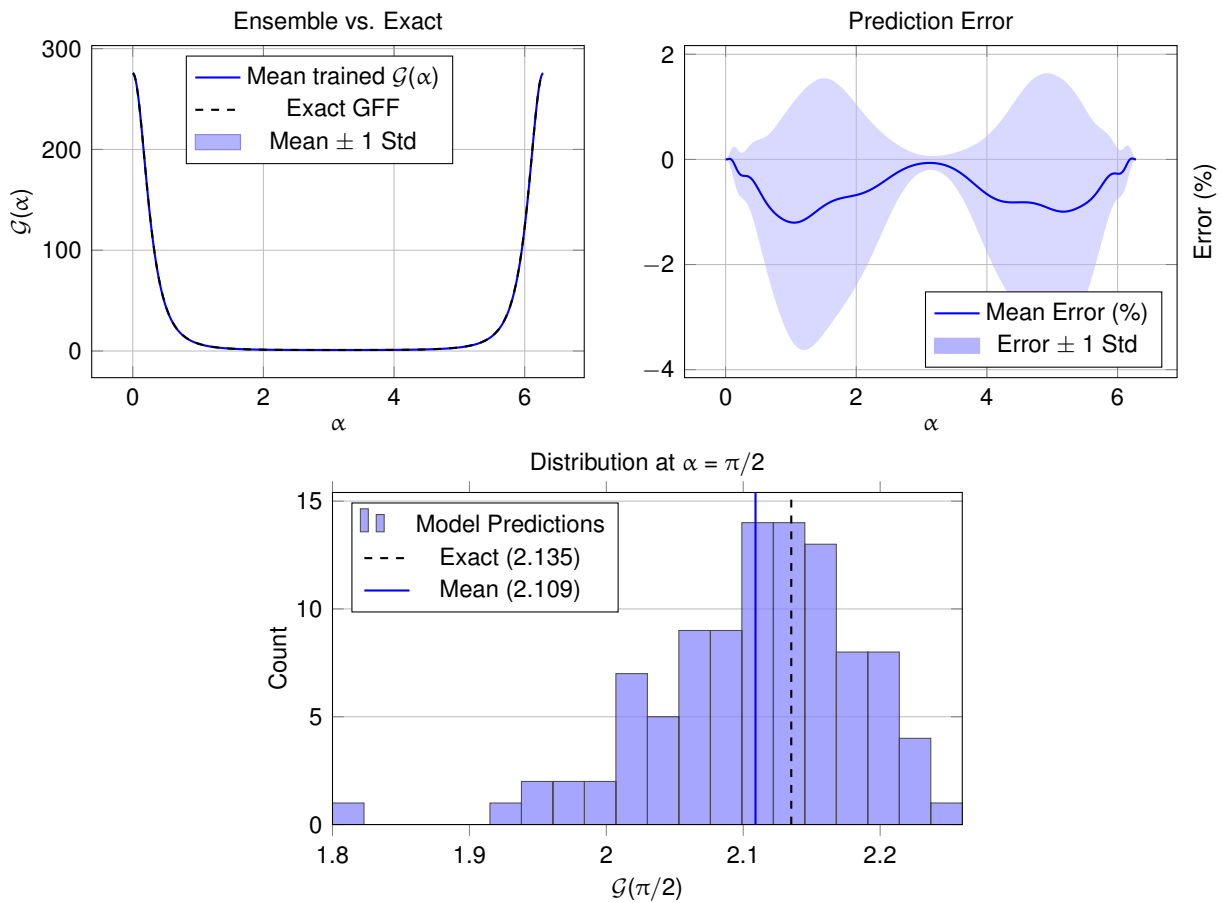

The exact thermal two-point function on $S^1_\beta \times \IR$ (for $\beta=1$) is
\begin{equation}
    \label{planeea}
    \GG(z,\bar z) = \sum_{n=-\infty}^\infty \left(\frac{z\bar z}{(n-z)(n-\bar z)}\right)^{\Delta_\phi}
    ~.
\end{equation}
As an example, we set again $\Delta_\phi=1.618$. For this case, results based on anchored neural network optimisation on a circle at $R=0.35$ are presented in Fig.\ \ref{fig:gffT_circle_summary}. The NN predictions continue to compare well with the exact correlator, with the largest deviations occurring in the vicinity of $\alpha=\frac{\pi}{2}$.

\subsection{3d Ising thermal two-point functions}

\begin{figure}[ht]
    \centering

    % Subfigure 1: Mean vs exact with std band
    \begin{subfigure}[b]{0.49\textwidth}
        \centering
        \begin{tikzpicture}
            \begin{axis}[
                width=\linewidth, height=6cm,
                xlabel={$\alpha$},
                ylabel={$\mathcal{G}(\alpha)$},
                title={Ensemble},
                grid=major,
                legend pos=north west, legend style={xshift=30pt},
                scaled y ticks=false,
                yticklabel style={
                    /pgf/number format/fixed,
                    /pgf/number format/precision=3,
                    font=\sansmath\sffamily\footnotesize,
                }
            ]

                % Mean
                \addplot [blue, thick]
                table [x=theta, y=mean_g, col sep=space]
                {plot_data/3d_Ising_finiteT_sigma_z00.15_circle_means.dat};
                \addlegendentry{Mean trained $\mathcal{G}(\alpha)$}
                % Uncertainty band
                \addplot [forget plot, name path=upper, draw=none]
                table [
                    x=theta,
                    y expr=\thisrow{mean_g}+\thisrow{std_g},
                    col sep=space
                ] {plot_data/3d_Ising_finiteT_sigma_z00.15_circle_means.dat};

                \addplot [forget plot, name path=lower, draw=none]
                table [
                    x=theta,
                    y expr=\thisrow{mean_g}-\thisrow{std_g},
                    col sep=space
                ] {plot_data/3d_Ising_finiteT_sigma_z00.15_circle_means.dat};

                \addplot [forget plot, fill=blue!30, fill opacity=0.5, draw=none]
                fill between [of=upper and lower];
                \addlegendimage{legend image code/.code={\fill[blue!30, draw=blue!50] (0cm,-0.1cm) rectangle (0.6cm,0.1cm);}}
                \addlegendentry{Mean $\pm$ 1 Std}

            \end{axis}
        \end{tikzpicture}
    \end{subfigure}%
    \hfill
    \begin{subfigure}[b]{0.49\textwidth}
        \centering
        \begin{tikzpicture}
            \begin{axis}[
                width=\linewidth, height=6cm,
                xlabel={$\alpha$},
                ylabel={Std},
                title={Standard Deviation},
                grid=major,
                ylabel style={at={(axis description cs:1.10,0.5)}, anchor=south},
                scaled y ticks=false,
                yticklabel style={
                    /pgf/number format/fixed,
                    /pgf/number format/precision=3,
                    font=\sansmath\sffamily\footnotesize,
                }
            ]

                % Mean
                \addplot [blue, thick]
                table [x=theta, y=std_g, col sep=space]
                {plot_data/3d_Ising_finiteT_sigma_z00.15_circle_means.dat};
                % Uncertainty band
            \end{axis}
        \end{tikzpicture}
    \end{subfigure}%
    \hfill

\begin{subfigure}[b]{0.60\textwidth}
    \centering
    \begin{tikzpicture}
	        \begin{axis}[
	            width=\linewidth, height=6cm,
	            xlabel={$\mathcal{G}(\pi/2)$},
	            ylabel={Count},
	            title={Distribution at $\alpha=\pi/2$},
	            ymin=0,
	            xmin=1.54, xmax=1.84,
	            ybar,
	            ymajorgrids=true,
	            xmajorgrids=false,
	            legend pos=north east,
	            scaled ticks=false,
		            xtick={1.54,1.58,1.62,1.66,1.70,1.74,1.78,1.82},
	            xticklabel style={
	                /pgf/number format/fixed,
	                /pgf/number format/precision=3,
	                font=\sansmath\sffamily\footnotesize,
            }
        ]

            % True histogram from raw samples
            \addplot [forget plot,
                fill=blue!50,
                draw=black,
                opacity=0.7,
	                hist={
	                    bins=20,
	                    data min=1.54,
	                    data max=1.84
	                }
	            ] table[y=pi2_prediction_g, col sep=space]
	            {plot_data/3d_Ising_finiteT_sigma_z00.15_circle_pi2_preds.dat};
            \addlegendimage{ybar,ybar legend,fill=blue!50, opacity  = 0.7, draw=black}
	            \addlegendentry{Model Predictions};
	            % Mean vertical line
	            \draw[blue, thick]
(axis cs:1.68806948,0) -- (axis cs:1.68806948,\pgfkeysvalueof{/pgfplots/ymax});
	
	\addlegendimage{legend image code/.code={\draw[blue, thick] (0cm,0cm)--(0.6cm,0cm);}}
	\addlegendentry{Mean ($1.688$)}
	        \end{axis}
	    \end{tikzpicture}
	\end{subfigure}
	    \caption{NN-predicted results for the 3d Ising thermal two-point function $\langle\sigma\sigma\rangle_{\beta}$ on the circle $z=0.5+0.35e^{i\alpha}$. The anchor point is fixed at $\alpha_{0} = \pi$ using \eqref{anchorcircleising}. As there are no analytic results to compare with, the top right figure instead plots the standard deviation of the NN prediction around the circle. The bottom plot presents a histogram of the NN predictions at $\alpha=\pi/2$ with mean $1.6881\pm 0.0531$.}
	    \label{fig:3d_Ising_ss_T_circle_summary}
\end{figure}
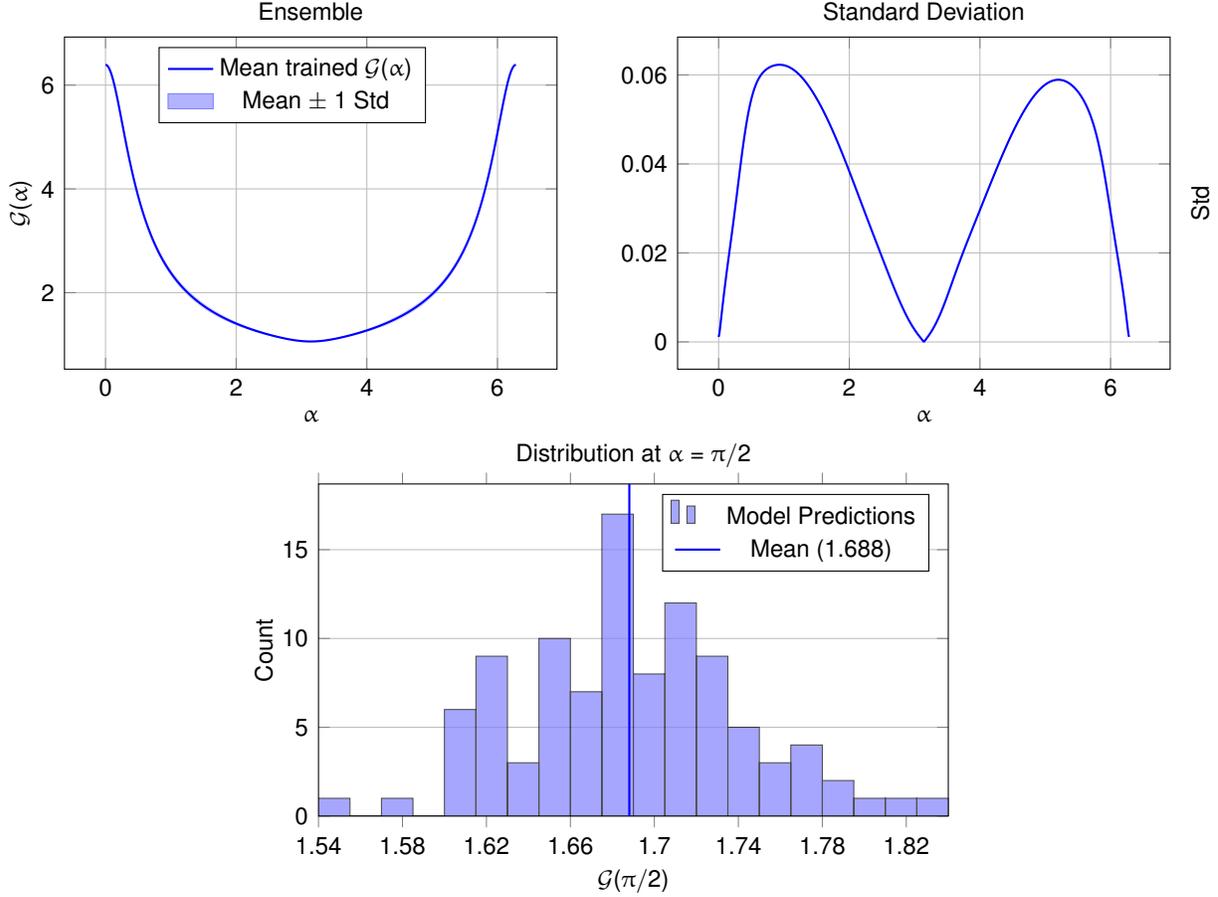

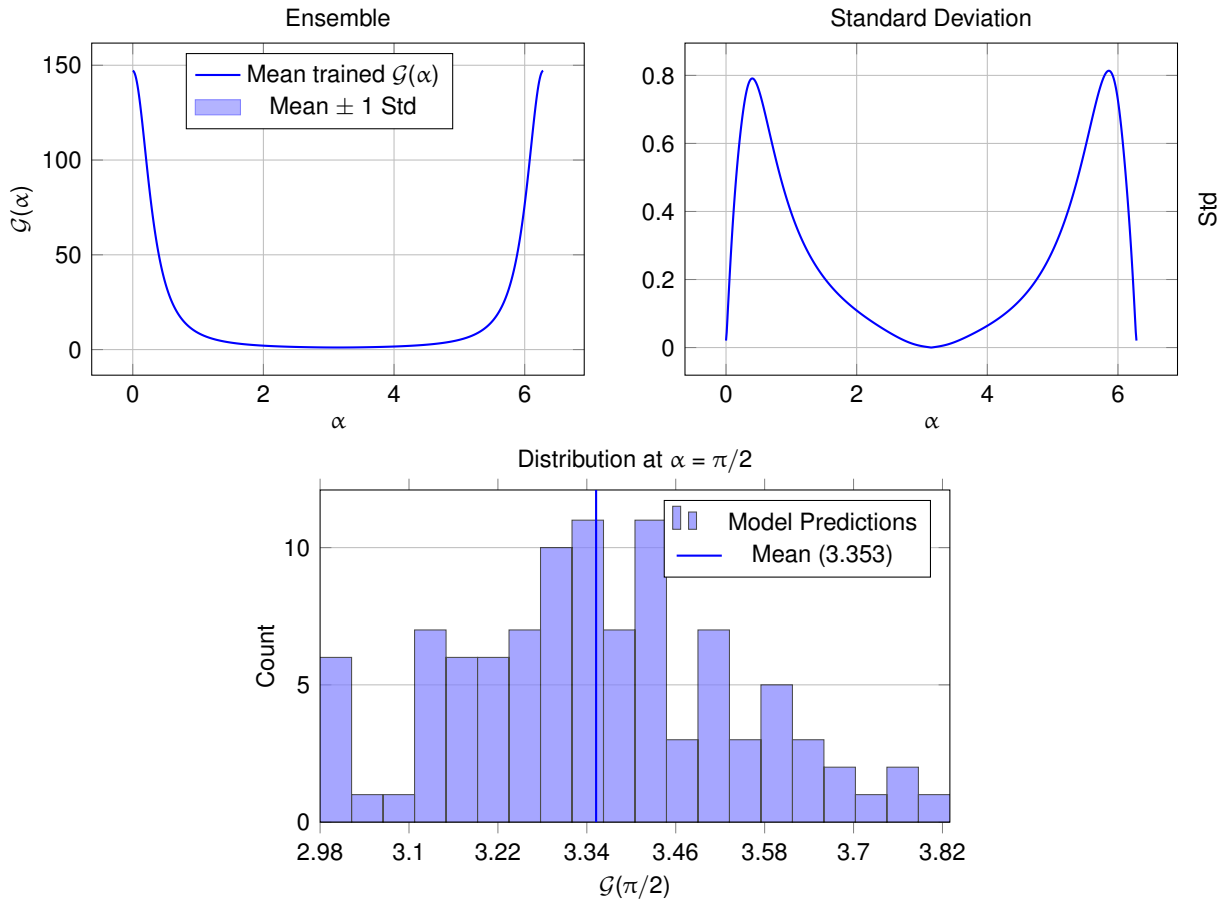
\begin{figure}[ht]
    \centering

    % Subfigure 1: Mean vs exact with std band
    \begin{subfigure}[b]{0.49\textwidth}
        \centering
        \begin{tikzpicture}
            \begin{axis}[
                width=\linewidth, height=6cm,
                xlabel={$\alpha$},
                ylabel={$\mathcal{G}(\alpha)$},
                title={Ensemble},
                grid=major,
                legend pos=north west, legend style={xshift=30pt},
                scaled y ticks=false,
                yticklabel style={
                    /pgf/number format/fixed,
                    /pgf/number format/precision=3,
                    font=\sansmath\sffamily\footnotesize,
                }
            ]

                % Mean
                \addplot [blue, thick]
                table [x=theta, y=mean_g, col sep=space]
                {plot_data/3d_Ising_finiteT_epsilon_z00.15_circle_means.dat};
                \addlegendentry{Mean trained $\mathcal{G}(\alpha)$}
                % Uncertainty band
                \addplot [forget plot, name path=upper, draw=none]
                table [
                    x=theta,
                    y expr=\thisrow{mean_g}+\thisrow{std_g},
                    col sep=space
                ] {plot_data/3d_Ising_finiteT_epsilon_z00.15_circle_means.dat};

                \addplot [forget plot, name path=lower, draw=none]
                table [
                    x=theta,
                    y expr=\thisrow{mean_g}-\thisrow{std_g},
                    col sep=space
                ] {plot_data/3d_Ising_finiteT_epsilon_z00.15_circle_means.dat};

                \addplot [forget plot, fill=blue!30, fill opacity=0.5, draw=none]
                fill between [of=upper and lower];
                \addlegendimage{legend image code/.code={\fill[blue!30, draw=blue!50] (0cm,-0.1cm) rectangle (0.6cm,0.1cm);}}
                \addlegendentry{Mean $\pm$ 1 Std}

            \end{axis}
        \end{tikzpicture}
    \end{subfigure}%
    \hfill
    \begin{subfigure}[b]{0.49\textwidth}
        \centering
        \begin{tikzpicture}
            \begin{axis}[
                width=\linewidth, height=6cm,
                xlabel={$\alpha$},
                ylabel={Std},
                title={Standard Deviation},
                grid=major,
                ylabel style={at={(axis description cs:1.10,0.5)}, anchor=south},
                legend pos=north west, legend style={xshift=50pt},
                scaled y ticks=false,
                yticklabel style={
                    /pgf/number format/fixed,
                    /pgf/number format/precision=3,
                    font=\sansmath\sffamily\footnotesize,
                }
            ]

                % Mean
                \addplot [blue, thick]
                table [x=theta, y=std_g, col sep=space]
                {plot_data/3d_Ising_finiteT_epsilon_z00.15_circle_means.dat};
            \end{axis}
        \end{tikzpicture}
    \end{subfigure}%
    \hfill
\begin{subfigure}[b]{0.60\textwidth}
    \centering
    \begin{tikzpicture}
	        \begin{axis}[
	            width=\linewidth, height=6cm,
	            xlabel={$\mathcal{G}(\pi/2)$},
	            ylabel={Count},
	            title={Distribution at $\alpha=\pi/2$},
	            ymin=0,
	            xmin=2.98, xmax=3.83,
	            ybar,
	            ymajorgrids=true,
	            xmajorgrids=false,
	            legend pos=north east,
	            scaled ticks=false,
	            xtick={2.98,3.10,3.22,3.34,3.46,3.58,3.70,3.82},
	            xticklabel style={
	                /pgf/number format/fixed,
	                /pgf/number format/precision=3,
	                font=\sansmath\sffamily\footnotesize,
            }
        ]

            % True histogram from raw samples
	            \addplot [forget plot,
	                fill=blue!50,
	                draw=black,
	                opacity=0.7,
	                hist={
	                    bins=20,
	                    data min=2.98,
	                    data max=3.83
	                }
	            ] table[y=pi2_prediction_g, col sep=space]
	            {plot_data/3d_Ising_finiteT_epsilon_z00.15_circle_pi2_preds.dat};
            \addlegendimage{ybar,ybar legend,fill=blue!50, opacity  = 0.7, draw=black}
            \addlegendentry{Model Predictions};
	            % Mean vertical line
	            \draw[blue, thick]
(axis cs:3.35252038,0) -- (axis cs:3.35252038,\pgfkeysvalueof{/pgfplots/ymax});

\addlegendimage{legend image code/.code={\draw[blue, thick] (0cm,0cm)--(0.6cm,0cm);}}
\addlegendentry{Mean ($3.353$)}
	        \end{axis}
	    \end{tikzpicture}
\end{subfigure}
		    \caption{NN-predicted results for the 3d Ising thermal two-point function $\langle\epsilon\epsilon\rangle_{\beta}$ on the circle $z=0.5+0.35e^{i\alpha}$. The anchor point is fixed at $\alpha_{0} = \pi$ using \eqref{anchorcircleising}. As there are no analytic results to compare with, the top right figure instead plots the standard deviation of the NN prediction around the circle. The bottom plot presents a histogram of the NN predictions at $\alpha=\pi/2$ with mean $3.353\pm 0.189$.}
		    \label{fig:3d_Ising_ee_T_circle_summary}
\end{figure}

As a final example, we consider the 3d Ising thermal two-point functions $\langle \sigma \sigma\rangle$, $\langle \epsilon \epsilon\rangle$ on the plane. There are no analytic or numerical results in the literature for these correlators to the best of knowledge, and the predictions we present here are genuinely new. For concreteness, we set again $R=0.35$ and use as anchor points the values of the correlators on the real line obtained in Sections \ref{sigsig} and \ref{epep}; from these runs we find that our anchor points on the circle can be taken as
\begin{equation}
    H_{\sigma}(z=0.15) = 0.06003\pm 0.00008,\quad H_{\epsilon}(z=0.15) = 0.09530\pm 0.00177.
    \label{anchorcircleising}
\end{equation}
The results for the $\langle \sigma \sigma\rangle_{\beta}$ correlator are summarised in Fig.\ \ref{fig:3d_Ising_ss_T_circle_summary} and the results for $\langle \epsilon \epsilon\rangle_{\beta}$ in Fig.\ \ref{fig:3d_Ising_ee_T_circle_summary}. In these cases, we present only the predictions without any comparison to known results.

\section{An alternative to NNs: Chebyshev--Tikhonov fits}
\label{chebyshev}

The discussion in Sections \ref{NNbias}, \ref{smooth} and the subsequent observations in a range of different examples, reinforce the key role of correlator smoothness. In Section \ref{smooth} we have already discussed how smoothness is captured by the structure of the Chebyshev spectrum in 1d correlators. Motivated by that analysis, we proceed to ask if it is possible to find an alternative efficient representation of 1d correlators in the interval $(0,1)$ based on Chebyshev polynomials that does not employ NNs. Along these lines, we employ a truncated basis of Chebyshev polynomials to set
\begin{equation}
    \label{chebTaa}
    \GG(z) = L(z) + H(z) \,, \qquad
    H(z) = z^a(1-z)^{-b} \sum_{n=0}^{N-1} c_n T_n(2z-1)\,.
\end{equation}
$N$ controls the order of the truncation of the Chebyshev expansion and the prefactor $z^{a}(1-z)^{-b}$ takes care of the leading order behaviour of $H(z)$ in the vicinity of the endpoints of the interval as before.

In this context, the implementation of the crossing equation \eqref{introac} together with the anchor point constraint on a grid becomes a linear least-squares, deterministic problem that can be solved trivially to produce crossing-symmetric functions satisfying the gap and anchor conditions at high degree. It should be clear by now, however, that these configurations will not be, in general, good approximations to the target physical correlator because the above scheme does not incorporate any mechanism that will enforce the required spectral decay, which played a crucial role in the previous analysis. To implement such a mechanism, it is natural to add an explicit penalty to the least-squares objective function of the form
\begin{equation}
    \label{chebTab}
    \lambda_{\rm reg} \sum_{n=0}^{N-1} \mu^{2n} |c_n|^2
    ~.
\end{equation}
In this expression, $\lambda_{\rm reg}$ is an overall positive weight and $\mu>1$ a parameter that enforces the exponential decay of higher Chebyshev coefficients. The degree of success of this type of fit relies on the proper choice of the parameters $N, \lambda_{\rm reg}$ and $\mu$. The optimal choice is not obvious, in general, unless we have a priori knowledge of the target function. In comparison, note that the NN representation has been useful because it operates generically without any case-specific finetuning. 

In what follows, we demonstrate two things: $(a)$ that a suitable tuning of the regularisation parameters can indeed fit the exact CFT correlators with excellent accuracy, reaffirming the role of spectral decay, and $(b)$ that a partially-informed fixed choice of regularisation parameters does not do as well, but can, nevertheless, produce decent approximations of CFT correlators. As a result, with suitable improvements, this approach may provide another useful tool in future studies of correlator smoothness in CFT.

For concreteness, we first display fits for a generalised free boson with $\Delta=1/4$ and for the Lee--Yang model in Fig.~\ref{fig:cheb_gfb_leeyang_two_panel}, using fixed decay parameters $\mu=1.2285$ and $\lambda_{\rm reg}=10^{-4}$. The same parameters produce sensible fits in other correlators, but not in these specific cases, fleshing out the fact that one needs to fine-tune them, in general, in a case-by-case basis. 

\begin{figure}[ht]
    \centering
    \begin{subfigure}[b]{0.49\textwidth}
        \centering
        \begin{tikzpicture}
            \begin{axis}[
                width=\linewidth,
                height=6cm,
                xlabel={$z$},
                ylabel={$\mathcal{G}(z)$},
                title={Untuned GFB Fit},
                xmin=0.01,
                xmax=0.9,
                grid=major,
                legend pos=north west,
            ]
                \addplot [black, dashed, thick] table [x=z, y=reference, col sep=tab] {plot_data/gfb_untuned.dat};
                \addlegendentry{Exact}
                \addplot [blue, thick] table [x=z, y=prediction, col sep=tab] {plot_data/gfb_untuned.dat};
                \addlegendentry{Chebyshev fit}
            \end{axis}
        \end{tikzpicture}
    \end{subfigure}\hfill
    \begin{subfigure}[b]{0.49\textwidth}
        \centering
        \begin{tikzpicture}
            \begin{axis}[
                width=\linewidth,
                height=6cm,
                xlabel={$z$},
                ylabel={$\mathcal{G}(z)$},
                title={Untuned Lee--Yang Fit},
                xmin=0.1,
                xmax=0.999,
                grid=major,
                ylabel style={at={(axis description cs:1.10,0.5)}, anchor=south},
                legend pos=north west,
            ]
                \addplot [black, dashed, thick] table [x=z, y=reference, col sep=tab] {plot_data/leeyang_untuned.dat};
                \addlegendentry{Exact}
                \addplot [blue, thick] table [x=z, y=prediction, col sep=tab] {plot_data/leeyang_untuned.dat};
                \addlegendentry{Chebyshev fit}
            \end{axis}
        \end{tikzpicture}
    \end{subfigure}
    \caption{Untuned Chebyshev--Tikhonov reconstructions of the generalised free boson correlator with $\Delta=\tfrac14$ and of the Lee--Yang correlator, compared against the corresponding exact correlators. The left panel uses the bosonic GFF parameters $\Delta=\tfrac14$, $a=\tfrac12$, $b=\tfrac12$, anchor point $z=0.3$, $N=20$, $\lambda_{\rm reg}=10^{-4}$, decay rate $1.2285$, and $z\in[0.01,0.9]$. The right panel uses the Lee--Yang parameters $\Delta=-\tfrac25$, $m=\tfrac23$, $a=-0.4$, $b=0.4$, anchor point $z=0.3$, $N=20$, $\lambda_{\rm reg}=10^{-4}$, decay rate $1.2285$, and $z\in[0.1,0.999]$.}
    \label{fig:cheb_gfb_leeyang_two_panel}
\end{figure}
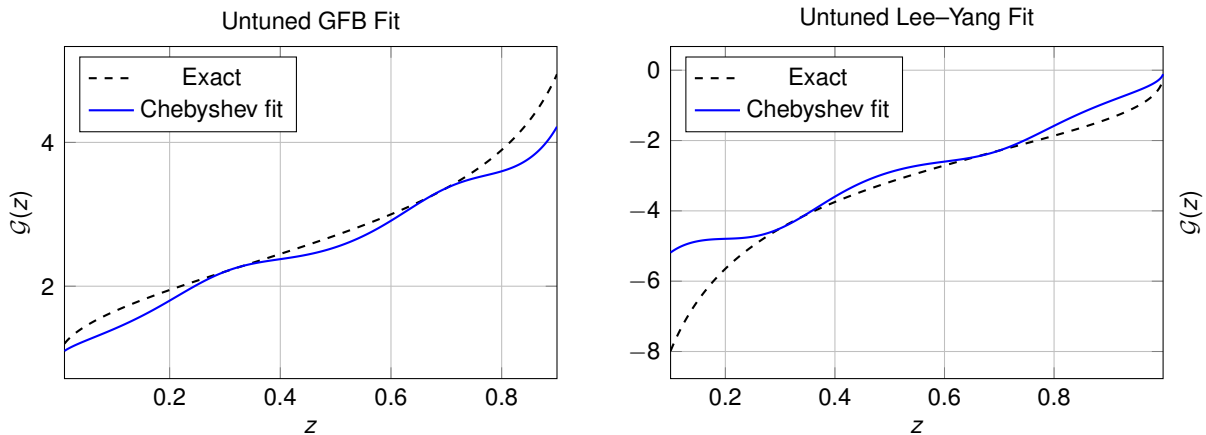

\begin{figure}[t!]
    \centering
    \begin{subfigure}[b]{0.49\textwidth}
        \centering
        \begin{tikzpicture}
            \begin{axis}[
                width=\linewidth,
                height=6cm,
                xlabel={$z$},
                ylabel={$\mathcal{G}(z)$},
                title={Untuned Chebyshev--Tikhonov Fit},
                xmin=0.01,
                xmax=0.95,
                grid=major,
                legend pos=north west,
            ]
                \addplot [black, dashed, thick] table [x=z, y=reference, col sep=tab] {plot_data/2d_untuned.dat};
                \addlegendentry{Exact}
                \addplot [blue, thick] table [x=z, y=prediction, col sep=tab] {plot_data/2d_untuned.dat};
                \addlegendentry{Chebyshev fit}
            \end{axis}
        \end{tikzpicture}
    \end{subfigure}\hfill
    \begin{subfigure}[b]{0.49\textwidth}
        \centering
        \begin{tikzpicture}
            \begin{axis}[
                width=\linewidth,
                height=6cm,
                xlabel={$z$},
                ylabel={$\mathcal{G}(z)$},
                title={Fine-Tuned Chebyshev--Tikhonov Fit},
                xmin=0.01,
                xmax=0.95,
                grid=major,
                ylabel style={at={(axis description cs:1.10,0.5)}, anchor=south},
                legend pos=north west,
            ]
                \addplot [black, dashed, thick] table [x=z, y=reference, col sep=tab] {plot_data/2d_exact_vs_chebyshev.dat};
                \addlegendentry{Exact}
                \addplot [blue, thick] table [x=z, y=prediction, col sep=tab] {plot_data/2d_exact_vs_chebyshev.dat};
                \addlegendentry{Chebyshev fit}
            \end{axis}
        \end{tikzpicture}
    \end{subfigure}
    \caption{Chebyshev--Tikhonov reconstruction of the 2d Ising $\langle \sigma\sigma\sigma\sigma\rangle$ correlator on the line, compared against the exact correlator. Both fits use $\Delta_\sigma=\tfrac18$, $m=3$, $N=20$, $\lambda_{\rm reg}=10^{-4}$, no anchor points, and the interval $z\in[0.01,0.95]$. The left panel is an untuned run with decay rate $1.2285$, while the right panel is a tuned run with decay rate $1.6$.}
    \label{fig:cheb_2d_sigma_two_panel}
\end{figure}
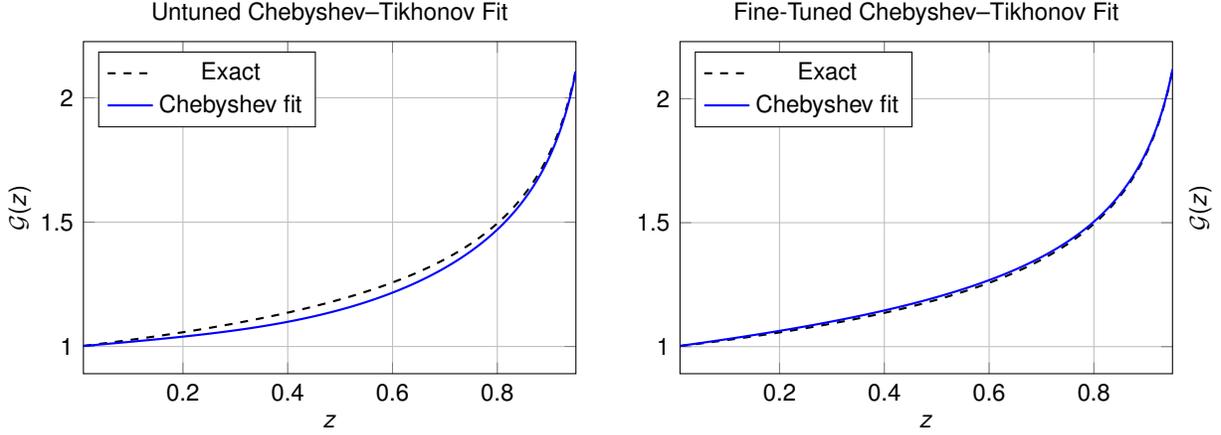

\begin{figure}[t!]
    \centering
    \begin{subfigure}[b]{0.49\textwidth}
        \centering
        \begin{tikzpicture}
            \begin{axis}[
                width=\linewidth,
                height=6cm,
                xlabel={$z$},
                ylabel={$\mathcal{G}(z)$},
                title={Untuned Chebyshev--Tikhonov Fit},
                xmin=0.01,
                xmax=0.9,
                grid=major,
                legend pos=north west,
            ]
                \addplot [black, dashed, thick] table [x=z, y=reference, col sep=tab] {plot_data/3d_sigma_noanchor.dat};
                \addlegendentry{Bootstrap}
                \addplot [blue, thick] table [x=z, y=prediction, col sep=tab] {plot_data/3d_sigma_noanchor.dat};
                \addlegendentry{Chebyshev fit}
            \end{axis}
        \end{tikzpicture}
    \end{subfigure}\hfill
    \begin{subfigure}[b]{0.49\textwidth}
        \centering
        \begin{tikzpicture}
            \begin{axis}[
                width=\linewidth,
                height=6cm,
                xlabel={$z$},
                ylabel={$\mathcal{G}(z)$},
                title={Fine-Tuned Chebyshev--Tikhonov Fit},
                xmin=0.01,
                xmax=0.9,
                grid=major,
                ylabel style={at={(axis description cs:1.10,0.5)}, anchor=south},
                legend pos=north west,
            ]
                \addplot [black, dashed, thick] table [x=z, y=reference, col sep=tab] {plot_data/3d_sigma_finetuned.dat};
                \addlegendentry{Bootstrap}
                \addplot [blue, thick] table [x=z, y=prediction, col sep=tab] {plot_data/3d_sigma_finetuned.dat};
                \addlegendentry{Chebyshev fit}
            \end{axis}
        \end{tikzpicture}
    \end{subfigure}
    \caption{Chebyshev--Tikhonov reconstruction of the 3d Ising $\langle \sigma\sigma\sigma\sigma\rangle$ correlator on the line, compared against the bootstrap reference data. Both fits use $\Delta_\sigma=0.51815$, $a=1.412625$, $b=1.0363$, $N=20$, $\lambda_{\rm reg}=10^{-4}$, no anchor points, and the interval $z\in[0.01,0.9]$. The left panel uses decay rate $1.2285$, while the right panel uses decay rate $5.0$.}
    \label{fig:cheb_3d_sigma_two_panel}
\end{figure}
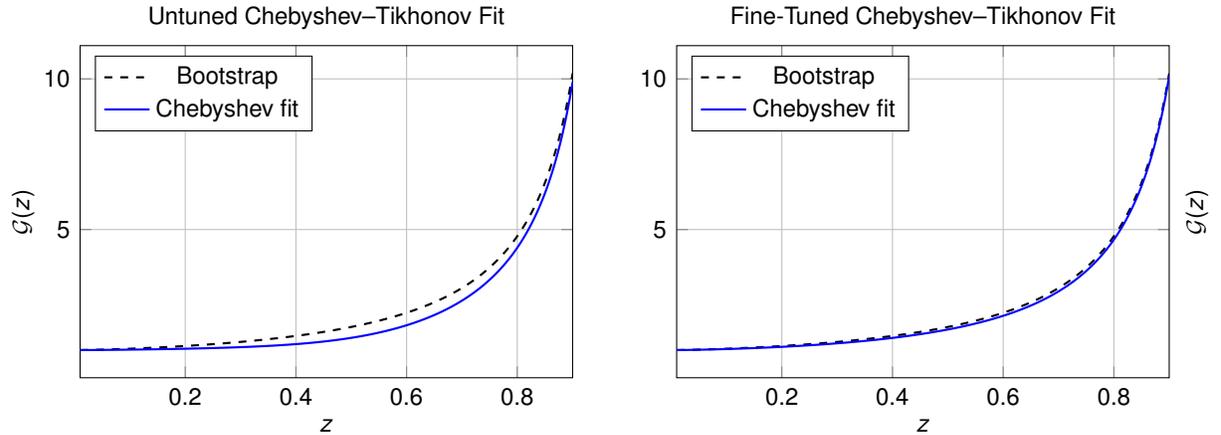

As a second set of examples, we consider the four-point function of the $\sigma$ field in the 2d and 3d Ising models. The corresponding plots are shown in Figs.~\ref{fig:cheb_2d_sigma_two_panel} and~\ref{fig:cheb_3d_sigma_two_panel}, respectively. In both figures, the left panel shows the result for the same untuned choice $\mu=1.2285$ as before, while the right panel shows the result for a finetuned value of $\mu$, for which the agreement is very accurate. We notice that the untuned fits are better approximations in these two cases compared to the previous examples. The novelty of the fits in Figs.~\ref{fig:cheb_2d_sigma_two_panel} and~\ref{fig:cheb_3d_sigma_two_panel} is that they involve fitting of the crossing equation \textit{without} an anchor point. This suggests that there may exist strategies based on smoothness, which produce excellent approximations of some conformal correlators on a line even without the input of anchor points.

\section{Outlook}
\label{outlook}

The results presented in this paper introduce the gradient-based optimisation of anchored neural networks as a practical and surprisingly effective method for the non-perturbative computation of CFT correlation functions on a line. Starting from a minimal set of input data---a scaling dimension, a spectral gap, and a single anchor value---a lightweight MLP trained on the crossing constraint converges to the physical correlator across a broad range of CFTs, spacetime dimensions and observables (vacuum four-point functions and thermal two-point functions) with percent-level accuracy. Even more surprisingly, there is evidence that it is possible to extend the approach beyond the diagonal kinematics, without the need for further input. 

To explain these observations, we emphasized the central role of correlator smoothness (a property never before quantified in CFT). For example, we observed that with appropriately finetuned spectral decay correlation functions on a line can be reproduced by a Chebyshev--Tikhonov fitting. Remarkably, the MLP approach achieves the same goal with less finetuning because of an unexpected alignment between the inherent spectral bias of gradient-based MLPs and the smoothness of CFT correlators. We close by highlighting several key aspects, open problems, and natural extensions of this work.

\paragraph{A variational principle for 1d CFT correlators?}
The deepest question raised by our findings is whether there exists a rigorously defined, \textit{theory-independent}, smoothness functional---perhaps a non-local functional analogous to the fractional Sobolev semi-norm---that is \emph{minimised} by physical CFT correlators among all crossing-symmetric functions with matching gap and anchor data. The evidence from Section~\ref{smooth} is encouraging but circumstantial. Establishing the existence of such a functional in CFT is expected to have profound implications on our understanding of conformal theories. It will improve our ability to perform efficient searches in the conformal bootstrap programme and cheap, non-perturbative computations of CFT correlators. The apparent independence of such a framework from unitarity (and positivity) constraints is an extra bonus.

\paragraph{From the line to the plane.}
Whether correlator smoothness can be a useful guide beyond the strict regime of diagonal kinematics is an interesting question that deserves further study. The concentric-circle strategy outlined in Section~\ref{plane} provides a concrete route towards reconstructing the full four-point function $\GG(z,\bar z)$ (on a disc) at zero temperature and the full two-point thermal correlators at non-zero spatial separation. By exploring further the validity and limitations of this approach one would hope to bring the method into closer contact with the full conformal bootstrap programme. A systematic study of the full correlators on the cross-ratio plane, including potential extensions to spinning correlators, is an important next step.

\paragraph{Mixed-correlator bootstrap.}
The treatment of the 2d Ising mixed $\sigma$-$\epsilon$ system (Section~\ref{ising:2d}) demonstrates that coupled crossing equations pose no fundamental obstacle to the approach. Applying the approach to more complicated systems of mixed correlators would be interesting. Additional examples along these lines are presented in Refs.~\cite{GKNS:2,GKNS:3}.

\paragraph{Anchor data and alternative approaches.}
A key ingredient of the methodology is the anchor point. This is perhaps the input that is most difficult to obtain from independent sources. In Section~\ref{thermal}, we obtained predictions for thermal 2-point functions in the 3d Ising model without explicit knowledge of an anchor point. In that particular application, an approximate anchor point was obtained close to the origin (at $z=0.05$) using the leading non-trivial OPE contribution, which combined a Monte Carlo prediction for the corresponding thermal OPE coefficient and information about the scaling dimension of the $\epsilon$ operator from previous independent bootstrap studies. In general, setting the anchor point very close to $z=0$ can be challenging (yielding, for example, significant variations in statistical runs). It is therefore interesting to develop further techniques that could efficiently exchange anchor points with more easily accessible low-dimensional OPE data. It would also be interesting to explore these questions in the context other strategies (e.g.\ strategies based on Chebyshev--Tikhonov fits).

\paragraph{OPE extraction.}
Our method is primarily designed to reconstruct CFT correlators. Once a correlator is obtained to reasonable numerical accuracy, a natural next step is to ask to what extent it can be used to constrain individual OPE data or to complement more traditional bootstrap approaches. Although extracting detailed spectral information from an approximate correlator is itself a nontrivial inverse problem, the correlator already contains substantial global information, including its crossing-symmetric analytic structure and its behaviour in different OPE limits. A particular direction is to combine the reconstruction with conformal dispersion relations \cite{Paulos:2020zxx}\footnote{For an implementation of a related strategy in the context of thermal CFT, see \cite{Niarchos:2025cdg}.}. Here, the double discontinuity is not taken as a separate input. Instead, once the neural network produces a candidate correlator $\mathcal G(z)$, one may test whether it is compatible with the expected dispersive representation, including the appropriate subtraction terms and asymptotic behavior. When the neural ansatz allows for controlled analytic continuation, this can be implemented by extracting the corresponding double discontinuity and checking consistency with the associated dispersion formula. In this way, the dispersion relation provides an additional nontrivial constraint on the reconstructed correlator, either as an \emph{a posteriori} check or, more ambitiously, as part of the loss function. Another natural step is to project the reconstructed correlator onto conformal blocks. Even if this inversion is only approximate, fitting $\mathcal G(z)$ by a truncated conformal block expansion over a suitable low-lying basis would already impose strong constraints, since a physically acceptable fit should remain stable under changes of fitting window, truncation order, and regularisation scheme, while yielding plausible low-lying dimensions and OPE coefficients. More broadly, this suggests a hybrid strategy in which the neural network first reconstructs a correlator consistent with crossing symmetry, asymptotic input, and anchor data, after which dispersion-based constraints and conformal block decomposition are used to further reduce the allowed space and extract approximate microscopic CFT data. We leave these interesting applications for future work \cite{GKNS:3}.

\paragraph{Companion papers.} An accompanying letter~\cite{GKNS:0} provides a concise summary of the key results of this work. Further applications will appear in two companion papers. The first one, \cite{GKNS:2}, applies the method to the 2d modular bootstrap on the torus and the Cardy condition on the annulus, reformulated as crossing equations for twist-field correlators in symmetric orbifold theories; examples include the Ising, Lee--Yang, and $SU(2)$ WZW models. The second companion paper, \cite{GKNS:3}, analyses (mixed) four-point correlators on the half-BPS Wilson line in 4d $\NN=4$ super-Yang-Mills, and holomorphic four-point functions of chiral quasi-primaries in the $\WW$-algebra arising from the chiral twist of half-BPS operators in the 6d $(2,0)$ SCFT.

\ack{We would like to thank A.~Kaviraj, S.~Pal, C.~Papageorgakis, A.~Stratoudakis, J.~ Thaler, and M.~Woolley for useful discussions, and J.~Barrat, D.~N.~Bozkurt, E.~Marchetto, A.~Miscioscia and E.~Pomoni for sharing their data on 3d Ising thermal correlators in Ref.\ \cite{Barrat:2025nvu}. Research presented in this work was initiated with and supported by an ``International Exchanges 2024 Global Round 1'' grant from the Royal Society (IES\textbackslash{}R1\textbackslash{}241082). The authors also thank Riken iTHEMS and the Yukawa Institute for Theoretical Physics at Kyoto University. Discussions during ``Progress of Theoretical Bootstrap'' were useful in completing this work. Numerical computations in this work have been largely performed on King's College London's CREATE~\cite{CREATE} computing cluster. Claude Code was used in this work for code development. KG is supported by the Royal Society under grant RF\textbackslash{}ERE\textbackslash{}231142. SK is supported by the UK's Engineering and Physical Sciences Research Council under grant EP/Z535035/1, through an EPSRC Doctoral Landscape Award. AS is supported by the Royal Society under grant URF\textbackslash{}R1\textbackslash211417 and by STFC under grant ST/X000753/1.}

\appendix

\section{Fractional Sobolev spaces and semi-norms}
\label{sobolev}

For the convenience of the reader, in this appendix we summarise a few useful mathematical concepts from the subject of fractional Sobolev spaces that underlie the smoothness analysis of Section~\ref{smooth}. Useful, pertinent material can also be found in many References, e.g.\ \cite{Adams:2003} and \cite{DiNezza:2012} (see also the following Wikipedia article \url{https://en.wikipedia.org/wiki/Sobolev\_space} for a first orientation to the subject).

\subsection{Definitions}

\paragraph{Definition: Weak Derivative.}
Consider a function $u$ that is integrable on any compact subset of its domain of definition. We say that $u$ is locally integrable and denote this by writing $u \in L^1_{\mathrm{loc}}(\Omega)$, where $\Omega \subseteq \IR^n$ is open. A function $v \in L^1_{\mathrm{loc}}(\Omega)$ is called the \textbf{weak derivative} of $u$ with respect to $x_i$, denoted $D_i u$ or $\partial u / \partial x_i$, if it satisfies
\[
\int_\Omega u \, \frac{\partial \phi}{\partial x_i} \, dx = -\int_\Omega v \, \phi \, dx
\]
for all test functions $\phi \in C^\infty(\Omega)$ with compact support.

More generally, for a multi-index $\alpha = (\alpha_1, \ldots, \alpha_n)$, the weak derivative $D^\alpha u$ is the function $v \in L^1_{\mathrm{loc}}(\Omega)$ satisfying
\[
\int_\Omega u \, D^\alpha \phi \, dx = (-1)^{|\alpha|} \int_\Omega v \, \phi \, dx
\]
for all $\phi \in C^\infty(\Omega)$ with compact support, where $|\alpha| = \alpha_1 + \cdots + \alpha_n$.

\hfill{$\blacksquare$}

\noindent
\textit{Remark.} This definition arises from integration by parts: if $u$ is classically differentiable, the boundary terms vanish for compactly supported $\phi$, and the classical derivative satisfies this relation. The weak formulation extends this to functions lacking pointwise differentiability.

\paragraph{Definition: Sobolev spaces.} 
Let $\Omega \subseteq \IR^n$ be an open set, $k \in \N_0$, and $1 \leq p \leq \infty$. The \textit{Sobolev space} $\Wkp(\Omega)$ is defined as
\[
\Wkp(\Omega) = \left\{ u \in \Lp(\Omega) : D^\alpha u \in \Lp(\Omega) \text{ for all } |\alpha| \leq k \right\}
\]
where $D^\alpha u$ denotes the weak derivative of $u$ corresponding to the multi-index $\alpha = (\alpha_1, \ldots, \alpha_n)$ with $|\alpha| = \alpha_1 + \cdots + \alpha_n$.

\hfill{$\blacksquare$}

\noindent
\textit{Remark.}
Sobolev spaces are also associated with the names of Slobodeckij and Aronszajn.

\paragraph{Definition: Sobolev Norm.}
The space $\Wkp(\Omega)$ is equipped with the norm
\[
\|u\|_{\Wkp} = 
\begin{cases} 
\displaystyle \left( \sum_{|\alpha| \leq k} \|D^\alpha u\|_{\Lp}^p \right)^{1/p} & \text{if } 1 \leq p < \infty \\[12pt]
\displaystyle \max_{|\alpha| \leq k} \|D^\alpha u\|_{L^\infty} & \text{if } p = \infty
\end{cases}
\]
\hfill{$\blacksquare$}

\noindent
\textit{Remark.}
For $p = 2$, it is customary to denote $\Hk(\Omega) \equiv W^{k,2}(\Omega)$. This is a Hilbert space with inner product
\[
\langle u, v \rangle_{\Hk} = \sum_{|\alpha| \leq k} \int_\Omega D^\alpha u \, \overline{D^\alpha v} \, dx
~.
\]

\paragraph{Definition: Fractional Sobolev semi-norm.} Consider a function $f$ on a bounded, smooth subset $\Omega$ of $\IR^n$. For $s\in (0,1)$ the \textbf{fractional Sobolev semi-norm} (or Gagliardo semi-norm) of $f$ is defined as
\begin{equation}
    \label{mathaa}
    [f]^p_{W^{s,p}(\Omega)} = \int_\Omega \int_\Omega dx_1 dx_2\,
    \frac{|f(x_1) - f(x_2)|^p}{|x_1-x_2|^{n+sp}}
    ~.
\end{equation}

\hfill{$\blacksquare$}

\noindent
\textit{Remark.}
This quantity can be viewed as a non-local measure of smoothness. It quantifies the difference of the values of $f$ at two points, penalised by an inverse power of the distance of the points. 

In the special case with $p=2$, $s=1/2$ one obtains the fractional Sobolev semi-norm
\begin{equation}
    \label{mathab}
    [f]^2_{W^{\frac{1}{2},2}(\Omega)} = \int_\Omega \int_\Omega dx_1 dx_2\,
    \frac{|f(x_1) - f(x_2)|^2}{|x_1-x_2|^{n+1}}\,.
\end{equation}

\subsection{Useful aside comments from harmonic analysis}
\label{harm}

\subsubsection{Harmonic extension on the double-slit geometry}

Consider a generic function $G(x)$ on $(0,1)$. It is guaranteed to have a harmonic extension $U(x,y)$ to the complex plane region
\begin{equation}
    \Upsilon = \IC\setminus\{(-\infty, 0) \cup (1,\infty)\}\,.
\end{equation}
This means that $U$ is such that
\begin{equation}
    \label{harm1aa}
    \nabla^2 U=0\,, \qquad U(x,0) = G(x)\,,\qquad x\in (0,1)\,.
\end{equation}
However, this harmonic extension is not unique, because the data on the cuts are not specified. Any of the harmonic extensions is a local minimum of the Dirichlet energy functional
\begin{equation}
    \label{harm1ab}
    E[f] = \int_\Upsilon \Big[ (\partial_x f)^2 + (\partial_y f)^2 \Big] dx\lsp dy\,.
\end{equation}
The global minimum of the Dirichlet energy is a distinguished harmonic extension, let us call it $U_*(x,y)$.

Now assume\footnote{This is relevant for scalar four-point correlation functions in unitary CFTs on the complex cross-ratio plane parametrised by $(z,\bar z)$.} that the function $G(x)$ on $(0,1)$ has an \textit{analytic (holomorphic) extension} on the double-slit complex plane $\Upsilon$ of the form
\begin{equation}
    \label{harmaa}
    G(x+iy) = u(x,y) + i v(x,y)\,.
\end{equation}
By construction, 
\begin{equation}
    \label{harmab}
    u(x,0) = G(x)\,, \qquad v(x,0) = 0\,,  \qquad x\in (0,1)\,.
\end{equation}
In addition, the Cauchy--Riemann conditions 
\begin{equation}
    \label{harmac}
    \partial_x u = \partial_y v\,, \qquad \partial_y u = - \partial_x v
\end{equation}
imply that both functions $u(x,y)$, $v(x,y)$ are harmonic
\begin{equation}
    \label{harmad}
    \nabla^2 u = \nabla^2 v = 0\,,
\end{equation}
and with suitable boundedness conditions at infinity one can express $G(x)$ in terms of the discontinuity on the cuts. 

The harmonic functions $u,v$ are \textit{local} minimizers of the Dirichlet energy functional subject to the boundary conditions \eqref{harmab} (for $u$ and $v$, respectively). That does not imply, however, that $u$ is necessarily the global minimum $U_*$. It may be different.

\subsubsection{Dirichlet energy and Sobolev semi-norm on \texorpdfstring{$\mathbb H$}{H}}

There is a beautiful identity in harmonic analysis that relates the Dirichlet energy of the harmonic (Poisson) extension $U_*(x,y)$ of the boundary data to the fractional Sobolev semi-norm \eqref{mathab}. In the case of the upper half plane $\mathbb H$ the precise statement can be formulated in the following manner.

Let $G(x)$ be a real function defined on the real line, which is viewed as the boundary of the upper half plane $\mathbb H$. Also, let $U_*(x,y)$ be the harmonic extension\footnote{In this case we supply boundary data on the whole real line, so the harmonic extension is unique.} of the function $G(x)$ on $\mathbb H$. By definition, $\nabla^2 U_*=0$ on $\mathbb H$ and $U_*(x,0)=G(x)$. The following identity holds
\begin{equation}
    \label{mathac}
    \int_{\mathbb H} \Big[ (\partial_x U_*)^2 + (\partial_y U_*)^2 \Big] dx\lsp dy 
    = \frac{1}{\pi} \int_\IR \int_\IR dx_1 dx_2\,
    \frac{|G(x_1) - G(x_2)|^2}{|x_1-x_2|^2}\,,
\end{equation}
where $C$ is some universal constant. Recall that the Dirichlet energy $E[U_*]$ on the LHS is already the (global) minimum over all the functions on $\mathbb H$ subject to the boundary condition on the real line given by the function $G(x)$.

\subsubsection{Parenthesis: the Bourgain--Brezis--Mironescu formula}

In 2001 Bourgain, Brezis and Mironescu (BBM), \cite{BourgainBrezisMironescu2001}, proved the following relation. For $G\in W^{1,p}(\Omega)$, $1\leq p<\infty$ and $\Omega \subset \IR^n$ bounded and smooth
\begin{equation}
    \label{mathaba}
    \lim_{s\to 1^-} (1-s) \int_\Omega \int_\Omega dx_1 dx_2\, \frac{|G(x_1) - G(x_2)|^p}{|x_1-x_2|^{n+sp}} = K_{n,p} \int_\Omega |\nabla G(x)|^p dx\,,
\end{equation}
where $K_{n,p}$ is a universal constant depending on the dimension $n$ and the exponent $p$. 

This formula also relates the Dirichlet energy to the fractional Sobolev semi-norm (its $s\to 1$ limit in particular), but this connection happens on the same space $\Omega$ and not in one dimension higher as in \eqref{mathac}.

We do not need the BBM formula in this paper, but could not resist mentioning it here as an interesting appearance of the fractional Sobolev semi-norm.

\subsubsection{Dirichlet energy and Sobolev semi-norm on the double-slit geometry}

Now, let us generalise the relation \eqref{mathac} to the double-slit geometry 
\begin{equation}
    \Upsilon = \IC\setminus\{(-\infty, 0) \cup (1,\infty)\}\,,
\end{equation}
where the two branch cuts on the real axis are excluded. This is more relevant for our purposes. The outcome is a relation of the form
\begin{equation}
    \label{mathba}
    \int_\Upsilon 
    \Big[ (\partial_x U_*)^2 + (\partial_y U_*)^2 \Big] dxdy 
    = \int \int_{(0,1)^2}dx_1 dx_2\, K(x_1,x_2) |G(x_1)-G(x_2)|^2 
    ~,
\end{equation}
with a modified kernel $K(x_1,x_2)$. The precise form of this kernel can be derived as follows. First, define 
\begin{equation}
    \label{mathbb}
    \Phi(z) = \sqrt{\frac{z}{z-1}}\,,
\end{equation}
which sends $(0,1)$ to the positive imaginary axis and $\Upsilon$ to the upper half plane $\mathbb H$. Then, we can use the upper-half plane result \eqref{mathac} to obtain the kernel
\begin{equation}
    \label{mathbd}
    K(x_1, x_2) = \frac{1}{\pi} 
    \frac{\sqrt{|\Phi'(x_1)| |\Phi'(x_2)|}}{|\Phi(x_1) - \Phi(x_2)|^2} = \frac{1}{2\pi} 
    \frac{(x_1x_2)^{-1/4}(1-x_1)^{-3/4}(1-x_2)^{-3/4}}{\left( \sqrt{\frac{x_1}{1-x_1}} - \sqrt{\frac{x_2}{1-x_2}} \right)^2}\,.
\end{equation}

\section{Fuzzy sphere implementation for 3d Ising CFT}
\label{fuzzy}
\subsection{Setup and finite size scaling analysis}
Here we summarise the fuzzy-sphere implementation used to compute the line-restricted four-point correlator $\langle \epsilon(0)\epsilon(z=\frac{1}{2})\epsilon(1)\epsilon(\infty)\rangle$ for the 3d Ising model, building on the methods developed in\cite{Zhu:2022gjc,Han:2023yyb}. To benchmark the quality of our results, we also extract the OPE coefficient $f_{\epsilon\epsilon\epsilon}$ and the two-point correlator $\langle\epsilon(0)\epsilon(z=\frac{1}{2})\rangle$, for which extremely precise bootstrap results exist, e.g. \cite{Simmons-Duffin:2016wlq}.

The standard, second-quantised Hamiltonian\footnote{We take two-flavour non-relativistic fermions on $S^2$ with monopole flux $4\pi s$. The lowest Landau level has $n_m=2s+1$ orbitals. We work at half filling. As $R\sim\sqrt{n_m}$ in the fuzzy-sphere regulator, subleading terms in $1/R$ encode finite-size effects.}  is represented as a matrix-product operator using \texttt{FuzzifiED} \cite{Zhou:2025liv}. We diagonalise the Hamiltonian using DMRG \cite{White:1992zz} to extract the lowest-energy scalar states in the $\mathbb{Z}_2$-even sector, $\{|g\rangle,|\epsilon\rangle\}$, and the $\mathbb{Z}_2$-odd sector, $\{|\sigma\rangle\}$. Local densities are defined as
\begin{equation}
    \hat{n}_{a} = \psi^{\dagger}(\Omega)\sigma_{a}\psi(\Omega)\,, \qquad a\in\{0,x,y,z\}\,,
\end{equation}
where $\psi_{i}(\Omega)$, $i\in\{\downarrow,\:\uparrow\}$, denotes the two fermion flavours, and $\sigma_a$ are Pauli matrices together with the identity operator. These operators admit a truncated expansion in spherical harmonics,
\begin{equation}
    \hat n_a(\Omega)
    =
    n_{m}\sum_{l=0}^{n_m-1}\sum_{m=-l}^{l}
    (\hat n_a)_{lm}\,Y_{lm}(\Omega)\,,
    \qquad
    (\hat n_a)_{lm}
    =
    \frac{1}{n_{m}}\int d\Omega\,Y^*_{lm}(\Omega)\,\hat n_a(\Omega)\,.
\end{equation}
We construct two microscopic realisations of a $\mathbb{Z}_2$-even scalar operator, collectively denoted $ \hat{\mathcal{O}}$ for the finite size scaling analysis. At finite $n_m$, $\hat{\mathcal{O}}$ admits an expansion in terms of $\mathbb{Z}_2$-even CFT operators,
\begin{equation}
    \hat{\mathcal{O}}(\Omega)
    = c_I(n_{m}) \hat I
    + \bigl[ c_{\epsilon}(n_{m})\,\hat\epsilon(\Omega) + \cdots \bigr]
    + \bigl[ c_{T}(n_{m})\,\hat T(\Omega) + \cdots \bigr]
    + \bigl[ c_{\epsilon'}(n_{m})\,\hat\epsilon'(\Omega) + \cdots \bigr]
    + \cdots,
\label{expansion}
\end{equation}
where $c_{X}(n_m)=\tilde c_{X} n_m^{-\Delta_{X}/2}$. For $(\hat{\mathcal O})_{00}$, only spin-$0$ CFT operators in \eqref{expansion} contribute. Hence,
\begin{align}
    \frac{
    \langle \epsilon| (\hat{\mathcal{O}})_{00} |\epsilon\rangle
    -
   \langle g| (\hat{\mathcal{O}})_{00}|g\rangle
    }{
    \langle \epsilon|(\hat{\mathcal{O}})_{00}|g\rangle
    } &\approx \frac{
\tilde c_{\epsilon} f_{\epsilon\epsilon\epsilon} n_m^{-\Delta_\epsilon/2}
+ \tilde c_{\square \epsilon} f_{\epsilon\epsilon\square \epsilon} n_m^{-(\Delta_\epsilon+2)/2}
+ \tilde c_{\epsilon'} f_{\epsilon\epsilon'\epsilon} n_m^{-\Delta_{\epsilon'}/2}+\cdots
}{
\tilde c_{\epsilon} n_m^{-\Delta_\epsilon/2}
+\cdots} \label{feee:finitesize1}
\\&\approx f_{\epsilon\epsilon\epsilon}+\left(\frac{\tilde{c}_{\square\epsilon}}{\tilde{c}_{\epsilon}}f_{\epsilon\epsilon\square \epsilon} \right)\frac{1}{n_m}+\left(\frac{\tilde{c}_{\epsilon'}}{\tilde{c}_{\epsilon}}f_{\epsilon\epsilon'\epsilon} \right)\frac{1}{n_m^{(\Delta_{\epsilon'}-\Delta_{\epsilon})/2}}+\cdots.\label{feee:finitesize2}
\end{align}
Similarly, we consider the partial wave expansion of the two-point and four-point correlators on $S^{2}$,
\begin{align}
    \langle \epsilon\epsilon\rangle(\theta;n_{m})
    &=
    \sum_{l=0}^{n_m-1}
    (2l+1)C^{(\epsilon\epsilon)}_lP_l(\cos\theta)\,,
    \\ \langle \epsilon\epsilon\epsilon\epsilon\rangle(\theta;n_{m})
    &=
    \sum_{l=0}^{n_m-1}
    (2l+1)\,C^{(\epsilon^4)}_lP_l(\cos\theta)\label{fuzzysphere:correxpans}\,,
\end{align}
where $C_l^{(\mathcal O_1\cdots \mathcal O_k)}$ is the $l^{\text{th}}$ partial-wave coefficient of the given correlator at fixed $n_m$. As $(\hat{\mathcal O})_{lm}$ transforms as a spin-$l$ representation of SO(3), the overlap $\langle g|(\hat{\mathcal O})_{l0}^\dagger(\hat{\mathcal O})_{l0}|g\rangle$ isolates the $l^{\text{th}}$ partial wave. Hence,
\begin{align}   
    &\hspace{-1cm}\frac{
    \langle g| (\hat{\mathcal{O}})_{l0}^\dagger (\hat{\mathcal{O}})_{l0} |g\rangle
    -
    \delta_{l0}\,|\langle g|(\hat{\mathcal{O}})_{00}|g\rangle|^2
    }{
    |\langle \varepsilon|(\hat{\mathcal{O}})_{00}|g\rangle|^2
    }\approx\nonumber
    \\&\approx\frac{n_{m}^{-\Delta_{\epsilon}}\left(|\tilde{c}_{\epsilon}|^{2}C_{l}^{(\epsilon\epsilon)}+ 2\operatorname{Re}[\tilde c_{ \epsilon}^{*}\tilde c_{\partial \epsilon}]C^{(\epsilon\partial \epsilon)}_{l} n_m^{-1/2}+|\tilde c_{\partial \epsilon} |^{2}C^{(\partial \epsilon\partial \epsilon)}_{l} n_m^{-1}+\cdots\right)+\cdots}{|\tilde{c}_{\epsilon}|^{2}n_{m}^{-\Delta_{\epsilon}}+\cdots}
    \\&\approx C_{l}^{(\epsilon\epsilon)}+\left(\frac{2\operatorname{Re}[\tilde{c}_{\epsilon}^{*}\tilde{c}_{\partial\epsilon}]C_{l}^{(\epsilon\partial\epsilon)}}{|\tilde{c}_{\epsilon}|^{2}}\right)\frac{1}{\sqrt{n_{m}}}+\left(\frac{
    |\tilde c_{\partial \epsilon} |^{2}}{|\tilde{c}_{\epsilon}|^{2}}C^{(\partial \epsilon\partial \epsilon)}_{l}\right)\frac{1}{n_{m}}+\cdots.
\end{align}
We can repeat this analysis for the four-point function by considering the overlap
\begin{align}
    &\hspace{-1cm}\frac{
    \langle \epsilon| (\hat{\mathcal{O}})_{l0}^\dagger (\hat{\mathcal{O}})_{l0} |\epsilon\rangle
    -
    \delta_{l0}
    \Bigl(
    2\,\langle g|(\hat{\mathcal{O}})_{00}|g\rangle
    \,\langle \epsilon|(\hat{\mathcal{O}})_{00}|\epsilon\rangle
    -
    |\langle g|(\hat{\mathcal{O}})_{00}|g\rangle|^2
    \Bigr)
    }{
    |\langle \epsilon|(\hat{\mathcal{O}})_{00}|g\rangle|^2} \approx\nonumber
    \\&\approx \frac{n_{m}^{-\Delta_{\epsilon}}\left(|\tilde{c}_{\epsilon}|^{2}C^{(\epsilon^{4})}_{l}+2\operatorname{Re}[\tilde{c}_{\epsilon}^{*}\tilde{c}_{\partial\epsilon}]C_{l}^{((\partial\epsilon)\epsilon^{3})}n_{m}^{-1/2}+|\tilde{c}_{\partial\epsilon}|^{2}C_{l}^{((\partial\epsilon)^{2}\epsilon^{2})}n_{m}^{-1}+\dots \right)+\dots}{|\tilde{c}_{\epsilon}|^{2}n_{m}^{-\Delta_{\epsilon}}+\cdots}
    \\&\approx C_{l}^{(\epsilon^{4})}+\left(\frac{2\operatorname{Re}[\tilde{c}_{\epsilon}^{*}\tilde{c}_{\partial\epsilon}]}{|\tilde{c}_{\epsilon}|^{2}}C_{l}^{((\partial\epsilon)\epsilon^{3})}\right)\frac{1}{\sqrt{n_{m}}}+\left(\frac{|\tilde{c}_{\partial\epsilon}|^{2}}{|\tilde{c}_{\epsilon}|^{2}}C_{l}^{((\partial\epsilon)^{2}\epsilon^{2})}\right)\frac{1}{n_{m}}+\cdots.
\end{align}
The finite-size scaling of these correlators at the crossing symmetric point to $\mathcal{O}(n_{m}^{-1})$ is then given by
\begin{align}
     \langle\epsilon\epsilon\rangle(\theta = \pi;n_{m}) &= \langle\epsilon\epsilon\rangle(\theta = \pi;n_{m}=\infty)+\frac{b_{\epsilon^{2}}}{\sqrt{n_{m}}}+\frac{c_{\epsilon^{2}}}{n_{m}}+\cdots,\label{fuzzysphere:scalee}
     \\ \langle\epsilon\epsilon\epsilon\epsilon\rangle(\theta = \pi;n_{m}) &= \langle\epsilon\epsilon\epsilon\epsilon\rangle(\theta = \pi;n_{m}=\infty)+\frac{b_{\epsilon^{4}}}{\sqrt{n_{m}}}+\frac{c_{\epsilon^{4}}}{n_{m}}+\cdots.\label{fuzzysphere:scaleeee}
\end{align}
We now discuss the construction of the two $\mathbb{Z}_{2}$-even operators whose continuum limit corresponds to the CFT operator $\hat{\epsilon}$.
\begin{enumerate}
\item[(i)] \textit{Bilocal implementation.}
We construct a bilocal proxy for the continuum $\hat{\epsilon}$ operator by convolving a local density with a rotationally invariant kernel,
$U(\Omega_a,\Omega_b)=U(\cos\theta_{ab})$, with truncated Legendre expansion \cite{Hu:2023xak}
\begin{equation}
  U(\Omega_a,\Omega_b)
  =
  \sum_{l=0}^{n_m-1}\frac{2l+1}{4\pi}\,\widetilde U_l\,P_l(\cos\theta_{ab})\,.
\end{equation}
At finite $n_m$, the coefficients $\widetilde U_l$ are fixed by matching the pseudopotentials $V_L$ of the fuzzy-sphere interaction via a finite $SU(2)$ recoupling transform. Defining the kernel-filtered density
\begin{equation}
  \hat{n}^{(U)}_{a}(\Omega)=\int d\Omega'\,U(\Omega,\Omega')\,\hat{n}_{a}(\Omega')\,,
\end{equation}
we take
\begin{equation}
   \hat{\mathcal{O}}(\Omega)=\hat{O}_\epsilon(\Omega)
  \equiv
  \hat n_0(\Omega)\,\hat n_0^{(U)}(\Omega)
  -
  \hat n_z(\Omega)\,\hat n_z^{(U)}(\Omega)
  +
  h\,\hat n_x(\Omega)\,.
\end{equation}
\item[(ii)] \textit{Local implementation.} We also consider the local density
\begin{equation}
    \hat{\mathcal{O}}(\Omega)=\hat n_{x}(\Omega)\,.
\end{equation}
\end{enumerate}

\subsection{Results for \texorpdfstring{$f_{\epsilon\epsilon\epsilon}$, $\langle\epsilon^{2}\rangle$ and $\langle\epsilon^{4}\rangle$}{f\_epsilon, <epsilon\^2> and <epsilon\^4>}}
\begin{table}[H]
\centering
\begin{tabular}{c|ccc|ccc}
\hline
& \multicolumn{3}{c|}{$\hat{O}_\epsilon$}
& \multicolumn{3}{c}{$\hat{n}_x$} \\
\hline
$n_m$
& $f_{\epsilon\epsilon\epsilon}$
& $\langle \epsilon\epsilon\rangle(\pi)$
& $\langle \epsilon\epsilon\epsilon\epsilon\rangle(\pi)$
& $f_{\epsilon\epsilon\epsilon}$
& $\langle \epsilon\epsilon\rangle(\pi)$
& $\langle \epsilon\epsilon\epsilon\epsilon\rangle(\pi)$ \\
\hline
8  & 2.010 & 0.2022 & 5.260 & 2.915 & 0.2465 & 8.104 \\
12 & 1.846 & 0.1724 & 4.785 & 2.462 & 0.1948 & 6.564 \\
16 & 1.770 & 0.1615 & 4.576 & 2.245 & 0.1761 & 5.886 \\
20 & 1.725 & 0.1558 & 4.455 & 2.115 & 0.1666 & 5.502 \\
24 & 1.696 & 0.1522 & 4.376 & 2.029 & 0.1609 & 5.252 \\
28 & 1.674 & -- & -- & 1.966 & 0.1570 & 5.075 \\
32 & 1.658 & -- & -- & 1.919 & 0.1542 & 4.944 \\
36 & 1.645 & -- & -- & 1.881 & 0.1521 & 4.841 \\
40 & -- & -- & -- & 1.851 & 0.1503 & 4.759 \\
\hline
\end{tabular}
\caption{
Fuzzy-sphere results for $f_{\epsilon\epsilon\epsilon}$, $\langle\epsilon\epsilon\rangle$ and $\langle\epsilon\epsilon\epsilon\epsilon\rangle$ at antipodal separation
$(r=1,\theta=\pi)$ using $\hat{O}_\epsilon$
and $\hat{n}_x$. Values omitted in the $\hat{O}_{\epsilon}$ column were omitted due to runtime/memory constraints.}
\label{tab:fuzzy-ising-epsilon}
\end{table}

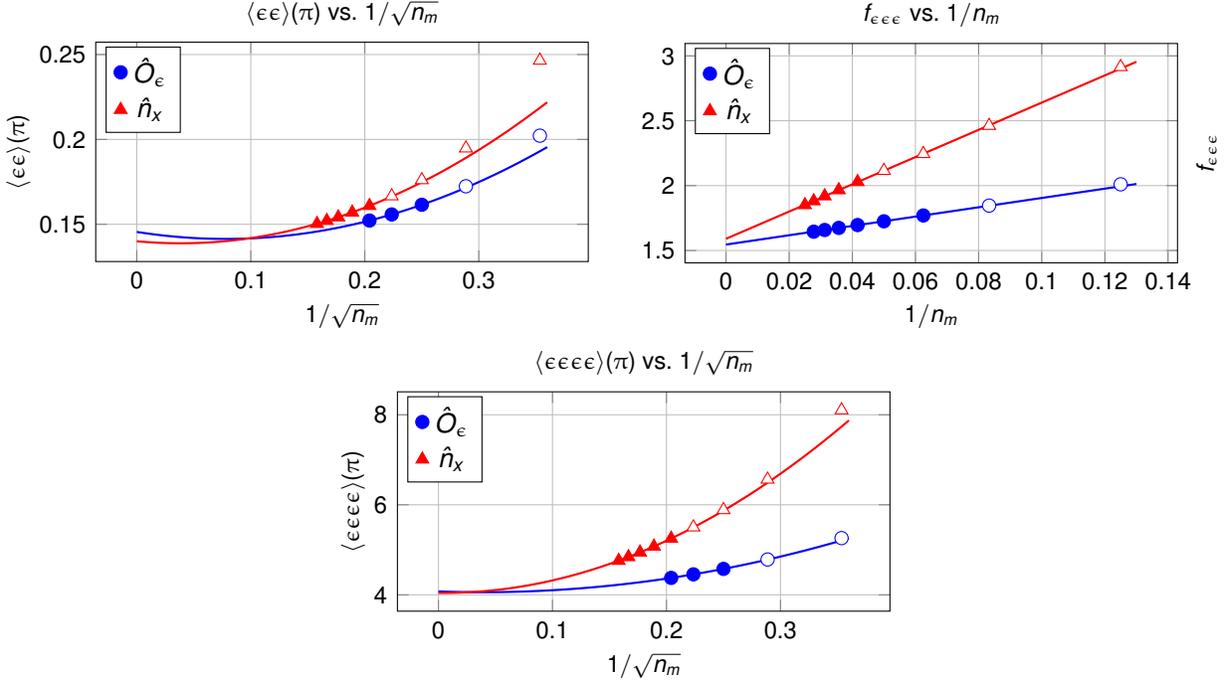
\begin{figure}[H]
\centering

% ===================== Top row =====================
\begin{subfigure}[b]{0.49\textwidth}
\centering
\begin{tikzpicture}
\begin{axis}[
    width=\linewidth, height=4.5cm,
    xlabel={$1/\sqrt{n_m}$},
    ylabel={$\langle \epsilon\epsilon \rangle(\pi)$},
    title={$\langle \epsilon\epsilon \rangle(\pi)$ vs.\ $1/\sqrt{n_m}$},
    grid=major,
    scaled x ticks=false,
    xticklabel style={
        /pgf/number format/fixed,
        /pgf/number format/precision=2
    },
    legend style={
        at={(0.02,0.98)},
        anchor=north west,
        fill=white,
        draw=black,
        font=\sansmath\sffamily\small
    }
]

% O_e data used in fit
\addplot[
    only marks,
    mark=*,
    mark size=2.5pt,
    mark options={draw=blue, fill=blue},
    restrict expr to domain={\thisrow{nm}}{16:100}
]
table[
    x expr=1/sqrt(\thisrow{nm}),
    y=Oe_ee_pi,
    col sep=space
] {plot_data/fuzzy_ising_epsilon.dat};
\addlegendentry{$\hat O_\epsilon$}

% O_e data not used in fit
\addplot[
    only marks,
    mark=*,
    mark size=2.5pt,
    mark options={draw=blue, fill=white},
    restrict expr to domain={\thisrow{nm}}{8:15}, forget plot
]
table[
    x expr=1/sqrt(\thisrow{nm}),
    y=Oe_ee_pi,
    col sep=space
] {plot_data/fuzzy_ising_epsilon.dat};

% n_x data used in fit
\addplot[
    only marks,
    mark=triangle*,
    mark size=2.8pt,
    mark options={draw=red, fill=red},
    restrict expr to domain={\thisrow{nm}}{24:100},
]
table[
    x expr=1/sqrt(\thisrow{nm}),
    y=nx_ee_pi,
    col sep=space
] {plot_data/fuzzy_ising_epsilon.dat};
\addlegendentry{$\hat n_x$}

% n_x data not used in fit
\addplot[
    only marks,
    mark=triangle*,
    mark size=2.8pt,
    mark options={draw=red, fill=white},
    restrict expr to domain={\thisrow{nm}}{8:23},
]
table[
    x expr=1/sqrt(\thisrow{nm}),
    y=nx_ee_pi,
    col sep=space
] {plot_data/fuzzy_ising_epsilon.dat};

% O_e fit
\addplot[
    blue,
    thick,
    domain=0:0.36,
    samples=200
]
{0.145509 - 0.105978*x + 0.679768*x^2};

% n_x fit
\addplot[
    red,
    thick,
    domain=0:0.36,
    samples=200
]
{0.140005 - 0.061667*x + 0.803214*x^2};

\end{axis}
\end{tikzpicture}
\end{subfigure}
\hfill
\begin{subfigure}[b]{0.49\textwidth}
\centering
\begin{tikzpicture}
\begin{axis}[
    width=\linewidth, height=4.5cm,
    xlabel={$1/n_m$},
    ylabel={$f_{\epsilon\epsilon\epsilon}$},
    title={$f_{\epsilon\epsilon\epsilon}$ vs.\ $1/n_m$},
    grid=major,
    ylabel style={at={(axis description cs:1.10,0.5)}, anchor=south},
    scaled x ticks=false,
    xticklabel style={
        /pgf/number format/fixed,
        /pgf/number format/precision=2
    },
    legend style={
        at={(0.02,0.98)},
        anchor=north west,
        fill=white,
        draw=black,
        font=\sansmath\sffamily\small
    }
]

% O_e data used in fit
\addplot[
    only marks,
    mark=*,
    mark size=2.5pt,
    mark options={draw=blue, fill=blue},
    restrict expr to domain={\thisrow{nm}}{16:100},
]
table[
    x expr=1/\thisrow{nm},
    y=Oe_feee,
    col sep=space
] {plot_data/fuzzy_ising_epsilon.dat};
\addlegendentry{$\hat O_\epsilon$}

% O_e data not used in fit
\addplot[
    only marks,
    mark=*,
    mark size=2.5pt,
    mark options={draw=blue, fill=white},
    restrict expr to domain={\thisrow{nm}}{8:15}, forget plot
]
table[
    x expr=1/\thisrow{nm},
    y=Oe_feee,
    col sep=space
] {plot_data/fuzzy_ising_epsilon.dat};

% n_x data used in fit
\addplot[
    only marks,
    mark=triangle*,
    mark size=2.8pt,
    mark options={draw=red, fill=red},
    restrict expr to domain={\thisrow{nm}}{24:100},
]
table[
    x expr=1/\thisrow{nm},
    y=nx_feee,
    col sep=space
] {plot_data/fuzzy_ising_epsilon.dat};
\addlegendentry{$\hat n_x$}

% n_x data not used in fit
\addplot[
    only marks,
    mark=triangle*,
    mark size=2.8pt,
    mark options={draw=red, fill=white},
    restrict expr to domain={\thisrow{nm}}{8:23}, forget plot
]
table[
    x expr=1/\thisrow{nm},
    y=nx_feee,
    col sep=space
] {plot_data/fuzzy_ising_epsilon.dat};

% O_e fit
\addplot[
    blue,
    thick,
    domain=0:0.13,
    samples=200
]
{1.54569 + 3.59118*x};

% n_x fit
\addplot[
    red,
    thick,
    domain=0:0.13,
    samples=200
]
{1.59025 + 10.49352*x};

\end{axis}
\end{tikzpicture}
\end{subfigure}
% ===================== Bottom row =====================
\begin{subfigure}[b]{0.49\textwidth}
\centering
\begin{tikzpicture}
\begin{axis}[
    width=\linewidth, height=4.5cm,
    xlabel={$1/\sqrt{n_m}$},
    ylabel={$\langle \epsilon\epsilon\epsilon\epsilon \rangle(\pi)$},
    title={$\langle \epsilon\epsilon\epsilon\epsilon \rangle(\pi)$ vs.\ $1/\sqrt{n_m}$},
    grid=major,
    scaled x ticks=false,
    xticklabel style={
        /pgf/number format/fixed,
        /pgf/number format/precision=2
    },
    legend style={
        at={(0.02,0.98)},
        anchor=north west,
        fill=white,
        draw=black,
        font=\sansmath\sffamily\small
    }
]

% O_e data used in fit
\addplot[
    only marks,
    mark=*,
    mark size=2.5pt,
    mark options={draw=blue, fill=blue},
    restrict expr to domain={\thisrow{nm}}{16:100},
]
table[
    x expr=1/sqrt(\thisrow{nm}),
    y=Oe_eeee_pi,
    col sep=space
] {plot_data/fuzzy_ising_epsilon.dat};
\addlegendentry{$\hat O_\epsilon$}

% O_e data not used in fit
\addplot[
    only marks,
    mark=*,
    mark size=2.5pt,
    mark options={draw=blue, fill=white},
    restrict expr to domain={\thisrow{nm}}{8:15}, forget plot
]
table[
    x expr=1/sqrt(\thisrow{nm}),
    y=Oe_eeee_pi,
    col sep=space
] {plot_data/fuzzy_ising_epsilon.dat};

% n_x data used in fit
\addplot[
    only marks,
    mark=triangle*,
    mark size=2.8pt,
    mark options={draw=red, fill=red},
    restrict expr to domain={\thisrow{nm}}{24:100},
]
table[
    x expr=1/sqrt(\thisrow{nm}),
    y=nx_eeee_pi,
    col sep=space
] {plot_data/fuzzy_ising_epsilon.dat};
\addlegendentry{$\hat n_x$}

% n_x data not used in fit
\addplot[
    only marks,
    mark=triangle*,
    mark size=2.8pt,
    mark options={draw=red, fill=white},
    restrict expr to domain={\thisrow{nm}}{8:23}, forget plot
]
table[
    x expr=1/sqrt(\thisrow{nm}),
    y=nx_eeee_pi,
    col sep=space
] {plot_data/fuzzy_ising_epsilon.dat};

% O_e fit
\addplot[
    blue,
    thick,
    domain=0:0.36,
    samples=200
]
{4.07524 - 0.88315*x + 11.54473*x^2};

% n_x fit
\addplot[
    red,
    thick,
    domain=0:0.36,
    samples=200
]
{4.04346 - 0.26314*x + 30.29017*x^2};

\end{axis}
\end{tikzpicture}
\end{subfigure}
\caption{Finite-size scaling of fuzzy sphere 3d Ising observables using $\{\hat O_\epsilon,\:\hat n_x\}$. Filled markers denote points included in the fits, while hollow denote points excluded. Best fit curves are recorded in Table \ref{tab:fuzzy-ising-epsilon-fit}.}
\label{fig:fuzzy_ising_scaling_compare}
\end{figure}

\begin{table}[H]
\centering
\begin{tabular}{c|c|c}
\hline
fit & $\hat{O}_\epsilon$
& $\hat{n}_x$ \\
\hline
$f_{\epsilon\epsilon\epsilon}$ & $1.54569 + 3.59118n_{m}^{-1}$ & $1.58458+10.67524n_{m}^{-1}$\\
$\langle \epsilon\epsilon\rangle(\pi)$ & $0.145509 - 0.105978n_{m}^{-1/2} + 0.679768n_{m}^{-1}$ & $0.140005 - 0.061667n_{m}^{-1/2} + 0.803214n_{m}^{-1}$ \\
$\langle \epsilon\epsilon\epsilon\epsilon\rangle(\pi)$ &$4.07524 - 0.88315n_{m}^{-1/2} + 11.54473n_{m}^{-1}$&$4.04346 - 0.26314n_{m}^{-1/2} + 30.29017n_{m}^{-1}$\\
\hline
\end{tabular}
\caption{
Fuzzy-sphere fits used in Fig. \ref{fig:fuzzy_ising_scaling_compare}. for $f_{\epsilon\epsilon\epsilon}$, $\langle\epsilon\epsilon\rangle$ and $\langle\epsilon\epsilon\epsilon\epsilon\rangle$ at antipodal separation using $\hat{O}_\epsilon$
and $\hat{n}_x$.}
\label{tab:fuzzy-ising-epsilon-fit}
\end{table}

\section{More 2d minimal models}
\label{app:moreminimal}

In this appendix we collect, as further illustration, the results of two more computations of the four-point function $\langle \phi_{1,2} \phi_{1,2} \phi_{1,2} \phi_{1,2} \rangle$ on a line in the 2d minimal models $\MM(7,8), \MM(13,14)$. 

\subsection{\texorpdfstring{$\MM(7,8)$}{M(7,8)}}
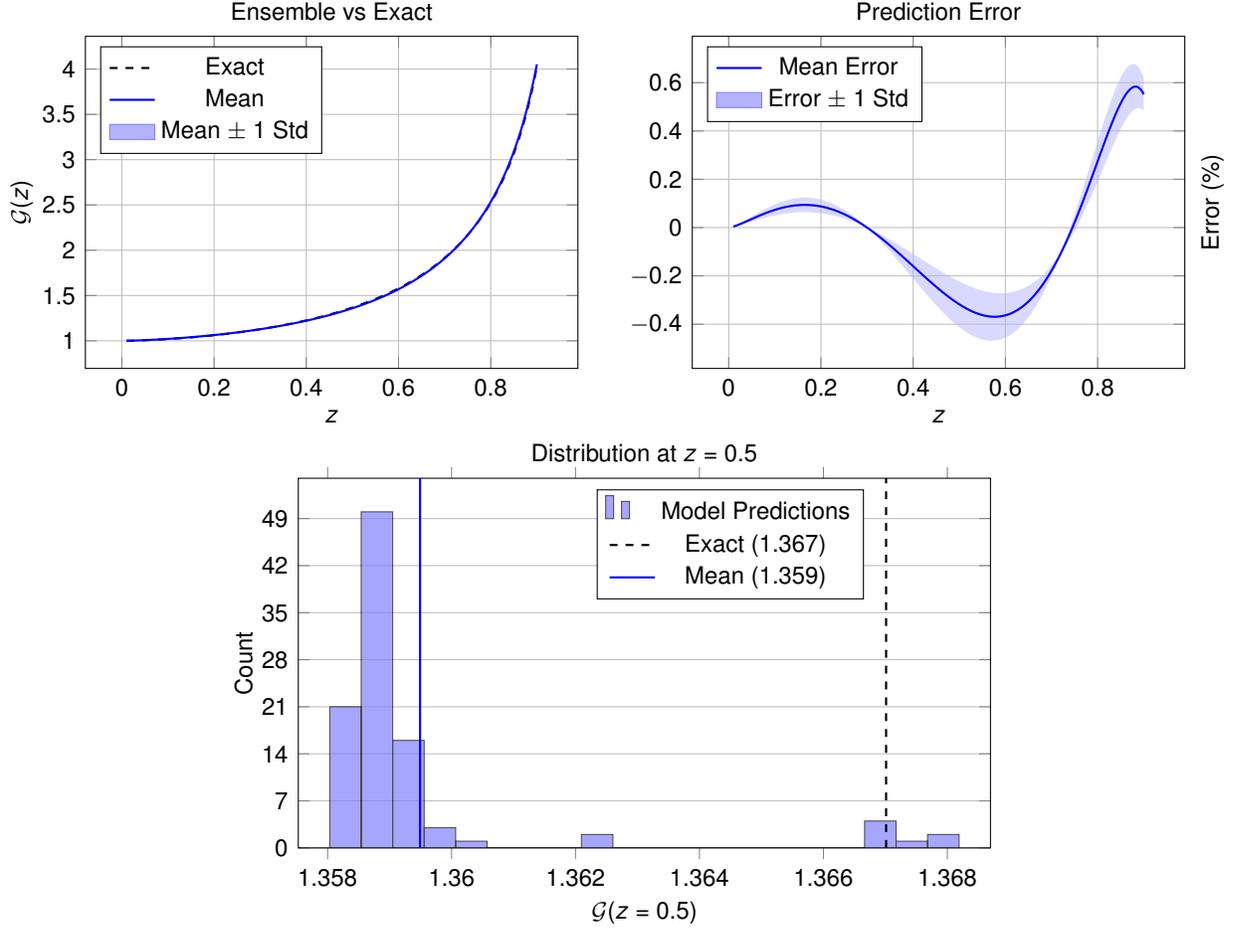
\begin{figure}[H]
    \centering
    % Subfigure 1: Comparison
    \begin{subfigure}[b]{0.49\textwidth}
        \centering
        \begin{tikzpicture}
            \begin{axis}[
                width=\linewidth, height=6cm,
                xlabel={$z$},
                ylabel={$\mathcal{G}(z)$},
                title={Ensemble vs Exact},
                grid=major,
                ytick distance=0.5,
                legend pos=north west,
            ]
                % Exact Solution
                \addplot [black, dashed, thick] table [x=z, y=Exact, col sep=space] {plot_data/mm_7_ensemble_comparison.dat};
                \addlegendentry{Exact}

                % Ensemble Mean
                \addplot [blue, thick] table [x=z, y=Mean, col sep=space] {plot_data/mm_7_ensemble_comparison.dat};
                \addlegendentry{Mean}

                % Uncertainty Band
                \addplot [forget plot, name path=upper, draw=none] table [x=z, y=Mean_plus_Std, col sep=space] {plot_data/mm_7_ensemble_comparison.dat};
                \addplot [forget plot, name path=lower, draw=none] table [x=z, y=Mean_minus_Std, col sep=space] {plot_data/mm_7_ensemble_comparison.dat};
                \addplot [forget plot, fill=blue!30, fill opacity=0.5, draw=none] fill between [of=upper and lower];
                \addlegendimage{legend image code/.code={\fill[blue!30, draw=blue!50] (0cm,-0.1cm) rectangle (0.6cm,0.1cm);}}
                \addlegendentry{Mean $\pm$ 1 Std}
            \end{axis}
        \end{tikzpicture}
        %\caption{}
        %\label{fig:mm_7_comparison}
    \end{subfigure}
    \hfill
    %
    % Subfigure 2: Percentage Error
    \begin{subfigure}[b]{0.49\textwidth}
        \centering
        \begin{tikzpicture}
            \begin{axis}[
                width=\linewidth, height=6cm,
                xlabel={$z$},
                ylabel={Error (\%)},
                ylabel style={at={(axis description cs:1.10,0.5)}, anchor=south},
                title={Prediction Error},
                grid=major,
                ytick distance=0.2,
                legend pos=north west,
            ]
                % Percentage Error
                \addplot [blue, thick] table [x=z, y=PctError, col sep=space] {plot_data/mm_7_percentage_error.dat};
                \addlegendentry{Mean Error}

                % Uncertainty Band
                \addplot [forget plot, name path=upper, draw=none] table [x=z, y=PctError_plus_PctStd, col sep=space] {plot_data/mm_7_percentage_error.dat};
                \addplot [forget plot, name path=lower, draw=none] table [x=z, y=PctError_minus_PctStd, col sep=space] {plot_data/mm_7_percentage_error.dat};
                \addplot [forget plot, fill=blue!30, fill opacity=0.5, draw=none] fill between [of=upper and lower];
                \addlegendimage{legend image code/.code={\fill[blue!30, draw=blue!50] (0cm,-0.1cm) rectangle (0.6cm,0.1cm);}}
                \addlegendentry{Error $\pm$ 1 Std}
            \end{axis}
        \end{tikzpicture}
        %\caption{}
        %\label{fig:mm_7_error}
    \end{subfigure}
    %
    %\vspace{1em}
    %
    % Subfigure 3: Histogram
    \begin{subfigure}[b]{0.65\textwidth}
        \centering
        \begin{tikzpicture}
            \begin{axis}[
                width=\linewidth, height=6.5cm,
                xlabel={$\mathcal{G}(z=0.5)$},
                ylabel={Count},
                title={Distribution at $z=0.5$},
                ybar,
                bar width=0.000508,
                xmin=1.357527, xmax=1.368696,
                ymin=0,
                ymajorgrids=true,
                xmajorgrids=false,
                ytick distance=7,
                legend pos=north east,
                legend style={xshift=-40pt},
                scaled ticks=false,
                xticklabel style={
                    /pgf/number format/fixed,
                    /pgf/number format/precision=3,
                    font=\sansmath\sffamily\footnotesize,
                }
            ]
                \addplot [fill=blue!50, draw=black, opacity=0.7] table [x=BinCenter, y=Count, col sep=space] {plot_data/mm_7_histogram_z0.500.dat};
                \addlegendentry{Model Predictions}
                \draw [black, dashed, thick] (axis cs:1.367012,\pgfkeysvalueof{/pgfplots/ymin}) -- (axis cs:1.367012,\pgfkeysvalueof{/pgfplots/ymax});
                \addlegendimage{legend image code/.code={\draw[black, dashed, thick] (0cm,0cm) -- (0.6cm,0cm);}}
                \addlegendentry{Exact ($1.367$)}
                \draw [blue, thick] (axis cs:1.359492,\pgfkeysvalueof{/pgfplots/ymin}) -- (axis cs:1.359492,\pgfkeysvalueof{/pgfplots/ymax});
                \addlegendimage{legend image code/.code={\draw[blue, thick] (0cm,0cm) -- (0.6cm,0cm);}}
                \addlegendentry{Mean ($1.359$)}
            \end{axis}
        \end{tikzpicture}
        %\caption{}
        %\label{fig:mm_7_hist}
    \end{subfigure}

    \caption{NN-predicted reduced four-point function $\GG(z)$ of the correlator $\langle \phi_{1,2}\phi_{1,2}\phi_{1,2}\phi_{1,2}\rangle$ for the 2d minimal model $\MM(7,8)$. These results are compared to the exact correlator \eqref{minaa}-\eqref{minac} with $m=7$. The NN predictions at $z=0.5$ is $1.359\pm 0.002$.}
    \label{fig:mm_7_summary}
\end{figure}

Results for the minimal model $\MM(7,8)$ are summarised in Fig.\ \ref{fig:mm_7_summary}. These results are based on 100 independent runs with MS training loss $(3.26\pm 0.639) \times 10^{-6}$. The disagreement with the exact analytic result is significantly below 1\% for most of the $z$-interval.

\subsection{\texorpdfstring{$\MM(13,14)$}{M(13,14)}}
Similar results for the minimal model $\MM(13,14)$ are summarised in Fig.\ \ref{fig:mm_13_summary}. These results are also based on 100 independent runs. The MS of the training loss was slightly better in this case, $(7.91\pm 0.160)\times 10^{-7}$, as well as the agreement with the exact analytic result. 

\begin{figure}[H]
    \centering
    % Subfigure 1: Comparison
    \begin{subfigure}[b]{0.49\textwidth}
        \centering
        \begin{tikzpicture}
            \begin{axis}[
                width=\linewidth, height=6cm,
                xlabel={$z$},
                ylabel={$\mathcal{G}(z)$},
                title={Ensemble vs Exact},
                grid=major,
                ytick distance=1,
                legend pos=north west,
            ]
                % Exact Solution
                \addplot [black, dashed, thick] table [x=z, y=Exact, col sep=space] {plot_data/mm_13_ensemble_comparison.dat};
                \addlegendentry{Exact}

                % Ensemble Mean
                \addplot [blue, thick] table [x=z, y=Mean, col sep=space] {plot_data/mm_13_ensemble_comparison.dat};
                \addlegendentry{Mean}

                % Uncertainty Band
                \addplot [forget plot, name path=upper, draw=none] table [x=z, y=Mean_plus_Std, col sep=space] {plot_data/mm_13_ensemble_comparison.dat};
                \addplot [forget plot, name path=lower, draw=none] table [x=z, y=Mean_minus_Std, col sep=space] {plot_data/mm_13_ensemble_comparison.dat};
                \addplot [forget plot, fill=blue!30, fill opacity=0.5, draw=none] fill between [of=upper and lower];
                \addlegendimage{legend image code/.code={\fill[blue!30, draw=blue!50] (0cm,-0.1cm) rectangle (0.6cm,0.1cm);}}
                \addlegendentry{Mean $\pm$ 1 Std}
            \end{axis}
        \end{tikzpicture}
        %\caption{}
        %\label{fig:mm_13_comparison}
    \end{subfigure}
    \hfill
    %
    % Subfigure 2: Percentage Error
    \begin{subfigure}[b]{0.49\textwidth}
        \centering
        \begin{tikzpicture}
            \begin{axis}[
                width=\linewidth, height=6cm,
                xlabel={$z$},
                ylabel={Error (\%)},
                ylabel style={at={(axis description cs:1.10,0.5)}, anchor=south},
                title={Prediction Error},
                grid=major,
                ytick distance=0.1,
                legend pos=north west,
                legend pos=north west,
            ]
                % Percentage Error
                \addplot [blue, thick] table [x=z, y=PctError, col sep=space] {plot_data/mm_13_percentage_error.dat};
                \addlegendentry{Mean Error}

                % Uncertainty Band
                \addplot [forget plot, name path=upper, draw=none] table [x=z, y=PctError_plus_PctStd, col sep=space] {plot_data/mm_13_percentage_error.dat};
                \addplot [forget plot, name path=lower, draw=none] table [x=z, y=PctError_minus_PctStd, col sep=space] {plot_data/mm_13_percentage_error.dat};
                \addplot [forget plot, fill=blue!30, fill opacity=0.5, draw=none] fill between [of=upper and lower];
                \addlegendimage{legend image code/.code={\fill[blue!30, draw=blue!50] (0cm,-0.1cm) rectangle (0.6cm,0.1cm);}}
                \addlegendentry{Error $\pm$ 1 Std}
            \end{axis}
        \end{tikzpicture}
        %\caption{}
        %\label{fig:mm_13_error}
    \end{subfigure}
    %
    %\vspace{1em}
    %
    % Subfigure 3: Histogram
    \begin{subfigure}[b]{0.65\textwidth}
        \centering
        \begin{tikzpicture}
            \begin{axis}[
                width=\linewidth, height=6.5cm,
                xlabel={$\mathcal{G}(z=0.5)$},
                ylabel={Count},
                title={Distribution at $z=0.5$},
                ybar,
                bar width=0.000061,
                xmin=1.421086, xmax=1.426162,
                ymin=0,
                ymajorgrids=true,
                xmajorgrids=false,
                legend pos=north east,
                legend style={xshift=-10pt},
                scaled ticks=false,
                xtick distance=0.002,
                ytick distance=2,
                xticklabel style={
                    /pgf/number format/fixed,
                    /pgf/number format/precision=3,
                    font=\sansmath\sffamily\footnotesize,
                }
            ]
                \addplot [fill=blue!50, draw=black, opacity=0.7] table [x=BinCenter, y=Count, col sep=space] {plot_data/mm_13_histogram_z0.500.dat};
                \addlegendentry{Model Predictions}
                \draw [black, dashed, thick] (axis cs:1.425932,\pgfkeysvalueof{/pgfplots/ymin}) -- (axis cs:1.425932,\pgfkeysvalueof{/pgfplots/ymax});
                \addlegendimage{legend image code/.code={\draw[black, dashed, thick] (0cm,0cm) -- (0.6cm,0cm);}}
                \addlegendentry{Exact ($1.426$)}
                \draw [blue, thick] (axis cs:1.421892,\pgfkeysvalueof{/pgfplots/ymin}) -- (axis cs:1.421892,\pgfkeysvalueof{/pgfplots/ymax});
                \addlegendimage{legend image code/.code={\draw[blue, thick] (0cm,0cm) -- (0.6cm,0cm);}}
                \addlegendentry{Mean ($1.422$)}
            \end{axis}
        \end{tikzpicture}
        %\caption{}
        %\label{fig:mm_13_hist}
    \end{subfigure}

    \caption{NN-predicted reduced four-point function $\GG(z)$ of the correlator $\langle \phi_{1,2}\phi_{1,2}\phi_{1,2}\phi_{1,2}\rangle$ for the 2d minimal model $\MM(13,14)$. These results are compared to the exact correlator \eqref{minaa}-\eqref{minac} for $m=13$. The NN prediction at $z=0.5$ is $1.4219\pm 0.0002$.}
    \label{fig:mm_13_summary}
\end{figure}

\bibliography{ancb1}

\end{document}